\documentclass[11pt]{article}
\usepackage{blindtext}
\usepackage{multirow}
\usepackage{graphicx}
\graphicspath{{images/}}
\usepackage{subfiles}
\usepackage{comment}
\usepackage{amssymb}
\usepackage{algorithm}
\usepackage{algpseudocode}
\usepackage{amsmath} 
\usepackage{array}
\usepackage{subcaption}
\usepackage{booktabs} 
\usepackage{longtable}
\usepackage{listings}
\usepackage{enumitem}
\usepackage{calc}
\usepackage[margin=1in]{geometry}
\newtheorem{thm}{Theorem}[section]

\newtheorem{definition}{Definition}

\newtheorem{assumption}{Assumption}
\newtheorem{example}{Example}
\newtheorem{remark}{Remark}

\usepackage{xcolor} 
\usepackage[utf8]{inputenc}
\usepackage{xr}
\definecolor{CardinalRed}{cmyk}{0,1,0.65,0.34} 
\definecolor{NavyBlue}{rgb}{0.0, 0.0, 0.5} 
\usepackage[colorlinks,linkcolor=NavyBlue,citecolor=NavyBlue,urlcolor=NavyBlue]{hyperref}
\usepackage{hyperref}

\usepackage{titlesec}
\titleformat{\section}
  {\normalfont\Large\bfseries}
  {Section~\thesection.}{0.5em}{}

\lstdefinestyle{mystyle}{
    backgroundcolor=\color{backcolour},   
    commentstyle=\color{codegreen},
    keywordstyle=\color{magenta},
    numberstyle=\tiny\color{codegray},
    stringstyle=\color{codepurple},
    basicstyle=\ttfamily\footnotesize,
    breakatwhitespace=false,         
    breaklines=true,    captionpos=b,                    
    keepspaces=true, numbers=left,                    
    numbersep=5pt, showspaces=false,                
    showstringspaces=false, showtabs=false, tabsize=2}
\makeatletter
\newcommand\longempty{}%
\newcommand\DoIfAndOnlyIfStandAlone{%
  \ifx\document\longempty
    \expandafter\@gobble
  \else
    \expandafter\@firstofone
  \fi
}%
\makeatother

\usepackage[natbibapa]{apacite}

\usepackage{listings}
\usepackage{xcolor}
\usepackage{colortbl}
\definecolor{codegreen}{rgb}{0,0.6,0}
\definecolor{codegray}{rgb}{0.5,0.5,0.5}
\definecolor{codepurple}{rgb}{0.58,0,0.82}
\definecolor{backcolour}{rgb}{0.95,0.95,0.92}

\usepackage{setspace}   
\usepackage{placeins}   
\newcommand{\E}{\mathbb{E}} 
\newcommand{\indep}{\perp\!\!\!\perp} 
\newcommand{\hvar}{\widehat{\operatorname{Var}}}
\newcommand{\hcov}{\widehat{\operatorname{Cov}}}
\newcommand{\HMX}{\textcolor{NavyBlue}{HMX (\citeyear{hainmueller2019much})}}

\newcommand{\hmx}{\textcolor{NavyBlue}{HMX (\citeyear{hainmueller2019much}) }}

\newtheorem{theorem}{Theorem}

\begin{document}


\title{\Large\bf A Practical Guide to Estimating Conditional Marginal Effects: Modern Approaches
\thanks{Jiehan Liu, PhD student, Department of Political Science, Stanford University. Email: \url{jiehanl@stanford.edu}. Ziyi Liu, PhD student, Haas School of Business, University of California, Berkeley. Email: \url{zyliu2023@berkeley.edu}. Yiqing Xu, Assistant Professor, Department of Political Science, Stanford University. Email: \url{yiqingxu@stanford.edu}. We are grateful to Justin Grimmer, Jens Hainmüller, and Jonathan Mummolo for helpful comments and suggestions. We thank the Series Editor, R. Michael Alvarez, for his guidance. We thank Ziyi Chen, Rivka Lipkovitz, Tianzhu Qin, Jinwen Wu, and Zehao Wang for excellent research assistance. Zehao and Tianzhu both contributed significantly to the implementation of the DML methods and the refinement of \texttt{interflex}. All errors are our own. We also thank the authors of the replicated studies for generously sharing their code and data. }
\\\bigskip}

\author{Jiehan Liu\\(Stanford)\and Ziyi Liu\\(Berkeley)\and Yiqing Xu\\(Stanford)}

\date{}

\maketitle

\begin{abstract}
\noindent This Element offers a practical guide to estimating conditional marginal effects—how treatment effects vary with a moderating variable—using modern statistical methods. Commonly used approaches, such as linear interaction models, often suffer from unclarified estimands, limited overlap, and restrictive functional forms. This guide begins by clearly defining the estimand and presenting the main identification results. It then reviews and improves upon existing solutions, such as the semiparametric kernel estimator, and introduces robust estimation strategies, including augmented inverse propensity score weighting with Lasso selection (AIPW-Lasso) and double machine learning (DML) with modern algorithms. Each method is evaluated through simulations and empirical examples, with practical recommendations tailored to sample size and research context. All tools are implemented in the accompanying \texttt{interflex} package for \texttt{R}.

\bigskip\noindent\textbf{Keywords:} causal inference, conditional marginal effects, double machine learning

\end{abstract}

\thispagestyle{empty}  
\clearpage
\newpage
\doublespacing


\setcounter{page}{1}
\abovedisplayskip=5pt
\belowdisplayskip=5pt


\begin{table}[!ht]
    \centering
     \caption{List of Acronyms and Their First Appearance}
    \begin{tabular}{l|l|c}  \hline    \hline
       Acronym  &  Meaning & First Appearance \\
       \hline
       AIPW & Augmented Inverse Propensity Weighting & Section~1 \\
       ATE & Average Treatment Effect & Section~1 \\
       BCG & Brambor--Clark--Golder (2006) & Section~1 \\
       CATE & Conditional Average Treatment Effect & Section~1 \\
       CME & Conditional Marginal Effect & Section~1 \\
       DGP & Data-Generating Process & Section~1 \\
       DML & Double/Debiased Machine Learning & Section~1 \\
       GLM & Generalized Linear Model & Section~4 \\
       HGB & Histogram Gradient Boosting & Section~4 \\
       HMX & Hainmueller--Mummolo--Xu (2019) & Section~1 \\
       IPW & Inverse Propensity Weighting & Section~1 \\
       IRM & Interactive Regression Model & Section~4 \\
       NN & Neural Network & Section~4 \\
       OLS & Ordinary Least Squares & Section~2 \\
       PDS & Post-Double Selection & Section~2 \\
       PLRM & Partially Linear Regression Model & Section~4 \\
       PO-Lasso & Partialing-Out Lasso & Section~3 \\
       RCT & Randomized Controlled Trial & Section~2 \\
       RF & Random Forest & Section~4 \\
       SUTVA & Stable Unit Treatment Value Assumption & Section~1 \\
       SVCM & Smooth Varying-Coefficient Model & Section~2 \\
       WLS & Weighted Least Squares & Section~2 \\
       \hline
    \end{tabular}

    \label{tab:acronyms}
\end{table}
\clearpage

   

\clearpage

\setcounter{tocdepth}{1}
\tableofcontents
\clearpage

\section{Introduction}

Understanding conditional relationships is central to social science research. The impact of a treatment—such as an experimental intervention, a policy, or an institution—on social, political, and economic outcomes often varies systematically across subgroups or contexts. Researchers are particularly interested in how the effect of a treatment \( D \) on an outcome \( Y \) changes with the value of a pre-determined covariate \( X \), known as the moderator, which is unaffected by \( D \). 

\begin{figure}[!htp]
\caption{Replicating \citet{huddy2015expressive}, Figure 2A}
\label{fig:huddy.original}
\centering
\includegraphics[width= 0.7\textwidth]{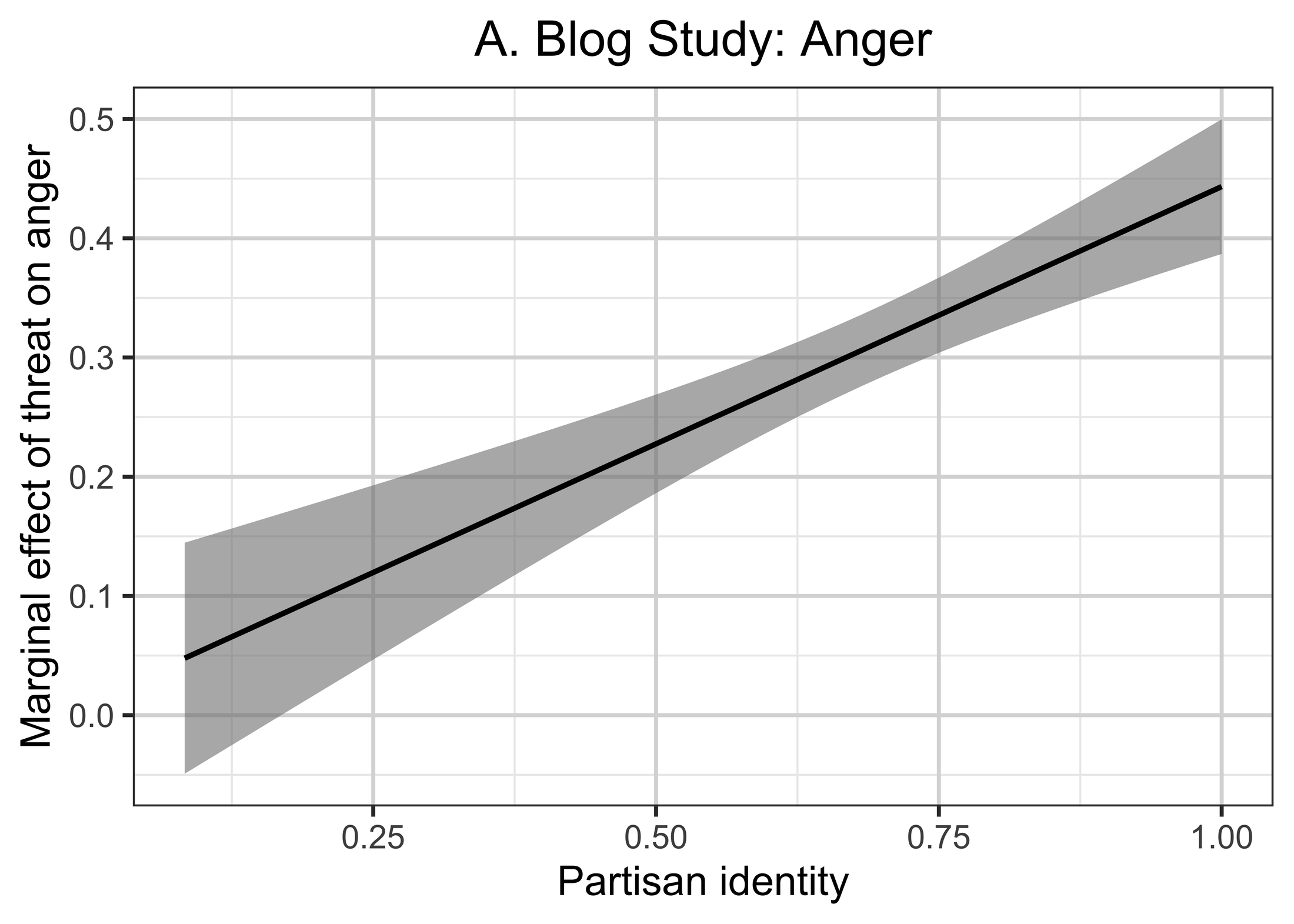}
\begin{minipage}{\linewidth}
\end{minipage}
\end{figure}

Methods for analyzing conditional relationships have evolved over time. Traditionally, researchers have relied on linear interaction models to probe these relationships. \citet{BCG2006} (Brambor, Clark, and Golder; hereafter, BCG 2006) introduced best practices for estimating and interpreting such models, including the widely used ``marginal effect plot'' to visualize these relationships when the moderator $X$ is continuous. Figure~\ref{fig:huddy.original}, reproduced from \citet{huddy2015expressive}, provides an example. The x-axis represents the moderator \( X \) (partisan identity), while the y-axis represents the effect of the treatment \( D \) (experimental partisan threat) on the outcome \( Y \) (anger). The gray ribbon indicates the 95\% (pointwise) confidence intervals.

Researchers typically draw several inferences from figures like this. First, they assess whether the effect of \( D \) on \( Y \) differs from zero by checking whether the horizontal zero line intersects the gray ribbon at a specific value of the moderator. Second, they identify regions where the gray ribbon lies entirely above or below zero (e.g., when partisan identity is larger than \( 0.2\)), interpreting these as conditions under which the theoretical effect is supported. Third, they examine whether the effect of \( D \) on \( Y \) changes monotonically with \( X \), inferring whether the moderator amplifies or weakens the treatment effect, either causally or descriptively.

More recently, \citet{hainmueller2019much} (Hainmueller, Mummolo, and Xu; hereafter, HMX 2019) highlighted the often-violated assumptions underlying these models, including the assumption of overlap and the reliance on linear interaction effects. To address these limitations, they introduced semiparametric kernel estimators, which relax functional form assumptions. Despite these advances, existing methods have several limitations. First, they are primarily motivated by an outcome-modeling perspective and often lack clear connections between estimation strategies and the underlying estimands of interest, leading to potential misinterpretation of estimated coefficients. Second, these approaches frequently rely on strong parametric assumptions. While the kernel estimator relaxes many of the functional form constraints of the linear estimator, it still imposes some structural constraints that may be violated in practice---for example, assuming that the higher order terms of the covariates, such as interactions, do not enter the outcome equation. Moreover, the overlap assumption is typically not critically evaluated. Third, existing methods have not systematically addressed important applications, such as those involving discrete outcomes.

This Element addresses these limitations through three key contributions. First, we provide precise definitions of the estimands of interest within the modern causal inference framework, clarifying the targets of identification and aligning estimation strategies---including parametric, semiparametric, and doubly robust machine-learning methods---with theoretical objectives. Second, we leverage recent advances in double/debiased machine learning (DML) estimators to present a flexible framework that accommodates a wide range of data-generating processes (DGPs) and use cases, including discrete and continuous treatments or outcomes, as well as potentially high-dimensional covariates, while ensuring valid inference for the targeted parameters. Finally, we offer practical recommendations for researchers based on evidence from simulations and empirical examples.

\subsection {Define the Problem}

We begin by defining our problem using the notations of the Neyman-Rubin potential outcomes framework \citep{neyman1923application, rubin1974estimating}, which provides the foundation for the rest of the Element. Consider a study with $n$ units, indexed by $i = 1, 2, \dots, n$. These units are independently and identically distributed (i.i.d.) from an infinite population denoted by $\mathcal{P}$. To simplify the discussion of uncertainty measures, we assume the existence of a super-population. This assumption is also standard in the DML literature.

For each unit $i$, we observe the quadruple $(Y_i, D_{i}, X_i, Z_i)$, where $Y_i$ denotes the outcome of interest, $D_{i}$ is the treatment indicator, and $V_i = (X_i, Z_i)$ represents a set of covariates not causally affected by the treatment and comprises two components: the moderator of interest $X_i \in \mathcal{X}$, which can be either discrete or continuous, and the remaining covariates used for identification $Z_i \in \mathcal{Z}$, which can be potentially high-dimensional. Within the scope of this Element, researchers are committed to studying the effect of $D$ on $Y$ conditional on $X$ (and only $X$), meaning that they do not select variables for analysis after the data have been collected. We formalize the data structure and the sampling procedure as follows:
\begin{assumption}Random Sampling.\label{assm:rs}
\begin{center}
$\{(Y_i, D_{i}, V_i), i = 1, 2, \cdots, n\}$ are i.i.d. sampled from the population $\mathcal{P}$.
\end{center}
\end{assumption}
Next, let's invoke the stable unit treatment value assumption (SUTVA) to define the potential outcomes for unit $i$. 
\begin{assumption} SUTVA.\label{assm:sutva}
\begin{center}
$Y_i(d_{1}, d_{2}, \cdots, d_{n}) = Y_i(d_{i})$
\end{center}
\end{assumption}
SUTVA states that unit $i$'s potential outcomes depend only on the treatment that the unit receives, not on what others receive, and that each treatment has a single version \citep{rubin1986comment, imbens2015causal}. 


As a starting point, let us consider the case in which the treatment is binary, where $D_{i} = 1$ indicates that unit $i$ receives the treatment condition, and $D_{i} = 0$ indicates the control condition. Under SUTVA and a binary treatment, unit $i$ has exactly two potential outcomes, denoted by $\{Y_{i}(d_{i}): d_{i} = 0, 1\}$. Therefore, we can link the observed outcome to the potential outcomes as follows:
\begin{equation}\label{eq:consistency}
Y_i = D_{i} Y_i(1) + (1 - D_{i}) Y_i(0).
\end{equation}
Equation~(\ref{eq:consistency}) is sometimes referred to as ``consistency'' in the causal inference literature; we avoid using this term to prevent confusion with the statistical concept of an estimator converging to the true value of a target parameter.

\paragraph{Conditional Marginal Effect} 

With a binary treatment, unit $i$'s individual treatment effect is defined as $\tau_{i} = Y_i(1) - Y_i(0)$, and the average treatment effect (ATE) of the population as $\tau_{ATE} = \mathbb{E}[Y_i(1) - Y_i(0)]$, where the expectation is taken over the population $\mathcal{P}$. Building on the ATE, the \emph{conditional average treatment effect} (CATE) captures the heterogeneous treatment effect given the value of  covariates $V_i$. 
\begin{equation}
\label{eq:CATE}
\tau(v) = \mathbb{E}[Y_i(1) - Y_i(0) \mid V_i = v],
\end{equation}
CATE measures the average treatment effect for specific groups defined by the value of $V_i$. In practice, estimating the CATE helps researchers better understand how the treatment effect varies across different population segments.

The main purpose of this Element is to examine how the treatment effect of $D$ on $Y$ changes given the value of $X$, a single moderator, and remaining covariates $Z$ used for identification. Therefore, we introduce the \emph{conditional marginal effect} (CME) as a special case of the CATE:
\begin{definition}[CME w/ Binary Treatments]\label{def.cme}
\begin{equation*}
\theta(x) = \E[Y_i(1) - Y_i(0) \mid X_i = x].
\end{equation*}
\end{definition}
Note that $\theta(x)$ is essentially an aggregated version of $\tau(v)$,  where we average out the effects of $Z_i$ while keeping $X_i = x$. Formally, this means:
$$\theta(x)=\mathbb{E}[\tau(x, Z_i) \mid X_i=x]=\int \tau(x, z) d F_{Z \mid X=x}(z).$$
in which the expectation integrates out the influence of $Z_i$, focusing solely on the variation introduced by $X_i$, hence the term ``marginal.''

Sometimes researchers are interested in comparing $\theta(x)$ at two (or more) different values of $x$. This estimand is referred to as \emph{effect modification} in the statistics literature \citep{tyler2009, bansak2020estimating}: 
$$\Delta \theta(x_{1}, x_{2}) = \theta(x_{1}) - \theta(x_{2}).$$ 
Importantly, effect modification is an associative estimand that describes treatment heterogeneity and should not be conflated with \emph{causal moderation}, which refers to the \emph{causal} effect of $X$ on the effect of $D$ on $Y$ (such as in a factorial experiment). In this Element, we do not discuss causal moderation because the moderator $X$ is usually not quasi-random or even manipulable.

\paragraph{A Toy Example}

To clarify the definition of $\theta(x)$, we provide a toy example with eight units, each assigned a survey weight, shown in Table~\ref{tab:detailed_data_effects}. Assume that, after weighting, these units represent the population of interest.
In this example, $(X_i, Z_i)$ comprises two discrete covariates, $D_i$ is a binary treatment indicator, and $Y_i(1)$ and $Y_i(0)$ denote the potential outcomes under treatment and control, respectively. The observed outcome $Y_i$ corresponds to the treatment actually received by each unit. $\tau_i$ represents the individual treatment effect for unit $i$. We arrange the rows of the table according to the values of $V_i$. Our objective is to define the CATE and CME.

\begin{table}[ht]
\centering
\caption{Toy Example}\vspace{-0.5em}
\label{tab:detailed_data_effects}
\begin{tabular}{ccccccccc}
\toprule
$i$ & $X_i$ & $Z_i$ & $Y_i(0)$ & $Y_i(1)$ & $\tau_i$ & $D_{i}$ & $Y_i$ & weight \\
\midrule
1  & 0 & 0 & \cellcolor{gray!30} 2 &  3& 1 & 1 & 3 & 1/2\\
2  & 0 & 0 &  2&  \cellcolor{gray!30} 0&  -2 & 0 & 2 & 1/2\\
3  & 0 & 1 &  \cellcolor{gray!30} 3&  7&  4 & 1 & 7 & 1/2\\
4  & 0 & 1 &  5&  \cellcolor{gray!30} 3&  -2 & 0 & 5 & 1/2\\
5  & 1 & 0 &  \cellcolor{gray!30} 10&  8& -2 & 1 & 8 & 2/3\\
6  & 1 & 0 &  4&  \cellcolor{gray!30} 1&  -3 & 0 & 4 & 2/3\\
7  & 1 & 1 &  \cellcolor{gray!30} 9&  9&  0 & 1 & 9 & 1/3\\
8  & 2 & 1 &  1& \cellcolor{gray!30} 0& -1 & 0 & 1 & 1\\ 
\bottomrule
\multicolumn{9}{p{0.5\textwidth}}{\textbf{Note:} Numbers in shaded cells represent counterfactuals and are not observed in a real dataset.}
\end{tabular}
\end{table}

\begin{example}(CATE)
Calculate $\tau(x, z)$ for $v = (0, 0)$:  
$$\tau(v) = \mathbb{E}[Y_i(1) - Y_i(0) \mid V_i = (0, 0)] = \frac{1 + (-2)}{2} = -0.5$$
\end{example}

\begin{example} (CME)
Calculate $\theta(x)$ for $x = 0$:  
\begin{align*}
& v = (0, 0): \quad \tau(v) = -0.5 & \Pr(Z_i = 0 \mid X_i = 0) = \frac{2}{4} = 0.5  \\
& v = (0, 1):  \quad  \tau(v) = 1 & \Pr(Z_i = 1 \mid X_i = 0) = \frac{2}{4} = 0.5
\end{align*}
\begin{align*}
\theta(0) &= \tau(0, 0)\cdot \Pr(Z_i = 0 \mid X_i = 0) + \tau(0, 1) \cdot \Pr(Z_i = 1 \mid X_i = 0) \\ 
& = (-0.5) \times 0.5 + (1) \times 0.5 = 0.25
\end{align*}
\end{example}

\begin{example} (Effect Modification) Calculate $\Delta\theta(x_{1}, x_{2})$ for $x_{1} = 0$, $x_{2} = 2$:
\begin{align*}
\theta(0) &= 0.25 \\
\theta(2) &= \tau((2,1))  = -1\\
\Delta\theta(0, 2) &= 0.25 - (-1) = 1.25
\end{align*}
\end{example}

The fundamental problem of causal inference states that for each unit, only one potential outcome can be observed at a time \citep{holland1986statistics}. For example, those numbers in shaded cells will never be observed. Therefore, given a sample, we cannot directly calculate the CATE or CME as demonstrated above. In the remainder of this Element, we will explore various estimation strategies that allow us to approximate these estimands under conditions of large sample sizes and when the necessary identifying assumptions are satisfied.


\subsection{Identification}

In this section, we focus on the identification and estimation of CME in the simplest scenario, where both the treatment $D$ and the moderator $X$ are discrete. While this setting is less common in the empirical literature, it provides a clear foundation for understanding the key concepts and results that will be applied throughout the Element.

We begin by presenting the identification results under two key assumptions: unconfoundedness and strict overlap. The goal of identification is to connect a statistical estimand, which can be estimated using observable data, to a meaningful causal estimand, such as the CME, which involves counterfactuals that are inherently unobservable. This connection is established through identifying assumptions. To simplify the discussion further, we assume that the treatment is binary.

\begin{assumption}[Unconfoundedness]\label{assm:unconf}
$$\{ Y_i(0), Y_i(1) \} \perp\!\!\!\perp D_{i} \mid V_i = v, \text{ for all } v \in \mathcal{V}.$$
\end{assumption}
The unconfoundedness assumption---also known as ignorability or selection on observables---is a key identifying assumption, though it is typically untestable. In randomized controlled trials, where the treatment assignment $D_{i}$ is randomized, this assumption is satisfied because the potential outcomes, $\{Y_i(0), Y_i(1)\}$, are independent of the randomly assigned $D_{i}$. Furthermore, in a randomized controlled trial, the probability of receiving the treatment for each participant is known, ensuring that this assumption holds by design.

In observational studies, however, unconfoundedness assumes that treatment assignment $D_{i}$ can be considered as good as randomized only after conditioning on a set of covariates $V_i = (X_i, Z_i)$, which capture all relevant confounders. This is equivalent to stating that within each cell defined by the vector $v$, the treatment is as-if randomly assigned. However, similar to a stratified randomized experiment, the probability of receiving the treatment may vary across cells. The key difference is that, in observational studies, researchers do not know what that probability is.

\begin{assumption}[Strict overlap]\label{assm:overlap}
$$\text{ for all } v \in \mathcal{V},\text{ }\exists \eta>0,\ \eta \leq \Pr(D_{i} = 1 \mid V_i = v) \leq 1-\eta,\ \text{with prob. }1.$$ 
\end{assumption}
In addition to unconfoundedness, ensuring the identifiability of the estimand across all values of the covariates requires the overlap assumption, also known as positivity or common support assumption. This assumption states that every unit must have a non-zero probability of being assigned to each treatment condition, i.e., $0 < \Pr(D_{i} = 1 \mid V_i = v) < 1$. Intuitively, this means that within each cell defined by $v$, there are both treated and control units, provided the sample size is sufficiently large. 

In Assumption~\ref{assm:overlap}, we require a slightly stronger condition than standard overlap, strict overlap. It says that the probabilities of treatment assignment given covariates, or propensity scores, must be uniformly bounded away from 0 and 1 by some positive constant $\eta$. Mechanically, this requirement avoids extreme values of propensity scores, thereby ensuring the identifiability of certain estimators, such as inverse propensity score weighting (IPW) and augmented inverse propensity score weighting (AIPW). It is also crucial for theoretical guarantees, including asymptotic normality of these estimators.

Now, we present two approaches for identifying the CME under unconfoundedness and strict overlap: (a) stratified difference-in-means and (b) IPW. When all covariates \( V_i \) are discrete, these two methods are numerically equivalent \citep{imbens2015causal}—see also the proof for the CME case in the Appendix. However, discussing both methods in this simple discrete setting is important, as they provide the foundation and intuition for more general cases. 

\subparagraph*{Stratified difference-in-means.} Denote \(\mu^{1}(v) = \E[Y_i \mid D_i = 1, V_i = v]\) and \(\mu^{0}(v) = \E[Y_i \mid D_i = 0, V_i = v]\), which represent the conditional mean outcomes under treatment and control conditions, respectively, given covariate values \(V_i = (X_i,Z_i)=(x,z)=v\). The limit of the stratified difference-in-means estimator can be written as:
\begin{align*}
\label{eq:cme_cate}
& \sum_{z} \{(\mu^{1}(v) - \mu^{0}(v)) \cdot \Pr(Z_{i} = z \mid X_i = x)\} \\
= \ &  \sum_{z} \{\tau(v)\cdot \Pr(Z_{i} = z \mid X_i = x)\} \qquad \text{(unconfoundedness \& overlap)}    \nonumber\\ 
=\ & \sum_{z} \{\left(\E[Y_i(1) - Y_i(0) \mid X_i = x,Z_i = z]\right)\Pr(Z_{i} = z \mid X_i = x)\}  \qquad (\text{CATE definition}) \nonumber\\ 
=\ & \E[Y_i(1) - Y_i(0) \mid X_i = x] \qquad (\text{tower rule}) \nonumber\\
=\ & \theta(x) 
\end{align*}
Under unconfoundedness and overlap, the stratified difference-in-means identifies the CME by taking a weighted average CATE across all strata defined by the values of $v$ based on the conditional probability \( \Pr(V_{i} = v \mid X = x) = \Pr(Z_{i} = z \mid X = x) \). The two conditional means, $\mu^{1}(v)$ and $\mu^{0}(v)$, are estimable by data given overlap and random sampling.

\subparagraph*{IPW.} As an alternative, we can identify the CME by weighting each observation by the inverse of the propensity score, which represents the probability of receiving the treatment given covariates, i.e., \(\pi(v) = \Pr(D_i=1 \mid V_i=v)\). The derivation follows:
\begin{align*}
& \E\Bigl[\tfrac{D_i\,Y_i}{\pi(V_i)} - \tfrac{(1 - D_i)\,Y_i}{1 - \pi(V_i)} \,\Big|\, X_i = x\Bigr]
= \E\Bigl[\tfrac{D_i\,Y_i}{\pi(V_i)}\,\Big|\, X_i = x\Bigr] - \E\Bigl[\tfrac{(1 - D_i)\,Y_i}{1 - \pi(V_i)} \,\Big|\, X_i = x\Bigr]\\
=\ & \E\Bigl[\frac{D_iY_i(1)}{\pi(V_i)} \Big|X_{i}=x\Bigr]-\E\Bigl[\frac{(1-D_i)Y_i(0)}{1-\pi(V_i)} \Big|X_{i}=x\Bigr]\\
=\ & \E[Y_i(1) \mid X_i=x] - \E[Y_i(0) \mid X_i=x] \quad (\text{unconfoundedness \& overlap}) \\
=\ & \E[Y_i(1) - Y_i(0) \mid X_i=x] = \theta(x) 
\end{align*}
This expression recovers the same mean difference \(\E[Y_i(1) - Y_i(0)\mid X_i=x]\) by appropriately weighting each observed \(Y_i\) with $X_{i} = x$, ensuring that the estimate accounts for differences in treatment assignment probabilities.


The discussion so far has focused on the simple scenario of binary treatments and discrete covariates. However, researchers are often interested in estimating and visualizing the CME when \( X \) is continuous. In fact, for the remainder of this Element, we assume \( X \) is a continuous variable, treating the case of discrete \( X \) as a special case. While the intuition behind the identification results remains largely the same, the definition of the CME requires additional care when the treatment is continuous, as we discuss below.

\subsection{Continuous Treatments}

Following \citet{hirano2004propensity}, we first adapt the potential outcomes framework to accommodate continuous treatments by considering \( d \) as an interval \(d = [d_0, d_1]\). This formulation allows us to explore the unit-level dose-response function. We assume that each unit \( i \)'s potential outcome function, \( Y_{i}(d) \), is twice continuously differentiable with respect to \( d \) for all \( d \in \mathcal{D} \), ensuring that the derivative \( \partial Y_i(d)/\partial d \) exists and is well defined. 

Moreover, we maintain Assumption~\ref{assm:rs} (random sampling) and Assumption~\ref{assm:sutva} (SUTVA). To link the observed outcome with the potential outcomes, we assume \( Y_i = Y_i(D_i) \), with $D_{i}$ being the observed dose for unit $i$. For each unit $i$, given a vector of covariates $Z_i$ and a moderator $X_i$, we define the dose–response function $\mu(d,x,z) = \mathbb{E}\left[Y(d) \mid X = x, Z = z\right]$, which describes how the expected potential outcome varies with the treatment level $d$, conditional on the covariates. Let $\mu_d(d,x,z) = \partial \mu(d,x,z)/\partial d$ denote its derivative with respect to $d$. The CME with a continuous treatment is defined as the expected value of $\mu_d(d,x,z)$ among units with $X=x$, where the expectation is taken over the joint distribution of both $D$ and $Z$:
\begin{definition}[CME with Continuous Treatments]\label{def:cme2}  
\begin{equation*}  
\theta(x)\;=\;\mathbb{E}\!\big[\,\mu_d(D,X,Z)\ \big|\ X=x\big].
\end{equation*}  
\end{definition} 
Intuitively, $\mu_d(D_i, X_i, Z_i)$ measures how the expected outcome for unit $i$ changes with a small increase in the treatment level around $D_i$, holding its covariates fixed. The CME, $\theta(x)$, averages these unit-specific rates of change across units with the same value of the moderator $x$. As before, we also need unconfoundedness and overlap assumptions to identify the CME. 


\bigskip In the context of a continuous treatment, the term “marginal effect” can be particularly ambiguous. In economics and political science, it often refers to the effect of an infinitesimal change in a regressor on the expected outcome, that is, the partial derivative of $\mathbb{E}[Y\mid X]$ with respect to a specific covariate. In this sense, “marginal” is analogous to “slope,” “trend,” or “velocity” \citep{arel2024interpret}. The term is especially common in nonlinear or discrete outcome models, where the marginal effect depends on the derivative of the link function. For example, in a probit model
\[
\Pr(Y=1\mid X)=\Phi(X\beta),
\]
where $\Phi(\cdot)$ is the standard normal CDF, the marginal effect of $x_k$ is
\[
\frac{\partial \Pr(Y=1\mid X)}{\partial x_k}
  = \phi(X\beta)\beta_k,
\]
with $\phi(\cdot)$ denoting the standard normal density. Unlike the linear case, this effect varies with $X$ because $\phi(X\beta)$ depends on the predicted probability. Researchers sometimes report the estimated marginal effect evaluated at the sample mean, $\phi(\bar X\hat\beta)\hat\beta_k$, as a key quantity of interest.

What adds to the confusion is that the two definitions coincide under the traditional linear interaction model, where the influence of $Z$ on $Y$ is additive and separated from those from $D$ and $Z$ (see details in the next section). However, our identification and estimation of the CME do not require linearity or additivity. In this Element, we follow the statistical tradition and use “marginal” to indicate that the expectation is taken over the joint distribution of all right-hand-side variables except $X$, integrating out the treatment $D$ and other covariates $Z$. 

\begin{assumption}[Unconfoundedness for continuous dose] 
$$Y_i(d) \perp\!\!\!\perp D_{i} \mid V_i = v, \quad \text{for all } d \in \mathcal{D} \text{ and } v \in \mathcal{V}$$
\end{assumption}
This assumption, introduced as ``weak unconfoundedness'' in \citet{hirano2004propensity}, requires only pairwise conditional independence between the treatment and each potential outcome. In contrast, the unconfoundedness in Assumption~\ref{assm:unconf} imposes joint independence across all potential outcomes. In other words, weak unconfoundedness assumes that, conditional on covariates, treatment assignment is independent of each potential outcome separately rather than of the entire set of potential outcomes jointly.

\begin{assumption}[Strict overlap for continuous $D$]
$$\quad f_{D|V}(d \mid v) > 0, \quad \text{for all } d \in \mathcal{D} \text{ and } v \in \mathcal{V}$$    
\end{assumption}
In the continuous treatment case, identifying the CME requires additional assumptions on functional form, which we will discuss in the subsequent sections along with estimation strategies.





\subsection {Approach and Organization}

Having defined the CME and the identification assumptions, the goal of this Element is to present estimation and inference strategies for common empirical settings. The methods we discuss progressively relax functional form restrictions---from linear and semiparametric outcome modeling to propensity-based, doubly robust, and DML approaches---typically at the cost of requiring larger datasets. We illustrate the ideas with empirical examples from political science.

We focus on unconfoundedness as the main identification strategy, which is most natural for cross-sectional data. Panel data settings rely on different identifying assumptions (e.g., parallel trends) and fall outside our scope, but we can accommodate fixed effects by first differencing outcomes or partialing them out using the Frisch--Waugh--Lovell (FWL) theorem.

The Element proceeds as follows. Section 2 reviews classic CME estimators, beginning with the linear interaction model and its practical limitations, including overlap concerns and misspecification. We then discuss a design phase to improve overlap and introduce diagnostic and semiparametric tools motivated by \HMX, including binning and the kernel estimator.

Section 3 shifts to propensity-based and doubly robust estimation. We introduce IPW and AIPW for continuous moderators, develop the orthogonal signals that underpin these approaches, and show how basis expansion and regularization yield AIPW-Lasso. We also extend the framework to continuous treatments via partially linear regression and a partialing-out strategy that constructs ``denoised'' variables before applying kernel or spline regression.

Section 4 presents the DML framework as an extension of AIPW to high-dimensional and nonlinear nuisance functions estimated by machine learning methods. We describe the roles of Neyman orthogonality, flexible learners, and cross-fitting, and apply DML to CME estimation with both binary and continuous treatments, including an application with a discrete outcome.

Section 5 reports Monte Carlo evidence comparing DML with the kernel estimator and AIPW-Lasso to assist researchers to choose methods suitable for their settings. Section 6 concludes with practical recommendations. All methods are implemented in the \texttt{interflex} package in \texttt{R}.


\clearpage

\section{The Classic Approaches}

In this section, we discuss classic approaches to estimating and visualizing the CME. We begin with the linear interaction model, which remains widely used in social science research. We introduce the variance estimator for the CME based on this model, which forms the basis for hypothesis testing. We also address the issue of multiple comparisons and explain how uniform confidence intervals can be constructed using a bootstrap procedure to mitigate this concern.  

Next, we examine two key challenges in applying the linear interaction model: violations of the overlap assumption (or lack of common support) and model misspecification (such as omitted interactions and nonlinearity). We discuss simple diagnostic tools and strategies to detect and address these issues.  

Finally, to relax functional form assumptions, we explore the kernel estimator proposed by \hmx and introduce several improvements, including a fully interacted specification and an adaptive kernel. Throughout this section, we emphasize the importance of clearly defining the estimand and explicitly stating identifying and modeling assumptions.

\subsection{Linear Interaction Model}

In the previous section, we considered the CME with a binary treatment $D \in \{0,1\}$, a discrete moderator $X$, and additional discrete covariates $Z$. Under the unconfoundedness assumption, the empirical estimand, which only involves observable quantities \citep{lundberg2021your}, of the CME at a particular value $X_i = x$ is:
\[
\E_z[\E[Y_i \mid D_{i}=1, X_i = x, Z_i = z]] - \E_z[\E[Y_i \mid D_{i}=0, X_i = x, Z_i = z]],
\]
where $\E[Y_i \mid D_{i}=d, X_i = x, Z_i = z]$, $d \in \{0, 1\}$ is a conditional expectation function that could take an arbitrary functional form. As discussed previously, when \( D \) is binary and the covariates are discrete and limited in number, it is feasible to estimate conditional expectations nonparametrically. However, as the number of covariates increases or continuous variables are introduced, nonparametric estimation becomes impractical. To address this, we can impose modeling assumptions that restrict the functional form of $\E[Y_i(d) \mid X_i, Z_i]$. Combining with the unconfoundedness assumption, the functional form assumption simplifies both estimation and interpretation.

The linear interaction model is the most commonly used approach in political science for studying marginal effects. We formally incorporate this widely used specification as a functional form assumption:
\begin{assumption}[Linear interaction model, or the linear estimator]\label{assm:linear}
$$Y_{i}(d) = \beta_0 + \beta_1 d + \beta_2 X_i  + \beta_3 (d \cdot X_i) +  Z_i^\top\beta_4 + \epsilon_i$$ 
where $\epsilon_i$ represents an idiosyncratic error.
\end{assumption}

Given Assumption~\ref{assm:linear}, the unconfoundedness assumption implies:
$$\{\beta_0 + \beta_1 d + \beta_2 X_i  + \beta_3 (d \cdot X_i) + Z_{i}^\top \beta_4  + \epsilon_i\} \indep D_{i} \mid X_{i}, Z_{i}$$
which further implies that $\epsilon_i \indep D_{i} \mid X_i, Z_i$. Combined with the model specification, it leads to the zero-conditional-mean assumption: $\E[\epsilon_i \mid D_{i}, X_i, Z_i] = 0$. Therefore, if the Assumption~\ref{assm:linear} is correct, the coefficients in the linear interaction model can be consistently estimated using Ordinary Least Squares (OLS) under unconfoundedness.

Given unconfoundedness, Assumption~\ref{assm:linear} (the linear interaction model) implies the following conditional expectation function for the observed outcome:
\begin{equation}\label{eq:linear}
\E[Y_i \mid D_{i}, X_i, Z_i] = \beta_0 + \beta_1 D_{i} + \beta_2 X_i  + \beta_3 (D_{i} \cdot X_i) +  Z_i^\top \beta_4
\end{equation}

\subparagraph*{The binary treatment case.} When $D_{i}$ is binary, under the unconfoundedness assumption and the Assumption~\ref{assm:linear}, Equation~(\ref{eq:linear}) is equivalent to the following conditional expectations:
\begin{align*}
\E[Y_i \mid D_{i} = 0, X_i, Z_i] &= \beta_0 + \beta_2 X_i + Z_{i}^\top \beta_4 \\
\E[Y_i \mid D_{i} = 1, X_i, Z_i] &= (\beta_0 + \beta_1) + (\beta_2 + \beta_3) X_i + Z_{i}^\top \beta_4 
\end{align*}
These equations imply linear relationships between \( X \) and the expected outcome, as well as between \( Z \) and the expected outcome. In addition, the coefficients for \( Z \) are constant across treatment values, and there are no interactions between \( X \) and \( Z \). The CME can be written as:
\begin{align*}
\theta(x) &= \E_z[\beta_0 + \beta_1 + \beta_2 x + \beta_3 \cdot x + Z_{i}^\top \beta_4 ] - \E_z[\beta_0 + \beta_2 x + Z_{i}^\top \beta_4 ]\\
&= (\beta_0 + \beta_1 + \beta_2 x + \beta_3 \cdot x + \beta_4^\top \E_z[Z_i]) - (\beta_0 + \beta_2 x +  \beta_4^\top \E_z[Z_i])\\
&= \beta_1 + \beta_3 x
\end{align*}
which is a linear combination of $\beta_1$ and $\beta_3$. Therefore, $\hat\theta(x) = \hat\beta_1 + \hat\beta_3 x$. This logic can be easily extended to discrete $D_{i}$.

\subparagraph*{The continuous treatment case.} 
In the previous section, we defined the CME for continuous treatments as the expected partial derivative of the dose-response function, i.e., $\mu(d,x,z) = \mathbb{E}\left[Y(d) \mid X = x, Z = z\right]$, where $\mu_d(d,x, z)=\partial \mu(d,x,z)/\partial d$, and $\theta(x)\;=\;\mathbb{E}\!\big[\,\mu_d(D,X,Z)\ \big|\ X=x\big]$. Under Assumption~\ref{assm:linear}, the potential outcome is linear in $d$ with a slope of $\beta_1 + \beta_3 X_i$. Consequently, the partial derivative with respect to $d$ is deterministic given $X=x$, and the CME simplifies to:
\begin{equation*}
\theta(x) = \beta_1 + \beta_3 x
\end{equation*}
This yields the same functional form as the binary treatment case. Unconfoundedness is crucial here because it ensures $\epsilon_i \indep D_{i} \mid X_i, Z_i$, allowing us to use OLS to consistently estimate the key coefficients $\beta_1$ and $\beta_3$.

\paragraph*{Inference and Hypothesis Testing} 

The above results suggest that, under the linear interaction model, the CME is a simple, linear function of $x$. For statistical inference, 95\%  \emph{pointwise} confidence intervals are commonly used in empirical research.  By 95\% pointwise confidence interval, we mean that at each fixed value of $x$, 95\% of such intervals would contain the true CME under repeated sampling. These intervals are calculated independently for each value of $x$, without considering the joint coverage probability across multiple values of $x$. 

We can use one of the following two approaches to construct the pointwise confidence intervals, both implemented in \texttt{interflex}: (i) analytical asymptotic variance based on normal approximation or (ii) bootstrapping, which involves repeatedly resampling the data and re-estimating $\hat\theta(x)$. Analytically, the variance of $\hat\theta(x)$ can be estimated as follows: 
\begin{equation*}
    \hvar(\hat\theta(x)) = \hvar(\hat{\beta}_1) + x^2 \hvar(\hat{\beta}_3) + 2x \hcov(\hat{\beta}_1, \hat{\beta}_3).
\end{equation*}
in which $\hvar(\hat{\beta}_1)$, $\hvar(\hat{\beta}_3)$ and $\hcov(\hat{\beta}_1, \hat{\beta}_3)$ are often obtained using the Eicker-Huber-White robust estimator or the cluster-robust estimator (if a cluster structure exists). 

In their seminal work, BCG (2006) recommend visualizing the CME, which they refer to as the marginal effects, using a plot where $x$ is placed on the x-axis and $\hat{\theta}(x)$, derived from the linear interaction model, is displayed on the y-axis. See Figure~1 in Section 1 for example. The plot also includes the 95\% pointwise confidence interval to provide a visual representation of the uncertainty associated with the estimates. We provide the details of the example and \texttt{R} code to replicate this figure below. 
\begin{example}
\label{ex:huddy2015}
\cite{huddy2015expressive} explores the effect of the expressive model of partisanship. Drawing on four survey studies, the authors argue that partisan identity drives campaign participation and strong emotional responses to ongoing campaign events. We replicate Figure 2A in the original paper, which suggests that strongly identified partisans are more likely to exhibit stronger emotional reactions---such as anger when threatened with electoral loss---compared to those with weaker partisan identities. The variables of interest include: 
\begin{itemize}
\item Outcome: level of anger (continuous, $\in [0, 1]$) 
\item Treatment: electoral loss threat or electoral win reassurance (binary, $\in \{0, 1\}$)
\item Moderator: strength of partisan identity (continuous, $\in [0, 1]$)
\end{itemize}
\end{example}

\begin{lstlisting}[language=R]
# R code excerpt
library(interflex) 

Y="totangry" # Anger 
D="threat" # Threat
X="pidentity" # Partisan Identity
Z <- c("issuestr2", "knowledge" , "educ" , "male" , "age10" )

## linear interaction model 
out.linear <- interflex(estimator =
"linear", data = df,  Y = Y, D = D, X = X, Z = Z, 
vartype = "bootstrap", na.rm = TRUE)

# plot the CME
plot(out.linear)
\end{lstlisting}

\begin{figure}[!ht]
\caption{Replicating \cite{huddy2015expressive} Figure 2A}
\label{fig:huddy}
\begin{minipage}{\linewidth}
\begin{center}
\hspace{-2em}\includegraphics[width=0.8\textwidth]{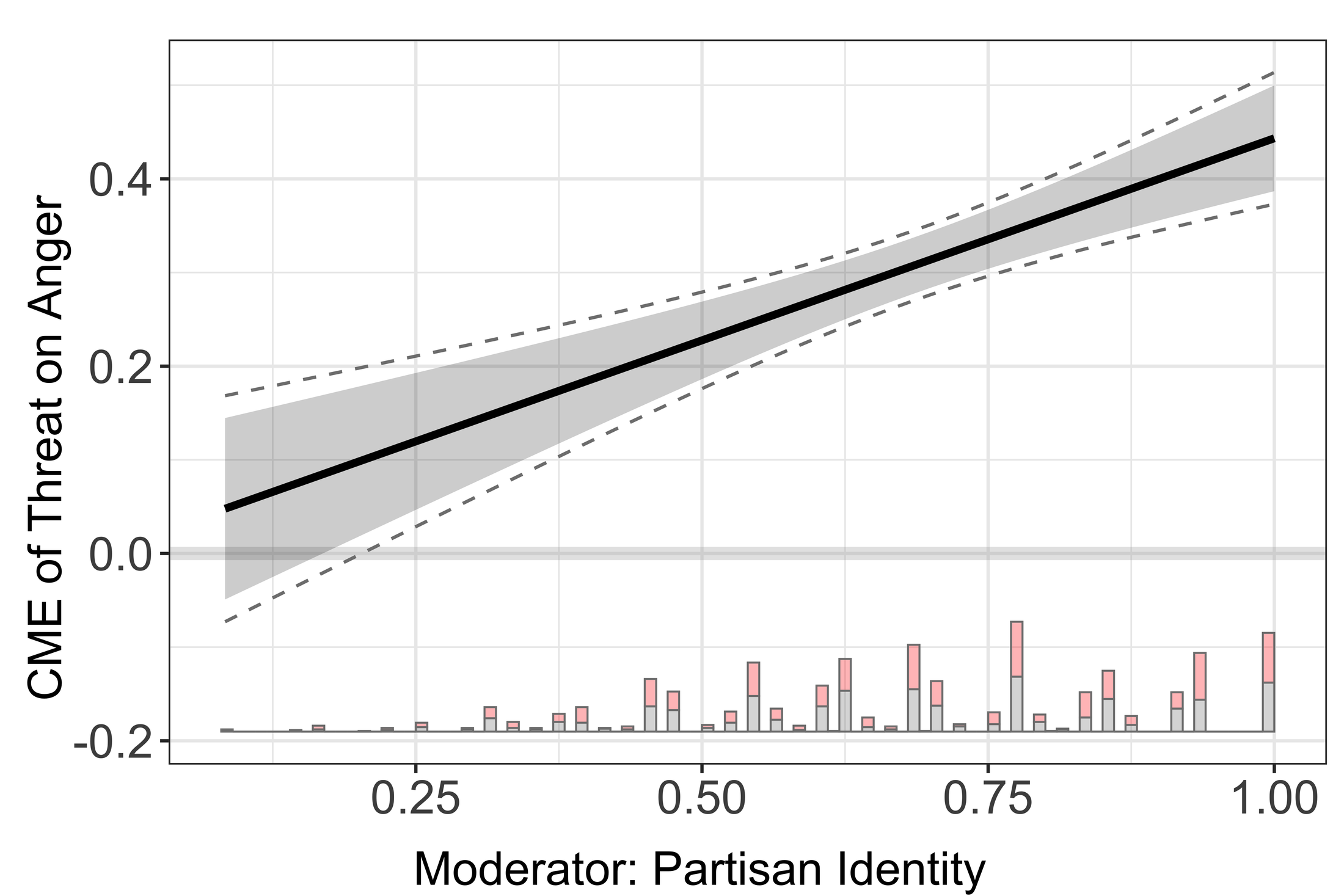}
\end{center}
{\footnotesize \emph{Notes:} The black line represents the CME estimates based on the linear interaction model. The shaded area represents 95\% pointwise confidence intervals. The dashed lines represent 95\% uniform confidence intervals. The histograms at the bottom of the figure depict the distributions of $X$ across treatment (pink) and control (gray) groups.}
\end{minipage}
\end{figure}




As mentioned in Section 1, researchers are often interested in testing three types of hypotheses:  
\begin{itemize}  
    \item $H_{0a}: \theta(x) = 0$ given a specific $x$. Researchers ask whether, at this particular value of $x$, the CME is statistically distinguishable from zero.  
    \item $H_{0b}: \Delta \theta(x_1, x_2) = 0$ or $\Delta \theta(x_1, x_2) > 0$ given specific values of $x_1$ and $x_2$. Researchers compare two CMEs and determine whether they are statistically distinguishable or whether one CME is larger than the other.  
    \item $H_{0c}: \theta(x) = 0$ (or $\theta(x) \gtrless 0$), $x \in \mathcal{X}$, a pre-specified set. Researchers investigate whether the CME across a set or region of $x$ values is jointly statistically distinguishable from zero or jointly larger or smaller than zero.  
\end{itemize}

To test $H_{0a}$, we can perform a $t$-test using $\hat\theta(x)$ and $\hvar(\hat\theta(x))$. This is equivalent to visually inspecting whether the pointwise confidence interval in the marginal effect plot intersects the horizontal zero line. For instance, in Figure~\ref{fig:huddy}, at $x = 0.5$, the pointwise confidence interval does not overlap with zero. This allows us to reject the null hypothesis that $\theta(0.5) = 0$ at the 5\% significance level.

To test $H_{0b}$, we first estimate $\Delta \theta(x_{1}, x_{2})$ using the following approach:
\begin{align*}
    \widehat{\Delta \theta}(x_{1}, x_{2}) &= \hat\theta(x_1) - \hat\theta(x_2) \\
    &= (\hat\beta_1 + \hat\beta_3 x_{1}) - (\hat\beta_1 + \hat\beta_3 x_{2}) \\
    &= \hat\beta_3 (x_{1} - x_{2}).
\end{align*}
The variance of this estimate is given by: $\hvar(\widehat{\Delta \theta}(x_{1}, x_{2})) = \hvar(\hat\beta_3)(x_1 - x_2)^2$. Because $(x_1 - x_2)$ is a constant, testing $H_{0b}$ is equivalent to testing whether $\beta_{3} = 0$. Indeed, researchers often interpret the statistical significance of $\hat\beta_{3}$, the coefficient of the interaction term, as evidence for heterogeneous treatment effects.

To test \( H_{0c} \), we need to construct uniform confidence intervals. Before doing so, we first discuss the issue of multiple comparisons.

\paragraph{Multiple Comparisons}  

\cite{esarey2018marginal} highlight a multiple comparisons issue associated with CME plots. They argue that BCG (2006)'s suggestion to use a visual test---examining whether the (pointwise) confidence interval includes zero at a given value of the moderator $X$ to conclude that the treatment $D$ and outcome $Y$ are statistically related at that value---is overly optimistic. This approach does not account for the multiple comparisons problem, which arises when making inferences across a range of moderator values, leading to inflated Type I error (or uncontrolled false positive rates). 

In the context of the linear interaction model, using an example with a binary moderator, \cite{esarey2018marginal} point out that the visual test is equivalent to conducting hypothesis testing twice, once in each subsample defined by $X$, with each test carrying its own probability of a false positive. This issue becomes even more pronounced when the treatment $D$ or the moderator $X$ is continuous, as the number of comparisons increases significantly, further compounding the likelihood of false positives. \cite{esarey2018marginal} recommend applying procedures that control the overall false discovery rate or familywise error rate, such as the sequential test procedure by \citet{benjamini1995controlling} or the joint $F$-test proposed by \citet{franzese2009modeling}. 

In the \texttt{interflex} package, we address the multiple comparisons issue in estimating CME by incorporating \emph{uniform} confidence intervals, also known as simultaneous confidence intervals. These intervals ensure a specified overall coverage level across all estimates, controlling the family-wise error rate and addressing the multiple comparisons problem. They are implemented using the sup-$t$ band approach introduced by \citet{montiel2017simultaneous} using bootstrapping, which is a simulation-based method that involves repeatedly resampling the data with replacement to estimate the sampling distribution of an estimator. The procedure takes the following steps:
\begin{enumerate}
    \item We discretize the pre-specified set of the moderator $X$, $\mathcal{X}$, into a grid of $k$ evaluation points, denoted as $\{x_1, x_2, \dots, x_k\}$. For each bootstrap replication $b \in \{1, \dots, B\}$, we compute the vector of CME estimates on this grid: $\hat{\theta}^{(b)} = (\hat{\theta}^{(b)}(x_1), \dots, \hat{\theta}^{(b)}(x_k))$.
    
    \item For each grid point $j \in \{1, \dots, k\}$, we can obtain the empirical quantiles from the bootstrap estimates: let $Q_{j,\zeta}$ and $Q_{j,1-\zeta}$ denote the $\zeta$-th and $(1-\zeta)$-th quantiles of the distribution $\{\hat{\theta}^{(b)}(x_j)\}_{b=1}^B$.
    
    \item We determine the largest $\zeta$ such that the rectangular band formed by these quantiles covers the entire true CME curve with probability at least $1-\alpha$:\footnote{The lower bound of $\zeta$, $\frac{\alpha}{2k}$, corresponds to the quantiles with the Bonferroni correction, while the upper bound, $\frac{\alpha}{2}$, corresponds to the quantile of the pointwise confidence interval.}
    \begin{equation*}
    \zeta^* = \sup\left\{\zeta \in \left[\frac{\alpha}{2k}, \frac{\alpha}{2}\right] \middle| \frac{1}{B} \sum_{b=1}^{B} \mathbf{1}\left( \forall j \in \{1,\dots,k\}: \hat{\theta}^{(b)}(x_j) \in [Q_{j, \zeta}, Q_{j, 1-\zeta}]\right) \geq 1 - \alpha \right\}
    \end{equation*}
    
    \item The uniform confidence intervals are constructed using this optimal $\zeta^*$, resulting in the band $[\hat{Q}_{j, \zeta^*}, \hat{Q}_{j, 1-\zeta^*}]$ at each grid point $x_j$.
\end{enumerate}

This method ensures that the confidence intervals are uniformly valid across all evaluation points in $\mathcal{X}$. In other words, the uniform confidence band covers the true CME function $\theta(x)$ simultaneously for all $x \in \mathcal{X}$ with probability at least $1-\alpha$.
Consequently, if the uniform confidence band excludes zero at any point $x \in \mathcal{X}$ (i.e., if the horizontal zero line does not lie entirely within the band), we can reject the joint null hypothesis $H_{0 c}: \theta(x)=0, \forall x \in \mathcal{X}$ at the $\alpha$ significance level.

In Figure~\ref{fig:huddy}, the dotted lines represent the uniform confidence intervals ($\mathcal{X}$ is set to the support of $X$), while the shaded area depicts the pointwise confidence intervals. The uniform confidence intervals are consistently wider, reflecting a more conservative approach to interval estimation that adjusts for multiple comparisons across the moderator's range. We observe that the lower bounds of the uniform confidence intervals cross the horizontal zero line around $x = 0.25$, suggesting that for values of the moderator $X$ greater than 0.25, the effect of the treatment becomes statistically distinguishable from zero at the 5\% level, indicating a significant treatment effect.\footnote{Using a uniform confidence band constructed over the entire support of $X$ to test hypotheses about a specific sub-region (e.g., $x > 0.25$) is statistically valid but conservative. If the research hypothesis had specified this sub-region \textit{a priori}, a narrower band could have been constructed specifically for that range, potentially increasing power \citep[see][for a detailed discussion on controlling false positive rates in this context]{esarey2018marginal}.} Therefore, a visual test using the uniform confidence intervals provides a practical method for performing hypothesis testing for $H_{0c}$.

\bigskip
Using the linear interaction model to estimate the CME presents several challenges, primarily lack of common support and model misspecification, as discussed in \HMX. In the remainder of this section, we illustrate these challenges with examples and propose several solutions to address them.

\subsection{Lack of Common Support} 

\hmx highlight that the overlap assumption is often violated in applications, particularly in observational studies, a problem commonly referred to as a ``lack of common support.'' Researchers using the linear interaction model typically report the CME by substituting \( x \) values into the conditional marginal effect formula, \( \hat\theta(x) = \hat\beta_{1} + \hat\beta_{3}x \), without checking for overlap. This occurs because regression produces estimates regardless of whether the overlap assumption holds.

When \( X \) is continuous, the overlap assumption requires two conditions: (1) a sufficient number of observations with values of the moderator \( X \) close to \( x_0 \), and (2) sufficient variation in the treatment \( D \) at or around \( x_0 \) for similar values of \( Z \). For example, if all data points near \( X = x_0 \) belong to the treatment group (\( D = 1 \)), overlap is violated. If either of these conditions is not met, CME estimates must rely on extrapolation based on the assumed functional form, extending the model to regions with little or no data. In such cases, the CME estimates become highly model-dependent and unreliable. We provide such an example below. 

\begin{example}
\label{ex:chapman2009}
\cite{chapman2009audience} examines the effect of international organization, specifically the authorizations granted by the U.N. Security Council, on public opinion of U.S. foreign policy. The authors use the size of rallies, which measures the short-term change in presidential approval ratings surrounding military disputes, as a proxy for support for foreign policy. We replicate the Figure 2 of the original paper, which suggests the effect of U.N. authorizations is conditional on similarity of preferences between the U.S and the U.N. Security Council: the public is more likely to favor policies that are explicitly approved by relatively conservative institutions, i.e., institutions with more heterogeneous preferences. The variables of interest include: 

\begin{itemize}
\item Outcome: size of rallies supporting the authorization (continuous, $\in [-16, 33]$) 
\item Treatment: granting of a U.N. authorization (binary, $\in \{0, 1\}$)
\item Moderator: preference distance between the U.S. and the Security Council (continuous, $\in [-1, 0]$)
\end{itemize}
\end{example}

\begin{lstlisting}[language=R]
# R code excerpt

library(interflex) 

Y="rally" # size of rallies
D="unauth" # U.N. authorization
X="S" # US affinity with UN Security Council
Z=c("priorpop","bipart","statemnt","nextelec","nytcov","buscon","revorg","wardumk","majopp","allies","war","SCappeal","regorgact","ordsever","hostlvl")

## linear estimator with uniform CI 
out.linear<-interflex(estimator = "linear",
    Y=Y,D=D,X=X,Z=Z, data=d, 
    vartype = "bootstrap", na.rm = TRUE)
plot(out.linear)

## diagnostic plot
out.raw <- interflex(estimator = "raw", data = d,Y=Y,D=D,X=X, na.rm = TRUE)
plot(out.raw)

## propensity score computation
V<-append(Z, X)
covariate_matrix <- d[V]
covariate_matrix[] <- lapply(covariate_matrix, function(x) as.numeric(as.character(x)))
covariate_matrix<-as.matrix(covariate_matrix)

cf<-causal_forest(covariate_matrix,  d$rally, d$unauth)
e.hat <- cf$W.hat
\end{lstlisting}

\begin{figure}[!th]
\centering
\caption{Replicating \cite{chapman2009audience} Figure 2} \label{fig:chapman}
\begin{subfigure}{0.7\textwidth}
  \centering
  \includegraphics[width=\textwidth]{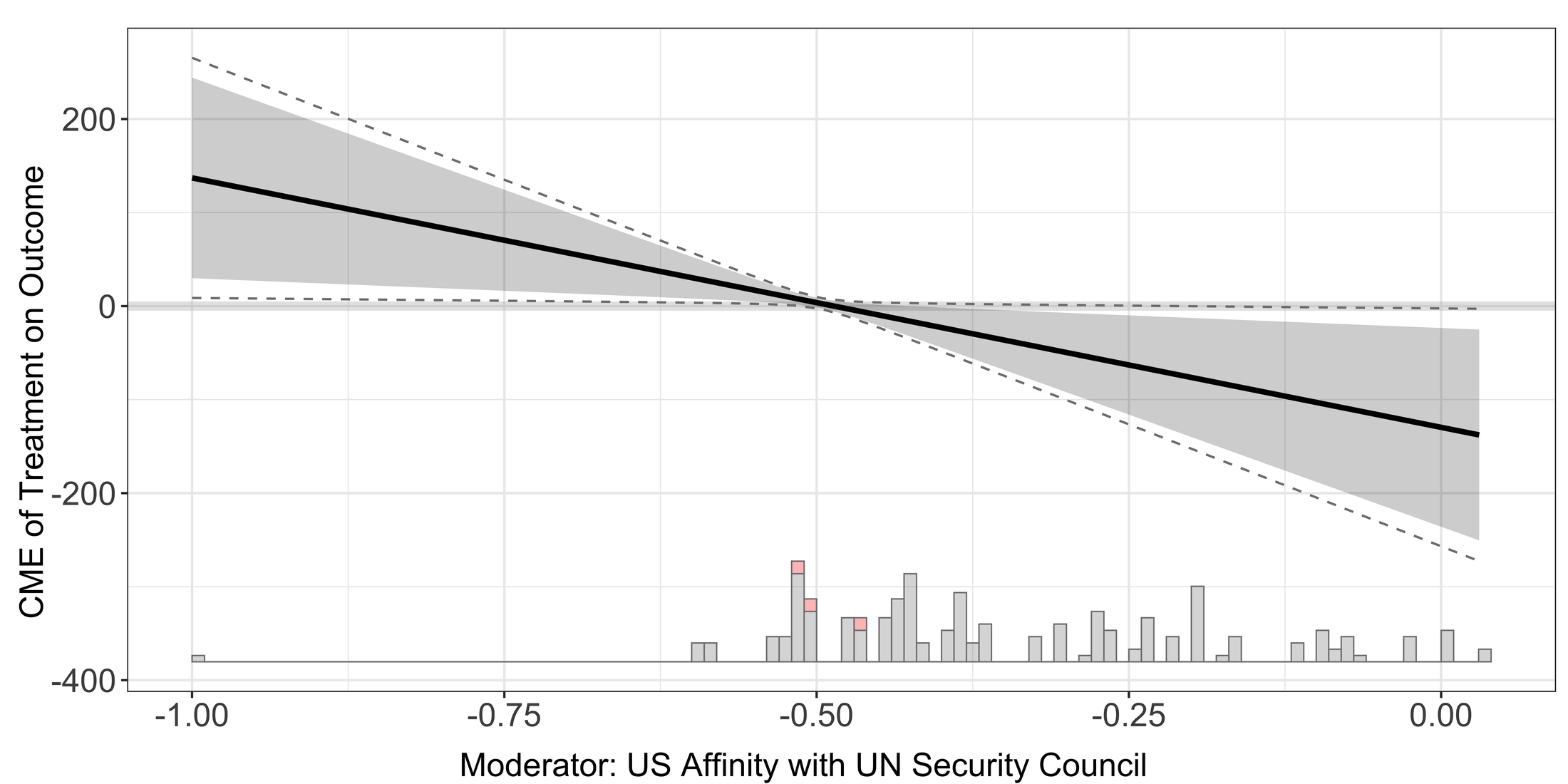}
  \caption{CME plot}
  \label{fig:chapman_cme}
\end{subfigure}
\vspace{1em}
\begin{subfigure}{0.45\textwidth}
  \centering
  \includegraphics[width=\textwidth]{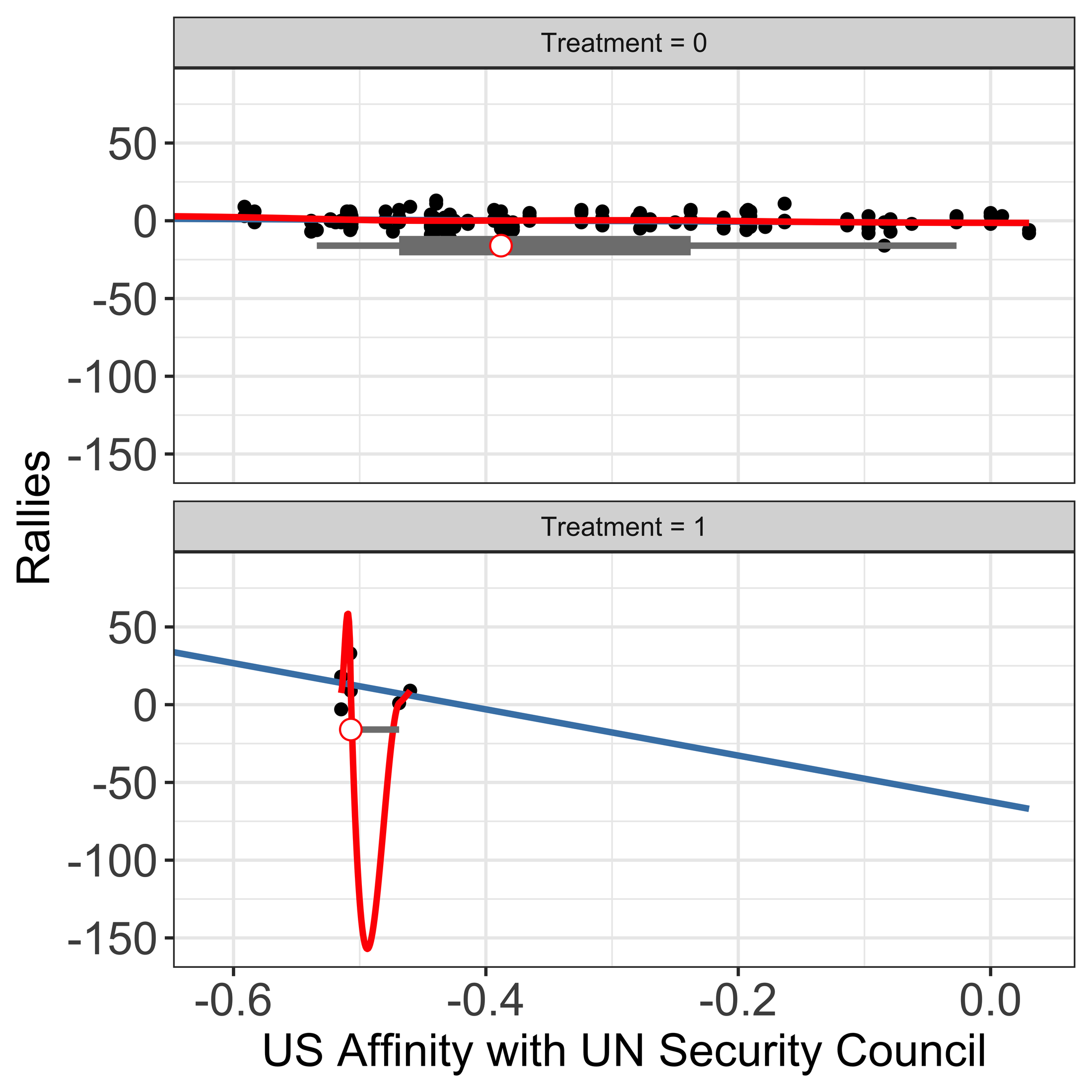}
  \caption{Outcome by treatment condition}
  \label{fig:chapman_raw}
\end{subfigure}
\hspace{1em}
\begin{subfigure}{0.45\textwidth}
  \centering
  \includegraphics[width=\textwidth]{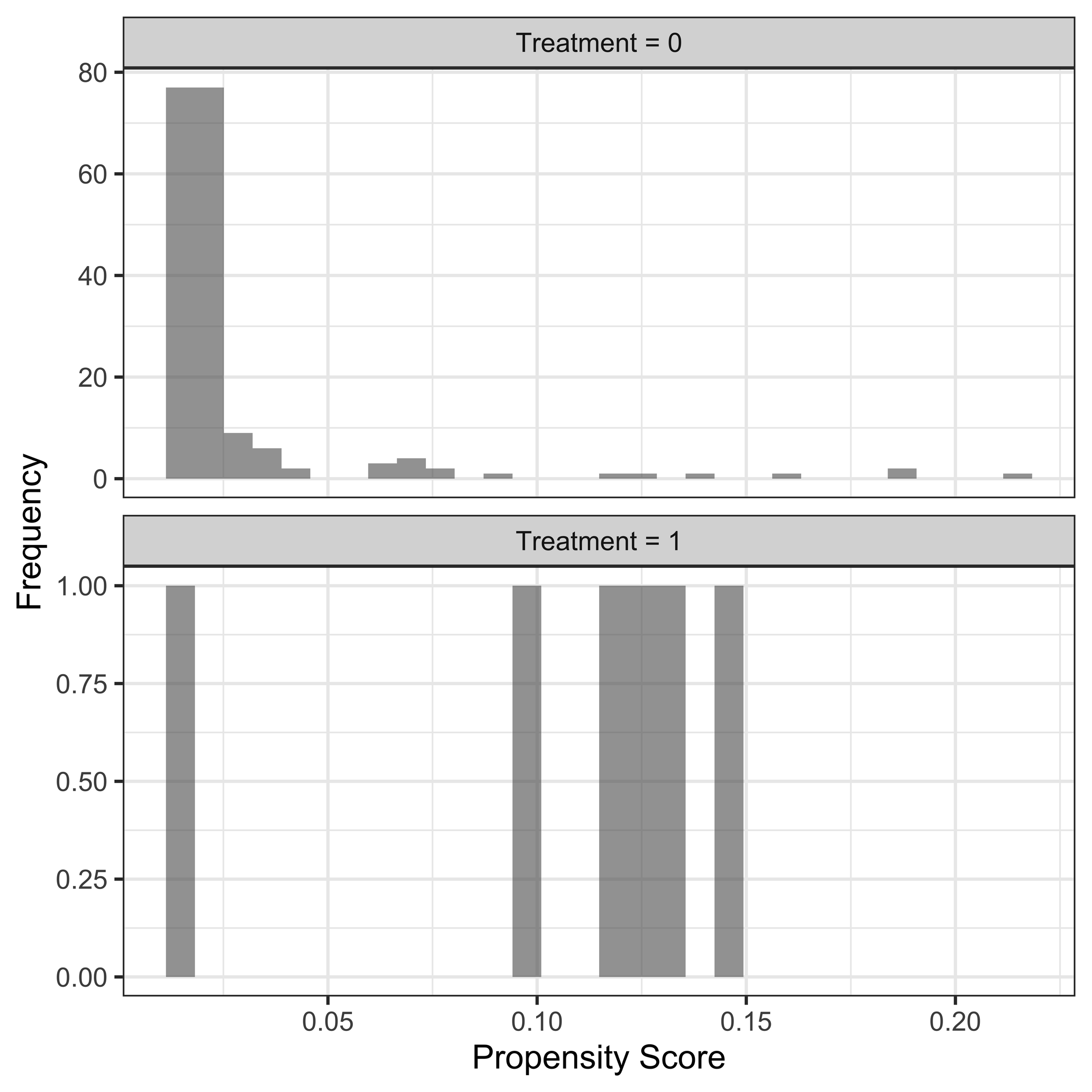}
  \caption{Propensity score}
  \label{fig:chapman_ps}
\end{subfigure}
\begin{minipage}{1\linewidth}
{\footnotesize \emph{Notes:} Treatment measures UN authorization, outcome is the size of rallies, measuring the short-term change in presidential approval ratings surrounding military disputes. (a) The CME plot. The black line represents the CME estimates based on a linear interaction model. The shaded area represents pointwise confidence intervals, while the dotted line indicates uniform confidence band, both of which are calculated via bootstrap. (b) Raw outcome plot by treatment condition; (c) Propensity score, estimated with \texttt{grf} package, by treatment condition.}
\end{minipage}
\end{figure}

Figure~\ref{fig:chapman}(a) presents the CME plot based on the linear interaction model. The authors interpret the positive CME for moderator values below $-0.5$ as evidence that the effect of U.N. authorization is larger when the U.S. policy position is far from those of Security Council members.

However, this finding appears to be entirely driven by extrapolation, and the lack of common support is easy to detect using simple plots. \hmx recommend (i) adding a histogram at the bottom of the CME plot and (ii) using a diagnostic plot, such as Figure~\ref{fig:chapman}(b), that displays the raw data by treatment condition. When the treatment is binary, we also recommend plotting the propensity score, a standard approach for assessing overlap \citep{heckman1999economics, imbens2025comparing}. 

Indeed, the histogram in Figure~\ref{fig:chapman}(a) reveals that there are almost no observations when the moderator is below \(-0.7\). Even more concerning is the lack of treated units (with U.N. authorization), which are concentrated in a narrow range around \(-0.5\). This indicates a severe lack of overlap in the regions where the author interprets the effect as positive. In fact, as shown in Figure~\ref{fig:chapman}(a), the downward slope of the CME estimates is almost entirely driven by six treated units through extrapolation. The propensity score plot in Figure~\ref{fig:chapman}(c) further confirms this issue, showing insufficient common support in these regions.

\paragraph*{Design Phase to Improve Overlap}

To address overlap concerns, we recommend incorporating a \emph{design phase} prior to any outcome analysis \citep[][Section 15]{imbens2015causal}. This recommendation reflects the broader view in causal inference that study design should conceptually precede outcome modeling, even in observational settings.

In observational studies, the design phase is a preparatory step in which researchers structure or preprocess the data \textit{before} examining outcomes. The objective is to approximate key features of a randomized controlled trial using only covariates and treatment assignment, without referencing the outcome variable. Concretely, researchers aim to construct a balanced sample, or subsamples, in which treated and control units are comparable in their covariate distributions. As \citet{imbens2015causal} emphasize, ``within this selected subsample, inferences are most robust and credible.''

A central purpose of the design phase is to diagnose and address lack of overlap. By examining the joint distribution of covariates and treatment, researchers can identify regions with insufficient support and trim or downweight those observations to avoid extrapolation beyond the data. Even when overlap formally holds, preprocessing can improve covariate balance and reduce residual confounding.

In practice, the design phase can be implemented using matching, trimming, or reweighting procedures. Outcome data should not be used at this stage, and hypothesis testing should be avoided, to prevent $p$-hacking and preserve the integrity of subsequent estimation.

The design phase does not substitute for identification or flexible modeling. It does not render unconfoundedness valid, nor does it address nonlinear or heterogeneous effects, which must be handled at the estimation stage. Empirical studies illustrate that such preprocessing can improve comparability across treatment groups and strengthen causal claims \citep{noble2024presidential}.

\begin{example}
\label{ex:Noble2024}
\citet{noble2024presidential} examines how Members of Congress in the United States leverage the president's symbolic power. The study argues that legislators—particularly those from the out-party—strategically reference the president to nationalize debates, shape constituent opinions, increase in-party approval, and reduce incentives for compromise. The analysis draws on a dataset of nearly two million floor speeches from 1973 to 2016. Here, we replicate the top panel of Figure 2 from the original paper, which shows that out-party legislators reference the president more frequently—and that this difference narrows as constituency support for the president increases. The variables of interest are:
\begin{itemize}
\item Outcome: Frequency of presidential references in Congressional speeches (continuous, $\in \left[0, 258\right]$) 
\item Treatment: Out-partisanship of lawmakers (binary, $\in \{0, 1\}$).
\item Moderator: District past presidential vote margin (continuous, $\in \left[-46.06, 46.99\right]$).
\end{itemize}
\end{example}

\begin{figure}[!ht]
\centering
\caption{Replicating \cite{noble2024presidential} Figure 2} \label{fig:noble}
\begin{subfigure}{\textwidth}
  \centering
  \begin{subfigure}{0.45\textwidth}
    \centering
    \includegraphics[width=\textwidth]{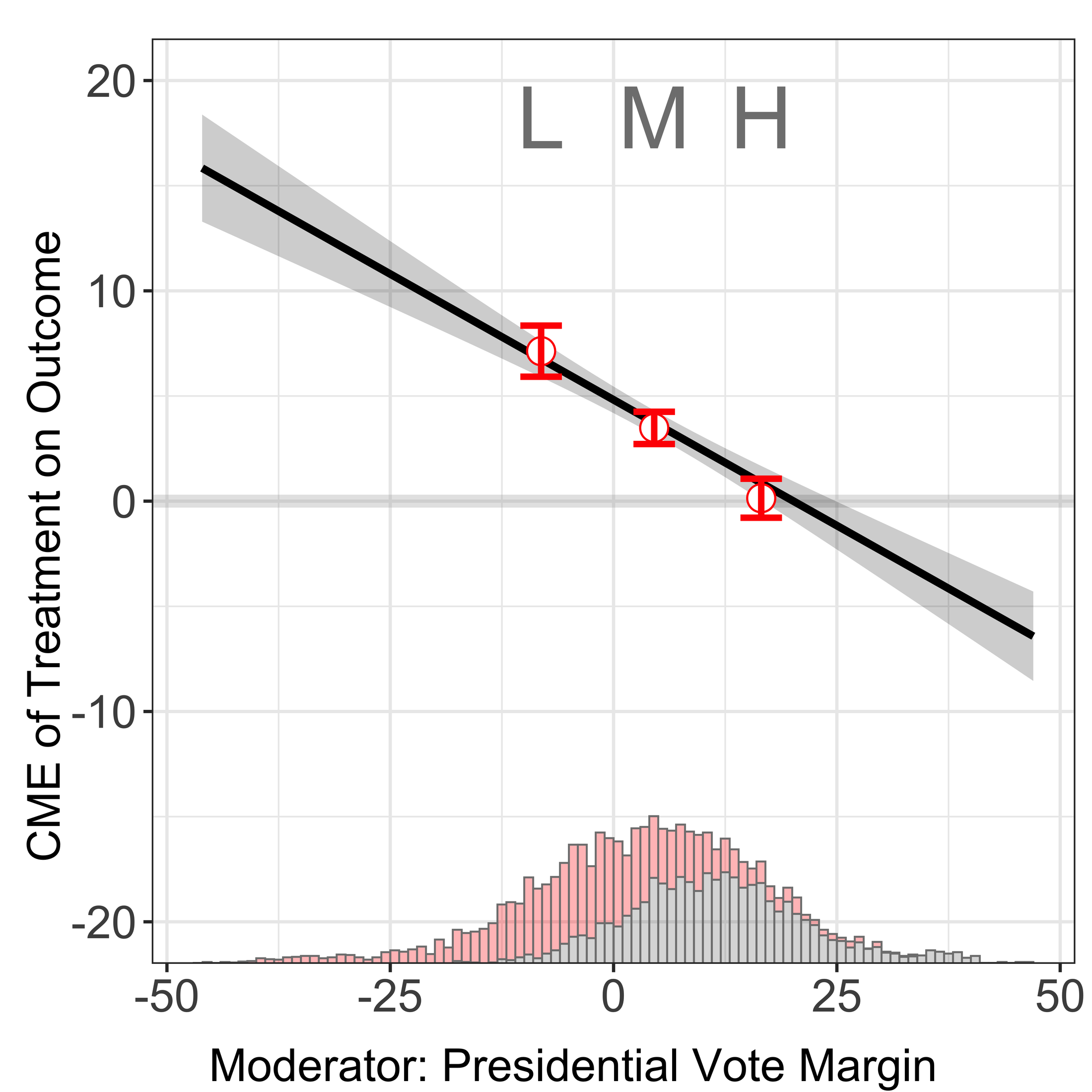}
  \end{subfigure}
  \hspace{1em}
  \begin{subfigure}{0.45\textwidth}
    \centering
    \includegraphics[width=\textwidth]{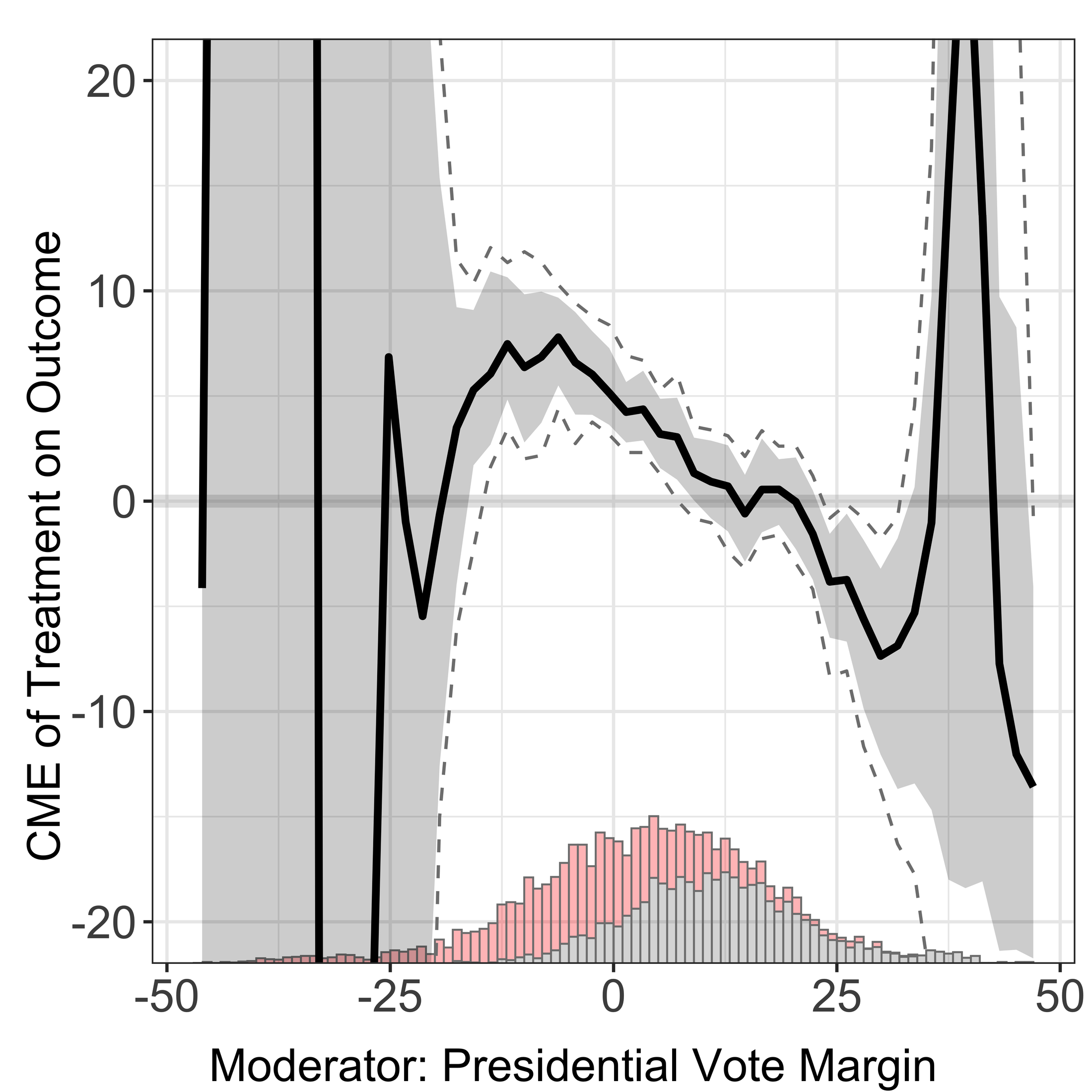}
  \end{subfigure}
  \caption{Full sample}
\end{subfigure}
\vspace{1em}
\begin{subfigure}{\textwidth}
  \centering
  \begin{subfigure}{0.45\textwidth}
    \centering
    \includegraphics[width=\textwidth]{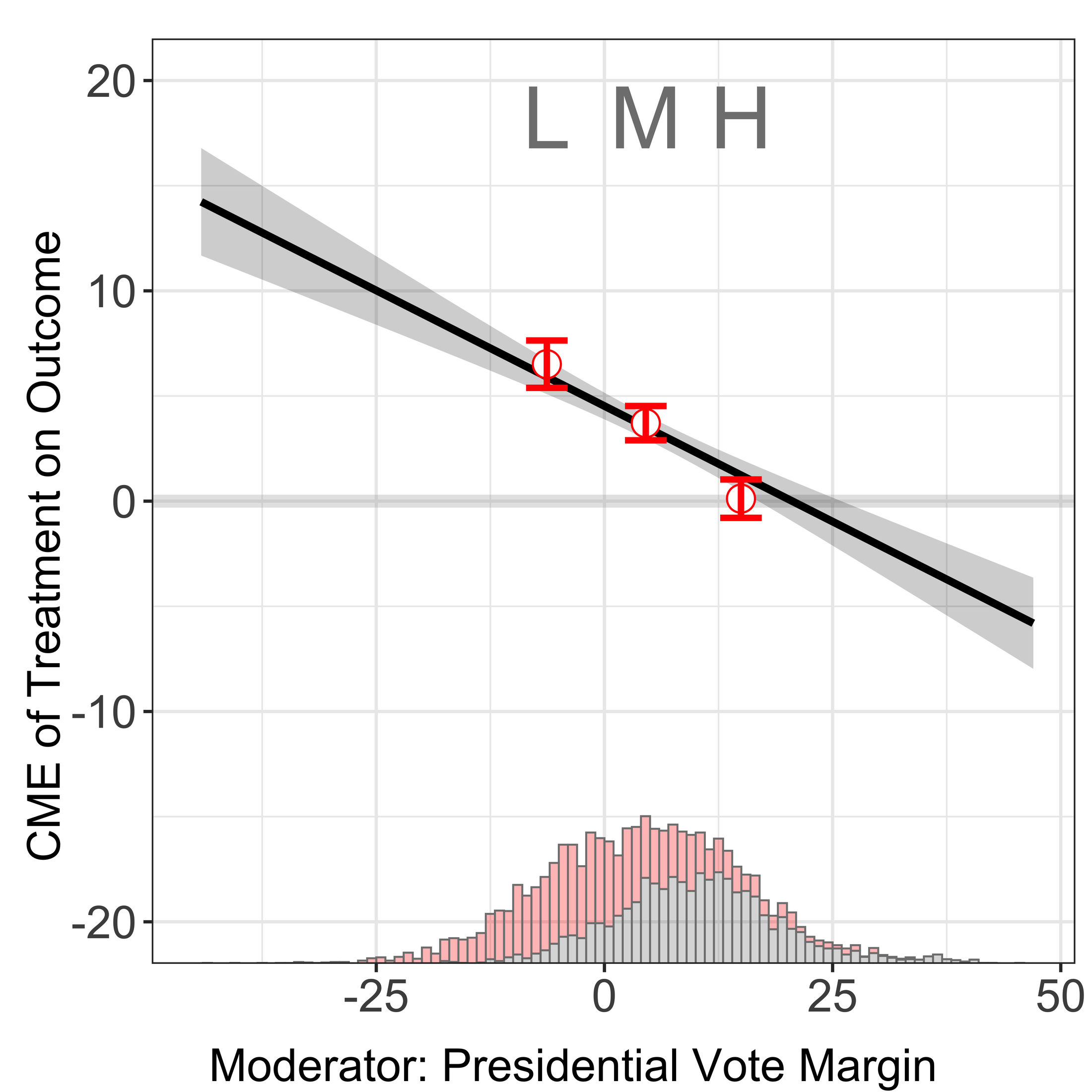}
  \end{subfigure}
  \hspace{1em}
  \begin{subfigure}{0.45\textwidth}
    \centering
    \includegraphics[width=\textwidth]{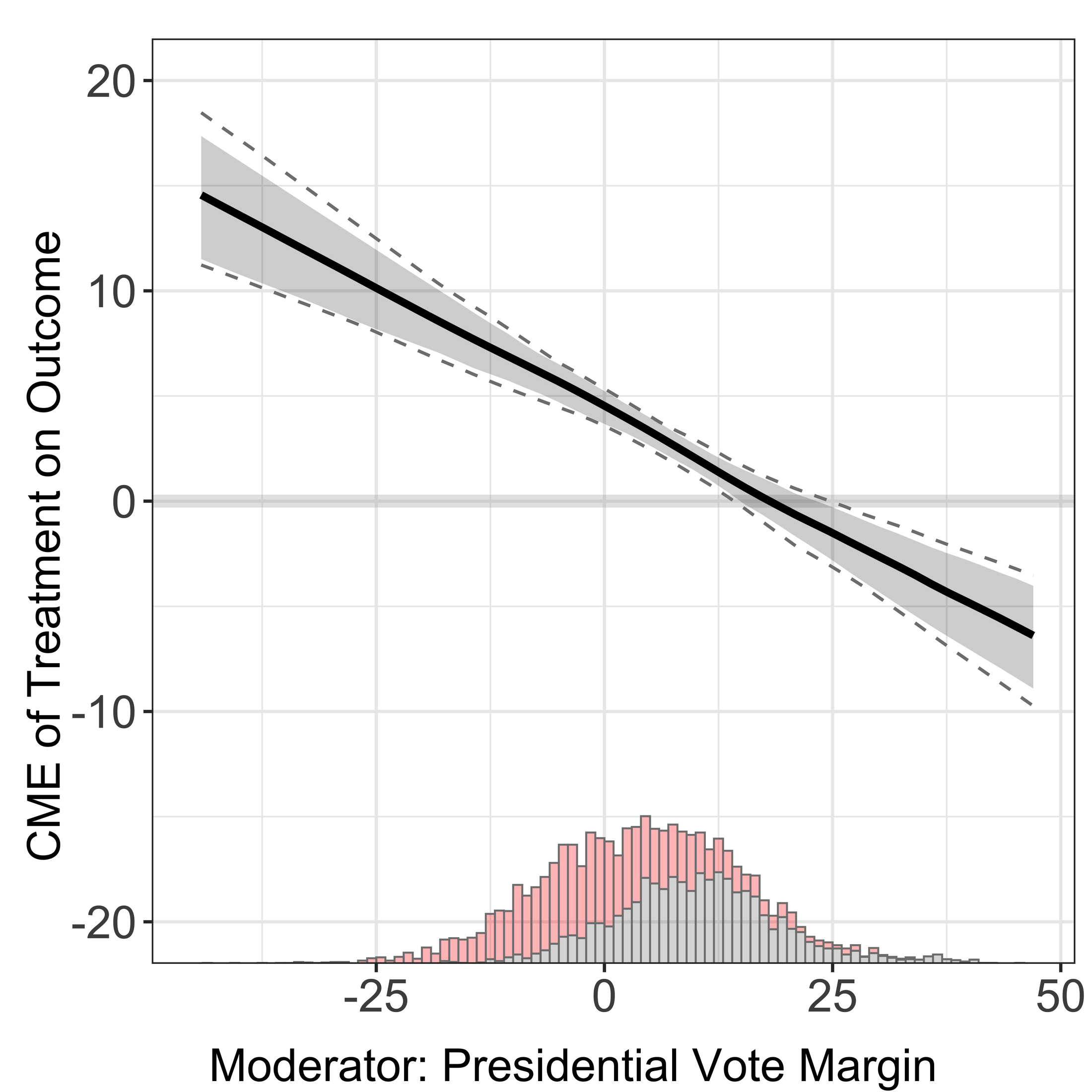}
  \end{subfigure}
  \caption{Trimmed sample}
\end{subfigure}
\begin{minipage}{1\linewidth}
{\footnotesize \emph{Notes:} Treatment $D$ is the out-partisanship of lawmakers, outcome $Y$ is the frequency of presidential references in congressional speeches. In each figure, the black dashed line represents the CME estimates; the red points (and bars) represent the binning estimates (and 95\% pointwise confidence intervals); the shaded area and dotted lines represent 95\% pointwise and uniform confidence intervals, respectively.}
\end{minipage}
\end{figure}

Using a large observational dataset, lack of overlap arises naturally due to partisan sorting in congressional districts: districts that overwhelmingly support the president tend to elect fewer out-party legislators, leading to imbalance across covariates—particularly the moderator. Figure~\ref{fig:noble}(a) shows CME estimates based on the linear estimator (left panel), overlaid with binning estimates, and the kernel estimator (right panel). At the bottom of each panel, we plot histograms of the moderator by treatment condition. The plots reveal limited common support at the extremes of the moderator (when $X \notin [-10, 20]$). Despite this, the linear estimator still produces estimates in these regions, which are not credible as they rely heavily on extrapolation. In contrast, the kernel estimator reflects this lack of support through very wide confidence intervals. The author is clearly aware of this issue and presents CME estimates over the range \(x \in [-4, 16]\), using the linear estimator, although the estimation is based on the full sample.

To ensure reasonable overlap, we trimmed the data based on the central 90\% percentile range of the propensity score $\hat \pi(V)=\Pr(D=1\mid X,Z)$. After trimming, both estimators exhibit more stable and nearly linear patterns, as shown in Figure~\ref{fig:noble}(a). The narrower confidence intervals and smoother trends reflect improved covariate balance between treated and control units in the trimmed sample. The resulting estimates support the original finding: as a constituency becomes more supportive of the president, the gap between out-party and in-party lawmakers in referencing the president in congressional speeches narrows. This example shows that data trimming during the design phase improves the credibility of the estimates and may even help justify simpler models.

\FloatBarrier

\subsection{Misspecification}

The linear interaction model makes rigid parametric assumption and can lead to misleading conclusions when misspecified. Below, we discuss two key scenarios where this occurs: (1) missing interaction terms between \( X \) and \( Z \) and (2) a nonlinear CME.
 
\paragraph{Missing Interaction Terms} 

\cite{blackwell2022reducing} and \cite{feigenberg2025omitted} highlight that the linear interaction model (as specified in Assumption~\ref{assm:linear}) may omit potential interactions between the moderator $X$ and other covariates $Z$, which are likely to be correlated in observational studies where neither $D$ nor $X$ is randomized. They refer to this specification as the single-interaction model, contrasting it with the fully-moderated model, where $X_i$ is interacted with all covariates:
\begin{equation*}
Y_{i}(d) = \beta_0 + \beta_1 d + \beta_2 X_i  + \beta_3 (d \cdot X_i) + Z_{i}^\top \beta_4  + (X_i \cdot Z_i)^\top \beta_5 + \epsilon_i.
\end{equation*}
The linear interaction model produces inconsistent estimates when (a) the treatment–moderator interaction is predictive of omitted interactions and (b) the omitted interactions significantly influence the outcome \citep{beiser2020problems}. This issue arises, for example, when the heterogeneous effect of \( D \) on \( Y \) depends not only on \( X \) but also on \( Z \), yet \( Z \) is included only as a level term.

To address this issue, \citet{blackwell2022reducing} recommend using a fully moderated model combined with a post-double-selection (PDS) procedure, originally introduced by \citet{belloni2014inference}. This method involves three steps: (i) applying Lasso to select interactions between \(X_i\) and \(Z_i\) that predict the outcome \(Y\), the treatment \(D\), and the interaction \(D \cdot X\); (ii) taking the union of the selected variables from each model; and (iii) running a post-Lasso regression using this union to estimate the causal effect. The key advantage of PDS-Lasso is that it mitigates regularization bias by using Lasso only for variable selection, while relying on unpenalized OLS in the final step. This preserves interpretability and guards against omitted variable bias without distorting parameter estimates.

The kernel estimator introduced in the next section is fully moderated and thus immune to this criticism, provided the number of additional covariates is limited. 

In Section 3, we introduce Lasso-based methods that are similar in spirit but differ in two key ways. First, they incorporate more flexible basis expansions of \(X\) and \(Z\), as well as the interaction of the basis expansions, to accommodate higher-order terms and interactions among covariates. Second, for binary treatments, we propose an augmented inverse propensity weighting (AIPW) estimator, and for continuous treatments, we use a partialing-out approach—both of which go beyond the post-Lasso regression strategy described here.
 
\paragraph{Nonlinear CME} 

As discussed earlier, while the linear interaction model allows the effect of \( D \) on \( Y \) to vary with \( X \), it imposes the restriction that \( \theta(x) = \beta_1 + \beta_3 x \). When \( D \) is continuous, this implies that the effect of \( D \) on \( Y \) changes with \( X \) at a constant rate, determined by \( \beta_3 \). However, in many empirical settings, this assumption is unrealistic, as treatment effects are rarely strictly linear in a covariate and may not even be monotonic. \hmx highlight this limitation and emphasize that violating this assumption can lead to misinterpretation of the CME.

\subparagraph{Binning estimator as a diagnostic tool.} To diagnose potential nonlinearity in the CME, \hmx propose the binning estimator, which partitions the range of the continuous moderator \( X \) into intervals and estimates the CME at an evaluation point within each interval. For simplicity, consider dividing \( X \) into three bins using cutoff points \( \delta_{1/3} \) and \( \delta_{2/3} \), representing the first and second terciles of the \( X \) distribution. Define three dummy variables:
$$G_{1,i} = \begin{cases}
1 & X_{i} < \delta_{1/3}\\
0 & \text{otherwise}
\end{cases}, \ 
G_{2,i} = \begin{cases}
1 & \delta_{1/3} \le X_{i} < \delta_{2/3}\\
0 & \text{otherwise}
\end{cases}, \ 
G_{3,i} = \begin{cases}
1 & X_{i} \ge \delta_{2/3}\\
0 & \text{otherwise}.
\end{cases}$$
Within each bin, we select an evaluation point \( x_j \) (such as the median of \( X \) values in that bin) and model the outcome \( Y \) as a piecewise linear function of \( X \), interacting with the treatment \( D \). We then estimate the following model using OLS:
\begin{equation*}
Y_{i} = \sum_{j=1}^{3}\left[\mu_j + \alpha_j D_{i} + \eta_j(X_{i} - x_j) + \beta_j(X_i - x_j)D_{i} \right] \cdot G_{j,i} + Z_{i}^\top\gamma + \epsilon_{i},
\end{equation*}
where $\mu_j, \alpha_j, \eta_j, \beta_j$ are unknown parameters to be estimated,  $Z$ represents additional covariates, and $\epsilon$ is an error term with $\mathbb{E}[\epsilon_{i} | X_{i} = x, D_{i}, Z_{i}] = 0$. The key insight is that at the chosen evaluation points \( x_j \), the CME simplifies to \( \theta(x_j) = \alpha_j \) since \( (X_i - x_j) = 0 \) at these points.

By allowing \( \alpha_j \) to vary across bins, the binning estimator accommodates flexible CME patterns across different segments of the moderator. Unlike the linear interaction model, which imposes a single global linear relationship on the CME, this piecewise approach relaxes the linearity assumption on the CME and can capture more nuanced patterns. At the same time, the binning estimator remains easier to implement than fully nonparametric or semiparametric methods, as it relies on standard regression techniques and well-established inference methods. Standard errors for \( \theta(x_j) \) are obtained from the regression output, eliminating the need for additional estimation steps or complex inference procedures.

We want to emphasize that the binning estimator should be used primarily as a diagnostic tool. With a fixed, coarse partition, the estimates $\alpha_j$ provide a ``local" test of the treatment effect at the chosen evaluation points $x_j$ (typically the bin medians), allowing us to visually assess whether the linear functional form is appropriate without imposing it globally.

\begin{example}
\label{ex:clark2006}
\cite{clark2006rehabilitating} examine the effect of the temporal proximity of presidential elections on the number of political parties in legislative elections. They argue that presidential elections, as the most important election in a polity, are most likely to exert the strongest influence when held concurrently with legislative elections. The direction of this ``coattails effect'' is moderated by the number of presidential candidates: temporal proximity to presidential elections reduces the number of parties contesting legislative elections when the number of presidential candidates is small, because parties that are not competitive in presidential races are disadvantaged.  We replicate Figure~2 in the original paper, the variables of interest are: 


\begin{itemize}
\item Outcome: number of parties competes in an election (continuous) 
\item Treatment: temporal proximity of presidential elections (continuous, $\in [0, 1]$)
\item Moderator Variable: number of presidential candidates. (continuous)
\end{itemize}
\end{example}

\begin{lstlisting}[language=R]
# R code excerpt
library(interflex) 

Y="enep1" #Effective number of electoral parties
D="proximity1" #Temporal proximity of presidential elections
X="enpres" #Number of presidential candidates
Z=c("uppertier", "logmag", "uppertier_eneg", "eneg", "logmag_eneg")

## diagnostic plot
cuts_raw <- c(0,1.64,3.28,4.92)
out.raw <- interflex(estimator = "raw", data = d, Y=Y,D=D,X=X, na.rm = TRUE, 
ncols = 4, cutoffs = cuts_raw)
plot(out.raw)

## binning estimator with uniform CI
cuts_bin <- c(0,0.1,3,4,7)
out.binning <- interflex(estimator = "binning", data = d, Y=Y,D=D,X=X,Z=Z, vartype ="bootstrap", nbins = 4,  cutoffs = cuts_bin) 
plot(out.binning)
\end{lstlisting}

\begin{figure}[!htb]
\centering
\caption{Replicating \cite{clark2006rehabilitating} Figure 2} \label{fig:clark}
\begin{subfigure}{0.8\textwidth}
  \centering
  \includegraphics[width=\textwidth]{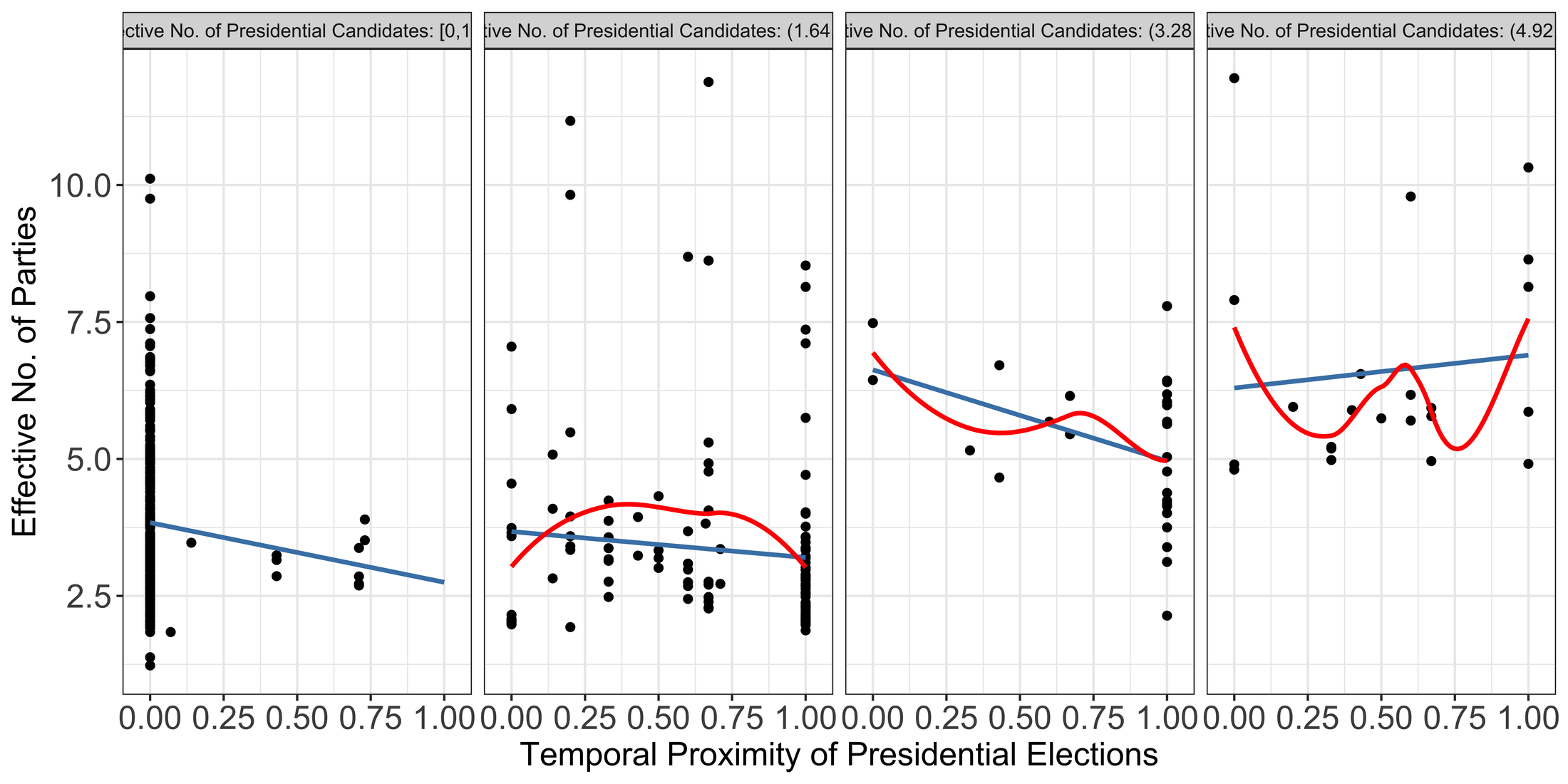}
  \caption{Raw outcome plot by moderator values}
  \label{fig:clark_raw}
\end{subfigure}
\vspace{1em}
\begin{subfigure}{0.8\textwidth}
  \centering
  \includegraphics[width=\textwidth]{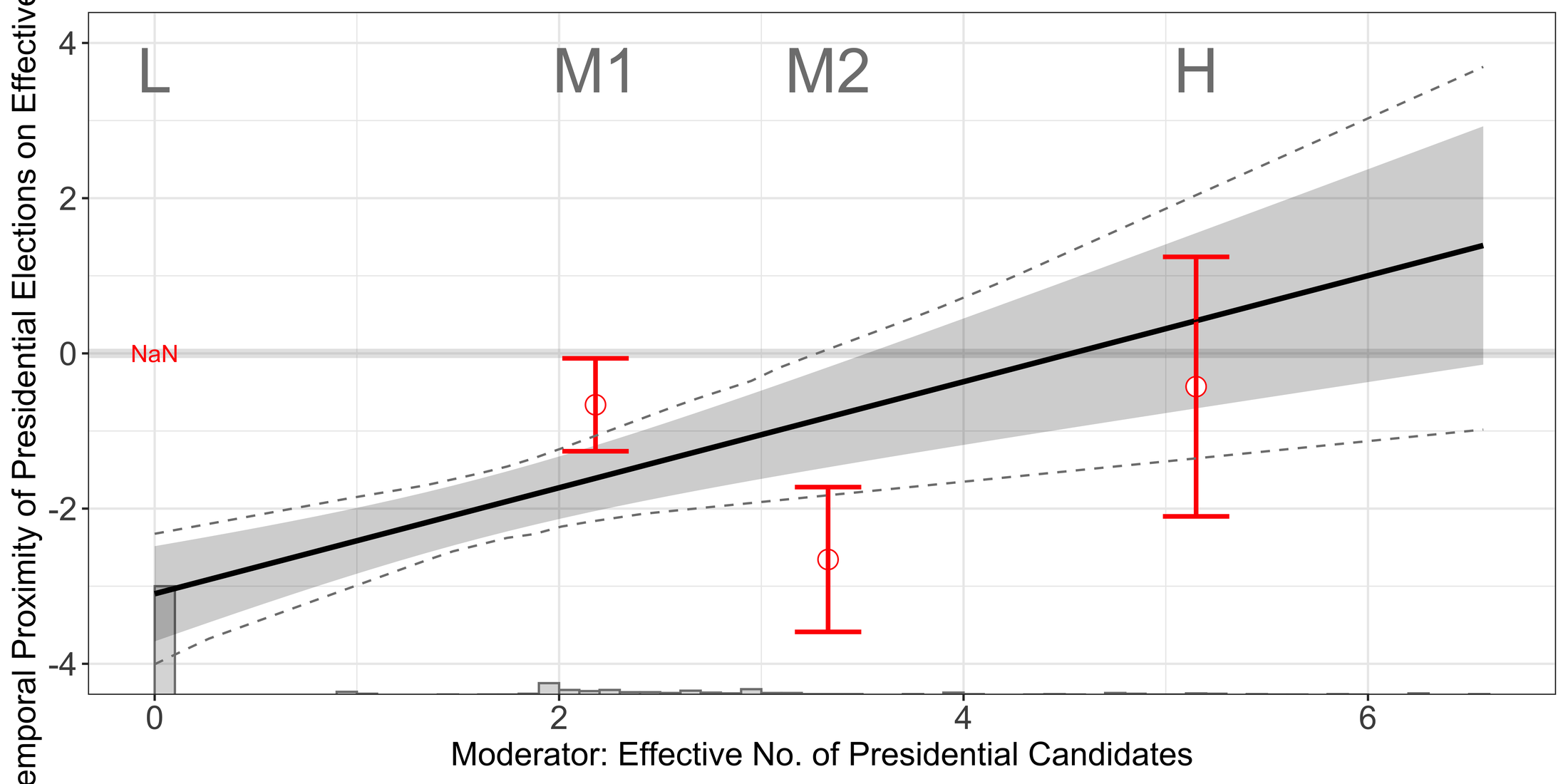}
  \caption{CME plot}
  \label{fig:clark_cme}
\end{subfigure}
\begin{minipage}{\linewidth}
  {\footnotesize\emph{Notes:} Treatment $D$ is presidential elections, outcome $Y$ is the effective number of electoral parties. (a) Plot of the raw outcome data in each bin of the moderator. (b) The CME estimates based on the linear interaction model (black line) and the binning estimator (red hollow circles). The gray ribbon and red bars show the 95\% pointwise confidence intervals, respectively. The dashed lines represent 95\% uniform confidence intervals of the linear interaction model.}
\end{minipage}
\end{figure}

Because a significant portion of the observations in the sample (59\%) have \( X = 0 \), as shown in Figure~\ref{fig:clark}(a), we discretize \( X \) into four bins: \( \{0\} \), \( (0, 3] \), \( (3, 4] \), and \( (4, 6.57] \). The last three bins contain approximately equal numbers of observations. Figure~\ref{fig:clark}(a) also reveals that the bivariate relationship between \( D \) and \( Y \) does not appear to increase with the moderator.

Figure~\ref{fig:clark}(b) compares the CME estimates from the linear interaction model and the binning estimator. Contrary to the study’s claim of a positive interaction based on the linear interaction model, the binning estimator indicates highly nonlinear effects. Specifically, the effect is negative and small in the second bin (\( X \in (0, 3] \)), turns negative and statistically significant in the third bin (\( X \in (3, 4] \)), and then approaches zero in the last bin (\( X \in (4, 6.57] \)). This pattern suggests that the linear interaction model is misspecified. 

Moreover, the treatment has no variation in the first bin (\( X = 0 \)), making the treatment effect non-identifiable at \( x = 0 \) due to a lack of common support. This further highlights the limitations of the linear interaction model in this case. Additionally, the uniform confidence interval for the second bin (M1) covers zero, meaning that we cannot reject the null hypothesis that \( \theta(x) < 0 \) for all \( x \leq 3 \) at the 5\% level, contradicting what the linear interaction model might have suggested.  

While we typically do not assume that the CME changes with \( X \) in a piecewise manner, this example demonstrates that the binning estimator is an easy-to-implement and powerful diagnostic tool for detecting potential nonlinearity in the CME. To achieve greater flexibility in modeling the functional form of the CME, we turn to more advanced methods, such as the kernel estimator, which we will discuss later in this section, and the augmented inverse propensity weighting (AIPW) and double/debiased machine learning (DML) estimators, which we will explore in the next two sections.



\subsection{The Kernel Estimator}

While parametric methods rely on strong and often unrealistic functional form assumptions, and nonparametric models avoid such assumptions but suffer from the ``curse of dimensionality,'' semiparametric methods offer a middle ground. They relax functional form restrictions by allowing some components of the model to be flexible while maintaining parametric structure in others. \hmx propose a semiparametric kernel estimator based on smooth varying-coefficient models (SVCM) \citep{hastie2017generalized} to estimate the CME. In the remainder of the Element, we refer to this method as the kernel estimator. We introduce two improvements: (i) allowing the coefficient on \(Z\) to vary with \(X\), making the model fully moderated, and (ii) incorporating an adaptive kernel bandwidth to improve estimation accuracy across different regions of the data distribution.

\begin{assumption}{Smooth varying-coefficient model , SVCM}\label{assm:svcm}
\[
 Y_i(d)   = f(X_i) + g(X_i) d + Z_i^\top \gamma(X_i) + \epsilon_i
\]
in which $f(\cdot), g(\cdot)$ and $\gamma(\cdot)$ are smooth functions. 
\end{assumption}
SVCM assumes that \( Y \) is a function of \( D \), \( X \), and \( Z \), while their impact on \( Y \) is modeled as an unspecified smooth function of \( X \). At each evaluation point $x_{0}$, each of them is approximated by a local linear regression:\footnote{\hmx assume a slightly more restrictive form of SVCM: $Y_i(d) = f(X_i) + g(X_i) d + Z_i^\top \gamma + \epsilon_i$, where \( \gamma \) does not vary with \( X \). Assumption~\ref{assm:svcm} represents a fully moderated version of their model.}
\begin{align*}
 f(X)_{|X-x_{0}|<\epsilon}  &= \mu(x_{0})+\eta(x_{0})(X-x_{0}) \\
 g(X)_{|X-x_{0}|<\epsilon}  &= \alpha(x_{0})+\beta(x_{0})(X-x_{0}) \\
 \gamma(X)_{|X-x_{0}|<\epsilon}  &= \rho(x_{0}) +\delta(x_{0})(X-x_{0}) 
\end{align*}

As before, given Assumption~\ref{assm:svcm}, the unconfoundedness assumption implies $\mathbb{E}[\epsilon_i | X_i, D_{i}, Z_i] = 0$. The CME becomes:
$$\theta(x)  = \mathbb{E}_Z\left[f(x) + g(x)  + Z_i^\top \gamma(x)\right] - \mathbb{E}_Z\left[f(x) + Z_i^\top \gamma(x)\right] = g(x).$$
when $D$ is binary. CME for a continuous $D$ similarly yields:
\[
\theta(x)  = \E_{Z}\left[ \frac{\partial (f(x) + g(x) d + Z_i^\top \gamma(x))}{\partial d}\right]  = g(x).
\]
Therefore, our goal is to obtain valid inference for \( g(x) \). In SVCM, the intercept \( f(x) \), treatment effect \( g(x) \), and control coefficients \( \gamma(x) \) are all flexible functions of the moderator \( X \), allowing for  nonlinear CME.

\hmx propose a kernel smoothing estimator for the CME with SVCM. For each evaluation point $x_0$ in the support of $X$, $\hat{f}(x_0)$, $\hat{g}(x_0)$, and $\hat{\gamma}(x_0)$ are estimated by minimizing the following weighted least-squares (WLS) objective function. With our two improvements, the kernel estimator that considers the influence of potentially all data points with greater emphasis on points closer to $x_0$ has the following weighted least-squares objective function: 
\[
\left( \hat{\mu}(x_0), \hat{\alpha}(x_0), \hat{\eta}(x_0), \hat{\beta}(x_0), \hat{\rho}(x_0),\hat{\delta}(x_{0}) \right) = \arg\min_{\tilde{\mu}, \tilde{\alpha}, \tilde{\eta}, \tilde{\beta}, \tilde{\rho},\tilde{\delta}} L(\tilde{\mu}, \tilde{\alpha}, \tilde{\eta}, \tilde{\beta}, \tilde{\rho},\tilde{\delta})
\]
in which:
\[
L = \sum_{i=1}^n \left\{ \left[ Y_i - \tilde{\mu} - \tilde{\alpha} D_i - \tilde{\eta}(X_i - x_0) - \tilde{\beta} D_i (X_i - x_0) - \tilde{\rho}^\top Z_i - \tilde{\delta}^\top Z_i(X_i-x_{0}) \right]^2 K \left( \frac{|X_i - x_0|}{h(x_0)} \right) \right\}
\]
Kernel smoothing averages data points using a kernel function, with weights determined by proximity to the estimation point. The kernel function, $K(.)$, assigns higher weights to observations closer to the target point \( x_0 \).  
With the above kernel estimator, \( \hat{f}(x_0) = \hat{\mu}(x_0) \) and \( \hat{g}(x_0) = \hat{\alpha}(x_0) \). Smoothing is performed using a Gaussian kernel \( K(\cdot) \), while the parameter of the bandwidth function \( h(x_0) \) is selected via least-squares cross-validation.  

\paragraph{Adaptive Kernel}

Classical bandwidth selection methods perform well only when the underlying density is approximately normal. They become problematic for long-tailed or multimodal distributions. A fixed-bandwidth kernel estimator performs well near the mode, where data density is high, but oversmooths in the tails, where observations are sparse, leading to biased estimates. To mitigate this issue, adaptive kernel density estimation adjusts the bandwidth based on the local density at each evaluation point, allowing for finer resolution in sparse regions while maintaining stability in denser areas.

With an adaptive kernel, the bandwidth \( h \) is no longer constant but varies with the location of the evaluation point \( x_0 \), denoted as \( h(x_0) \).  We take the following density-based adaptation \citep{silverman2018density}.  \begin{equation*}
    h(x_0) = h_0 \times \left( \frac{\bar{\rho}^{GM}}{\rho(x_0)} \right)^{\frac{1}{2}} 
\end{equation*}
where $h_0$ is a global baseline bandwidth, $\rho(x_0)$ denotes the local design density at $x_0$, and $\bar{\rho}^{\mathrm{GM}}$ is the geometric mean of $\left\{\rho\left(x_i\right)\right\}$ over the sample. In practice, we first obtain a pilot estimate of $\rho(\cdot)$ (e.g., via kernel density estimation) and then plug it into the adaptation rule to produce bandwidths $h(x_0)$ for the kernel regression. This construction makes $h(x_0)$ inversely proportional to the local design density: in regions with sparse data, the bandwidth increases, providing broader smoothing across $X$; in denser regions, it decreases, enabling more localized smoothing and preserving finer features. The baseline bandwidth $h_0$ is chosen by 10-fold least-squares cross-validation. The sample is randomly partitioned into ten folds. For each fold $k$, we fit the estimator on the remaining nine folds and compute predictions $\widehat{m}^{(-k)}_{h^{(-k)}(x_i)}(x_i)$ for the held-out observations $i \in I_k$, where $h^{(-k)}(x_i)$ denotes the adaptive bandwidth computed from the training data for that fold. The cross-validation criterion is
\[
\mathrm{CV}(h_0)
= \frac{1}{n} \sum_{k=1}^{10} \sum_{i \in I_k}
\bigl[ Y_i - \widehat{m}^{(-k)}_{h^{(-k)}(x_i)}(x_i) \bigr]^2,
\]
and the selected $h_0$ minimizes $\mathrm{CV}(h_0)$, thereby balancing bias and variance.

A common issue in semiparametric regression is boundary bias, where estimates near the edges of the data range tend to be less accurate due to fewer observations. As recommended by \citet{fan1995local}, \hmx address this problem by incorporating two additional terms, \( \eta(X - x_0) \) and \( \beta D(X - x_0) \), which capture the first partial derivative of \( Y \) with respect to \( X \) at each \( x_0 \). With a fully moderated version, $\delta Z (X-x_0)$ is also included for each covariate $Z$. These terms adjust for the local slope of the response surface, helping to reduce bias at the boundaries of the support of \( X \), where data are sparse and estimation uncertainty is higher.

\paragraph{Inference}

Once we obtain the CME estimates through kernel-weighted least squares, the next step is to estimate the standard errors of the estimated parameters. This can be done using two approaches: (i) analytically, via the robust variance estimator, and (ii) empirically, using bootstrapping.

We can estimate pointwise variance analytically using the robust (sandwich) covariance matrix. The local linear kernel regression estimates \( \hat{\alpha}(x_0) \) and \( \hat{\beta}(x_0) \) are obtained as coefficients from a weighted least squares regression. Consequently, the variance of these estimates can be obtained directly from the robust variance-covariance matrix of the WLS estimator, without requiring additional Taylor expansions.

The confidence interval for \( \hat{\theta}(x) = \hat{\alpha}(x) \) is calculated as: $CI: \hat{\theta} \pm t \times SE(\hat{\theta})$, where \( t \) is the critical value from the \( t \)-distribution for a 95\% confidence interval, and \( SE(\hat{\theta}) \) is the standard error of the estimate.

An alternative approach is bootstrapping. Given a dataset \( S \) with \( n \) observations, we generate \( B \) bootstrap samples \( S_1, S_2, \dots, S_B \) by sampling with replacement from \( S \). For each bootstrap sample \( S_b \), we compute the kernel estimates and extract the CME \( \hat{\theta}_b = \hat{\alpha}_b \). The standard deviation of these bootstrap estimates approximates the standard error:
\[
SE(\hat{\theta}) \approx \sqrt{\frac{1}{B-1} \sum_{b=1}^{B} (\hat{\theta}_b - \bar{\theta}_B)^2}
\]
where \( \bar{\theta}_B \) is the mean of the bootstrap estimates.


Furthermore, we extend the bootstrap approach to construct uniform confidence intervals, similar to those used with the linear interaction model. Again, we implement the bootstrap-based sup-$t$ band method, ensuring valid inference across the entire range of \( X \).

\bigskip

\begin{figure}[!ht]
\centering
\caption{Applying the Kernel Estimator}\label{fig:kernel}
\begin{subfigure}{0.45\textwidth}
  \centering
  \includegraphics[width=\textwidth]{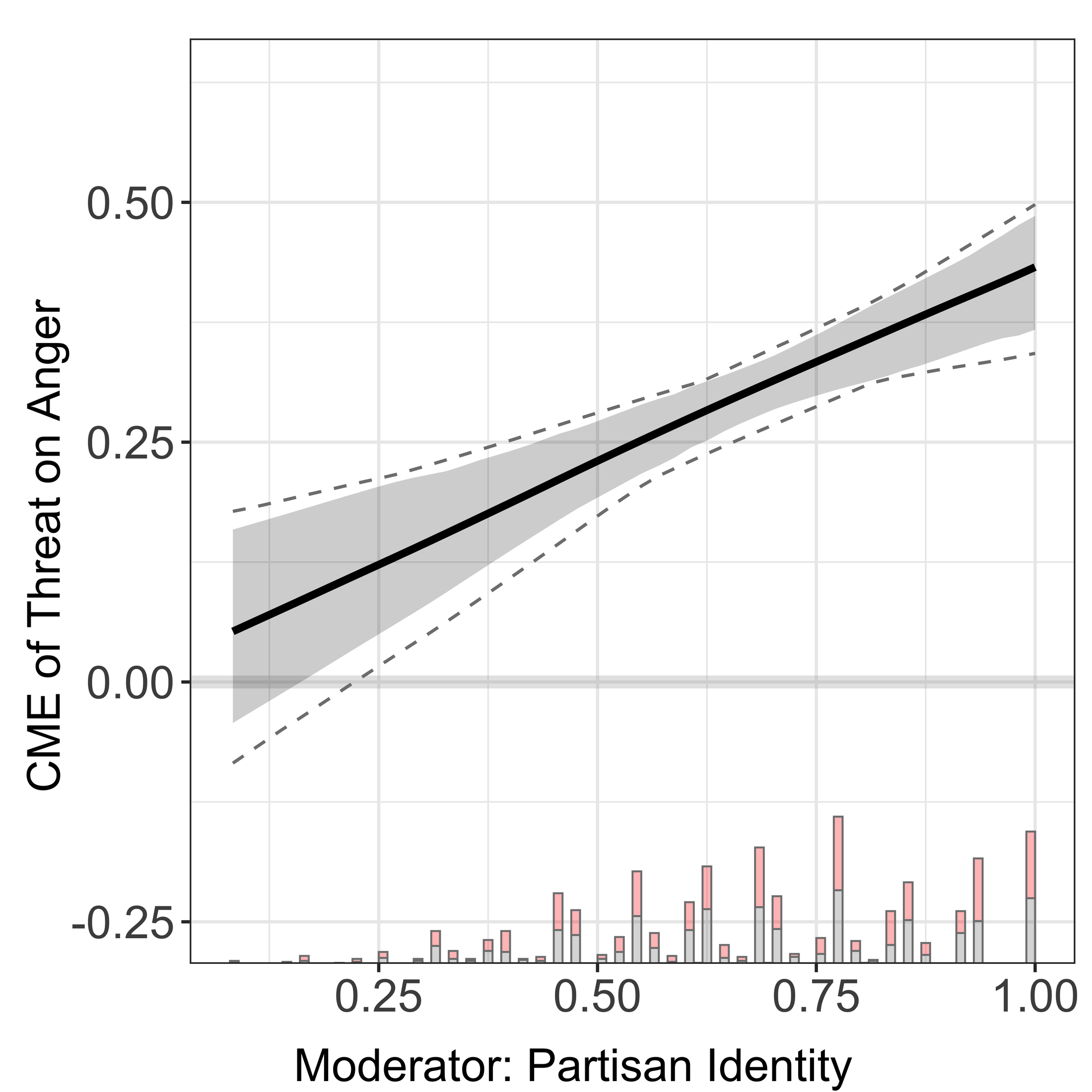}
  \caption*{(a)}
  \label{fig:huddy_kernel}
\end{subfigure}
\hspace{1em}
\begin{subfigure}{0.45\textwidth}
  \centering
  \includegraphics[width=\textwidth]{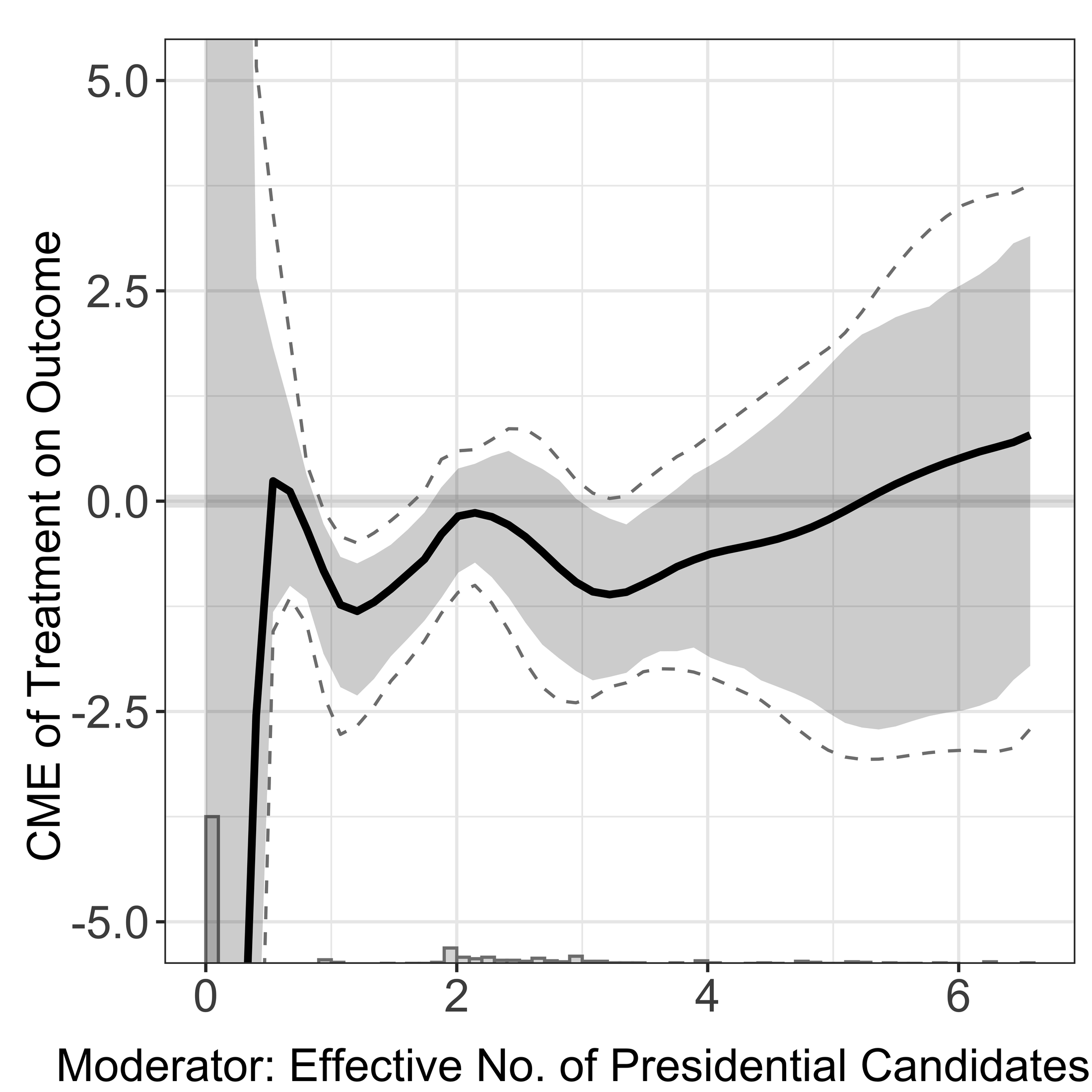}
  \caption*{(b)}
  \label{fig:clark_kernel}
\end{subfigure}
\vspace{1em}
\begin{minipage}{\linewidth}
{\footnotesize \emph{Notes:} The above figures replicate Figure 2A from \cite{huddy2015expressive} (a) and Figure 2 from \cite{clark2006rehabilitating} (b). In each figure, the black dashed line represents the CME estimates based on the kernel estimator; shaded area and dotted lines represent pointwise and uniform confidence intervals, respectively, via nonparametric bootstrapping.}
\end{minipage}
\end{figure}

We apply the kernel estimator to Examples~\ref{ex:huddy2015} and \ref{ex:clark2006}. The CME estimates from both examples are shown in Figure~\ref{fig:kernel}. Using data from \citet{huddy2015expressive}, the kernel estimator yields CME estimates that are nearly identical to those from the linear interaction model, suggesting that Assumption~\ref{assm:linear} is reasonable in this setting. 

\begin{lstlisting}[language=R]
# R code excerpt

library(interflex) 
### Huddy et. al 2015 
Y="totangry" #Anger
D="threat" #Threat
X="pidentity" #Partisan Identity
Z<-c("issuestr2", "pidstr2",  "knowledge" , "educ" , "male" , "age10" ) 

out.kernel1 <- interflex(estimator = "kernel", 
            Y=Y,D=D,X=X,data=d,
            vartype = "bootstrap", ## uniform CI
            full.moderate = TRUE, 
            na.rm = TRUE)
plot(out.kernel1)

### Clark and Golder 2006
Y="enep1" #Effective number of electoral parties
D="proximity1" #Temporal proximity of presidential elections
X="enpres" #Number of presidential candidates"
Z=c("uppertier" ,"logmag" , "uppertier_eneg" ,"eneg", "logmag_eneg")

out.kernel2 <- interflex(estimator = "kernel", 
            Y=Y,D=D,X=X,data=d,
            vartype = "bootstrap", ## uniform CI
            full.moderate = TRUE, na.rm = TRUE)
plot(out.kernel2)
\end{lstlisting}

In contrast, using data from \citet{clark2006rehabilitating}, the CME estimates differ substantially from those based on the linear interaction model. The first thing we notice is the absence of variation in the treatment variable when the moderator value is zero, indicating that the treatment effect is unidentifiable at this point due to lack of common support. This is not captured by the linear interaction model, but with kernel estimator, the confidence intervals blow up as the moderator approaches zero, given that there is no variation in the treatment variable at this point. 

The kernel estimator also picks up on nonlinearity in the CME masked by the linear interaction model. Contrary to the anticipated positive interaction suggested by the study, the CME estimates are insignificant when $X$ is close to 0, shifts to a negative and statistically significant effect as the moderator ranges from approximately $[0.5, 2]$, and then reverts nearly to indistinguishable from zero when $X$ is bigger than 2.

In both examples, the patterns revealed by the kernel estimator are broadly consistent with the findings from the binning estimator, which we recommend as a diagnostic tool.

\subsection{Summary}

This section covered classic CME estimation: the linear interaction model with bootstrap-based uniform confidence intervals to address multiple comparisons; diagnostics for lack of common support and misspecification (raw outcome and propensity-score plots, the binning estimator), together with a design-phase remedy; and a semiparametric kernel estimator extended with a fully moderated specification and adaptive bandwidth.


\clearpage

\section{AIPW and Extensions}

So far, our analysis has centered on modeling \(\mathbb{E}[Y(d)\mid X,Z]\) with parametric or semiparametric specifications and using unconfoundedness to estimate the CME. These approaches, however, do not directly incorporate the treatment assignment mechanism and can be restrictive in how they handle rich covariates \(Z\), especially when nonlinearities and interactions matter.

This section addresses these limitations using propensity-score weighting and its augmented form. We begin with the binary-treatment case, introduce augmented inverse probability weighting (AIPW), and highlight its \textit{double robustness}: the CME estimate remains consistent if either the outcome model or the propensity-score model is correctly specified \citep{knaus2022double, kennedy2023towards}. We then extend the same logic to continuous treatments using a residualization-based identification strategy \citep{nie2021quasi}. The AIPW and ``partial-out'' frameworks provide the conceptual foundation for the double/debiased machine learning (DML) approach introduced in the next section.

\subsection{Outcome, IPW and AIPW Signals for the CME}

This section formalizes how to estimate the CME for a binary treatment using three complementary identification strategies---outcome modeling, inverse probability weighting (IPW), and augmented IPW (AIPW). We establish identification for each approach and discuss their robustness and variance trade-offs. Each method maps the data for each observation into a corresponding \emph{signal}. Under the respective model assumptions, the conditional expectation of these signals given \(X=x\) equals the CME, \(\theta(x)\).

In Section~1, we have defined the conditional mean outcome by treatment condition \(\mu^{d}(v) = \mathbb{E}[Y_i \mid D_i = d, V_i = v]\) for \(d \in \{0,1\}\), and the propensity score \(\pi(v) = \Pr(D_i = 1 \mid V_i = v)\). We showed that, when \(D\) is binary and all covariates in \(V\) are discrete, under unconfoundedness and overlap assumptions, the stratified difference-in-means identifies the CME:
\[
\sum_v \left\{ \left( \mu^{1}(v) - \mu^{0}(v) \right) \cdot \Pr(V = v \mid X = x) \right\} = \theta(x),
\]
which naturally extends to cases where \(Z\) includes continuous variables by replacing the sum with an integral:
\[
\int \left( \mu^{1}(v) - \mu^{0}(v) \right) \cdot f_{V \mid X}(v \mid x) \, dv = \theta(x),
\]
where \(f_{V \mid X}(v \mid x)\) denotes the conditional density of \(V\) given \(X = x\). We can therefore use expectation notation to cover both cases:
\begin{equation}\label{eq:outcome}
\mathbb{E}\!\left[\mu^{1}(V_i) - \mu^{0}(V_i) \mid X_i = x\right] = \theta(x).
\end{equation}
Moreover, in Section~1 we also showed that the IPW estimator identifies the CME:
\begin{equation}\label{eq:ipw}
\mathbb{E}\!\left[\left.\frac{D_i Y_i}{\pi(V_i)}  - \frac{(1-D_i) Y_i}{1 - \pi(V_i)}  \right|\, X_i = x\right] = \theta(x).
\end{equation}
Both expectations are taken over the conditional distribution of \(V\) given \(X\).

Building on the strengths of both outcome modeling and IPW, \citet{robins1994estimation} introduced the AIPW estimator, which combines the outcome model and the propensity-score model to improve robustness and efficiency in estimating the CME. The AIPW estimator for the CME can be constructed as
\begin{align}
& \mathbb{E}\!\left[\mu^{1}(V_i) - \mu^{0}(V_i)\mid X_i = x \right] \;+\nonumber \\[2pt]
& \mathbb{E}\!\left[\frac{D_i}{\pi(V_i)}\bigl(Y_i - \mu^{1}(V_i)\bigr) - \frac{1 - D_i}{1 - \pi(V_i)}\bigl(Y_i - \mu^{0}(V_i)\bigr) \,\Big|\, X_i = x \right] \;=\; \theta(x). \label{eq:aipw}
\end{align}
The proof in the appendix shows that the equality holds (i.e., AIPW is consistent) when either the outcome models \(\mu^{1}(V_i)\) and \(\mu^{0}(V_i)\) or the propensity-score model \(\pi(V_i)\) is correctly specified, but not necessarily both. This property is known as \emph{double robustness}.

Asymptotically, beyond double robustness, AIPW exhibits lower variance than IPW when both the outcome and propensity-score models are correctly specified. \citet{robins1994estimation} show that AIPW attains the smallest asymptotic variance within the class of inverse probability–weighted estimators. However, when only the outcome model(s) are correctly specified, AIPW remains consistent but can have higher variance than a purely outcome-based estimator.

In finite samples, AIPW does not necessarily dominate the outcome model or IPW. \citet{li2019addressing} demonstrate via simulations that its finite-sample performance depends heavily on overlap in propensity scores. When overlap is poor---that is, many estimated propensity scores \(\hat{\pi}(V_i)\) lie near 0 or 1, producing very large weights---AIPW can perform worse than an outcome-based estimator.

Motivated by this finite-sample concern, we adopt propensity-score clipping to mitigate instability from extreme weights. IPW and AIPW estimators can be sensitive when the estimated propensity score \(\hat{\pi}(V_i)\) is close to 0 or 1. To address this, we replace \(\hat{\pi}(V_i)\) with
\[
\tilde{\pi}(V_i)=\min \!\left\{\max \!\bigl(\hat{\pi}(V_i), \alpha\bigr), 1-\alpha\right\},
\]
for a small threshold \(\alpha>0\). Clipping reduces variance and improves finite-sample stability, at the cost of a small bias when scores are modified \citep{cole2008constructing}. In the remainder of this section, we set \(\alpha = 10^{-2}\) in our implementations.

\subparagraph*{Signals for the CME.}
We now introduce \emph{signals}, defined as per-observation transformations whose conditional expectations equal the CME.
We define outcome, IPW, and AIPW signals as
\begin{align*}
 \Lambda_i^{(\mathrm{outcome})} &= \mu^{1}(V_i)  - \mu^{0}(V_i),\\
 \Lambda_i^{(\mathrm{ipw})} &= \frac{D_i}{\pi(V_i)}\,Y_i - \frac{1 - D_i}{1 -\pi(V_i)}\,Y_i,\\
 \Lambda_i^{(\mathrm{aipw})} &= \mu^{1}(V_i) - \mu^{0}(V_i) +
     \frac{D_i}{\pi(V_i)}\bigl(Y_i-\mu^{1}(V_i)\bigr) -
    \frac{1-D_i}{1-\pi(V_i)}\bigl(Y_i-\mu^{0}(V_i)\bigr).
\end{align*}
Therefore, each of Equations~(\ref{eq:outcome})--(\ref{eq:aipw}) implies an identification result for a corresponding signal under the correctly specified model:\footnote{By “correctly specified,” we mean that the working model coincides with the true data generating process on the support of \(V\): for all \(v\), \(\mu^{d}(v)=\mathbb{E}[Y(d)\mid V=v]\) for \(d\in\{0,1\}\) and/or \(\pi(v)=\Pr(D=1\mid V=v)\), with positivity \(0<\pi(v)<1\). For estimated models \(\hat\mu^d,\hat\pi\), this can be read as large-sample convergence to the truth.}
\begin{align*}
\text{Equation~(\ref{eq:outcome}) } &\Rightarrow\quad  \mathbb{E}\!\left[\Lambda_i^{(\mathrm{outcome})} \mid X_i = x\right] = \theta(x)\quad \text{if }\mu=(\mu^1,\mu^0) \text{ is correctly specified}, \\
\text{Equation~(\ref{eq:ipw}) } &\Rightarrow\quad \mathbb{E}\!\left[\Lambda_i^{(\mathrm{ipw})} \mid X_i = x\right] = \theta(x)\quad \text{if }\pi \text{ is correctly specified}, \\
\text{Equation~(\ref{eq:aipw}) } &\Rightarrow\quad \mathbb{E}\!\left[\Lambda_i^{(\mathrm{aipw})} \mid X_i = x\right] = \theta(x)\quad \text{if either }\mu=(\mu^1,\mu^0) \text{ or }\pi \text{ is correctly specified}.
\end{align*}
In other words, taking the conditional expectation of any of the outcome, IPW, or AIPW signals given \(X_i = x\), under their respective modeling assumptions, identifies the CME.

\subparagraph*{Signals for the CME on the treated.}
Although we focus primarily on the CME (a special case of the conditional average treatment effect, CATE), the same logic extends to other estimands. A commonly used estimand is the conditional average treatment effect on the treated (CATT), \(\mathbb{E}[Y_i(1)-Y_i(0) \mid D_i=1, V_i=v]\). Analogously, restricting attention to heterogeneity in the moderator yields the conditional marginal effect on the treated (CMET),
\[
\mathrm{CMET}(x)=\mathbb{E}\!\left[Y_i(1)-Y_i(0) \mid D_i=1, X_i=x\right].
\]
Outcome, IPW, and AIPW signals for the CMET are
\begin{align*}
 \Xi_i^{(\mathrm{outcome})} &= \frac{Y_i-\mu^{0}(V_i)}{\Pr(D_i=1\mid X_i)}\,D_i, \\
 \Xi_i^{(\mathrm{ipw})} &= \frac{Y_i\bigl(D_i-\pi(V_i)\bigr)}{\Pr(D_i=1\mid X_i)\bigl(1-\pi(V_i)\bigr)}, \\
 \Xi_i^{(\mathrm{aipw})} &= \frac{1}{\Pr(D_i=1\mid X_i)}\bigl(Y_i-\mu^{0}(V_i)\bigr)\!\left(D_i-\frac{\pi(V_i)(1-D_i)}{1-\pi(V_i)}\right).
\end{align*}
Here, \(\Pr(D_i=1\mid X_i)\) is the probability of treatment given the moderator; it can be estimated parametrically or nonparametrically.\footnote{It is worth noting that \(\Pr(D_i=1\mid X_i)\) is the marginal propensity score, which is conditioned only on $X$, and needs to be estimated separately from the full propensity score $\pi\left(V_i\right)$, which is conditioned on $V = (X, Z)$.} Under unconfoundedness and overlap, we establish in the appendix that under correctly specified models:
\[
\mathbb{E}\!\left[\Xi_i^{(\mathrm{outcome})}\mid X_i=x\right]
=\mathbb{E}\!\left[\Xi_i^{(\mathrm{ipw})}\mid X_i=x\right]
=\mathbb{E}\!\left[\Xi_i^{(\mathrm{aipw})}\mid X_i=x\right]
= \mathrm{CMET}(x).
\]
The AIPW signal possesses this property when either the outcome model $\mu^{0}(\cdot)$ or the propensity-score model $\pi(\cdot)$ is correctly specified. Notably, estimation of \(\mathrm{CMET}(x)\) does not require modeling the treated potential-outcome function \(\mu^{1}(\cdot)\).

\subsection{Estimation Strategies}
This subsection is slightly more technical than the preceding material. Readers new to weighting and doubly robust methods can first skim the three-step ``recipe'' in estimating CME using AIPW and the worked examples, then return to details as needed; those already familiar with IPW and AIPW can treat this as a compact reference.

Implementing the outcome, IPW, and AIPW estimators for the CME proceeds in three steps. First, estimate the propensity score \(\Pr(D = 1 \mid X, Z)\) and the outcome models \(\mathbb{E}[Y \mid D=0, X, Z]\) and \(\mathbb{E}[Y \mid D=1, X, Z]\). We refer to these as \emph{nuisance parameters}: they are not of direct substantive interest but are essential for identifying and estimating the CME.

Second, given the estimated outcome models \(\hat{\mu}^{1}(\cdot)\) and \(\hat{\mu}^{0}(\cdot)\) and the estimated propensity score \(\hat{\pi}(\cdot)\), construct the outcome, IPW, or AIPW \emph{signals} according to Equations~(\ref{eq:outcome})--(\ref{eq:aipw}) and the corresponding signal definitions. For each observation, obtain its constructed signal \(\hat{\Lambda}_{i}\) under each estimator.

Finally, estimate the CME as a function of \(X\) using the constructed signals. When \(X\) is discrete, compute subgroup means by averaging the signals within each level of \(X\). When \(X\) is continuous, regress the signals \(\{\hat{\Lambda}_{i}\}\) on a flexible representation of \(X\) (e.g., a B\mbox{-}spline basis expansion) to obtain a smooth estimate of the CME. Alternatively, we can use the semiparametric kernel regression, as introduced in Section~2, to estimate this relationship.

\paragraph{Steps 1 and 2: Constructing Signals}

We begin by discussing how to construct signals from data using outcome modeling, IPW, and AIPW. 

\subparagraph{Signals from outcome modeling.}
With the outcome-modeling approach, we can fit two separate linear regressions for data with \(D_i = 1\) and \(D_i = 0\). Specifically, we regress \(Y_i\) on \((1, V_i)\) \emph{separately} within each treatment group, yielding the following estimates:
\[
\hat{\mu}^{0}(V_i)  = \hat\beta_{0}^{0} + V_i \hat{\beta}_{V}^{0}\;,\qquad
\hat{\mu}^{1}(V_i) = \hat\beta_{0}^{1} + V_i \hat{\beta}_{V}^{1}\;,
\]
where \((\hat\beta_{0}^{0}, \hat{\beta}_{V}^{0\prime})\) and \((\hat\beta_{0}^{1}, \hat{\beta}_{V}^{1\prime})\) are the OLS coefficients from these two regressions. For each observation \(i\), we then construct the signal from the outcome model:
\[
\hat{\Lambda}_i^{(\mathrm{outcome})} = \hat{\mu}^{1}(V_i) - \hat{\mu}^{0}(V_i) = (\hat\beta_{0}^{1} - \hat\beta_{0}^{0}) + V_i (\hat{\beta}_{V}^{1} - \hat{\beta}_{V}^{0})\,,
\]
which represents the predicted treatment effect for observation \(i\).

\subparagraph*{IPW signals.}
For the IPW estimator, we first fit a propensity-score model using a logistic regression of \(D\) on \(V\). For each observation \(i\), the estimated propensity score is
\[
\hat{\pi}(V_i)
= \widehat{\Pr}\bigl(D_i = 1 \mid V_i\bigr)
= \frac{\exp\bigl(\hat{\gamma}_0 + V_i \hat{\gamma}_V\bigr)}
{1 + \exp\bigl(\hat{\gamma}_0 + V_i \hat{\gamma}_V\bigr)}\,,
\]
where \((\hat{\gamma}_0, \hat{\gamma}_V^{\prime})\) are coefficients from the logistic regression.\footnote{As discussed previously, we can clip the estimated propensity score to \([\alpha, 1-\alpha]\) with \(\alpha = 0.01\) to improve the stability of the estimator.}

For each observation \(i\), we then construct the IPW signals, which are the IPW-adjusted outcomes:
\[
\hat{\Lambda}_i^{(\mathrm{ipw})}
= \frac{D_i}{\hat{\pi}(V_i)}\,Y_i - \frac{1 - D_i}{1 - \hat{\pi}(V_i)}\,Y_i.
\]
This adjustment reweights the treated and control outcomes by \(\frac{1}{\hat{\pi}(V_i)}\) and \(-\frac{1}{1 - \hat{\pi}(V_i)}\), respectively:
\[
\hat{\Lambda}_i^{(\mathrm{ipw})}
= \frac{Y_i}{\hat{\pi}(V_i)}\,
\quad \text{if } D_i=1,
\qquad
\hat{\Lambda}_i^{(\mathrm{ipw})}
= -\frac{Y_i}{1 - \hat{\pi}(V_i)}\,
\quad \text{if } D_i=0.
\]
This reweighting creates a pseudo-population in which covariates \(V_i\) are (asymptotically) balanced between treated and control groups via the propensity score, ensuring that a simple difference in means in this pseudo-population can recover the treatment effect. Specifically, among treated units, those with smaller estimated propensity scores receive more weight, while among control units, those with higher estimated propensity scores receive more weight. This adjustment corrects for selection bias induced by \(V\) by aligning covariate distributions across treatment groups.

\subparagraph*{AIPW signals.}
The AIPW estimator combines elements of both outcome modeling and IPW. We first fit two linear regression models, \(\hat{\mu}^{1}(V_{i})\) and \(\hat{\mu}^{0}(V_{i})\), as well as the propensity-score model \(\hat{\pi}(V_{i})\). We then construct the signal as follows:
\[
\hat{\Lambda}_i^{(\mathrm{aipw})}
=
\hat{\mu}^{1}(V_i)
-
\hat{\mu}^{0}(V_i)
+
\frac{D_i}{\hat{\pi}(V_i)}\bigl(Y_i-\hat{\mu}^{1}(V_i)\bigr)
-
\frac{1-D_i}{1-\hat{\pi}(V_i)}\bigl(Y_i-\hat{\mu}^{0}(V_i)\bigr).
\]
Note that
\begin{align*}
\hat{\Lambda}_i^{(\mathrm{aipw})}
&=
\hat{\mu}^{1}(V_i)-\hat{\mu}^{0}(V_i)
+
\frac{1}{\hat{\pi}(V_i)}\bigl(Y_i-\hat{\mu}^{1}(V_i)\bigr)
&& \text{if } D_i=1, \\
\hat{\Lambda}_i^{(\mathrm{aipw})}
&=
\hat{\mu}^{1}(V_i)-\hat{\mu}^{0}(V_i)
-
\frac{1}{1 - \hat{\pi}(V_i)}\bigl(Y_i-\hat{\mu}^{0}(V_i)\bigr)
&& \text{if } D_i=0.
\end{align*}
For both treated and control units, the first term, \(\hat{\mu}^{1}(V_i) - \hat{\mu}^{0}(V_i)\), represents the predicted treatment effect for unit \(i\) based on the outcome model. The second term, \(\frac{1}{\hat{\pi}(V_i)}\bigl(Y_i-\hat{\mu}^{1}(V_i)\bigr)\) for treated units and \(\frac{1}{1 - \hat{\pi}(V_i)}\bigl(Y_i-\hat{\mu}^{0}(V_i)\bigr)\) for control units, adjusts for discrepancies between observed outcomes and their outcome modeling-based predictions by reweighting \emph{residuals from the outcome model}. Similar to the IPW estimator, treated units with smaller estimated propensity scores receive more weight, while control units with larger estimated propensity scores are weighted more heavily. However, instead of reweighting the raw outcome \(Y_i\), AIPW reweights the residuals \(\bigl(Y_i-\hat{\mu}^{D_{i}}(V_i)\bigr)\), which helps reduce sensitivity to misspecification in the outcome model.

Fixed effects can be included in outcome models to control for unobserved unit-invariant or time-invariant confounders; however, we do not recommend incorporating fixed effects into the propensity-score estimation. Since propensity-score models typically use nonlinear specifications (e.g., logit or probit), including a large number of fixed effects can lead to the incidental parameters problem \citep{neyman1948consistent}. This problem arises when the number of fixed-effect parameters increases with the number of observations, resulting in biased and inconsistent maximum likelihood estimates when the number of observations per group is small.

Next, we use a simulated example to illustrate the advantages of IPW and AIPW over a purely outcome-modeling approach. In this example, the CME is nonlinear, but the researcher misspecifies the outcome model as a linear interaction model. Fortunately, the researcher correctly specifies the propensity-score model. This scenario highlights the limitations of relying solely on outcome modeling, particularly when using rigid parametric models. Later, we introduce basis expansions to relax parametric assumptions and enhance estimation flexibility.

\begin{example}[A Simulated Sample with a Misspecified Outcome Model]
The sample consists of 1{,}000 observations, with a moderator \(X_i\) uniformly distributed on \([-2, 2]\) and two covariates \(Z_{i1}\) and \(Z_{i2}\) uniformly distributed on \([0, 1]\). Let \(V_i = [X_i, Z_{i1}, Z_{i2}]\) be the set of all confounders. Treatment is assigned with probability
\[
\Pr(D_i = 1|X_i,Z_i) = \frac{\exp(0.5 X_i + 0.5 Z_{i1})}{1+\exp(0.5 X_i + 0.5 Z_{i1})}.
\]
The outcome \(Y\) is generated via
\[
Y_i = 1 + X_i + D_i - X_i^2\,D_i + Z_{i1} + \epsilon_i,
\]
where \(\epsilon_i \sim \mathcal{N}(0,1)\). Unconfoundedness holds: \(D_i \indep (Y_i(0), Y_i(1)) \mid V_i\). The CME is nonlinear in \(x\), i.e., \(\theta(x) = 1 - x^2\). Covariate \(Z_{i2}\) is redundant since it does not directly affect \(D_i\) or \(Y_i\). However, the researcher correctly specifies the propensity-score model but misspecifies the outcome model as
\[
Y_i = 1 + X_i + D_i + X_iD_i + Z_{i1} + \epsilon_i.
\]
\end{example}

\begin{figure}[!th]
\caption{Constructed Signals}
\begin{subfigure}[b]{0.45\textwidth}
    \centering
    \includegraphics[width=\textwidth]{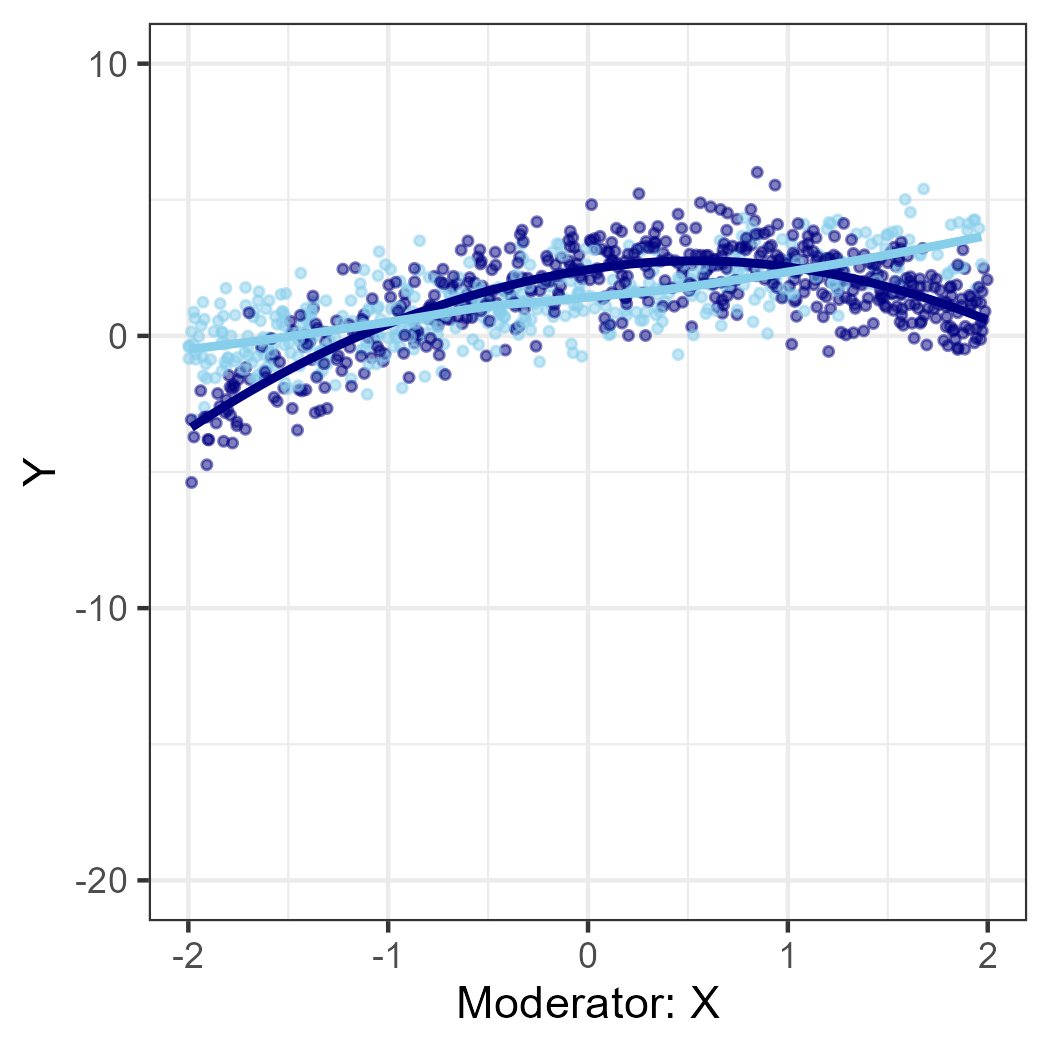}
    \caption{\hspace{0.5em} Raw Outcomes}
\end{subfigure}
\hspace{0.02\textwidth}
\begin{subfigure}[b]{0.45\textwidth}
    \centering
    \includegraphics[width=\textwidth]{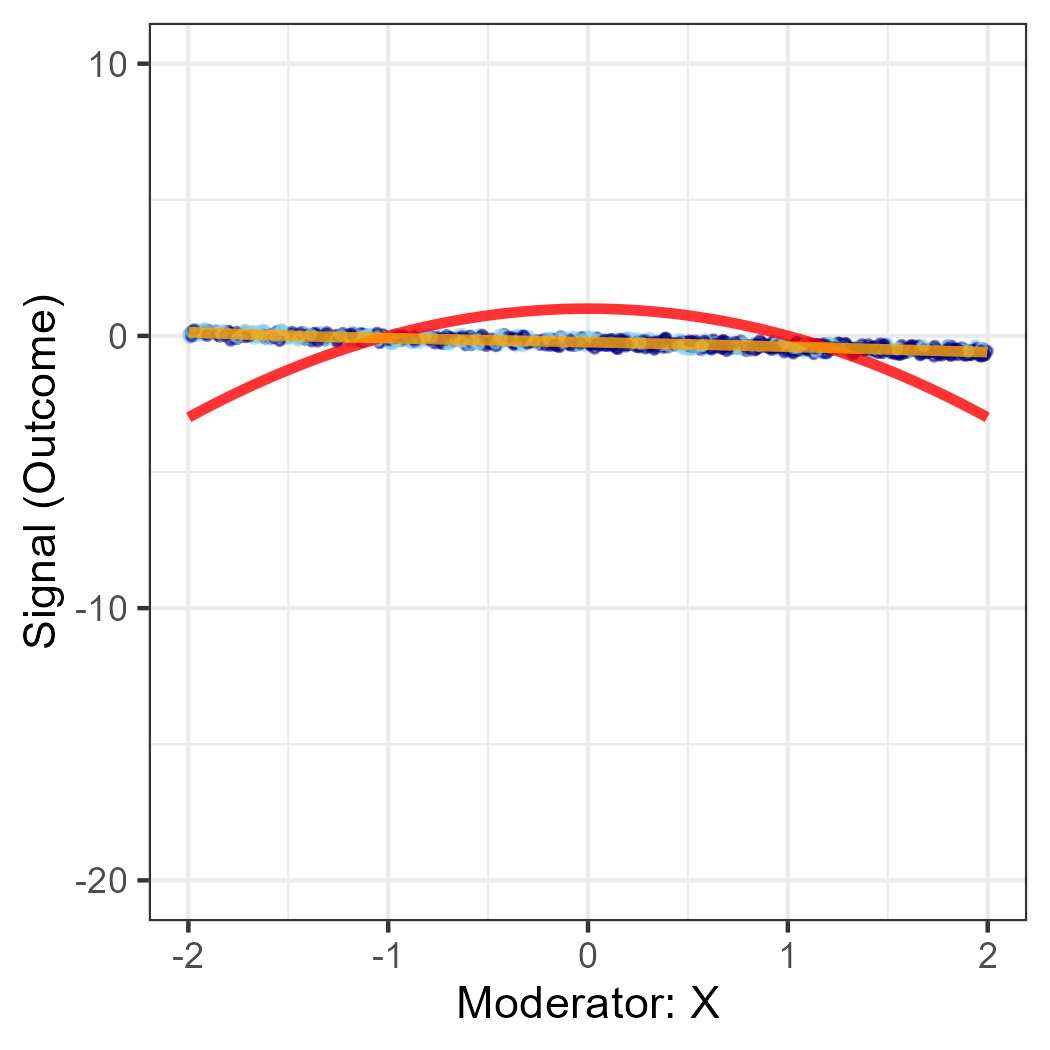}
    \caption{\hspace{0.5em} Outcome Model Signals \& \(\hat\theta(x)\)}
\end{subfigure}\\
\begin{subfigure}[b]{0.45\textwidth}
    \centering
    \includegraphics[width=\textwidth]{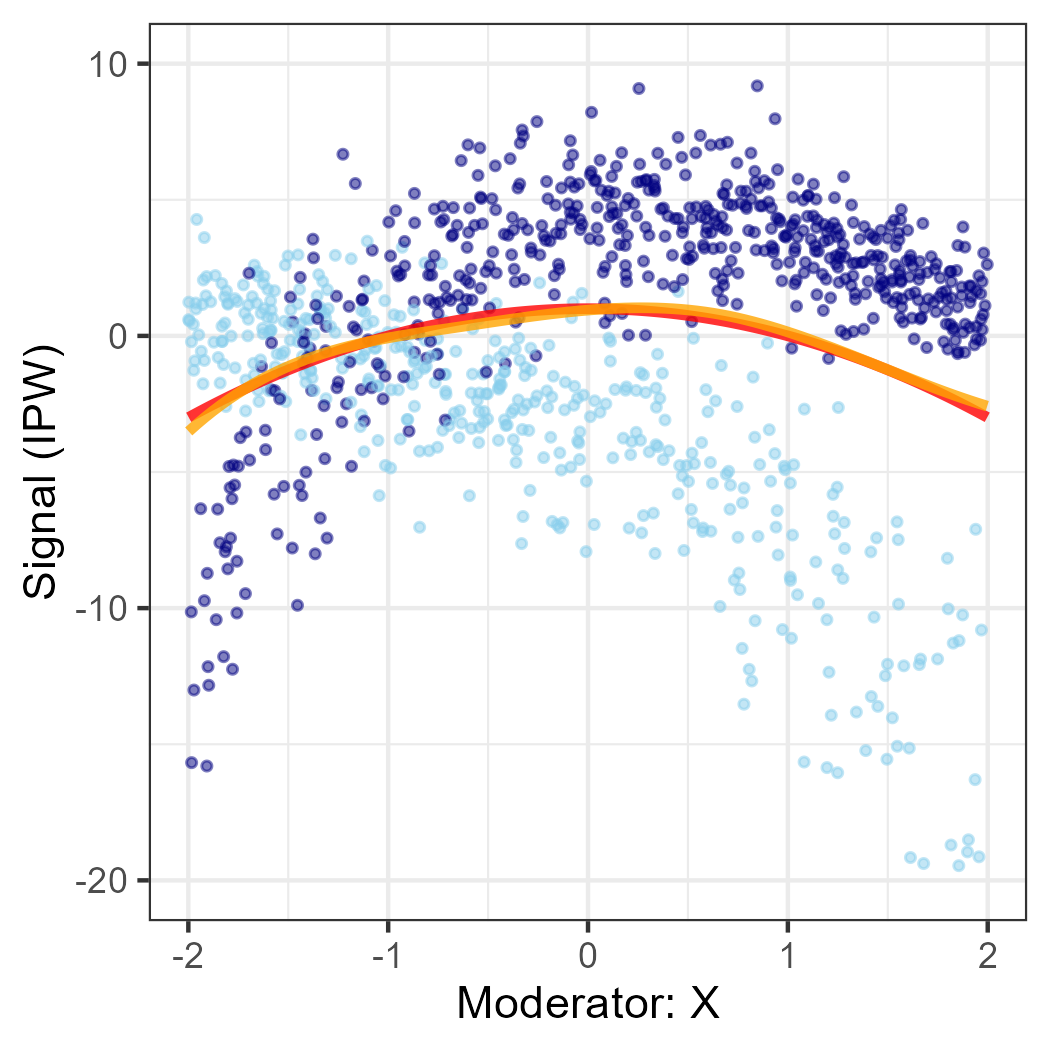}
    \caption{\hspace{0.5em} IPW Signals \& \(\hat\theta(x)\)}
\end{subfigure}
\hspace{0.02\textwidth}
\begin{subfigure}[b]{0.45\textwidth}
    \centering
    \includegraphics[width=\textwidth]{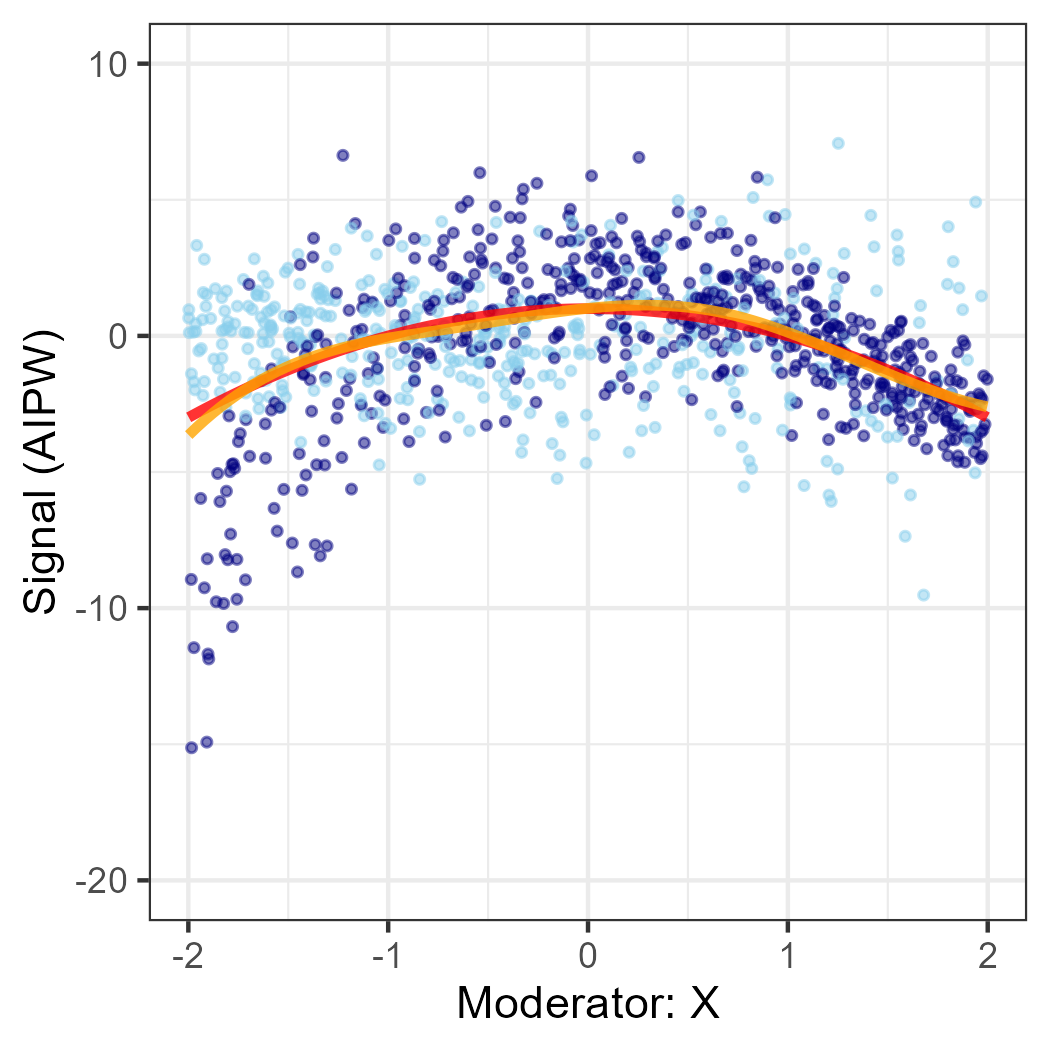}
    \caption{\hspace{0.5em} AIPW Signals \& \(\hat\theta(x)\)}
\end{subfigure}
\label{fig:AIPW_signal}\\
{\footnotesize \textit{Note}: In (a), the dark blue and light blue dots represent outcome values from the treatment and control groups, respectively; the dark blue and light blue lines represent the LOESS fits for the two groups. In (b)–(d), the dark blue and light blue dots represent signals of the observations in the treatment and control groups, respectively; the red line represents the true CME, while the orange line represents the CME estimates based on the signals after smoothing.}
\end{figure}

Figure~\ref{fig:AIPW_signal}(a) plots the raw outcome \(Y_i\) against the moderator \(X_i\), using dark blue for treated units (\(D_i=1\)) and light blue for control units (\(D_i=0\)). In the treated group, the relationship between \(X_i\) and \(Y_i\) is nonlinear.

Figure~\ref{fig:AIPW_signal}(b) plots the signals from the outcome model, \(\hat{\Lambda}_i^{(\mathrm{outcome})}\). Because the model is misspecified as linear, the signals—representing predicted treatment effects—are linear in \(X\). Here, we fit a B-spline regression to these signals to obtain a smooth estimate of the CME with respect to $X$ (the orange line). These misspecified signals lead to misleading CME estimates, which are also linear in \(X\) and differ from the true CME shown in red.

Figure~\ref{fig:AIPW_signal}(c) plots the signals from the IPW estimator, \(\hat{\Lambda}_i^{(\mathrm{ipw})}\). Compared to the corresponding raw outcomes, treated observations (dark blue) with small \(X\) values are relatively upweighted, as are control observations (light blue) with large \(X\) values. This occurs because \(D\) (and the estimated propensity score) is positively correlated with \(X\). Since the propensity-score model is correctly specified, this adjustment mitigates selection bias and helps reveal the true CME after smoothing.

In Figure~\ref{fig:AIPW_signal}(d), we plot the AIPW signals, \(\hat{\Lambda}_i^{(\mathrm{aipw})}\). They behave similarly to the IPW signals and successfully recover the true CME after smoothing even when the outcome model is misspecified, illustrating the doubly robust property. Notably, their spread is much smaller than that of IPW because the misspecified outcome model still retains some predictive power for the outcome. Moreover, they exhibit fewer extreme values—particularly for control units—than the IPW signals.

\paragraph{Step 3: Smoothing}

So far, we have glossed over the smoothing step, which reduces the signals to a one-dimensional CME along \(X\); we now formally discuss it. Once we obtain the signals \(\hat{\Lambda}_i\) from outcome modeling, IPW, or AIPW, the final step is to approximate the CME by fitting a flexible function \(f(X)\) that represents the conditional mean of these signals given \(X\). When the moderator \(X\) is discrete, this reduces to averaging the signals within each group defined by \(X\). For a continuous moderator, we consider two approaches for estimating \(f(X)\): one using kernel regression, which parallels the kernel estimator discussed in Section~2, and another using B\mbox{-}spline regression.

\subparagraph{Kernel Regression}

One can employ the kernel-weighted local polynomial smoother (LOESS). For each evaluation point \(x_0\), LOESS constructs a locally weighted regression of \(\hat{\Lambda}\) on \(X\) within a neighborhood around \(x_0\). Users can specify a \texttt{span} parameter, \(\alpha\), which determines the size of this local neighborhood as a fraction of the overall range of \(X\). Let
\[
W_{i}(x_{0};\alpha) \;=\; K \!\left(\frac{|X_i - x_0|}{\alpha\,\mathrm{range}(X)}\right),
\]
where \(K(\cdot)\) is a kernel weight function (often the tricube weight in standard LOESS). One then solves a weighted least squares problem:
\[
\bigl(\hat{\beta}_0(x_0), \hat{\beta}_1(x_0)\bigr)
\;=\;
\operatorname*{arg\,min}_{\beta_0,\,\beta_1}
\sum_{i=1}^n
\Bigl[
\hat{\Lambda}_i
-
\beta_0
-
\beta_1\bigl(X_i - x_0\bigr)
\Bigr]^2
\,W_{i}(x_{0};\alpha),
\]
where \(\hat{\Lambda}_i\) denotes the signals. The coefficient \(\hat{\beta}_0(x_0)\) represents the local intercept at \(x_0\), while \(\hat{\beta}_1(x_0)\) captures the local slope in the neighborhood of \(x_0\). Therefore, the fitted function at \(x_0\) is
\[
\hat{f}(x_0) \;=\; \hat{\beta}_0(x_0).
\]
This approach ensures that data points closer to \(x_0\) receive higher weight, while those farther away have less influence. If a higher-order polynomial fit is desired (e.g., quadratic), include additional terms such as \(\beta_2 (X_i - x_0)^2\), \(\beta_3 (X_i - x_0)^3\), etc., with the kernel weights \(W_{i}(x_0;\alpha)\) unchanged.

To select an optimal \(\alpha\), we employ \(K\)-fold cross-validation. For each candidate \(\alpha\) in a predefined grid, compute the mean squared error (MSE) over the folds:
\[
\operatorname{MSE}(\alpha)
\;=\;
\frac{1}{K}
\sum_{k=1}^{K}
\frac{1}{n_k}
\sum_{i \in \mathrm{Fold}_k}
\Bigl(\hat{\Lambda}_i - \hat{f}_{-k}(X_i;\alpha)\Bigr)^2,
\]
where \(\hat{f}_{-k}\) denotes the LOESS fit \emph{excluding} observations in the \(k\)-th fold, and \(n_k\) is the number of observations in that fold. We then choose
\[
\alpha^* \;=\; \operatorname*{arg\,min}_\alpha \operatorname{MSE}(\alpha),
\]
and refit LOESS on the entire dataset using \(\alpha^*\). Predicting over the evaluation grid yields the final estimate \(\hat{f}(x)\), which in turn provides the CME at each \(x\).

\subparagraph*{B-Spline Regression}
A B\mbox{-}spline is a piecewise polynomial function that ensures smoothness at the junctions of polynomial segments (the \emph{knots}). In our setting, for the signals \(\hat{\Lambda}_i\), we fit a linear model of the form
\[
\hat{\Lambda}_i
=
\beta_0
+
\sum_{j=1}^{K}\beta_j\,B_j(X_i)
+
\epsilon_i,
\]
where \(B_j(\cdot)\) is the \(j\)-th B\mbox{-}spline basis function of chosen degree (\texttt{spline\_degree}) and dimension (\texttt{spline\_df}). By selecting an adequate number of knots and a suitable polynomial degree,\footnote{In implementation, we set \texttt{spline\_degree} to 3 (cubic spline) and \texttt{spline\_df} to 6; users may choose \texttt{spline\_degree} and \texttt{spline\_df} based on other criteria such as AIC or BIC.} the B\mbox{-}spline expansions \(\{B_j(X)\}_{j=1}^K\) capture nonlinearities in the signal as a function of \(X\). After estimating \(\{\beta_j\}\) via least squares, we predict \(\hat{f}(x)\) over an evaluation grid to obtain \(\theta(x)\). Formally,
\[
\hat{f}(x) \;=\; \hat{\beta}_0 + \sum_{j=1}^{K}\hat{\beta}_j\,B_j(x).
\]

\begin{figure}[!th]
\caption{Kernel vs.\ B\mbox{-}Spline Regressions for Smoothing}
\begin{subfigure}[b]{0.47\textwidth}
    \centering
    \includegraphics[width=\textwidth]{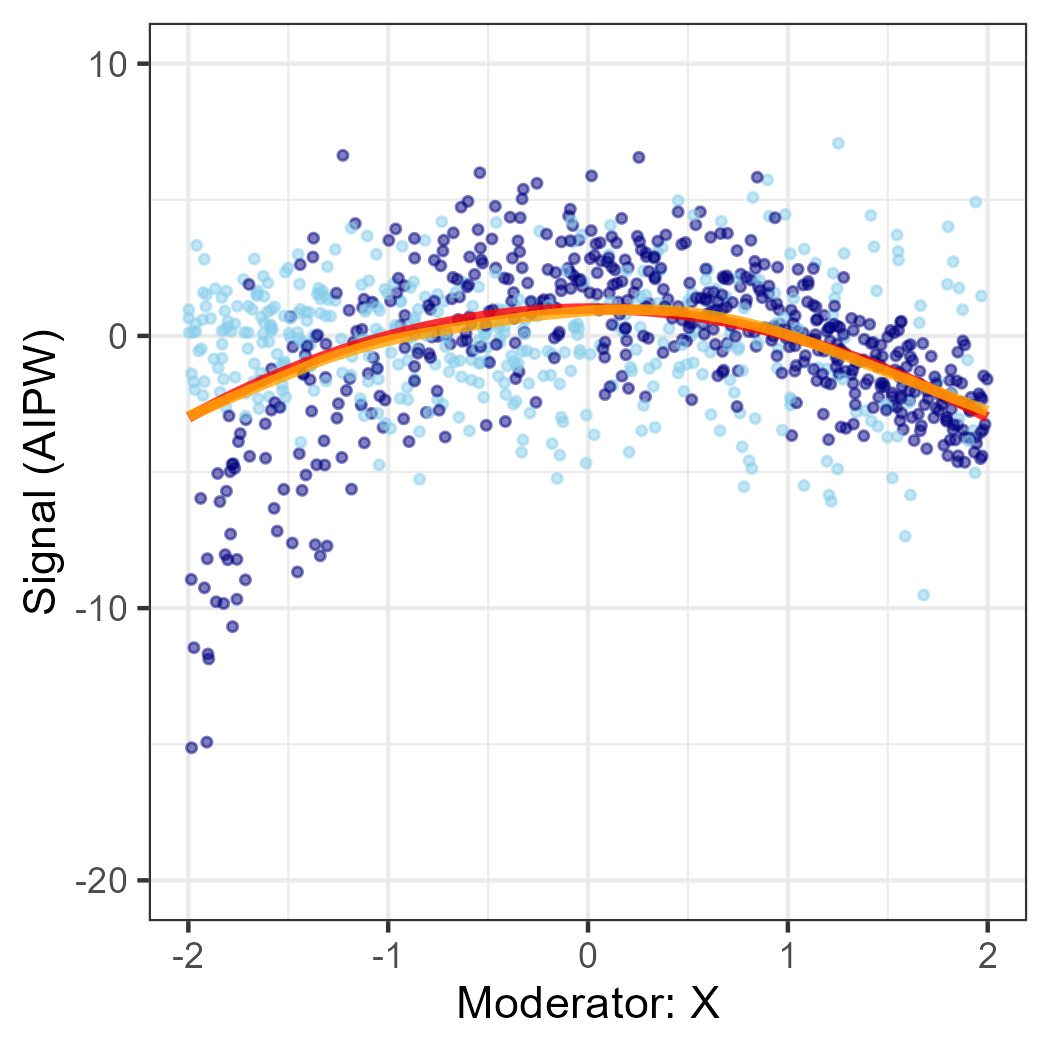}
    \caption{\hspace{1em} Kernel Regression}
\end{subfigure}
\hspace{0.02\textwidth}
\begin{subfigure}[b]{0.47\textwidth}
    \centering
    \includegraphics[width=\textwidth]{figures/chp3_example1_aipw.png}
    \caption{\hspace{1em} B\mbox{-}Spline Regression}
\end{subfigure}
\label{fig:AIPW_cme_construct}\\
{\footnotesize \textit{Note}: Dark blue and light blue dots represent AIPW signals for the treatment and control groups, respectively. The red line represents the true CME, while the orange line represents the estimated CME.}
\end{figure}

Both kernel and B\mbox{-}spline regression are flexible modeling strategies for capturing nonlinear relationships, but they differ in how smoothness is regulated and how tuning parameters (knot configuration vs.\ \(\alpha\)-span) are selected. B\mbox{-}spline regression employs a polynomial basis with continuity constraints at knot boundaries, whereas kernel/LOESS methods adaptively weight data within localized neighborhoods to achieve smooth fits. In either case, the estimated function \(\hat{f}(x)\) captures the shape of \(\hat{\Lambda}\) as a function of \(X\), delivering pointwise CME estimates across the range of the moderator. Figure~\ref{fig:AIPW_cme_construct} compares these two approaches (orange lines) in practice, showing that they generally produce similar fits when applied to \(\hat{\Lambda}_i^{(\mathrm{aipw})}\). In all subsequent CME estimations in this section, we continue to use B-spline regression to obtain the CME from the constructed signals due to its computational efficiency.

\FloatBarrier

\paragraph{Basis Expansion and Variable Selection}

To increase flexibility in the nuisance models, one can replace the raw covariates \(V=(X,Z)\) with a richer set of transformations \(\psi(V)\), such as low-degree polynomials or B\mbox{-}spline bases, and include selected interaction terms. This enlarges the space of outcome and propensity-score specifications, allowing \(\mathbb{E}[Y\mid D,X,Z]\) and \(\Pr(D=1\mid X,Z)\) to better approximate nonlinearities, threshold effects, and heterogeneous slopes that are poorly captured by linear terms. In practice, it is usually better to use flexible but parsimonious bases (e.g., cubic splines with a small number of knots placed at quantiles) rather than high-degree polynomials \citep{de1978practical, hastie2009elements}.

Allowing richer bases comes with a cost: the number of regressors can grow quickly, especially once interactions are added, which increases variance and makes conventional OLS or logit unstable. Regularization addresses this problem. We therefore pair basis expansion with Lasso-based variable selection in the outcome and propensity-score models, and often refit selected variables using unpenalized estimators (post-selection Lasso) to improve prediction \citep{belloni2012sparse, belloni2013least}. Simulations in the Online Appendix (Section B) illustrate the main pattern: basis expansions can substantially improve outcome and AIPW performance when the DGP is nonlinear, while IPW can remain sensitive to near-violations of overlap; adding Lasso selection stabilizes signals and improves CME estimation when many expanded terms are redundant. For completeness, the Online Appendix (Section B) also provides full implementation details, including the step-by-step algorithm.
\vspace{2em}

With this in place, we return to a unified workflow. Outcome modeling, IPW, and AIPW all follow the same three steps: (1) estimate nuisance models; (2) construct signals; and (3) smooth the signals over \(X\). In what follows, we refer to the AIPW estimator paired with basis expansion and post-Lasso selection as \emph{AIPW-Lasso} (and analogously \emph{outcome-Lasso} and \emph{IPW-Lasso}), and summarize the procedure in Algorithm~\ref{algo:aipw-lasso}.

\begin{algorithm}[!ht]
\footnotesize
\caption{AIPW-Lasso}\label{algo:aipw-lasso}
\begin{algorithmic}[1]
\State \textbf{Inputs:}
\State \quad Data: \( \{(X_i, Z_i, D_i, Y_i)\} \) for \(i=1,\dots,n\)
\State \quad\quad \(D_i \in \{0,1\}\) is a binary treatment
\State \quad\quad \(V_i := (X_i, Z_i)\) are covariates, of which \(X_i\) is the moderator
\State \quad\quad \(Y_i\) is the outcome
\State \quad Choice of basis functions \(\psi(\cdot)\) and smoothing method
\State \textbf{Outputs:}
\State \quad Estimated function \( \hat{\theta}(x) = \mathbb{E}[Y_i(1) - Y_i(0) \mid X_i = x]\)

\State \textbf{1) Estimate outcome model coefficients}
\State \quad * Using treated units \(\{i: D_i = 1\}\):
\State \quad \quad - Run Lasso regression with basis expansion \(\psi(V_i)\) to select active set \(\widehat{\mathcal{A}}_1\)
\State \quad \quad - Run OLS on selected covariates \(\widehat{\mathcal{A}}_1\) to obtain coefficients \(\hat{\beta}_1\)
\State \quad * Using control units \(\{i: D_i = 0\}\):
\State \quad \quad - Run Lasso regression with basis expansion \(\psi(V_i)\) to select active set \(\widehat{\mathcal{A}}_0\)
\State \quad \quad - Run OLS on selected covariates \(\widehat{\mathcal{A}}_0\) to obtain coefficients \(\hat{\beta}_0\)

\State \textbf{2) Estimate propensity-score coefficients}
\State \quad * Using all units:
\State \quad \quad - Run logit Lasso regression with basis expansion \(\psi(V_i)\) to select active set \(\widehat{\mathcal{A}}_p\)
\State \quad \quad - Run unpenalized logit regression on \(\widehat{\mathcal{A}}_p\) to obtain coefficients \(\hat{\gamma}\)

\State \textbf{3) Predict nuisance parameters for all observations}
\State \quad * For all \(i=1,\dots,n\), using the estimated coefficients and basis \(\psi(V_i)\):
\State \quad \quad \(\hat{\mu}^{1}(V_i) \leftarrow \text{predict using } \hat{\beta}_1\)
\State \quad \quad \(\hat{\mu}^{0}(V_i) \leftarrow \text{predict using } \hat{\beta}_0\)
\State \quad \quad \(\hat{\pi}(V_i) \leftarrow \text{predict using } \hat{\gamma}\)

\State \textbf{4) Compute AIPW signals}
\State \quad \( \hat{\Lambda}_i =
\hat{\mu}^{1}(V_i) - \hat{\mu}^{0}(V_i) + \frac{D_i}{\hat{\pi}(V_i)} (Y_i - \hat{\mu}^{1}(V_i)) - \frac{1 - D_i}{1 - \hat{\pi}(V_i)} (Y_i - \hat{\mu}^{0}(V_i)) \)

\State \textbf{5) Project AIPW signals onto \(X\) to obtain \( \hat{\theta}(x) \)}
\State \quad Regress \(\hat{\Lambda}_i\) on \(X_i\), using kernel or B\mbox{-}spline regression (with cross-validation for hyperparameter selections) \(\Rightarrow \hat{\theta}(x)\).

\State \textbf{Return:}
\State \quad The function \( \hat{\theta}(x) \) = estimated CME as a function of \(x\).
\end{algorithmic}
\end{algorithm}

\subsection{Inference}

Constructing confidence intervals analytically for the CME is challenging due to the multi-step nature of the estimation procedure. Our aim is to capture uncertainty arising from two main sources: (1) the estimation of the outcome and propensity-score models, which includes uncertainty from Lasso-based variable selection; and (2) the smoothing step that derives the CME from the resulting signals. We do not account for uncertainty from tuning or model-selection decisions—such as the choice of knots in the B-spline expansion, the selection of the Lasso penalty parameter \(\lambda\), or the choice of the span parameter in kernel regression.

To approximate the sampling distribution of the CME estimates \(\hat{\theta}(x)\), we implement a nonparametric bootstrap procedure that replicates the main estimation pipeline. In each bootstrap iteration, we resample the data with replacement and re-run the Lasso variable-selection step—using the same fixed penalty parameter as in the original estimation—to identify the active covariates. We then refit the outcome and propensity-score models using OLS and logistic regression on the selected covariates. The CME is subsequently estimated via kernel regression, using the span parameter fixed at its original cross-validated value. The B-spline expansion (including knot placement and polynomial degree) is also held fixed across bootstrap iterations.

Pointwise confidence intervals for \(\theta(x)\) at each evaluation point \(x\) can be constructed using either a percentile-based method or a normal approximation based on the empirical bootstrap variance. Similarly, uniform confidence intervals that ensure simultaneous coverage across the entire range of \(X\) can be constructed following the approaches outlined in previous sections.

\begin{figure}[!th]
 \caption{Estimated CME with Simulated Data: Kernel vs.\ Post-Lasso Methods}
 \label{fig:AIPW_inference}
 \begin{subfigure}[b]{0.47\textwidth}
     \centering
     \includegraphics[width=\textwidth]{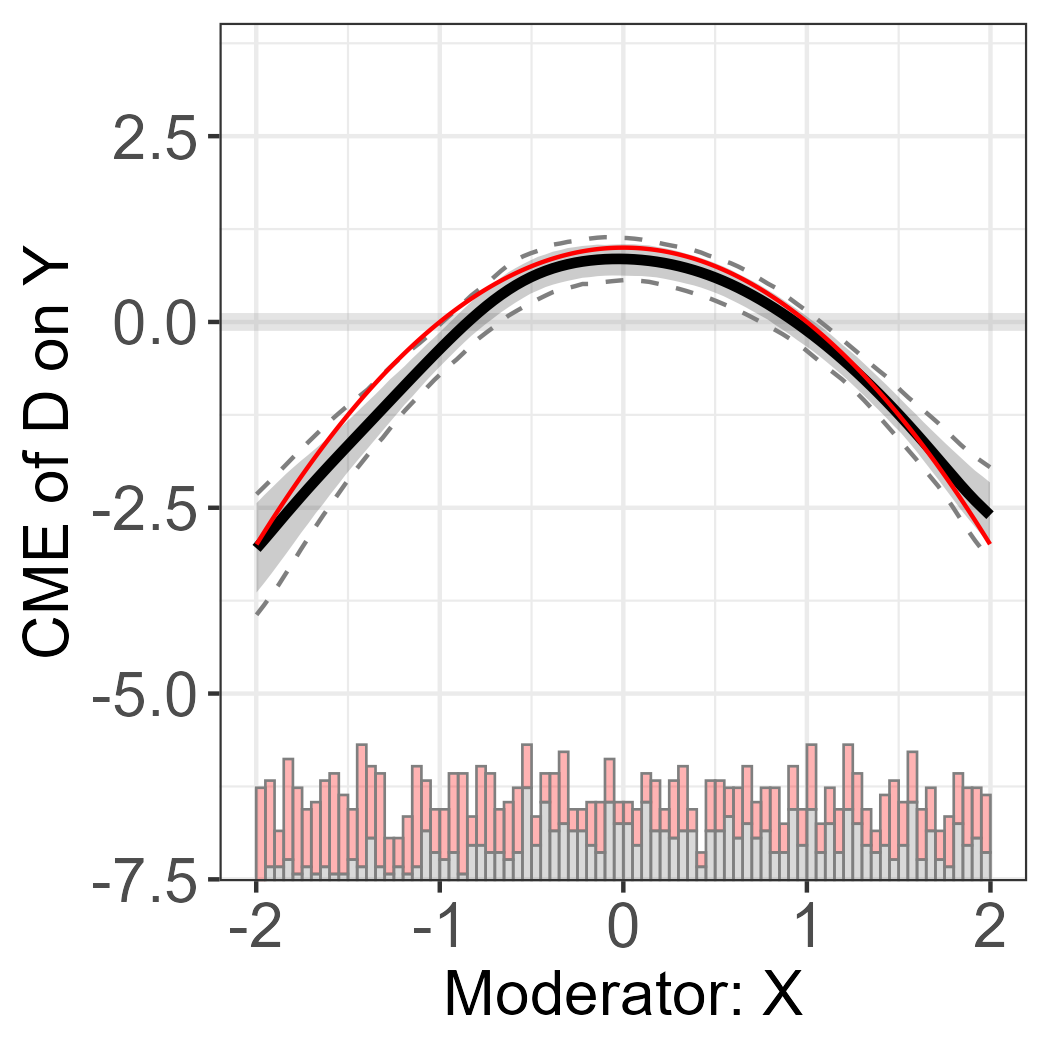}
     \caption{\hspace{1em} Kernel Estimator}
 \end{subfigure}
 \hspace{0.02\textwidth}
 \begin{subfigure}[b]{0.47\textwidth}
     \centering
     \includegraphics[width=\textwidth]{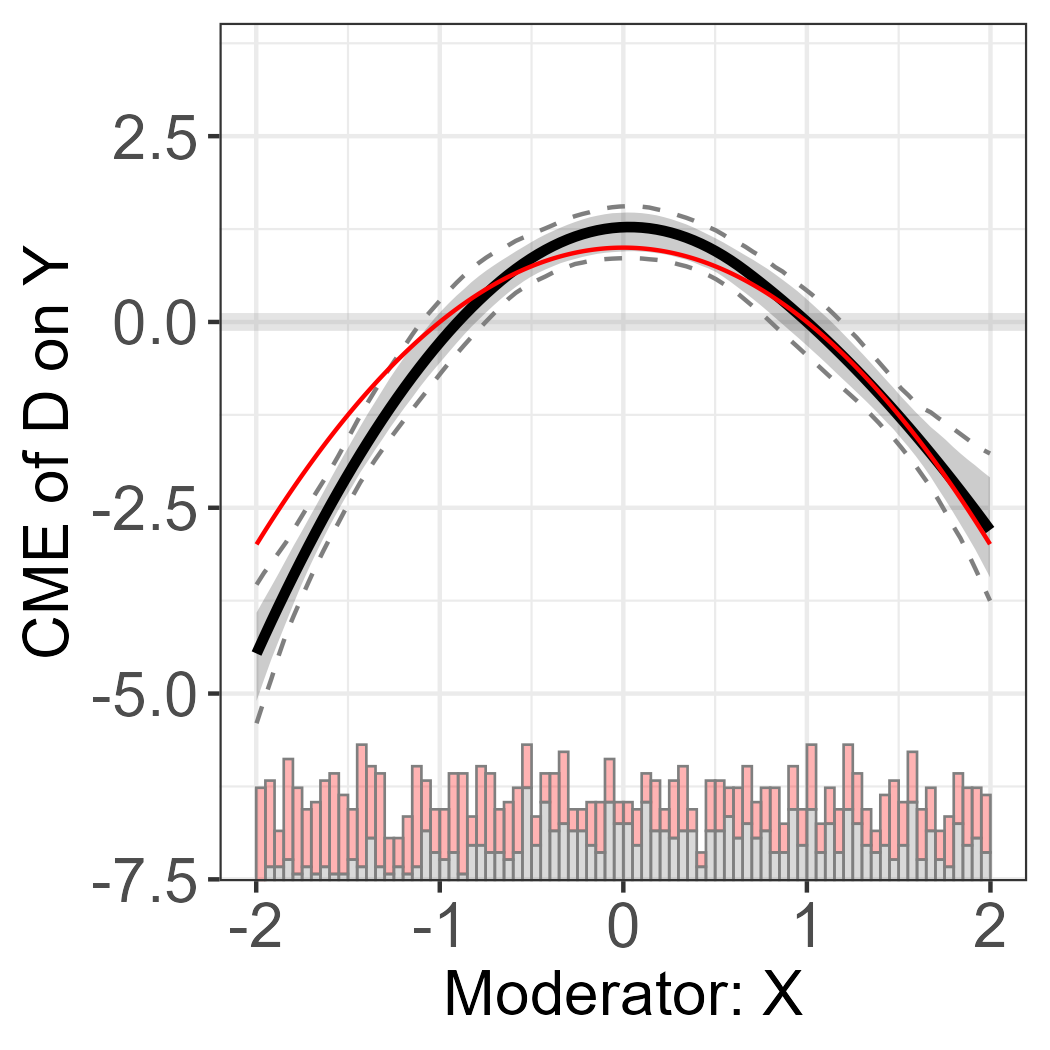}
     \caption{\hspace{1em} Outcome-Lasso}
 \end{subfigure}
 \\ 
 \begin{subfigure}[b]{0.47\textwidth}
     \centering
     \includegraphics[width=\textwidth]{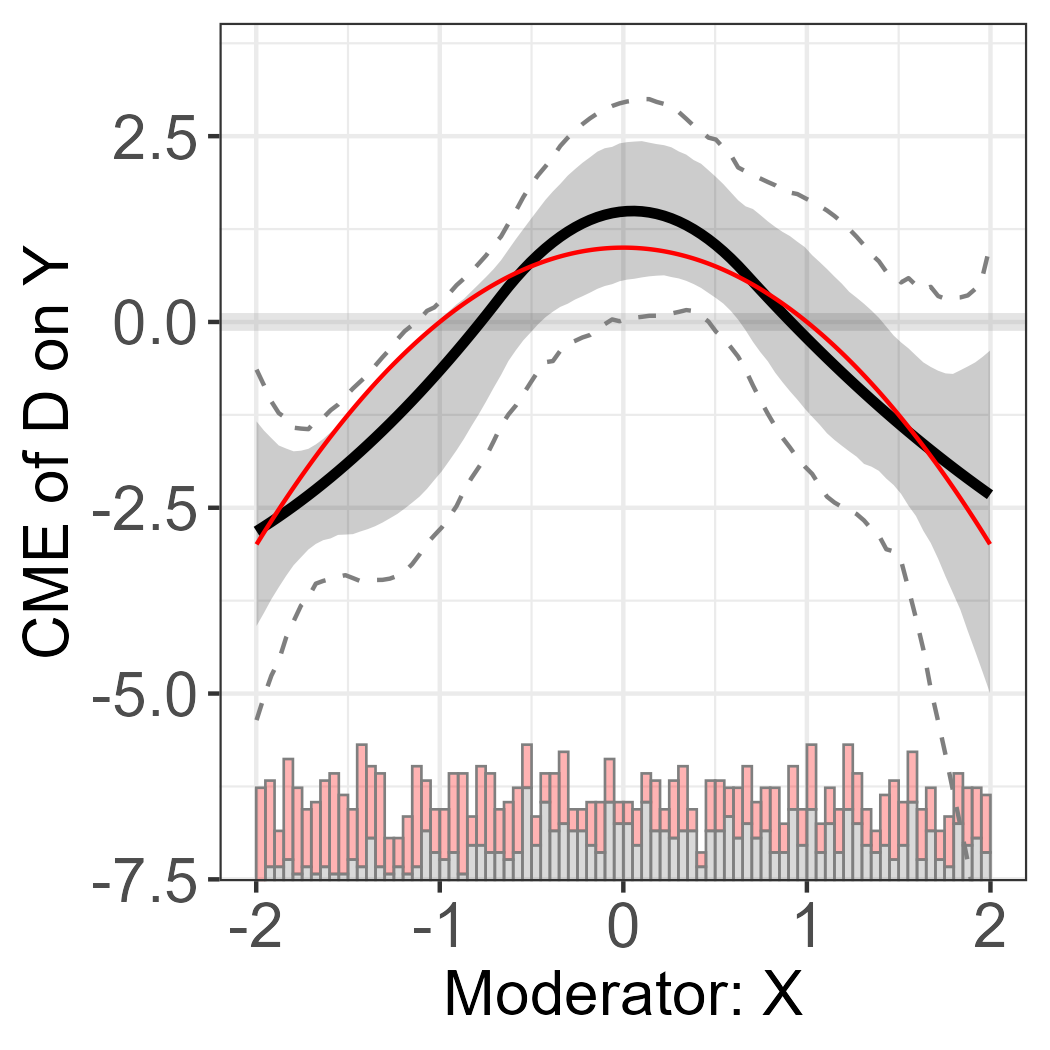}
     \caption{\hspace{1em} IPW-Lasso}
 \end{subfigure}
 \hspace{0.02\textwidth}
 \begin{subfigure}[b]{0.47\textwidth}
     \centering
     \includegraphics[width=\textwidth]{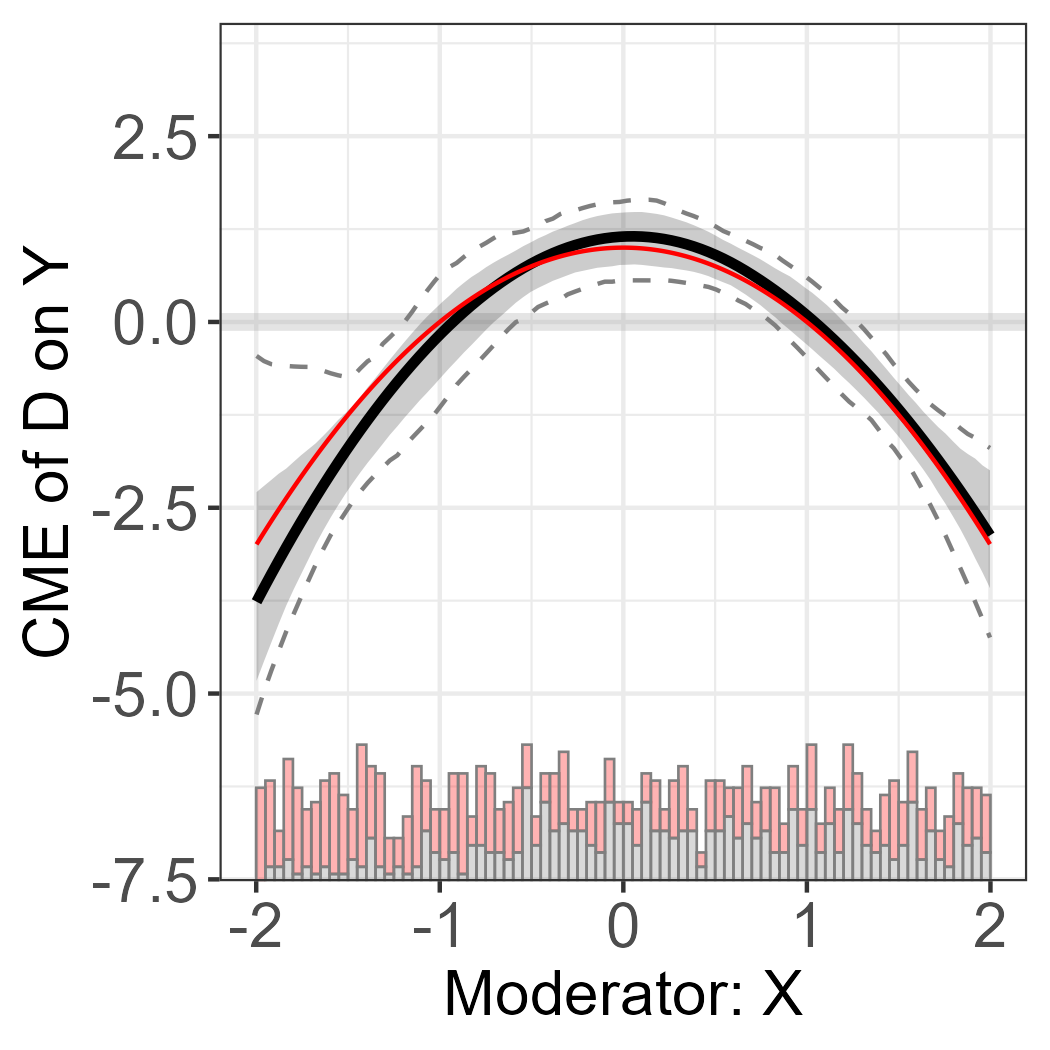}
     \caption{\hspace{1em} AIPW-Lasso}
 \end{subfigure} \vspace{-0.5em}
 \\
{\footnotesize \textit{Note}: In each figure, the red line represents the true CME; the black line represents the estimated CME. The shaded area and the dashed lines represent the 95\% pointwise and uniform confidence intervals, respectively. The histograms at the bottom of the figure depict the distributions of \(X\) across treatment (pink) and control (gray) groups.}
\end{figure}

In Figure~\ref{fig:AIPW_inference}, we display the estimated CME curves along with pointwise and uniform confidence intervals on the simulated data described in Section~B of the Online Appendix (``Many Covariates and Complex Nonlinear Relationships''), using the kernel estimator introduced in Section~2, the outcome-modeling approach (post-Lasso), the IPW-Lasso estimator, and the AIPW-Lasso estimator. The outcome-modeling approach yields an estimated CME that deviates slightly from the true CME when \(X\) is relatively small. The IPW-Lasso estimator produces noisier estimates and wider confidence intervals. The AIPW-Lasso estimator provides an estimated CME that more closely aligns with the truth than the other post-Lasso estimators. Interestingly, the kernel estimator performs fairly well, too.

\FloatBarrier
\bigskip

Finally, we apply four methods, including the kernel estimator, outcome-Lasso, IPW-Lasso, and AIPW-Lasso, to the data from \textbf{Example~\ref{ex:Noble2024}}. The last three methods all employ basis expansion and post-selection-Lasso regression. The code snippet below demonstrates how to implement all three procedures using the \texttt{interflex} package:

\begin{lstlisting}[language=R]
D <- "opp" 
Y <- "pres_ref" 
X <- "pres_vote_margin"
Z <- c("prev_vote","majority","leadership","seniority","tot_speech")
d <- d[which(d[,X]<=16.5 & d[,X]>=-4.1),]

## Use AIPW Signals
est.aipw<-interflex(estimator='lasso',signal = 'aipw',
                    data = d, nboots = 2000,
                    Y=Y,D=D,X=X, Z = Z,
                    treat.type = "discrete",
                    reduce.dimension = 'kernel',
                    na.rm = TRUE)

## Use IPW Signals
est.ipw<-interflex(estimator='lasso',signal = 'ipw',
                       data = d, nboots = 2000,
                       Y=Y,D=D,X=X, Z = Z,
                       treat.type = "discrete",
                       reduce.dimension = 'kernel',
                       na.rm = TRUE)

## Use Outcome Modeling Signals
est.outcome<-interflex(estimator='lasso',signal = 'outcome',
                       data = d, nboots = 2000,
                       Y=Y,D=D,X=X, Z = Z,
                       treat.type = "discrete",
                       reduce.dimension = 'kernel',
                       na.rm = TRUE)
\end{lstlisting}

\begin{figure}[!th]
\caption{Replicating \citet{noble2024presidential} Figure 2: Kernel vs Post-Lasso-Methods}\label{fig:noble_aipw}
\begin{subfigure}[b]{0.47\textwidth}
    \centering
    \includegraphics[width=\textwidth]{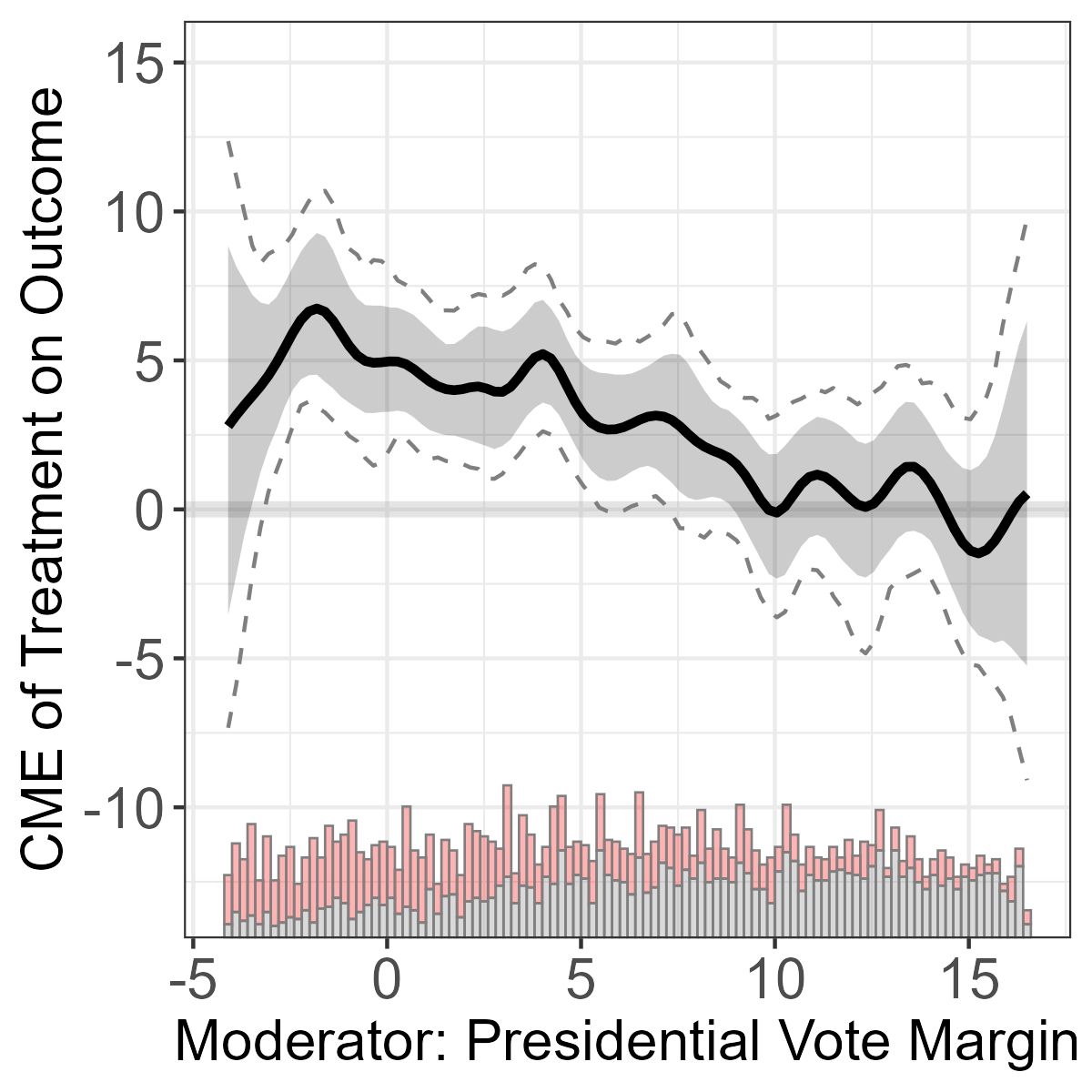}
    \caption{\hspace{1em} Kernel Estimator}
\end{subfigure}
\hspace{0.02\textwidth}  
\begin{subfigure}[b]{0.47\textwidth}
    \centering
    \includegraphics[width=\textwidth]{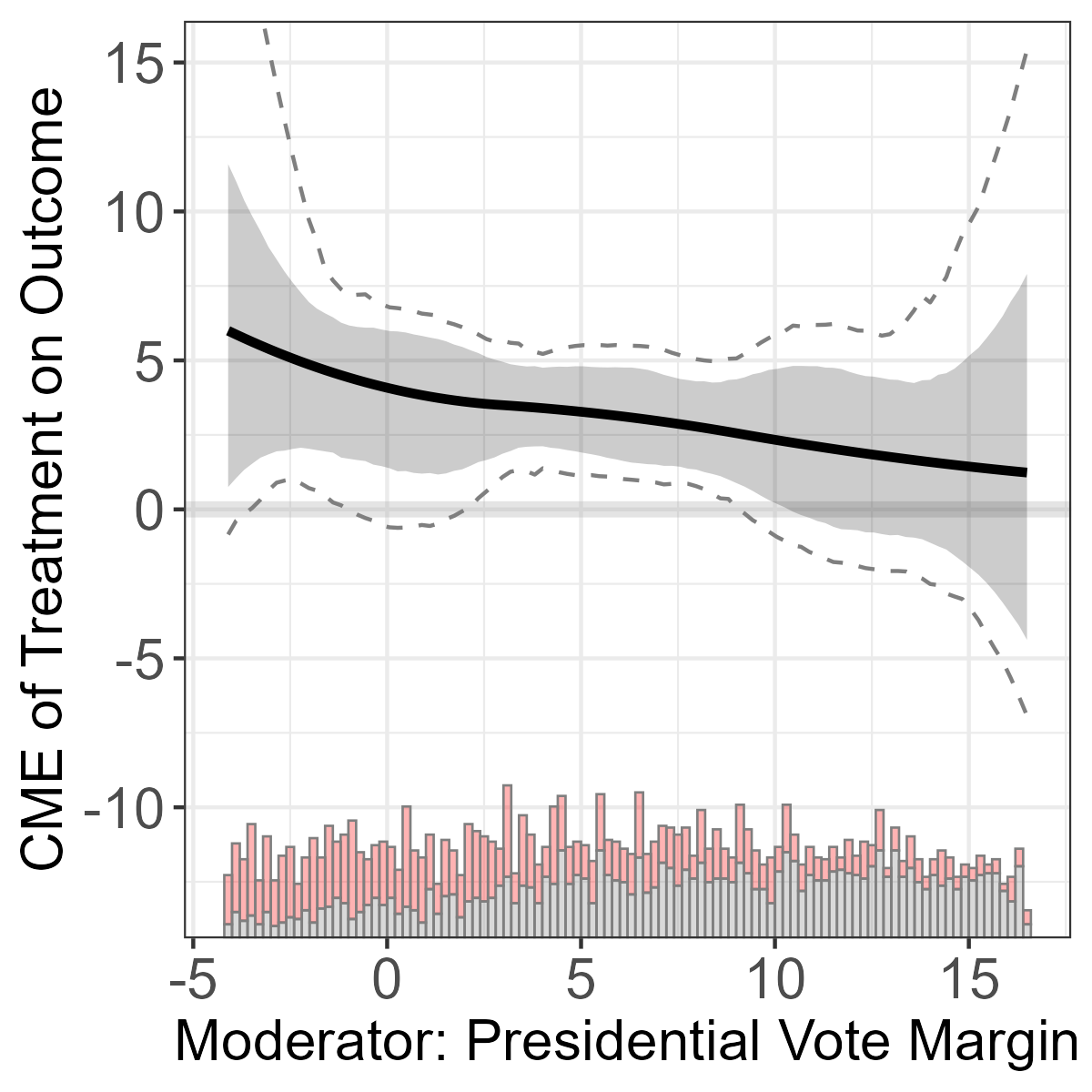}
    \caption{\hspace{1em} Outcome-Lasso}
\end{subfigure}\\ 
\begin{subfigure}[b]{0.47\textwidth}
    \centering
    \includegraphics[width=\textwidth]{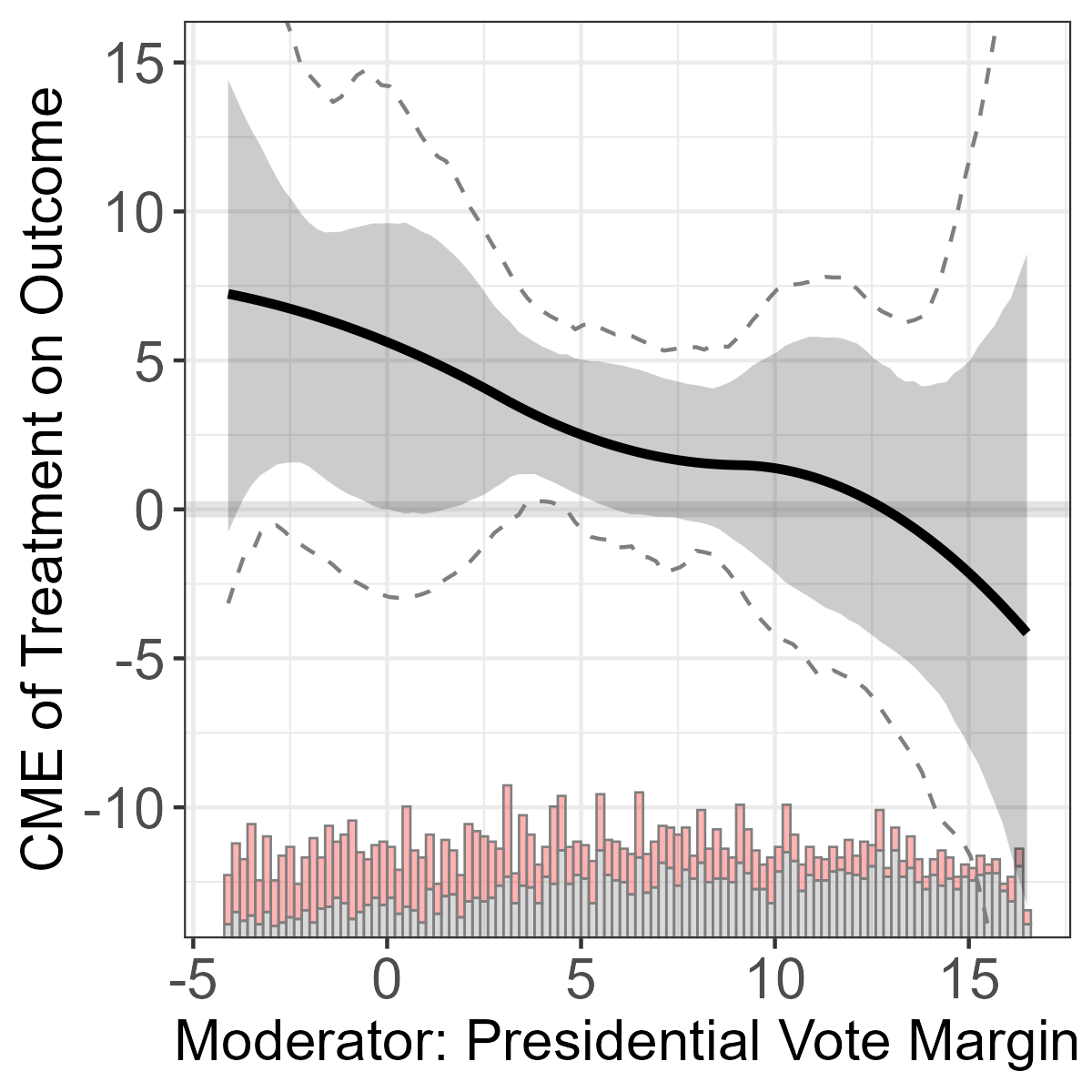}
    \caption{\hspace{1em} IPW-Lasso}
\end{subfigure}
\hspace{0.02\textwidth}  
\begin{subfigure}[b]{0.47\textwidth}
    \centering
    \includegraphics[width=\textwidth]{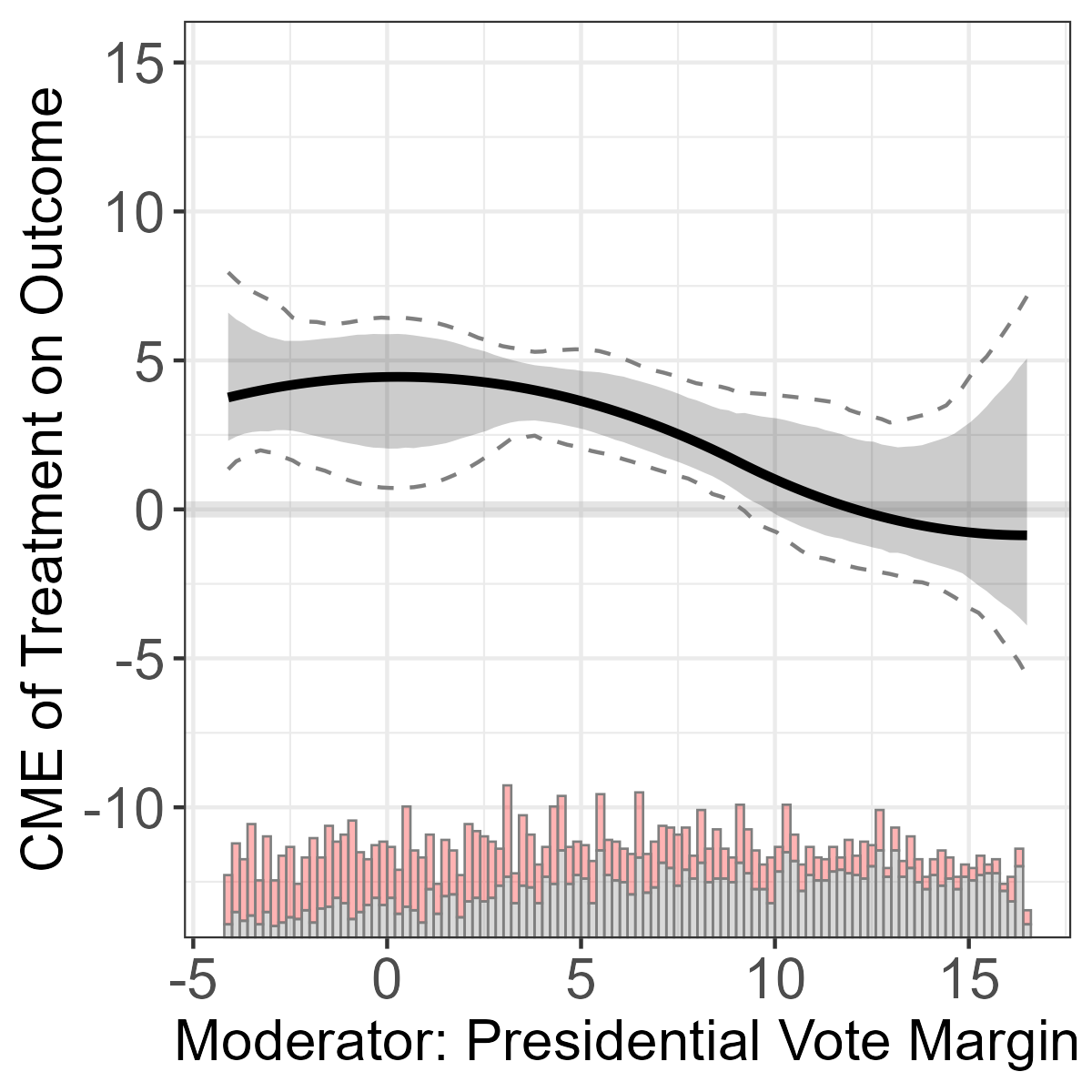}
    \caption{\hspace{1em} AIPW-Lasso}
\end{subfigure}
\\
{\footnotesize \textit{Note}: The treatment is out-partisanship of lawmakers, the outcome is frequency of presidential references in Congressional speeches. In each figure, the black line represents the estimated CME. The shaded area and the dashed lines represent the pointwise and uniform confidence intervals, respectively. The histograms at the bottom of the figure depict the distributions of $X$ across treatment (pink) and control (gray) groups.}
\end{figure}

In Figure~\ref{fig:noble_aipw}, we display the estimated CME along with their pointwise and uniform confidence intervals. The gray band represents the pointwise confidence intervals, while the dashed lines indicate the uniform confidence intervals that ensure simultaneous coverage across all values of \(x\). The uniform intervals are wider because they capture the uncertainty of the entire estimated curve rather than at individual points. Overall, the figure demonstrates that the main conclusion from \citet{noble2024presidential} remains robust across various estimators. The kernel estimator selects a large bandwidth via cross-validation, resulting in a nearly linear estimated CME. The outcome modeling approach also shows a decreasing CME, although the rate of decrease is smaller than that estimated by the kernel method. The IPW estimator exhibits a clear decreasing pattern but comes with noisy estimates and wide confidence intervals. Finally, the AIPW estimator produces narrower confidence intervals than the outcome modeling approach and generally shows a weakly decreasing trend; however, when the local vote margin is not lopsided (i.e., \(-5 \leq X \leq 5\)), the estimated CME does not appear strictly monotonic.

\subsection{Continuous Treatment}

We now consider a continuous treatment $D$ and focus on the conditional marginal effect
\[
\operatorname{CME}(x) \;=\; \E\!\left[\partial_d \mu(D,X,Z)\,\middle|\,X=x\right],
\]
where $\mu(d,x,z):=\E[Y(d)\mid X=x,Z=z]$ and
$\partial_d\mu(d,x,z):=\partial \mu(d,x,z)/\partial d$. For ease of identification, we adopt a partially linear regression model (PLRM), first introduced by \citet{robinson1988root}. This model assumes the potential outcome $Y(d)$ is a linear function of $d$, but the coefficient on $d$ can vary with $X$.

\begin{assumption}[Partially Linear Regression Model with $X$ as the Moderator]\label{assm:plrm}
Let $V\equiv(X,Z)$. There exist functions $\theta:\mathcal{X}\to\mathbb{R}$,
$g:\mathcal{V}\to\mathbb{R}$, and $m_D:\mathcal{V}\to\mathbb{R}$ such that
\begin{align*}
Y_i(d) &= \theta(X_i)\,d + g(V_i) + \zeta_i, \qquad \E[\zeta_i\mid V_i]=0,\\
D_i &= m_D(V_i) + \varepsilon_i, \qquad \E[\varepsilon_i\mid V_i]=0.
\end{align*}
\end{assumption}

Under this assumption, the observed outcome is $Y_i = Y_i(D_i) = \theta(X_i)D_i + g(V_i) + \zeta_i$. Under Assumption~\ref{assm:plrm}, taking conditional expectations yields
\[
\mu(d,x,z)=\E[Y(d)\mid X=x,Z=z]=\theta(x)d+g(x,z),
\]
so $\partial_d\mu(d,x,z)=\theta(x)$ and therefore $\operatorname{CME}(x)=\theta(x)$. Our estimation target is the function $\theta(x)$ itself, which captures the marginal effect of $D$ on $Y$, and the model assumes this effect varies \textit{only} with $X$.

To elucidate the identification strategy, we use the ``partialing out" approach. Note that:
\[
\mathbb{E}[Y_i \mid V_i] = \mathbb{E}[\theta(X_i)\,D_i \mid V_i] + \mathbb{E}[g(V_i)\mid V_i] + \mathbb{E}[\zeta_i\mid V_i].
\]
By our assumptions, $\mathbb{E}[g(V_i)\mid V_i] = g(V_i)$, $\mathbb{E}[\zeta_i\mid V_i] = 0$ (by Assumption \ref{assm:plrm}), and $\mathbb{E}[\theta(X_i)\,D_i \mid V_i] = \theta(X_i)\,\mathbb{E}[D_i \mid V_i]$ since $X_i$ is a component of $V_i$. We also have $\mathbb{E}[D_i\mid V_i] = m_D(V_i)$. Substituting these yields:
\[
\mathbb{E}[Y_i \mid V_i] = \theta(X_i)\,m_D(V_i) + g(V_i).
\]
Now, define the residualized outcome $\tilde{Y}_i \equiv Y_i - \mathbb{E}[Y_i\mid V_i]$ and residualized treatment $\tilde{D}_i \equiv D_i - \mathbb{E}[D_i\mid V_i]$. We have
\begin{align*}
\tilde{Y}_i &= \left(\theta(X_i)\,D_i + g(V_i) + \zeta_i\right) - \left(\theta(X_i)\,m_D(V_i) + g(V_i)\right) \\
&= \theta(X_i)\,(D_i - m_D(V_i)) + \zeta_i \\
&= \theta(X_i)\,\tilde{D}_i + \zeta_i.
\end{align*}
In this final equation, $\tilde{Y}_i$ and $\tilde{D}_i$ represent the residualized outcome and treatment after removing the effects of all covariates in $V_i$. \citet{chernozhukov2015post} gives a thorough discussion of this ``partialing out" procedure. Intuitively, we are only removing the confounding variation in $D$ attributable to $V = (X,Z)$; the CME is the slope of $Y$ with respect to the remaining, orthogonalized part of $D$, indexed by $X$.

In practice, we estimate $\mathbb{E}[Y\mid V_i]$ and $\mathbb{E}[D_i\mid V_i]$ using flexible methods—such as basis expansions combined with post-selection Lasso—to “denoise” $Y_i$ and $D_i$, yielding $\hat{\tilde{Y}}_i$ and $\hat{\tilde{D}}_i$. We then estimate the CME by a varying-coefficient regression of $\hat{\tilde{Y}}_i$ on $\hat{\tilde{D}}_i$ with coefficient $\theta\left(X_i\right)$, relying on the key property that $\zeta_i$ remains orthogonal to $\tilde{D}_i$ (i.e., $\mathbb{E}[\zeta_i\mid \tilde{D}_i] = 0$).

For implementation, one approach is to fit the following B-spline regression:
\[
\hat{\theta} \in \arg\min_{\theta\in\Theta_{\mathrm{bs}}}
\frac{1}{n}\sum_{i=1}^n\bigl(\hat{\tilde{Y}}_i-\theta(X_i)\,\hat{\tilde{D}}_i\bigr)^2.
\]
Under the sieve representation, $\theta(x)\approx \beta^T \phi(x)$, where $\phi(x)$ is a set of B-spline basis functions.

Alternatively, one can use a kernel-based local linear estimator introduced in Section 2. From our derivation, the population model for the residuals is:
\[
\tilde{Y}_i = \theta(X_i)\tilde{D}_i + \zeta_i,\quad \mathbb{E}[\zeta_i \mid X_i, \tilde{D}_i] = 0.
\]
To estimate $\theta(X=x_0)$, we can use a local linear approximation for $\theta(x)$:
\[
\theta(X_i) \approx \alpha(x_0) + \beta(x_0)(X_i - x_0) \quad \text{for } X_i \text{ near } x_0.
\]
The local minimization problem is then:
\[
\left(\hat{\alpha}(x_0), \hat{\beta}(x_0)\right) =
\arg\min_{\tilde{\alpha},\,\tilde{\beta}}\,
\sum_{i=1}^n \left[\hat{\tilde{Y}}_i - \left(\tilde{\alpha} + \tilde{\beta}(X_i - x_0)\right)\hat{\tilde{D}}_i\right]^2 K\left(\frac{X_i - x_0}{h(x_0)}\right).
\]
The estimate of the CME at $x_0$ is $\hat{\theta}(x_0) = \hat{\alpha}(x_0)$. Researchers can select the bandwidth function $h(x_{0})$ via cross-validation. We summarize this partial-out Lasso (PO-Lasso) procedure in Algorithm~\ref{algo:po-lasso}, which is modified based on \citet{chernozhukov2015post}.

\begin{algorithm}[!ht]
\footnotesize
\caption{PO-Lasso for CME}\label{algo:po-lasso}
\begin{algorithmic}[1]
\State \textbf{Inputs:}
\State \quad Data: \( \{(X_i, Z_i, D_i, Y_i)\} \) for \(i=1,...,n\)
\State \quad\quad \(D_i\) is a continuous treatment
\State \quad\quad \(V_i := (X_i, Z_i)\) are covariates, in which \(X_i\) is the moderator
\State \quad\quad \(Y_i\) is the outcome
\State \quad Choice of basis functions (e.g., B-splines) $\psi(V_i)$
\State \textbf{Outputs:}
\State \quad Estimated function \( \hat{\theta}(x) = \widehat{\operatorname{CME}}(x)\)

\State \textbf{1) Basis Expansion}
\State \quad * Conduct basis expansion of $V_i$, obtain $\psi(V_i)$.

\State \textbf{2) Partial out covariates from the outcome (Post-Lasso)}
\State \quad * Run a Lasso regression of $Y_i$ on $\psi(V_i)$ to select the active set of basis functions, $\widehat{\mathcal{A}}_y$.
\State \quad * Run an OLS regression of $Y_i$ on the selected basis functions indexed by $\widehat{\mathcal{A}}_y$.
\State \quad * Compute fitted values $\hat{m}_Y(V_i) = \hat{\mathbb{E}}[Y_i \mid V_i]$ from this OLS regression.
\State \quad * Compute residuals $\hat{\tilde{Y}}_i = Y_i - \hat{m}_Y(V_i)$ for all $i$.

\State \textbf{3) Partial out covariates from the treatment (Post-Lasso)}
\State \quad * Run a Lasso regression of $D_i$ on $\psi(V_i)$ to select the active set of basis functions, $\widehat{\mathcal{A}}_d$.
\State \quad * Run an OLS regression of $D_i$ on the selected basis functions indexed by $\widehat{\mathcal{A}}_d$.
\State \quad * Compute fitted values $\hat{m}_D(V_i) = \hat{\mathbb{E}}[D_i \mid V_i]$ from this OLS regression.
\State \quad * Compute residuals $\hat{\tilde{D}}_i = D_i - \hat{m}_D(V_i)$ for all $i$.

\State \textbf{4) Estimate $\theta(x)$ using residuals}
\State \quad * Estimate the varying-coefficient model $\hat{\tilde{Y}}_i = \theta(X_i)\hat{\tilde{D}}_i + \zeta_i$ using the estimated residuals $(\hat{\tilde{Y}}_i, \hat{\tilde{D}}_i)$ and the moderator $X_i$.
\State \quad * This estimation can use B-splines (interacting $\hat{\tilde{D}}_i$ with basis of $X_i$) or local linear regression (regressing $\hat{\tilde{Y}}_i$ on $\hat{\tilde{D}}_i$ with kernel weights).

\State \textbf{Return:}
\State \quad The function \( \hat{\theta}(x) \) = estimated CME as a function of \(x\).
\end{algorithmic}
\end{algorithm}

We provide a simulated example below.

\begin{example}[Complex DGP with a Continuous Treatment]\label{ex:sim.cont}
We simulate a sample of \(1000\) observations. The moderator \(X_i \sim \mathrm{Unif}[-2,2]\). Four additional covariates \(\{Z_{i1},Z_{i2},Z_{i3},Z_{i4}\}\) are sampled from $\mathcal{N}(0,1)$. The continuous treatment $D_i$ is generated as:
\[
D_i = \underbrace{0.5 Z_{i1} + X_i^2}_{m_D(V_i)} + \varepsilon_i.
\]
where $\varepsilon_i\sim \mathcal{N}(0,1)$. The outcome $Y_i$ is generated as:
\[
Y_i = 1 + 1.5 X_i + D_i - D_i X_i^2 + 2 X_i \exp(1 + Z_{i1}) + 2\,\mathbf{1}\{Z_{i2} > 0\} Z_2 + \zeta_i.
\]
with $\zeta_i\sim \mathcal{N}(0,1)$. We can rewrite the outcome equation to match the PLRM form $Y_i = \theta(X_i)D_i + g(V_i) + \zeta_i$:
\[
Y_i = \underbrace{(1 - X_i^2)}_{\theta(X_i)} D_i + \underbrace{\left(1 + 1.5 X_i + 2 X_i \exp(1 + Z_{i1}) + 2\,\mathbf{1}\{Z_{i2} > 0\} Z_2\right)}_{g(V_i)} + \zeta_i.
\]
The true CME is therefore $\theta(x) = 1 - x^2$. The challenge comes from $g(V_i)$, a highly complex, non-linear function of $X$ and $Z$ that acts as a confounder. 
\end{example}

\begin{figure}[!th]
\caption{CME Estimates for Example 4: Continuous Treatment}
\begin{subfigure}[b]{0.47\textwidth}
    \centering
    \includegraphics[width=\textwidth]{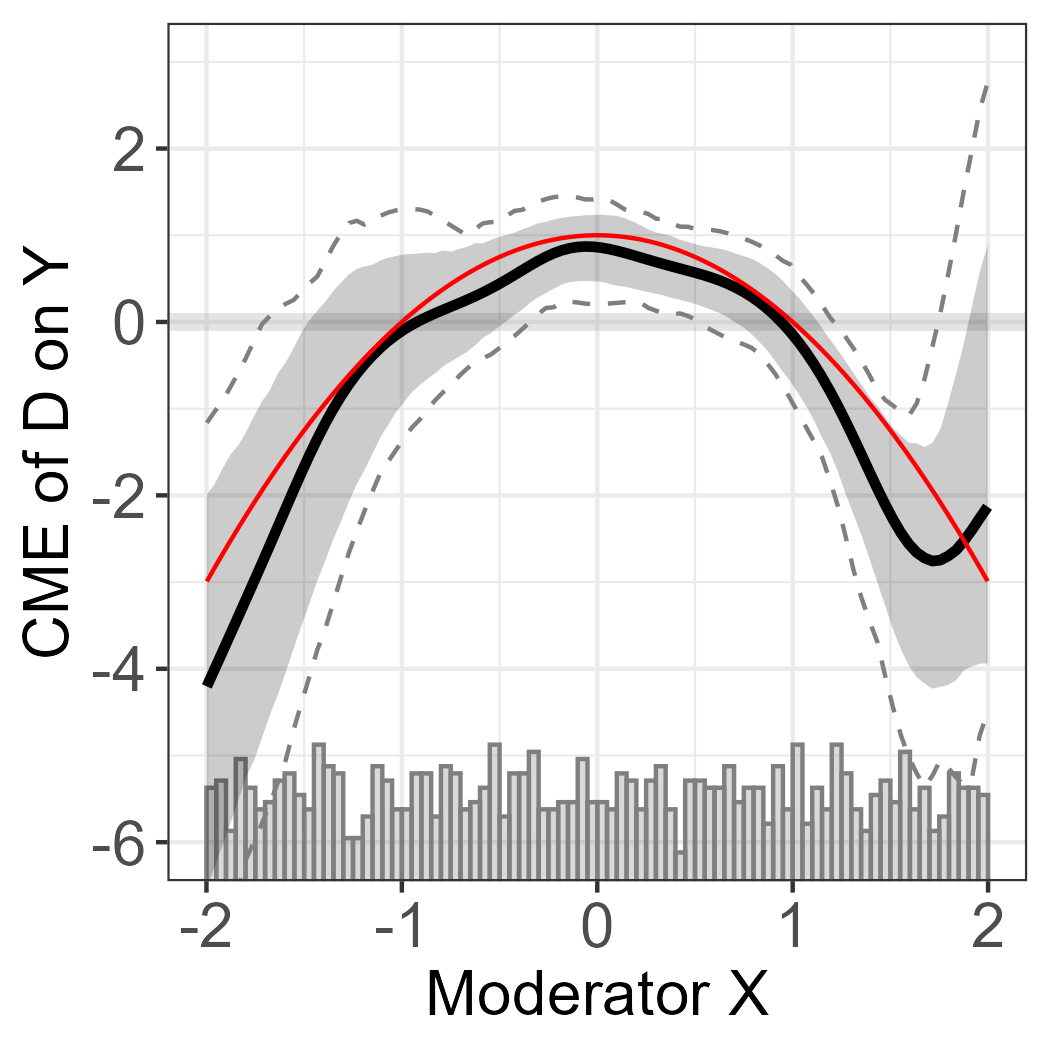}
    \caption{\hspace{0.5em} Kernel Estimator}
\end{subfigure}
\hspace{0.02\textwidth}  
\begin{subfigure}[b]{0.47\textwidth}
    \centering
    \includegraphics[width=\textwidth]{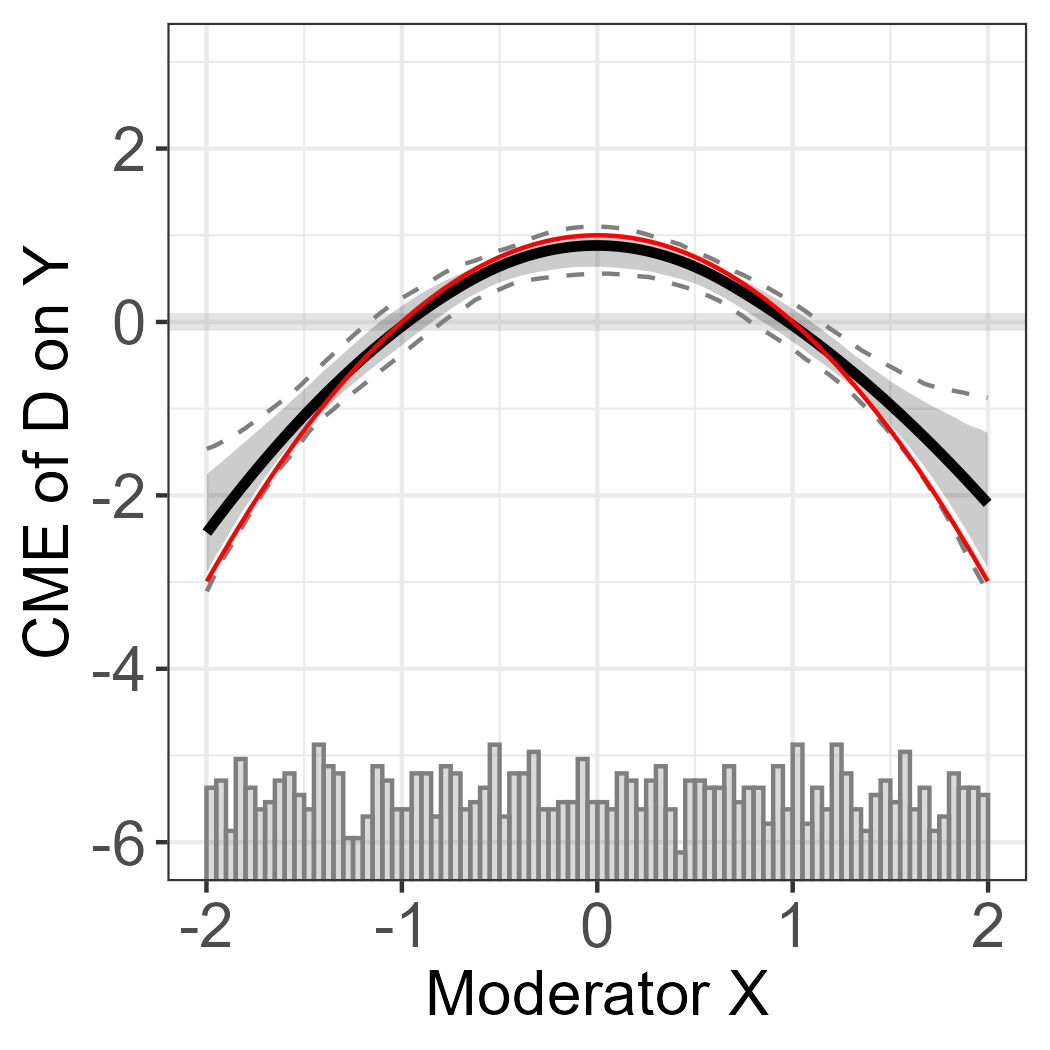}
    \caption{\hspace{0.5em} PO-Lasso}
\end{subfigure}
\label{fig:denoise_kernel}\\
{\footnotesize \textit{Note}: In each figure, the red line represents the true CME; the black line represents the estimated CME. The shaded area and the dashed lines represent the 95\% pointwise and uniform confidence intervals, respectively. The histograms at the bottom of the figure depict the distributions of $X$.}
\end{figure}

In Figure~\ref{fig:denoise_kernel}, we compare two estimation strategies using data from Example~\ref{ex:sim.cont}. In panel (a), we apply the kernel estimator discussed in Section 2. Because it cannot effectively accommodate the strong nonlinearity in $Z_i$ (i.e., the $g(V_i)$ term), the resulting black curve diverges from the true effect (red), especially in the tails of $X_i$. In contrast, panel (b) shows the result from PO-Lasso using kernel regression for smoothing. Here, the black curve closely aligns with the true effect (red) across most of the domain. This illustrates how partialing out can correct for distortions caused by nonlinearities and covariate interactions in the DGP, yielding a substantially more accurate estimate of the CME.

Finally, we illustrate PO-Lasso using an empirical example from political science. 

\begin{example}
\label{ex:adiguzel2023}
\cite{adiguzel2023out} examines the impact of the incumbent's public health provision on electoral outcomes. The authors use a reform in Turkey that significantly altered the geographic distribution of health clinics. Here, we replicate Figure 4A in the study. The authors find that reduced congestion levels significantly increased the AKP's vote share, especially in poorer communities which exhibited a more substantial response to the improvements in health care access. The variables of interest are:
\begin{itemize}
\item Outcome: Change in the vote share for the incumbent AKP party (continuous, $\in [-33.9, 39.5]$) 
\item Treatment: Change in congestion (continuous, $\in [-28.8, 20.8]$); congestion is measured by the number of patients per doctor at the nearest clinic, reducing congestion indicating improved service quality
\item Moderator: Logarithm of median property value (continuous, $\in[-0.2, 7.7]$)
\end{itemize}
\end{example}

We apply the kernel estimator from Section 2 and the PO-Lasso estimator. We use the B-spline regression in the last step of estimating $\theta(x)$ as it is less time consuming. The code snippet below demonstrates how to implement the PO-Lasso estimator using the \texttt{interflex} package. We trim the data to exclude moderator values where the distribution is sparse.

\begin{lstlisting}[language=R]
D <- "Dpop1000_asm_dr_3"
Y <- "Dakp_3"
X <- "lograyic09"
Z <- c('Duniversity_3', 'Dodr_3', 'Dydr_3', 'Dpopulation_3', 'Dhospital_pri_3', 'Dhospital_pub_3')

df<-df %>% dplyr::filter(lograyic09 <=7) %>%
  dplyr::filter(lograyic09 >=3)

est.lasso<-interflex(estimator = 'lasso', data = df, nboots = 2000,
                    Y = Y, D = D, X = X, Z = Z, na.rm = TRUE)

\end{lstlisting}

\begin{figure}[!th]
\caption{Replicating \citet{adiguzel2023out} Figure 4A}
\begin{subfigure}[b]{0.45\textwidth}
    \centering
    \includegraphics[width=\textwidth]{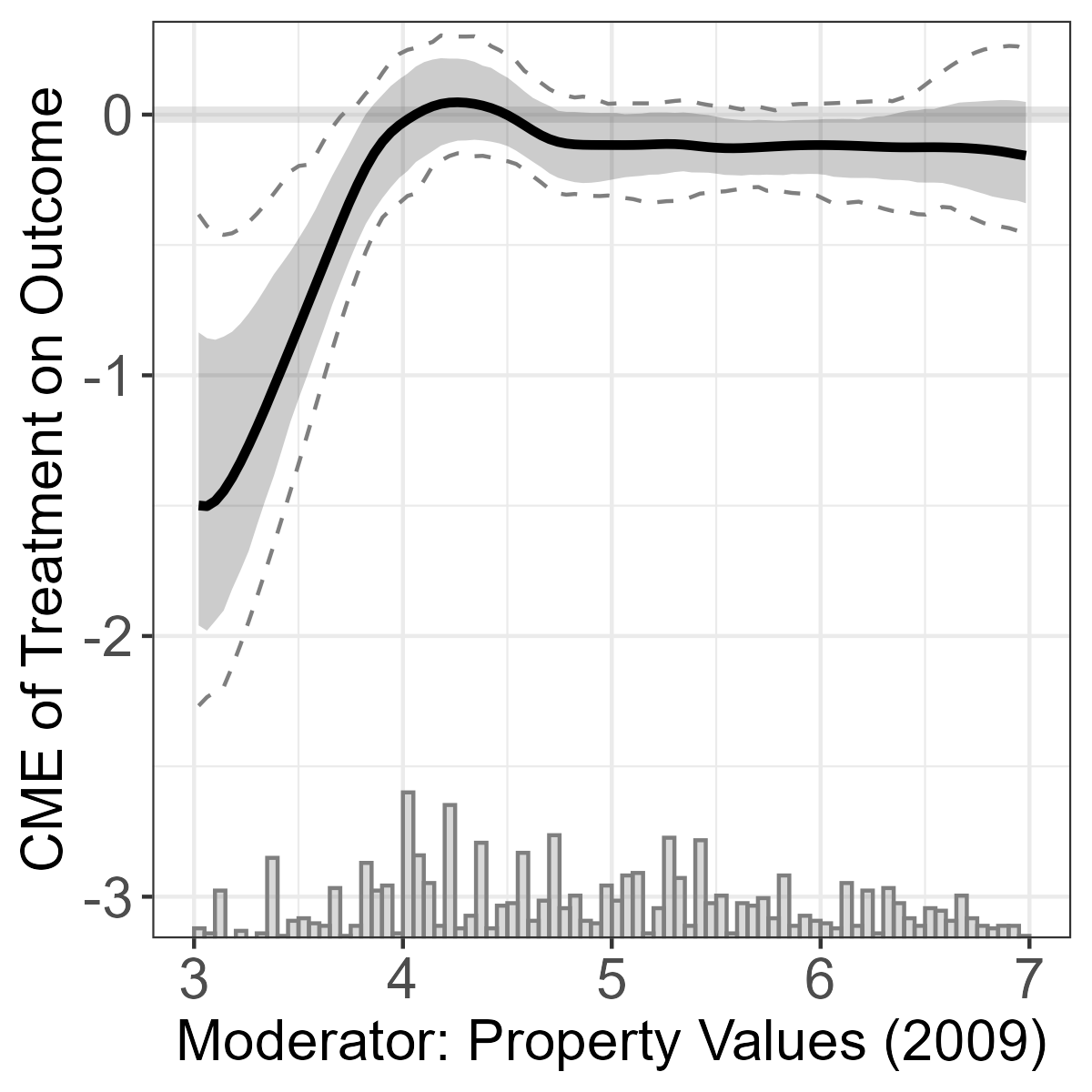}
    \caption{\hspace{0.5em} Kernel Estimator}
\end{subfigure}
\hspace{0.02\textwidth}  
\begin{subfigure}[b]{0.45\textwidth}
    \centering
    \includegraphics[width=\textwidth]{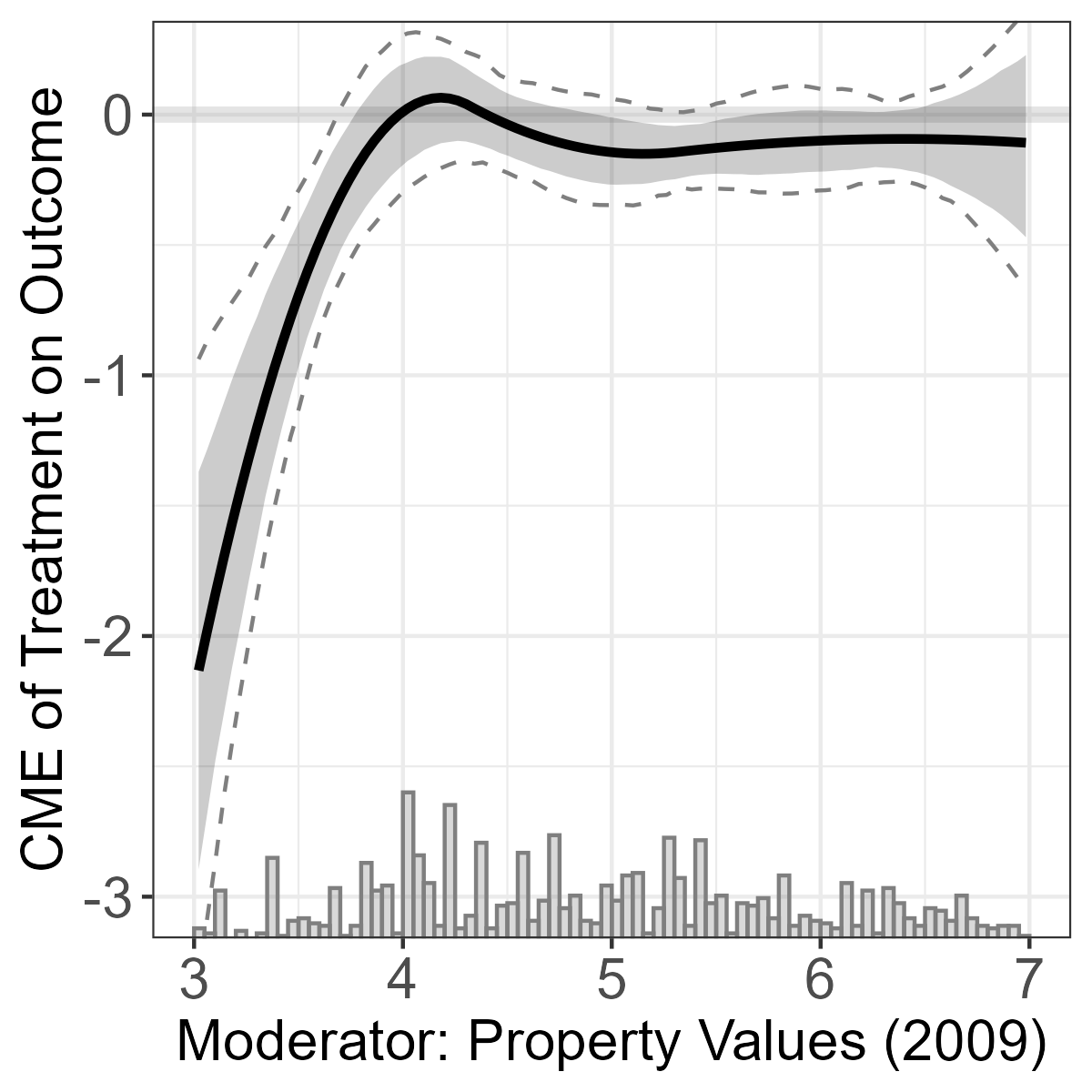}
    \caption{\hspace{0.5em} PO-Lasso}
\end{subfigure}
\label{fig:Adiguzel_denoise}\\
{\footnotesize \textit{Note}: The treatment is the change in congestion level, the outcome is the change
in the vote share for the incumbent AKP party. In each figure, the black line represents the estimated CME. The shaded area and the dashed lines represent the 95\% pointwise and uniform confidence intervals, respectively. The histograms at the bottom of the figure depict the distributions of the moderator.}
\end{figure}

In Figure~\ref{fig:Adiguzel_denoise}, we display the estimated CME along with their 95\% pointwise and uniform confidence intervals. The gray band represents pointwise intervals, while the dashed lines indicate uniform intervals that ensure simultaneous coverage across all values of \(x\). Both the kernel estimator and PO-Lasso yield broadly consistent findings: poorer neighborhoods with \(X < 4\) respond more strongly to improvements in health care, while the effect is essentially nonexistent in other areas. Although the main argument in \citet{adiguzel2023out} remains supported, the more flexible estimators reveal a more nuanced pattern than the original linear model.

\subsection{Summary}

This section developed two doubly robust estimators. AIPW combines outcome modeling and propensity-score weighting, remaining consistent when either model is correctly specified and reducing variance relative to IPW alone. Pairing AIPW with basis expansion and post-Lasso selection gives AIPW-Lasso, which handles high-dimensional covariates; we use a nonparametric bootstrap to capture uncertainty across both stages. For continuous treatments, an analogous partialing-out strategy yields PO-Lasso. Both methods provide the conceptual bridge to the DML framework.


\clearpage

\section{Double Machine Learning}

In previous sections, we introduced inverse probability weighting (IPW) and augmented inverse probability weighting (AIPW) estimators for the CME. With AIPW, we emphasized a general principle: construct an estimating equation that combines an outcome model and a treatment model so that moderate misspecification in either component does not generate first-order bias in the target. This section makes that principle explicit by introducing Neyman orthogonality and showing that the AIPW signal is orthogonal. Neyman orthogonality is the central property behind modern doubly robust estimation and high-dimensional inference. Intuitively, it ensures that small errors in nuisance functions do not translate into first-order bias in the final CME estimate.

Throughout this Element, the primary estimand is the CME function $\theta(x)$, where $x$ indexes a single moderator $X$. Identification and estimation rely on nuisance functions $\eta_0(V)$ defined on the full covariate set $V=(X,Z)$. These components include the propensity score $\Pr(D=1\mid V)$, outcome regressions $\E[Y\mid D=d,V]$, and, depending on the setting, additional conditional moments such as variances or density derivatives. In applications with many covariates or strong nonlinearity, parametric and semiparametric specifications may be too rigid to approximate $\eta_0$ accurately. Flexible learners, such as random forests (RF), neural nets (NN), histogram gradient boosting (HGB), address this by reducing functional-form assumptions, but they typically converge more slowly and depend on regularization to control complexity. Regularization reduces variance but induces bias in nuisance estimates; a naive plug-in approach can therefore destroy the $\sqrt{n}$-rate needed for asymptotic normality and valid inference for $\theta(\cdot)$ \citep{chernozhukov2018double,semenova2021debiased}.

Double/debiased machine learning (DML), also called ``double machine learning'' or ``orthogonalized machine learning,'' provides a formal framework for using machine learning nuisance estimates while preserving $\sqrt{n}$-inference. The modern DML framework is developed in \citet{chernozhukov2018double} for fixed-dimensional parameters and extended by \citet{semenova2021debiased} to function-valued targets such as the CME. In its canonical form, DML proceeds in two stages. In the first stage, one estimates the nuisance functions using flexible methods with sample splitting and cross-fitting, so that nuisance fits are obtained on training folds and evaluated on separate holdout folds. In the second stage, one constructs a Neyman-orthogonal score (or ``signal'') using these cross-fitted nuisance estimates and solves the associated moment condition to obtain the parameter of interest. Orthogonalization targets regularization bias, while cross-fitting mitigates overfitting bias; together they yield $\sqrt{n}$-consistent, asymptotically normal estimators under mild rate and regularity conditions \citep{chernozhukov2018double}.

When the estimand is a function rather than a scalar, an additional smoothing step is required. After forming orthogonal signals, we recover $\theta(x)$ by projecting the signal onto a basis in $X$ whose dimension may grow with the sample size, paralleling the smoothing step introduced earlier but now applied to cross-fitted orthogonal pseudo-outcomes \citep{semenova2021debiased}. Conceptually, orthogonality and cross-fitting handle nuisance estimation and remove first-order bias, while the final projection ``collapses out'' the remaining covariates $Z$ and returns the CME as a function of $X$.

The remainder of the section proceeds as follows. We first introduce the general DML setup and its key ingredients—orthogonality, high-quality nuisance learning, and cross-fitting—allowing $\theta_0$ to be either finite-dimensional or function-valued. We then specialize to CME estimation under the interactive regression model (binary treatment) and the partially linear regression model (continuous treatment), detailing how orthogonal scores and cross-fitting are implemented in each case and how the final smoothing step recovers $\theta(x)$. We conclude with empirical illustrations using the \texttt{interflex} package.

\subsection{Key Ingredients of DML}

The general DML framework includes three key ingredients summarized in Table~\ref{tab:dml_components}, : (i) Neyman orthogonality, (ii) high-quality machine learning methods, and (iii) sample-splitting and cross-fitting strategies. We first lay out the necessary assumption for the properties of DML estimator to hold for fixed dimensional (scalar) parameter, then lay out the regularity condition to build the pointwise and uniform asymptotic properties of the DML estimator for infinite-dimensional parameter. 


\begin{table}[ht]
\centering
\caption{Key Ingredients of DML}
\label{tab:dml_components}
\begin{tabular}{p{0.28\textwidth}|p{0.65\textwidth}}
\hline\hline
\textbf{Component} & \textbf{Purpose} \\
\hline
Neyman orthogonal score & Ensures that the estimation of the parameter of interest $\theta_0$ is robust against estimation biases of nuisance parameters $\hat{\eta}$. \\
\hline
High-quality machine learning method & Ensures the estimated nuisance parameter, $\hat{\eta}$, converges to the true parameter, $\eta$, at a sufficiently fast rate, specifically $o_p(n^{-1/4-\delta})$ for some $\delta \geq 0$. \\
\hline
Sample splitting and cross-fitting & Utilizes sample splitting to mitigate overfitting bias by independently estimating $\eta$ and $\theta$ using different data subsets.  \\
\hline
\end{tabular}
\end{table}

\paragraph{Neyman Orthogonality}

In the previous section, we saw that AIPW constructs an estimator for the CME that remains consistent if either the outcome model or the propensity model is correctly specified. This double robustness arises primarily from Neyman orthogonality: the AIPW signals are constructed so that small errors in nuisance functions induce only second-order biases, preventing first-order bias in the CME estimate. Recall that, in the binary-treatment setting, the AIPW signals takes the form:
\begin{equation}
\label{eq:aipw.signal}
\Lambda_i = \mu^1(V_i) - \mu^0(V_i)
+ \frac{D_i}{\pi(V_i)} \bigl(Y_i - \mu^1(V_i)\bigr)
- \frac{1 - D_i}{\,1 - \pi(V_i)\,} \bigl(Y_i - \mu^0(V_i)\bigr),
\end{equation}
where $\mu^d(V_i)$ estimates the outcome model $\mathbb{E}[Y_i(d) \mid V_i]$ and $\pi(V_i)$ estimates the propensity score model $\Pr(D_i=1 \mid V_i)$. The key property is that if $\hat{\mu}^d(\cdot)$ and $\hat{\pi}(\cdot)$ are slightly misspecified around their true values, the first-order bias in the final estimation for signal $\mathbb{E}[\Lambda_i]$ still cancels out. This insensitivity to small modeling errors reflects Neyman orthogonality. The score function derived from the AIPW signal satisfies Neyman orthogonality:
\begin{equation*}
\left.\frac{\partial}{\partial \eta}
\Bigl(\mathbb{E}[\Lambda_i(\eta)]\Bigr)
\right|_{\eta = \eta_0}= 0,
\end{equation*}
where \(\eta_0\) denotes the true nuisance functions \((\mu^{0}_{0}, \mu^{1}_{0},\, \pi_0)\). This derivative is taken with respect to local perturbations of true \(\eta\). We provide proofs in the Appendix showing that the partial derivative with respect to the nuisance functions is zero, meaning that small errors in estimating \(\eta\) do not affect the target parameter \(\theta(\cdot)\) to the first order. In the AIPW formula, Neyman orthogonality underlies the double robustness property: even if one nuisance function is misspecified, the main parameter \(\theta(\cdot)\) remains consistently estimated.

DML builds on the idea of Neyman orthogonality. When estimating the target function \(\theta_0\) in the presence of high-dimensional nuisance functions \(\eta\), a key concern is the potential influence of regularization bias in \(\eta\) on the estimation of \(\theta_0\). Neyman orthogonality, introduced in \citet{neyman1959optimal} and \citet{neyman1979c}, offers a structured approach to reduce the sensitivity of the final estimator to certain classes of errors in the nuisance function estimates. We begin with a generic score $\psi(W_i, \theta, \eta)$, where $W_i =(Y_i,D_i,X_i,Z_i)$ represents data, to specify a population moment condition:
\begin{equation*}
\mathbb{E}[\psi(W_i; \theta_0, \eta_0)] = 0,
\end{equation*}

\noindent where $\theta_0$ denotes the true target parameter and $\eta_0$ denotes the true nuisance parameters. Neyman orthogonality requires the score function to satisfy the following condition:
\begin{definition}[Neyman orthogonality] \label{def:neyman-ortho}
\begin{equation*}
\frac{\partial}{\partial \eta} \, \mathbb{E}[\psi(W_i; \theta_0, \eta)] \bigg|_{\eta = \eta_0} = 0
\end{equation*}
\end{definition}
In other words, infinitesimal changes around \(\eta_0\) do not alter the main moment condition to first order. This is exactly what we observed in the classic AIPW formula: each component involving \(\mu^1\), \(\mu^0\), and \(\pi\) is constructed so that small estimation errors cancel out. The score function is linear in the parameter of interest as well.

The DML framework in \citet{chernozhukov2018double} is first developed for a scalar $\theta_0$ and then extended to function-valued targets such as the CME. We state the scalar case here and return to the function-valued extension in the ``Estimation of Function-Valued Parameters'' subsection below. Let $\psi(W; \theta, \eta)$ be a score function, where $\theta$ is a scalar parameter.

\begin{assumption}[Linear scores with approximate Neyman Orthogonality]\label{assump:approx_ortho}
Suppose:

\renewcommand\labelenumi{(\theenumi)}
\begin{enumerate}
\item \label{assump-A}%
Identification via moment condition. The true parameter $\theta_0$ solves $\mathbb{E}\bigl[\psi(W;\theta_0,\eta_0)\bigr] = 0.$

\item \label{assump-B}%
Linearity in $\theta$. The score is linear in $\theta$. Formally, there exist functions $\psi^a(\cdot)$ and $\psi^b(\cdot)$ such that
\[
\psi(W;\theta,\eta) = \psi^a(W;\eta)\,\theta + \psi^b(W;\eta).
\]

\item \label{assump-C}%
Smoothness in $\eta$. The map $\eta \mapsto \mathbb{E}\bigl[\psi(W;\theta,\eta)\bigr]$ is twice Gateaux-differentiable in a neighborhood around the true $\eta_0$.

\item \label{assump-D}%
(Near) Neyman Orthogonality. Small perturbations $\eta - \eta_0$ have only a second-order effect on the moment condition. Formally, for a sequence $\delta_n \to 0$,
\[
\sup_{\eta \in \mathcal{T}_n} \left\|\partial_{\eta} \mathbb{E}[\psi(W;\theta_0,\eta)]\big|_{\eta=\eta_0}\,[\eta - \eta_0]\right\| \le \delta_n\,n^{-1/2}.
\]

\item Identification condition. The matrix $J_0 = \mathbb{E}\bigl[\psi^a(W;\eta_0)\bigr]$ has singular values bounded away from 0 and $\infty$. This is a standard invertibility requirement to identify $\theta_0$.
\end{enumerate}
\end{assumption}

In practice, constructing the score function begins with a naive moment function that would correctly identify \(\theta_0\) if \(\eta_0\) were known exactly. This function is then adjusted—or ``de-biased''—to satisfy the orthogonality condition. We will show this process in more details in the context of CME estimation with binary and continuous treatments in the following subsections.

\paragraph{High-quality Machine Learning Method} 

The double robustness property of the DML approach ensures resilience to mild errors in \(\hat{\eta}_0\). The key requirement is that the estimated nuisance parameter \(\hat{\eta}\) converges to the true parameter \(\eta_0\) at a sufficiently fast rate, specifically \(o_p(n^{-1/4 - \delta})\) for some \(\delta \geq 0\): $\|\hat\eta-\eta_0\|=o_p(n^{-1/4-\delta})$, meaning that the nuisance estimation error vanishes in probability faster than $n^{-1/4-\delta}$. 

\begin{assumption}
[Score regularity and quality of ML estimator]\label{assump:ml_quality}
Suppose the following conditions hold with high probability:

\renewcommand\labelenumi{(\theenumi)}
\begin{enumerate}
\item \label{assump-ml-A}%
Good neighborhood for $\hat{\eta}$. With high probability, the machine learning nuisance estimator $\hat{\eta}$ falls into some set $T_N$, which is a neighborhood around $\eta_0$. Formally, $\hat{\eta}\in T_N$, where $T_N$ includes $\eta_0$ and is constrained by conditions below.

\item \label{assump-ml-B}%
Bounded score moments. There exists an integer $q\ge2$ and constant $c_1$ such that for all $\eta\in T_N$,
\[
\bigl(\mathbb{E}\bigl[\|\psi(W;\theta_0,\eta)\|^q\bigr]\bigr)^{1/q}
\;\le\; c_1,
\quad \bigl(\mathbb{E}\bigl[\|\psi^a(W;\eta)\|^q\bigr]\bigr)^{1/q}
\;\le\; c_1.
\]
This ensures $\psi$ and its derivatives have finite moments, allowing standard limit laws to apply.

\item \label{assump-ml-C}%
Controlled rates. For $\eta\in T_N$, define quantities (such as \(r_N\), \(r'_N\), \(\lambda_N\), and \(\lambda'_N\)) capturing how much $\psi$ and $\psi^a$ can change when $\eta$ deviates slightly from $\eta_0$. We assume these differences remain small, typically on the order of $\delta_N \to 0$ as $N$ grows. Concretely:
\[
\sup_{\eta \in T_N}
\left\|
\mathbb{E}[\psi^a(W;\eta)]
-
\mathbb{E}[\psi^a(W;\eta_0)]
\right\|
\;\le\;
\delta_N,
\]
and similarly for $\psi(W;\theta_0,\eta)$. This guarantees ``smoothness'' in $\eta$ and that small estimation errors in $\eta$ do not blow up the score.

\item Non-degenerate variance. The variance/covariance of $\psi$ under $\eta_0$ is bounded away from zero. Formally, all eigenvalues of
\[
\mathbb{E}\bigl[
\psi(W;\theta_0,\eta_0)\,
\psi(W;\theta_0,\eta_0)^\top
\bigr]
\]
exceed a positive constant $c_0$. This ensures the asymptotic variance is well-defined.
\end{enumerate}
\end{assumption}

This condition can be satisfied when \(\eta_0\) is effectively approximated using methods such as decision trees or advanced neural network architectures \citep{chernozhukov2018double}. The \texttt{interflex} package implements three machine learning approaches: NN, RF, and HGB.

It is important to note that no single machine learning algorithm performs best across all scenarios. Therefore, no algorithm strictly dominates in all applications; performance varies widely depending on factors such as the specific task and the dimensionality of the dataset. In Table~\ref{tab:ml_techniques}, we outline some broadly accepted pros and cons of each machine learning method.

\begin{longtable}{@{}p{0.10\textwidth}>{\raggedright\arraybackslash}p{0.42\textwidth}>{\raggedright\arraybackslash}p{0.42\textwidth}@{}}
    \caption{Comparison of Machine Learning Techniques} \label{tab:ml_techniques} \\
    \toprule
    \textbf{Method} & \textbf{Pros} & \textbf{Cons} \\
    \midrule
    \endfirsthead
    \toprule
    \textbf{Method} & \textbf{Pros} & \textbf{Cons} \\
    \midrule
    \endhead
    \midrule
    \bottomrule
    \endlastfoot
    \textbf{NN} & \small
    • Excel in processing large, complex datasets, particularly those involving unstructured forms such as images, text, and audio. \newline
    • Use multiple layers to capture intricate patterns and dependencies within data, facilitating high levels of data abstraction. & \small
    • Require careful tuning of many hyperparameters (architecture, learning rate, regularization) to achieve optimal performance. \newline
    • Require significantly more data than traditional algorithms to generalize well, making them unsuitable for small datasets.
    \\ \hline
    \textbf{RF} & \small
    • Effective for classification and regression, leveraging bagging to reduce variance and control overfitting. \newline
    • Generally robust against overfitting compared to single decision trees and requires less hyperparameter tuning than boosting. & \small
    • Performance can suffer with high-dimensional sparse data, as random feature selection may fail to identify informative splits. \newline
    • Cannot extrapolate trends beyond the training data range and struggles with linear relationships compared to parametric models.
    \\ \hline
    \textbf{HGB} & \small
    • Efficiently handles large datasets by binning continuous features, offering faster training speeds than standard gradient boosting. \newline
    • Natively handles missing values and categorical features without extensive preprocessing. \newline
    • Effective for imbalanced datasets by focusing on hard-to-classify examples. & \small
    • More sensitive to noise and outliers than Random Forests, as the algorithm iteratively corrects errors, potentially overfitting to anomalies. \newline
    • Requires more careful tuning of hyperparameters (learning rate, tree depth) than Random Forests to prevent overfitting.
    \\
\end{longtable}

\paragraph{Sample Splitting and Cross-Fitting}

Sample splitting serves as a crucial technique alongside the use of orthogonal score functions and machine learning methods for estimating nuisance components. Overfitting bias can arise when the nuisance parameters are estimated on the same sample used to estimate the parameter of interest \(\theta\). Sample splitting helps mitigate this bias by ensuring independence between the data used to estimate the nuisance functions (\(\eta_0\), training data) and the data used to estimate the causal parameter (\(\theta_0\), holdout data).

Cross-fitting, as an efficient form of data splitting, maximizes the use of available data by alternating the roles of training and holdout sets in a cross-validated framework and combining the estimates from multiple splits to improve efficiency. The process can be outlined as follows:
\begin{enumerate}
\item Randomly partition the dataset into $K$ folds, $(I_k)^K_{k=1}$, where each fold has $|I_k| = n/K$ observations and $I^c_k = \{1, \ldots, n\} \setminus I_k$ denotes the complement set used for training.
\item For each fold $k$, estimate the nuisance parameters $\hat{\eta}_{0,k}$ using the data in $I^c_k$.\footnote{Strictly speaking, to preserve the fold-level independence (i.e., $\hat{\eta}_{0,k}$ being a function only of the training sample), any hyperparameter tuning---such as selecting the Lasso penalty via cross-validation---should be carried out using only observations in $I_k^c$, typically through an ``inner'' cross-validation restricted to $I_k^c$, followed by refitting the nuisance learner on all of $I_k^c$ at the chosen hyperparameters. In practice, many implementations tune hyperparameters globally (e.g., cross-validate once on the full sample and reuse the selected hyperparameters across folds) to reduce computational burden.}
\item Compute the fold-specific estimator $\hat{\theta}_{0,k}$ using the empirical average over observations in $I_k$, and average the results across all $K$ folds to obtain the final estimate $\hat{\theta}_0$.
\end{enumerate}

The cross-fitting procedure leverages the full dataset while enhancing the robustness of the estimation process by reducing dependence on any single data partition.

\bigskip

The three key ingredients, namely Neyman orthogonality, the high-quality machine learning method, and sample-splitting and cross-fitting strategy,  when effectively integrated, lead to the robust properties of the canonical DML estimator. 

\begin{thm}[Properties of the DML] Suppose that Assumption \ref{assump:approx_ortho} and Assumption \ref{assump:ml_quality} hold. In addition, suppose that $\delta_n \ge n^{-1/2}$ for all $n \ge 1$. Then the DML estimator $\hat{\theta}_0$ concentrates in a $1/\sqrt{n}$ neighborhood of $\theta_0$ and is approximately linear and centered Gaussian \citep{chernozhukov2018double}:

\begin{equation}
\label{eq:asy_norm}
\sqrt{n}\, \sigma^{-1} (\hat{\theta}_0 - \theta_0) = \frac{1}{\sqrt{n}} \sum_{i=1}^{n} \bar{\psi}(W_i) + O_P(\rho_n) \leadsto N(0, 1),
\end{equation}

\noindent uniformly over $P \in \mathcal{P}_n$, where the size of the remainder term obeys

\begin{equation}
\label{eq:remainder}
\rho_n := n^{-1/2} + r_n + r'_n + n^{1/2} \lambda_n + n^{1/2} \lambda'_n \leq \delta_n,
\end{equation}

\noindent $\bar{\psi}(\cdot) := -\sigma^{-1} J_0^{-1} \psi(W; \theta_0, \eta_0)$ is the influence function, and the approximate variance is:

\begin{equation}
\label{eq:covariance}
\sigma^2 := J_0^{-1} \mathbb{E}_P[\psi(W; \theta_0, \eta_0)\psi(W; \theta_0, \eta_0)^\top] (J_0^{-1})^\top.
\end{equation}
\end{thm}

Equation~(\ref{eq:asy_norm}) states that \(\hat{\theta}_0\) is asymptotically normal with mean \(\theta_0\) and variance \(\sigma^2 / n\). Equation~(\ref{eq:remainder}) defines the remainder term, which captures the cumulative effect of small errors in the final asymptotic expansion. Intuitively, the terms \(r_n\), \(r'_n\), \(\lambda_n\), and \(\lambda'_n\) reflect the quality of nuisance function estimation and the degree to which orthogonality mitigates the influence of these errors on the target parameter. Equation~(\ref{eq:covariance}) gives the asymptotic variance, where \(J_0\) denotes the derivative of the score function in $\theta$, capturing its sensitivity to deviations of \(\theta\) from \(\theta_0\).

Taken together, the theorem establishes that the DML estimator for the scalar parameter \(\theta_0\) achieves \(\sqrt{n}\)-rate convergence and is approximately normally distributed. Both the convergence rate and the asymptotic distribution hold uniformly over a broad class of underlying data-generating processes.

\paragraph{Estimation of Function-Valued Parameters}

When \(\theta_0\) is a function of \(x \in \mathcal{X}\), such as the CME, we need an additional smoothing step to handle the infinite-dimensional target. The standard approach is to project \(\theta(x)\) onto a finite but growing set of basis functions. For instance, with B-spline expansions, we write
\[
\theta_0(x) \approx p(x)^\top \beta_0,
\]
where \(p(x) \in \mathbb{R}^{d}\) is a vector of \(d\) basis functions (e.g., splines or polynomials). After constructing the orthogonal signal \(\hat{\Lambda}_i\) using cross-fitting, we regress \(\hat{\Lambda}_i\) on \(p(X_i)\) to obtain the coefficients \(\hat{\beta}\). The final estimator of the function takes the form \(\hat{\theta}(x) = p(x)^\top \hat{\beta}\). Allowing the dimension \(d\) to increase with the sample size improves flexibility but requires additional regularity conditions.

Extending \(\sqrt{N}\)-inference from scalar to function-valued parameters requires controlling three sources of error. First, the sieve must approximate \(\theta_0(x)\) sufficiently well so that approximation bias vanishes as \(d\) increases. Second, the number of basis functions \(d=d_N\) cannot grow too quickly relative to \(N\); otherwise, variance from estimating many coefficients dominates the signal. Third, the cross-fitted nuisance estimators must converge sufficiently fast, and the orthogonal score must have well-behaved moments, so that nuisance estimation errors remain second-order after orthogonalization \citep{semenova2021debiased}. Under these conditions, the estimator \(\hat{\theta}(x)\) remains \(\sqrt{N}\)-consistent and asymptotically normal pointwise. Stronger restrictions on basis growth and tail behavior allow uniform confidence bands over \(x \in \mathcal{X}\). The formal regularity conditions are reported in Online Appendix Section C. 

With these elements in place, we now turn to the construction of orthogonal scores in concrete causal models. We begin with the binary treatment case, where the DML estimator builds on the familiar AIPW signal and implements cross-fitting to ensure valid inference for \(\theta(x)\).

\subsection{Binary Treatment}

In the case of binary treatment, \citet{semenova2021debiased} show that the Neyman-orthogonal signal for the CME can be constructed using the following expression:
\begin{equation}
\label{eq:neyman_score_irm}
\Lambda_{i}(\eta) = \mu(1, V_i) - \mu(0, V_i) + \frac{D_i(Y_i - \mu(1, V_i))}{\pi(V_i)} - \frac{(1 - D_i)(Y_i - \mu(0, V_i))}{1 - \pi(V_i)}
\end{equation}
where \(\mu(d, v) = \E[Y_i \mid D_i=d, V_i = v]\) is the (observed) outcome model and \(\pi(v) = \Pr(D_i = 1 \mid V_i = v)\) is the propensity score. We denote the true value of these nuisance parameters as \(\eta_0(v) := \{\pi_0(v), \mu_0(1, v), \mu_0(0, v)\}\).

Equation~(\ref{eq:neyman_score_irm}) shares the same form as the AIPW signal in Equation~(\ref{eq:aipw.signal}), which is orthogonal with respect to the nuisance parameter $\eta_0(v)$. This AIPW-style Neyman score is first-order insensitive to small estimation errors in \(\hat{\mu}(\cdot)\) and \(\hat{\pi}(\cdot)\), preserving the orthogonality property. It also provides the foundation for establishing both pointwise and uniform asymptotic theory for the resulting DML estimator for the CME \(\theta(x)\). \citet{bonvini2023flexibly} propose an alternative signal that allows the influence of \(X\) on \(Y\) to depend partially on other covariates in \(Z\); however, we retain the standard signal to target the CME \(\theta(x)\) as defined in \citet{semenova2021debiased}.

Under the unconfoundedness assumption introduced in Section 1 (Assumption~\ref{assm:unconf}), the CME can be expressed as the conditional expectation of this observable signal. Specifically, when \( Y_i(d) \indep D_i \mid V_i = v \), we have
\[
\E [\Lambda_i(\eta_0) \mid X_i = x ] = \theta(x).
\]
In the DML framework, we do not require the functional forms of the nuisance models to be known parametrically. Instead, we rely on machine learning models to consistently estimate the nuisance components as the sample size becomes sufficiently large.

With binary treatment, we make the following modeling assumption.
\begin{assumption}[Interactive Regression Model, IRM]
We assume the potential outcomes and treatment assignment satisfy:
\begin{align*}
Y_i(d) & = g_0(d, V_i) + u_i, & \E [u_i &\mid V_i] = 0  \\
D_i & = m_0(V_i) + \epsilon_i, & \E [\epsilon_i &\mid V_i ] =0
\end{align*}
where $g_0(d, v) := \E[Y(d) \mid V_i=v]$ is the structural conditional mean of the potential outcome, and $m_0(v) := \Pr(D_i=1 \mid V_i=v)$ is the propensity score.
\end{assumption}

Under the unconfoundedness assumption ($Y(d) \perp D \mid V$), the structural function $g_0(d, v)$ coincides with the observational regression function $\mu_0(d, v) := \E[Y \mid D=d, V=v]$. This identification result justifies using machine learning methods to estimate $\mu_0$ from observational data.

This model is termed ``interactive'' because it allows the potential outcomes \(Y_i(1)\) and \(Y_i(0)\) to follow arbitrarily different functional forms in \(V_i\), unlike partially linear models which assume the treatment effect is constant or independent of controls. With the IRM, the CME is given by:
\[\theta(x) = \E [g_0(1, x, Z) - g_0(0, x, Z) \mid X = x ].\]
The IRM is general, encompassing both the linear interaction model and the smooth varying coefficient model (SVCM) as special cases. Specifically, the linear interaction model assumes
\[
\E[Y_i(d) \mid X_i, Z_i] = \beta_0 + \beta_1 d + \beta_2 X_i + \beta_3 (d \cdot X_i) + \beta_4 Z_i,
\]
which constrains \(g_0(\cdot)\) to be linear. Similarly, the SVCM assumes
\[
\E[Y_i(d) \mid X_i, Z_i] = f(X_i) + q(X_i) d + Z_i \gamma(X_i),
\]
which allows for smooth variation over \(X\).

In the binary treatment case, we first estimate the nuisance functions using machine learning methods with cross-fitting. We then construct the AIPW-style Neyman-orthogonal signal for each unit \(i\). If \(X\) is discrete, the CME can be recovered by taking sample averages of \(\{\hat{\Lambda}_i\}\) within each group \(X_i = x\). If \(X\) is continuous, we regress \(\hat{\Lambda}_i\) on basis functions of \(X\) (as described in the previous section) to estimate the curve \(\theta(x)\).


\begin{algorithm}[!ht]
\footnotesize
\caption{DML Estimation of CME (Binary \(D\))}
\begin{algorithmic}[1]
\State \textbf{Inputs:}
\State \quad Data: \( \{(X_i, Z_i, D_i, Y_i)\} \) for \(i=1,...,n\)
\State \quad\quad \(D_i \in \{0,1\}\) is binary treatment
\State \quad\quad \(V_i := (X_i, Z_i)\) are controls (\(X_i\) is the moderator)
\State \quad\quad \(Y_i\) is the outcome
\State \quad Number of folds \(K\) (e.g., \(K=5\))
\State \quad Choice of basis functions \(p(X)\) (e.g., B-splines) for the final projection
\State \textbf{Outputs:}
\State \quad Estimated function \( \hat{\theta}(x) = \widehat{\E}[Y(1) - Y(0) \mid X = x]\)

\State \textbf{1) Partition data into \(K\) folds}
\State \quad Randomly split the indices \(\{1,...,n\}\) into \(K\) disjoint sets: \(I_1, ..., I_K\).
\State \quad Let \(I_k\) be the ``holdout" set and \(I_k^c = \{1,...,n\} \setminus I_k\) be the ``training" set.

\State \textbf{2) Estimate nuisance functions (cross-fitting)}
\For{\(k = 1\) to \(K\)}
\State \quad (a) On the training set \(I_k^c\):
\State \quad\quad * Fit outcome models: \(\hat{\mu}_1(v) \approx \E[Y \mid D=1, V=v]\), \(\hat{\mu}_0(v) \approx \E[Y \mid D=0, V=v]\)
\State \quad\quad * Fit propensity model: \(\hat{\pi}(v) \approx \Pr(D=1 \mid V=v)\)
\State \quad (b) For each observation \(i\) in the holdout set \(I_k\):
\State \quad\quad * Compute predictions: \( \hat{\mu}_1^{(-k)}(V_i), \hat{\mu}_0^{(-k)}(V_i), \hat{\pi}^{(-k)}(V_i) \)
\EndFor

\State \textbf{3) Compute AIPW-style Neyman-orthogonal signal}
\For{each fold \(k\), for each \(i\) in \(I_k\)}
\State \quad Construct the pseudo-outcome \(\Lambda_i\):
\State \quad \( \Lambda_i = \hat{\mu}_1^{(-k)}(V_i) - \hat{\mu}_0^{(-k)}(V_i) \)
\State \quad \quad \quad \( + \frac{D_i (Y_i - \hat{\mu}_1^{(-k)}(V_i))}{\hat{\pi}^{(-k)}(V_i)} - \frac{(1 - D_i) (Y_i - \hat{\mu}_0^{(-k)}(V_i))}{1 - \hat{\pi}^{(-k)}(V_i)} \)
\EndFor

\State \textbf{4) Project signal onto \(X\) to obtain \( \hat{\theta}(x) \)}
\State \quad Run a regression (e.g., OLS) of the signal \(\Lambda_i\) on the basis functions \(p(X_i)\):
\State \quad \( \hat{\beta} = \arg\min_{b} \sum_{i=1}^n \left( \Lambda_i - p(X_i)^\top b \right)^2 \)
\State \quad Construct estimator: \(\hat{\theta}(x) = p(x)^\top \hat{\beta}\)

\State \textbf{Return:}
\State \quad The function \( \hat{\theta}(x) \) = estimated CME.
\end{algorithmic}
\label{algorithm:binary}
\end{algorithm}

\bigskip
We now demonstrate how to implement this procedure using the \texttt{interflex} package. Using data from \textbf{Example~\ref{ex:Noble2024}} based on \citet{noble2024presidential}, we compare the results from AIPW-Lasso to those from the DML estimators. 

\begin{lstlisting}[language=R]
# R code excerpt

library(interflex) 
D="opp" # Out-partisanship of Lawmakers 
Y="pres_ref" # Frequency of Presidential References
X="pres_vote_margin" #Presidential Vote Margin
Z=c("prev_vote","majority","leadership","seniority","tot_speech")

## For code to generate plot (a), see last section 

## DML estimator with neural network machine learning method 
out.dml.nn<-interflex(estimator='DML', 
              data = d, model.y="nn", model.t = "nn",
              Y=Y,D=D,X=X, Z = Z, 
              treat.type = "discrete", na.rm = TRUE,
              vartype = "bootstrap", parallel = TRUE)
plot(out.dml.nn)

## DML estimator with random forests machine learning method 
out.dml.rf<-interflex(estimator='DML', 
              data = d, model.y="rf", model.t = "rf",
              Y=Y,D=D,X=X, Z = Z, 
              treat.type = "discrete", na.rm = TRUE,
              vartype = "bootstrap", parallel = TRUE)
plot(out.dml.rf)

## DML estimator with histogram gradient boosting machine learning method 
out.dml.hgb<-interflex(estimator='DML', 
              data = d, model.y="hgb", model.t = "hgb",
              Y=Y,D=D,X=X, Z = Z, 
              treat.type = "discrete", na.rm = TRUE,
              vartype = "bootstrap", parallel = TRUE)
plot(out.dml.hgb)
\end{lstlisting}

\begin{figure}
\caption{Replicating \cite{noble2024presidential} Figure 2: AIPW-Lasso vs DML Estimators}\label{fig:noble.dml}
\centering
\begin{subfigure}[b]{0.45\textwidth}
    \centering
    \includegraphics[width=\textwidth]{figures/chp3_noble_2024_aipw.png}
    \subcaption{AIPW-Lasso}
\end{subfigure}
\begin{subfigure}[b]{0.45\textwidth}
    \centering
    \includegraphics[width=\textwidth]{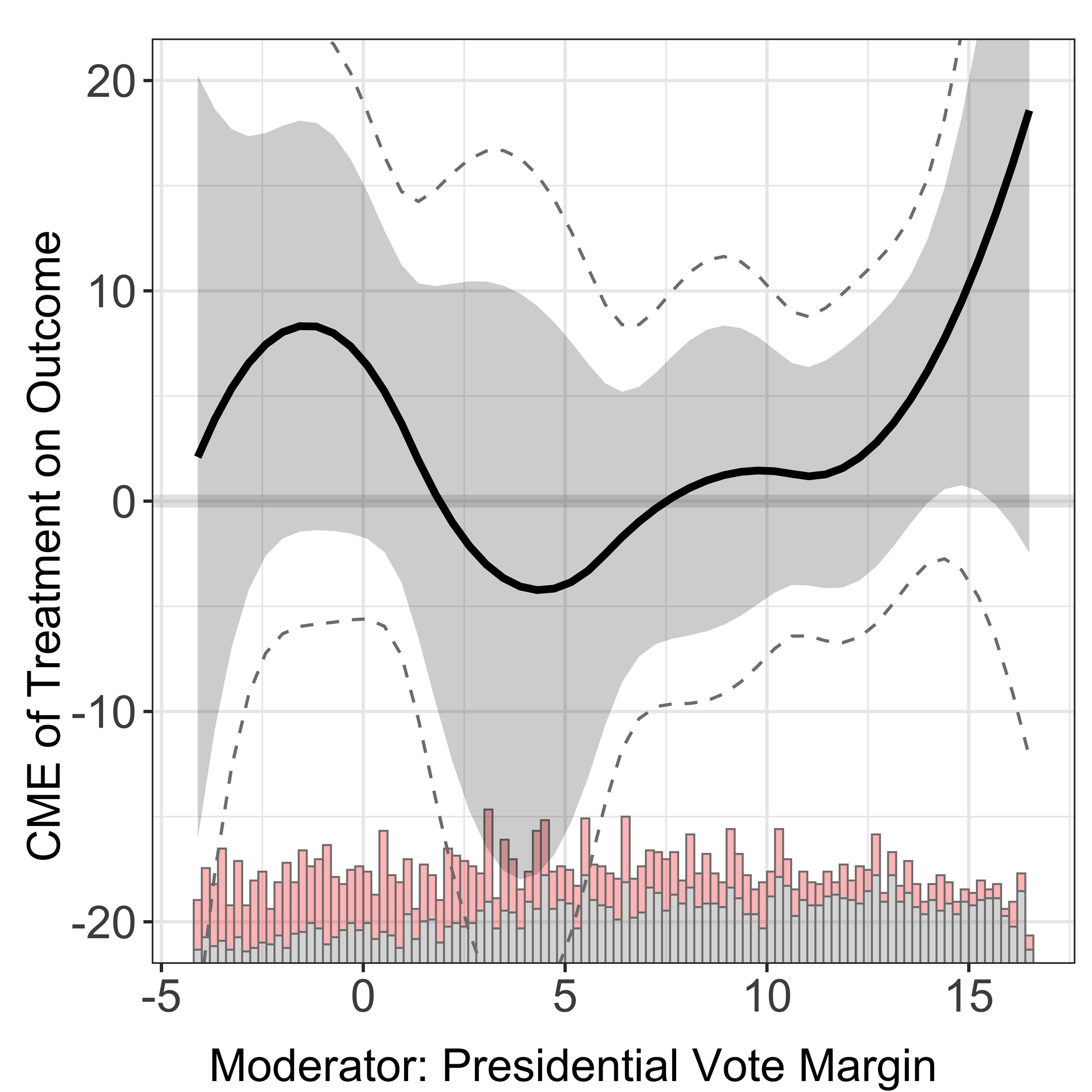}
    \subcaption{DML-NN}
\end{subfigure}
\begin{subfigure}[b]{0.45\textwidth}
    \centering
    \includegraphics[width=\textwidth]{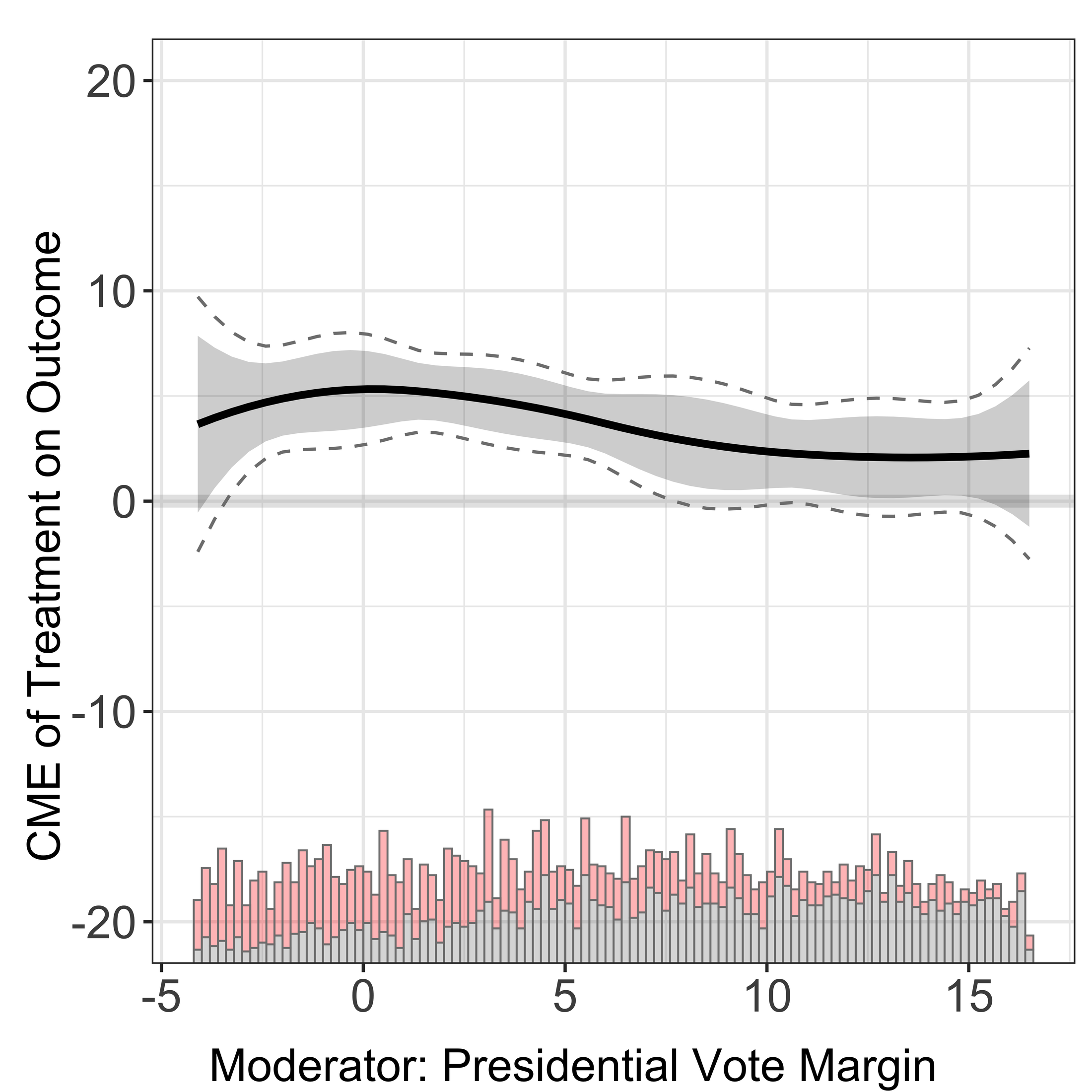}
    \subcaption{DML-RF}
\end{subfigure}
\begin{subfigure}[b]{0.45\textwidth}
    \centering
    \includegraphics[width=\textwidth]{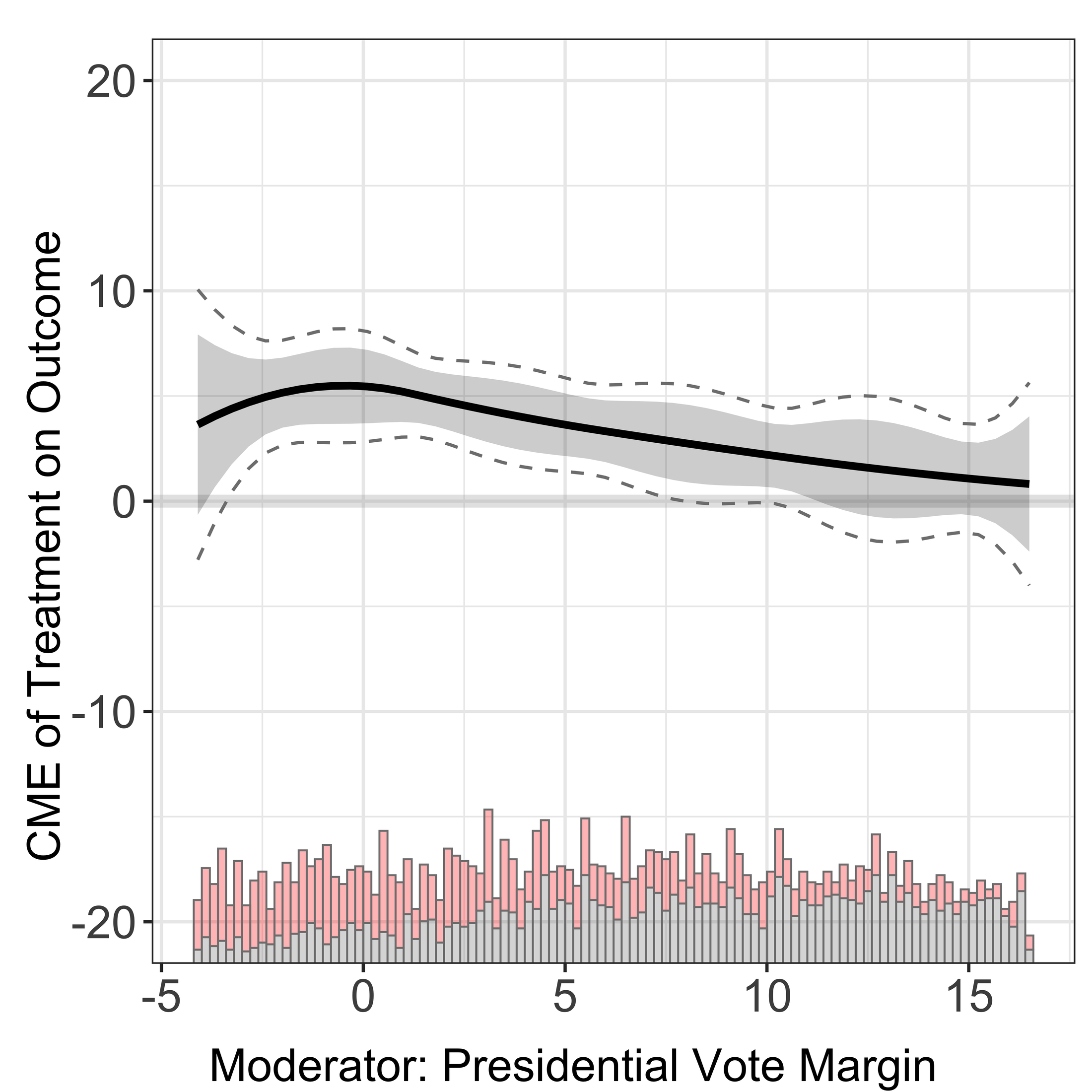}
    \subcaption{DML-HGB}
\end{subfigure}
\begin{minipage}{\linewidth}
{\footnotesize \emph{Notes:} The treatment $D$ is out-partisanship of lawmakers, the outcome $Y$ is frequency of presidential references in Congressional speeches. (a) CME estimates from AIPW-Lasso; (b)-(d) CME estimates from the DML estimator, with nuisance parameter estimated with NN, RF, and HGB, respectively. The gray ribbon show the 95\% pointwise confidence intervals, respectively. The dashed lines represent 95\% uniform confidence intervals.}  
\end{minipage}
\end{figure}

Figure~\ref{fig:noble.dml} compares CME estimates from the AIPW-Lasso estimator (a) and the DML estimators (b)–(d). As before, we trim the sample based on the moderator \(X\), restricting it to the central 95\% quantile range to improve overlap. AIPW-Lasso yields monotonically decreasing CME estimates in \(x\), with fairly narrow confidence intervals. The DML estimators produce broadly similar patterns but with much wider confidence intervals. Substantively, both methods suggest that the presidential-reference treatment has a modestly positive effect on the outcome in areas of low constituency partisanship. AIPW-Lasso shows a statistically significant treatment effect for \(X \in [-5, 10]\), while the DML estimators, despite greater uncertainty, indicate a significant effect for out-party legislators when \(X \in [0, 5]\).

\clearpage

\subsection{Continuous Treatment}

In the general case of continuous treatment, \citet{semenova2021debiased} derive a Neyman-orthogonal score that is robust to small estimation errors in the nuisance parameters $\eta_0 = \{\mu_0(d,v),\, f_{D\mid V}(d\mid v)\}$. Here, $\mu_0(d,v) = \E[Y \mid D = d, V = v]$ represents the (observed) outcome model, and $f_{D\mid V}$ is the conditional density of the treatment. The general Neyman-orthogonal signal takes the form:
\[
\Lambda_{i}(\eta) = -\partial_d \log f_{D\mid V}(D_i\mid V_i)\,[Y_i-\mu(D_i,V_i)] + \partial_d\mu(D_i,V_i).
\]
Under standard unconfoundedness and overlap assumptions, the CME is identified as the conditional expectation of this signal:
\begin{equation*}
\E\!\big[\Lambda_i(\eta_0)\mid X_i=x\big]\;=\;\E\!\big[\partial_d \mu_0(D_i,V_i)\mid X_i=x\big]\;=\;\theta(x).
\end{equation*}
We provide the proof in the Appendix.

While theoretically robust, this general approach requires fully nonparametric estimation of the conditional density \(f_{D \mid V}\) and its derivative, which can be highly unstable in high dimensions. To address this practical challenge, we adopt the Partially Linear Regression Model (PLRM) \citep{robinson1988root} as our framework.

\begin{assumption}[Partially Linear Regression Model, PLRM]
\label{assump:plrm}
We assume the outcome and treatment assignment satisfy:
\begin{align*}
Y_i(d) &= \theta(X_i) \cdot d + g(V_i) + \epsilon_i, \quad \E[\epsilon_i \mid V_i] = 0 \\
D_i &= m(V_i) + \xi_i, \quad \E[\xi_i \mid V_i] = 0
\end{align*}
where $Y_i(d)$ is the potential outcome; $\theta(X_i)$ is the conditional marginal effect of interest; and $m(V_i) := \E[D_i \mid V_i]$ is the conditional mean of the treatment.
\end{assumption}

This assumption imposes a key restriction: the outcome \(Y_i\) must depend linearly on the treatment \(D_i\), although the slope \(\theta(X_i)\) is allowed to vary arbitrarily with the moderator \(X_i\).

It is worth clarifying what estimand the PLRM-based DML estimator targets relative to the general continuous-treatment CME in Definition~\ref{def:cme2}. Definition~\ref{def:cme2} sets $\theta(x) = \E[\mu_d(D,X,Z)\mid X=x]$, where $\mu_d$ is the partial derivative of a generally nonlinear dose-response function and the expectation averages over both $D$ and $Z$. Under the PLRM, the dose-response function is linear in $d$, so $\mu_d(D,X,Z) = \theta(X)$ deterministically, and the derivative-averaging in Definition~\ref{def:cme2} collapses to $\theta(X)$ itself. When the PLRM holds, the two targets coincide. When the PLRM is misspecified, the DML estimator based on the Robinson score does not recover the derivative-average in Definition~\ref{def:cme2}; instead it recovers a best-linear projection of $\mu_d$ onto the residualized treatment (in the sense of \citealt{robinson1988root}), weighted by the conditional variance of $D$ given $V$. This projection remains a useful and interpretable summary of how the effect of $D$ varies with $X$, but readers targeting the derivative-average specifically should either (i) verify the PLRM is plausible in their setting, or (ii) use the general density-based score of \citet{semenova2021debiased} above, at the cost of estimating $f_{D\mid V}$ and its derivative.

This linearity assumption enables us to replace the complex density-based score with the ``partialing out'' transformation. By defining the residuals \(\tilde{Y}_i = Y_i - \E[Y_i \mid V_i]\) and \(\tilde{D}_i = D_i - \E[D_i \mid V_i]\), the relationship simplifies to:
\[
\tilde{Y}_i = \theta(X_i) \tilde{D}_i + \epsilon_i.
\]
This corresponds to a Neyman-orthogonal moment condition based solely on conditional means \(\eta = \{g(V), m(V)\}\):
\[
\psi(W; \theta, \eta) = \bigl( (Y - g(V)) - \theta(X)(D - m(V)) \bigr) (D - m(V)).
\]
This score allows us to estimate the CME without estimating densities, making it robust to small estimation errors in \(g(V)\) and \(m(V)\) while satisfying DML requirements. However, this stability comes at the cost of the linearity assumption. If the true relationship involves higher-order effects (e.g., \(Y\) depends on \(D^2\)), the PLRM is misspecified, and the estimator may fail to capture the true marginal effect.

Traditional parametric methods can be viewed as restricted forms of this partially linear model. For instance, if \(\theta(\cdot)\) is assumed to be constant, we recover the classic Robinson model. If \(\theta(X)\) is linear (i.e., \(\theta(X) = \beta_1 + \beta_3 X\)) and the baseline \(g(V)\) is linear in controls, we recover a standard linear interaction model. The semiparametric smooth varying coefficient model (SVCM) is also a special case where \(\theta(X)\) is smooth but the baseline effect of controls is restricted to be linear, i.e., \(g(V) = f(X) + Z\gamma(X)\).

The key innovation in this framework is the integration of this model with DML. Unlike the PO-Lasso approach discussed in the previous section, which typically trains and evaluates residuals on the same sample, DML employs cross-fitting. We partition the data into folds, training the high-dimensional nuisance functions \(\hat{m}(V)\) and \(\hat{g}(V)\) on training folds and evaluating the orthogonal residuals on the holdout fold. This procedure ensures that the nuisance estimation errors (converging at \(o_P(n^{-1/4})\)) do not bias the final inference. We provide pseudo-code below to clarify this process.

\begin{algorithm}[!ht]
\footnotesize
\caption{DML Estimation of CME (Continuous \(D\))}
\begin{algorithmic}[1]
\State \textbf{Inputs:}
\State \quad Data: \( \{(X_i, Z_i, D_i, Y_i)\} \) for \(i=1,...,n\)
\State \quad\quad \(D_i \in \mathbb{R}\) is a continuous treatment
\State \quad\quad \(V_i := (X_i, Z_i)\) are covariates (\(X_i\) is the moderator)
\State \quad\quad \(Y_i\) is the outcome
\State \quad Number of folds \(K\) (e.g., \(K=5\))
\State \quad Choice of basis functions \(p(X)\) (e.g., B-splines)
\State \textbf{Outputs:}
\State \quad Estimated function \( \hat{\theta}(x) = \widehat{\E}[\partial_d Y(d)\mid X_i=x]\)

\State \textbf{1) Partition data into \(K\) folds}
\State \quad Randomly split the indices \(\{1,...,n\}\) into \(K\) disjoint sets: \(I_1, ..., I_K\).
\State \quad Let \(I_k\) be the ``holdout" set and \(I_k^c = \{1,...,n\} \setminus I_k\) be the ``training" set.

\State \textbf{2) Estimate nuisance functions (cross-fitting)}
\For{\(k = 1\) to \(K\)}
\State \quad (a) On the training set \(I_k^c\):
\State \quad\quad * Fit outcome model: \( \hat{g}(v) \approx \E[Y \mid V=v] \)
\State \quad\quad * Fit treatment mean model: \( \hat{m}(v) \approx \E[D \mid V=v] \)
\State \quad (b) For each observation \(i\) in the holdout set \(I_k\):
\State \quad\quad * Compute predictions: \( \hat{g}^{(-k)}(V_i), \hat{m}^{(-k)}(V_i) \)
\EndFor

\State \textbf{3) Form orthogonal residuals}
\For{each fold \(k\), for each \(i\) in \(I_k\)}
\State \quad Residualize Outcome: \( \tilde{Y}_i = Y_i - \hat{g}^{(-k)}(V_i) \)
\State \quad Residualize Treatment: \( \tilde{D}_i = D_i - \hat{m}^{(-k)}(V_i) \)
\EndFor
\State \quad Collect the full set of residuals \( \{ (\tilde{Y}_i, \tilde{D}_i, X_i) \} \) for \( i=1,\dots,n \).

\State \textbf{4) Project residuals onto \(X\) to obtain \( \hat{\theta}(x) \)}
\State \quad To estimate the varying coefficient \(\theta(X)\), regress \(\tilde{Y}_i\) on the interaction \(\tilde{D}_i \cdot p(X_i)\):
\State \quad \( \hat{\beta} = \arg\min_{b} \sum_{i=1}^n \left( \tilde{Y}_i - \tilde{D}_i \cdot (p(X_i)^\top b) \right)^2 \)
\State \quad Construct estimator: \(\hat{\theta}(x) = p(x)^\top \hat{\beta}\)

\State \textbf{Return:}
\State \quad The function \( \hat{\theta}(x) \) = estimated CME curve.
\end{algorithmic}
\label{algorithm:continuous}
\end{algorithm}
\FloatBarrier

We provide an R code excerpt to implement the estimation using the \texttt{interflex} package. Using data from \textbf{Example~\ref{ex:adiguzel2023}}, we compare the results from the PO-Lasso estimator with those from the DML estimators. 

\begin{lstlisting}[language=R]
# R code excerpt

library(interflex) 

D="Dpop1000_asm_dr_3" #Change in Congestion
Y="Dakp_3" #Change in AKP vote share
X="lograyic09" #Property values (2009)
Z=c('Duniversity_3', 'Dodr_3', 'Dydr_3', 'Dpopulation_3', 'Dhospital_pri_3', 'Dhospital_pub_3')

## For code to generate plot (a), see last section 

## DML estimator with neural network machine learning method 
out.dml.nn<-interflex(estimator='DML', 
              data = d, model.y="nn", model.t = "nn",
              Y=Y,D=D,X=X, Z = Z, treat.type = "continuous", na.rm = TRUE, 
              vartype = "bootstrap",parallel = TRUE)
plot(out.dml.nn)

## DML estimator with random forests machine learning method 
out.dml.rf<-interflex(estimator='DML', 
              data = d, model.y="rf", model.t = "rf",
              Y=Y,D=D,X=X, Z = Z, treat.type = "continuous", na.rm = TRUE, 
              vartype = "bootstrap",parallel = TRUE)
plot(out.dml.rf)

## DML estimator with histogram gradient boosting machine learning method 
out.dml.hgb<-interflex(estimator='DML', 
              data = d, model.y="hgb", model.t = "hgb",
              Y=Y,D=D,X=X, Z = Z,treat.type = "continuous", na.rm = TRUE, 
              vartype = "bootstrap",parallel = TRUE)
plot(out.dml.hgb)
\end{lstlisting}


\begin{figure}
\caption{Replicating \cite{adiguzel2023out}\\
PO-Lasso vs DML Methods}\label{fig:Adiguzel2003}
\centering
\begin{subfigure}[b]{0.4\textwidth}
    \centering
    \includegraphics[width=\textwidth]{figures/chp3_Adiguzel_denoise.png}
    \subcaption{PO-Lasso}
\end{subfigure}\hspace{2em}
\begin{subfigure}[b]{0.4\textwidth}
    \centering
    \includegraphics[width=\textwidth]{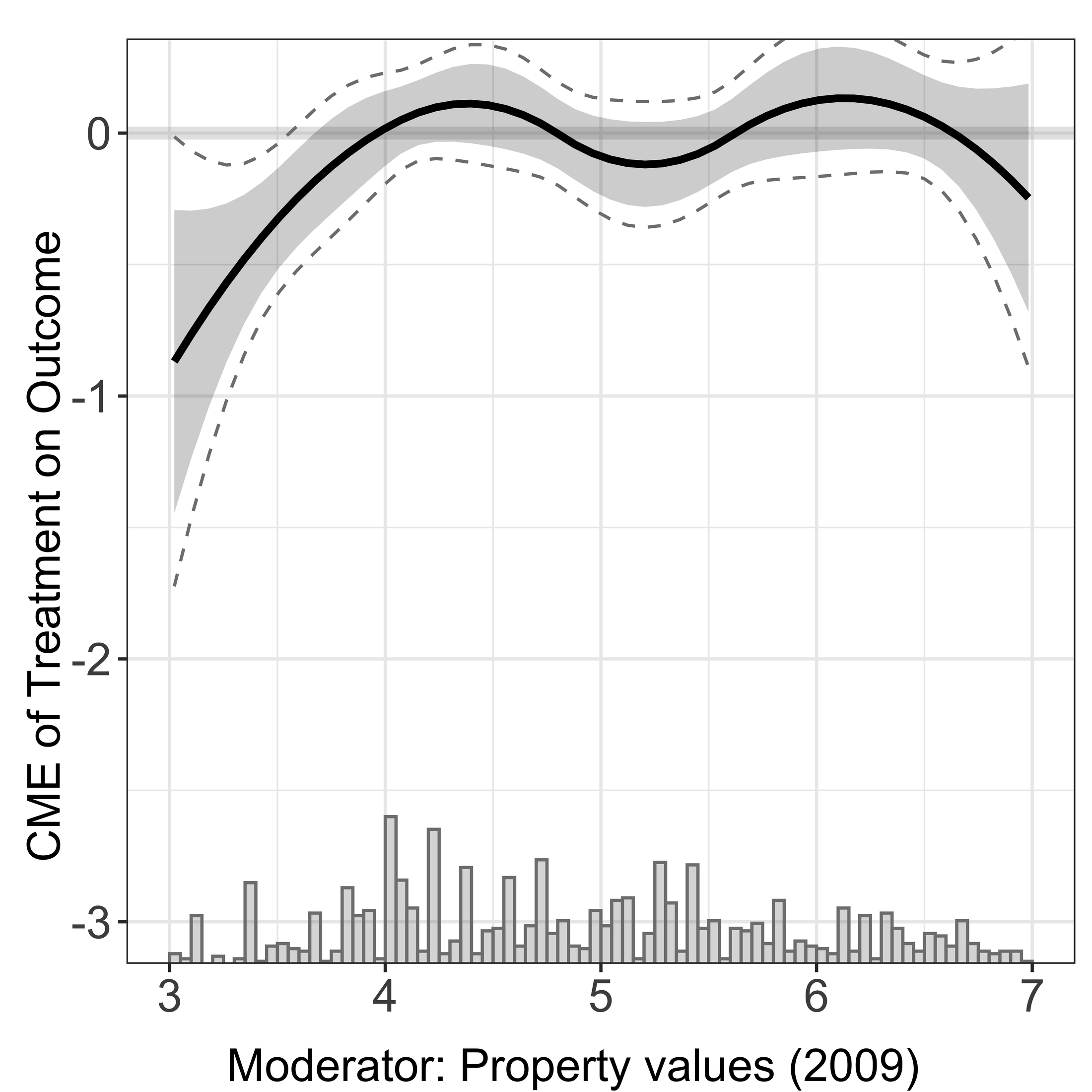}
    \subcaption{DML-NN}
\end{subfigure}\\
\begin{subfigure}[b]{0.4\textwidth}
    \centering
    \includegraphics[width=\textwidth]{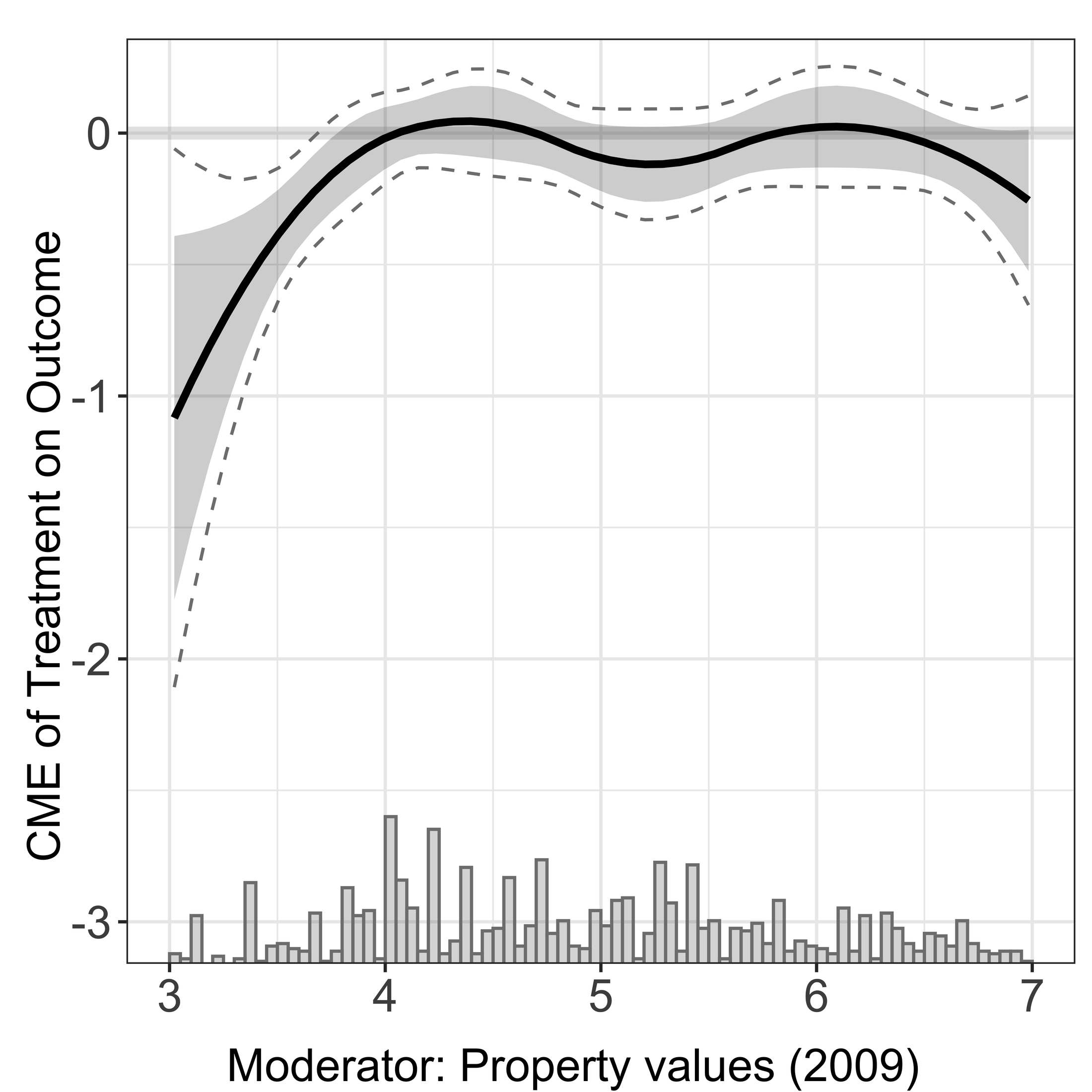}
    \subcaption{DML-RF}    
\end{subfigure}\hspace{2em}
\begin{subfigure}[b]{0.4\textwidth}
    \centering
    \includegraphics[width=\textwidth]{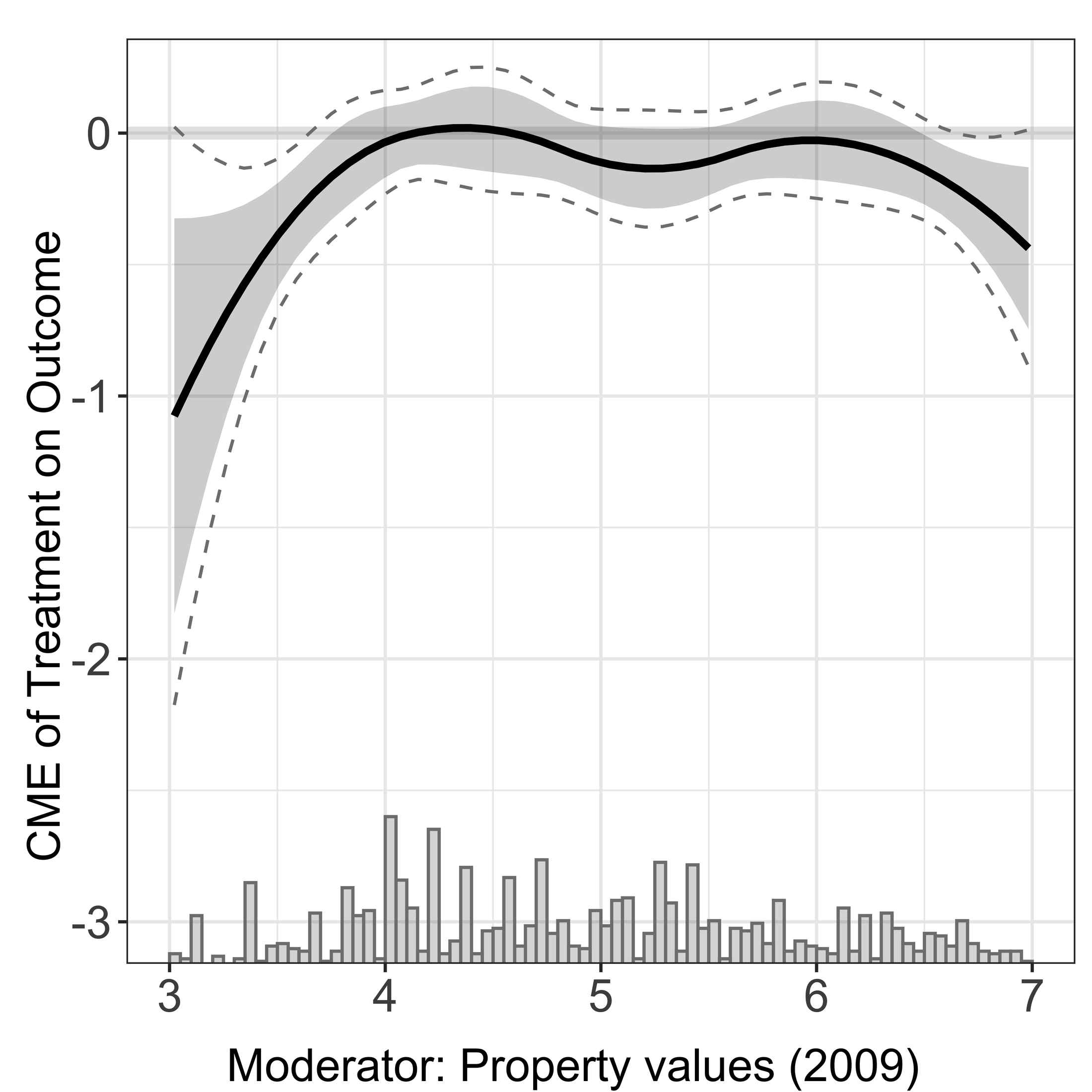}
    \subcaption{DM-HGB}
\end{subfigure}
\begin{minipage}{\linewidth}
{\footnotesize \emph{Notes:} The treatment $D$ measures change in congestion level, the outcome $Y$ measures change in the vote share for the incumbent AKP party. (a) CME estimates from PO-Lasso; (b)-(d) CME estimates from the DML estimator, with nuisance parameter estimated with NN, RF, and HGB, respectively. The gray ribbon show the 95\% pointwise confidence intervals, respectively. The dashed lines represent 95\% uniform confidence intervals.}  
\end{minipage}
\end{figure}

Figure~\ref{fig:Adiguzel2003} displays CME estimates of the effect of changes in congestion on AKP vote share across communities with varying median property values. The estimation sample is trimmed to the range \(3 \leq \log(\text{property value}) \leq 7\) due to sparse observations outside that interval. We have previous shown the CME estimates from PO-Lasso in the last section, now displayed in panel (a). Panels (b)-(d) show that the DML estimator produces similar patterns, with slightly bigger estimates when $X$ is small. They also yield wider confidence intervals.
\subsection{Inference}

Under the DML regularity conditions, the DML estimator $\hat{\theta}(x) = p(x)^\top \hat{\beta}$ is $\sqrt{N}$-consistent and asymptotically normal pointwise:
\begin{equation*}
    \sqrt{N}\,\big(\hat{\theta}(x)-\theta_0(x)\big)\ \leadsto\ N\!\big(0,\ \sigma^2(x)\big).
\end{equation*}

Let $q_i$ denote the vector of regressors used in the final projection step (e.g., $q_i = p(X_i)$ for binary treatment or $q_i = \tilde{D}_i \cdot p(X_i)$ for continuous treatment). Let $s_i$ be the orthogonal signal (the pseudo-outcome $\hat{\Lambda}_i$ or $\tilde{Y}_i$). Finally, let $u_i = s_i - q_i^\top\beta_0$ denote the second-stage residual.

We define the population design matrix $J_0 := \mathbb{E}[q_i q_i^\top]$ and the score variance matrix $\Omega_0 := \mathbb{E}[q_i q_i^\top u_i^2]$. The influence function for the scalar parameter $\theta(x) = p(x)^\top \beta_0$ is given by $\psi_i(x) = p(x)^\top J_0^{-1} q_i u_i$. Consequently, the asymptotic sandwich variance is:
\begin{equation*}
    \sigma^2(x) \;=\; p(x)^\top\,J_0^{-1}\,\Omega_0\,J_0^{-1}\,p(x).
\end{equation*}
This extends the scalar DML variance formula ($J_0^{-2}\mathbb{E}[\psi^2]$) to the estimation of a function via projection. Using cross-fitting, we construct the plug-in estimators:
\begin{equation*}
    \hat{J} \;=\; \frac{1}{N}\sum_{k=1}^{K}\sum_{i\in I_k} q_i q_i^\top,
    \qquad
    \hat{\Omega} \;=\; \frac{1}{N}\sum_{k=1}^{K}\sum_{i\in I_k} q_i q_i^\top \hat{u}_i^2,
\end{equation*}
where $\hat{u}_i$ are the empirical residuals. The pointwise variance estimator at $x$ is:
\begin{equation*}
    \hat{\sigma}^2(x) \;=\; p(x)^\top\,\hat{J}^{-1}\,\hat{\Omega}\,\hat{J}^{-1}\,p(x).
\end{equation*}
Let $\hat{\sigma}(x) = \sqrt{\hat{\sigma}^2(x)}$. An approximate $(1-\alpha)$ pointwise confidence interval for the CME at $x$ is:
\begin{equation*}
    \left[\,\hat{\theta}(x)\ \pm\ \Phi^{-1}(1-\alpha/2)\ \frac{\hat{\sigma}(x)}{\sqrt{N}}\,\right].
\end{equation*}

To obtain a confidence band that is valid simultaneously over a set $\mathcal{X}$, we apply a Gaussian multiplier bootstrap \citep{chernozhukov2013gaussian} to the projected influence function. Let $\{\xi_i^{\,b}\}_{i=1}^N$ be i.i.d. standard normal bootstrap weights, independent of the data. For each bootstrap draw $b=1,\dots,B$, we compute the process:
\begin{equation*}
    t^{*,b}(x) \;=\; \frac{1}{\sqrt{N}}\sum_{k=1}^{K}\sum_{i\in I_k} \xi_i^{\,b}\, \frac{p(x)^\top \hat{J}^{-1} (q_i \hat{u}_i)}{\hat{\sigma}(x)}.
\end{equation*}
Let $c_{1-\alpha}$ be the empirical $(1-\alpha)$-quantile of the maximal statistic $\{\sup_{x\in\mathcal{X}}|t^{*,b}(x)|\}_{b=1}^B$. The resulting simultaneous $(1-\alpha)$ confidence band is:
\begin{equation*}
    \left[\,\hat{\theta}(x)\ \pm\ c_{1-\alpha}\ \frac{\hat{\sigma}(x)}{\sqrt{N}}\,\right] \qquad \text{for all } x\in \mathcal{X}.
\end{equation*}

\subsection{Discrete Outcome}

Finally, we explore the application of AIPW and DML estimators to models with discrete outcomes. Traditional parametric approaches for binary or count data typically rely on link functions—such as logit, probit, Poisson, or negative binomial—to capture nonlinearity between predictors and the outcome. This introduces two key challenges. First, it becomes difficult to disentangle the nonlinearity imposed by the link function from the intrinsic nonlinearity of the CME function $\theta(x)$. Second, when the treatment \(D\) is continuous, the marginal effect \(\frac{\partial \mathbb{E}[Y]}{\partial D}\) depends on both \(X\) and \(D\) through the link function derivative, requiring an additional averaging step to recover the CME.

When the treatment is binary, the link function is naturally handled by the nuisance learners (e.g., using logistic regression for the propensity score and outcome model). However, when the treatment is continuous, the standard DML framework relying on the PLRM does not explicitly model the link function. In this setting, DML serves as a linear approximation strategy—analogous to the Linear Probability Model—treating the composite structure \(\mathbb{E}[Y|D,V]\) as a single nuisance component while preserving Neyman orthogonality for the target parameter.

\paragraph{Link Functions in Discrete Outcome Models}

In parametric generalized linear model (GLM) approaches, the conditional mean \(\mu = \mathbb{E}[Y \mid D, V]\) is connected to a predictor \(m(D, V)\) via an inverse link function \(g^{-1}(\cdot)\):
\[
\mathbb{E}[Y \mid D, V] = g^{-1}\bigl(m(D, V)\bigr).
\]
Common choices for the link function \(g(\cdot)\) include:
\begin{itemize}
    \item \textbf{Logit Link (Binary):} \(\mu = \frac{1}{1+\exp\{-m(D, V)\}}\).
    \item \textbf{Probit Link (Binary):} \(\mu = \Phi\bigl(m(D, V)\bigr)\), where \(\Phi\) is the standard normal CDF.
    \item \textbf{Poisson/Log Link (Count):} \(\mu = \exp\{m(D, V)\}\).
\end{itemize}
Classically, the predictor is modeled as a linear interaction:
\[
m(D,X,Z) = \beta_0 + \beta_1 D + \beta_2 X + \beta_3 (D \times X) + \gamma^\top Z.
\]

\paragraph{Challenges in Estimating the CME}

For a binary treatment, the CME is simply the difference in expectations, $\mathbb{E}[Y(1) - Y(0) \mid X]$. However, for a continuous treatment, the CME is defined as the expected partial derivative of the dose-response function. Under the composite model \(\mathbb{E}[Y\mid D=d, V=v] = g^{-1}(m(d,v))\), the chain rule yields:
\[
\frac{\partial}{\partial d} \mathbb{E}[Y\mid d, v] \;=\; g^{-1\,\prime}\!\big(m(d,v)\big) \cdot \frac{\partial m(d,v)}{\partial d}.
\]
Using the linear specification for $m(\cdot)$ above, this becomes:
\[
\frac{\partial \mathbb{E}[Y]}{\partial d} = g^{-1\,\prime}\!\big(\beta_0+\beta_1 d+\beta_2 x+\beta_3 d x + \dots\big)\cdot(\beta_1+\beta_3 x).
\]
Consequently, the CME at $X=x$ requires averaging this derivative over the distribution of the treatment $D$ and other covariates $Z$:
\[
\theta(x) \;=\; \mathbb{E}\!\left[\, g^{-1\,\prime}\!\big(m(D,x,Z)\big)\cdot(\beta_1+\beta_3 x)\ \Big|\ X=x\right].
\]
This highlights the complexity: the marginal effect depends on the baseline probability or count (via $g^{-1\,\prime}$), meaning that even with a linear latent predictor, the observed marginal effect varies with $D$.

\paragraph{DML Strategy for Discrete Outcomes}

The DML framework addresses discrete outcomes by concentrating on the estimation of nuisance functions rather than on specific link functions.

With a binary treatment, the target $\mathbb{E}[Y(1) - Y(0) \mid X]$ is identified using the standard AIPW signal. The adjustment lies in the choice of learners. The propensity score $\pi(V)$ is estimated using a classification method, such as regularized logistic regression or a random forests classifier. The outcome regressions $\mu(d,V)$ are also estimated using classification learners so that predicted values remain within the unit interval. The orthogonal score and subsequent projection onto basis functions of $X$ proceed as in the continuous-outcome case.

With a continuous treatment, the target becomes $\mathbb{E}\left[\left.\frac{\partial Y}{\partial D}\right\rvert X = x\right]$. Even if the data-generating process follows a GLM, such as Poisson, we implement DML through the partially linear regression model (PLRM). In this formulation, the estimator recovers a best linear approximation to the conditional mean function and yields an interpretable average partial effect without requiring direct estimation of density derivatives. We estimate $\mathbb{E}[Y\mid V]$ using flexible probabilistic regressions and $\mathbb{E}[D\mid V]$ using standard regressors, form residuals $\tilde{Y}$ and $\tilde{D}$, and regress $\tilde{Y}$ on $\tilde{D}$ interacted with basis functions of $X$. Although classification methods may be used to respect outcome bounds in nuisance estimation, the final projection step remains a linear regression of residuals, as implied by the orthogonality of the Robinson score.

\subsection{Summary}

DML extends the double robustness of AIPW to high-dimensional and nonlinear nuisance functions. Neyman orthogonality removes first-order bias from regularization, while cross-fitting controls overfitting, yielding $\sqrt{n}$-consistent and asymptotically normal CME estimates with flexible learners (NN, RF, HGB). The framework covers binary treatments via an AIPW-style score, continuous treatments via partially linear residualization, and discrete outcomes by adapting the nuisance learners while leaving the orthogonal projection unchanged.


\clearpage

\section{Monte Carlo Evidence}

In prior sections, we introduced the linear interaction, kernel, augmented inverse probability weighting with Lasso (AIPW-Lasso), and double/debiased machine learning (DML) estimators for the CME. We now compare these approaches using Monte Carlo simulations in a binary treatment setting. We consider three data-generating processes of increasing complexity. The first features linear covariate effects, allowing a benchmark comparison when only the CME is nonlinear. The second introduces nonlinear effects in additional covariates. The third incorporates high-order interactions and discontinuities in the nuisance functions to stress-test flexible learners. Section D of the Online Appendix supplements these exercises with an additional simulation on cross-validated tuning for DML learners.

\subsection{Simulation Study 1: Linear Covariate Effects}

In the first simulation study, the covariate affects the outcome linearly. In this setting, the kernel estimator (with an appropriately chosen bandwidth) can approximate the true marginal effect accurately. Our simulations show that, with sufficiently large sample sizes, DML methods perform comparably to the kernel estimator with cross-validated bandwidth. However, for smaller samples, DML methods can be more sensitive to noise, often resulting in underfitting of their machine learning models. 

To properly evaluate these methods, we define the outcome \(Y_i\) as
\[
Y_i = \theta(X_i)D_i + g(V_{i}) + \varepsilon_i,
\]
where \(\theta(X_i)\) is the CME of \(D_i\) on \(Y_i\), \(g(V_{i}) = \mathbb{E}[Y_i \mid D_i = 0, V_i]\) is the model for the untreated counterfactual outcome, and \(\varepsilon \sim \mathcal{N}(0, 1)\) is the i.i.d. error term. The covariates \(V_i = [X_i, Z_i]\) are i.i.d drawn from a uniform distribution: $X_i, Z_i$ are i.i.d drawn from $\textrm{Unif}(-\sqrt{3}, \sqrt{3})$. The CME is quadratic in $x$: $\theta(x) = x^2$, and $V_i$ enters the outcome model \(g(V_{i})\) linearly: $g(V_{i}) = 1 + X_i + 0.5Z_i$.

The propensity score \(\Pr[D_i = 1 \mid V_i]\) follows a logistic model:
\[
\pi(V_{i}) = \Pr[D_i = 1 \mid V_i] 
= \frac{\exp(0.5X_i + 0.5Z_i)}{1 + \exp(0.5X_i + 0.5Z_i)} =  \textrm{logit}^{-1}(0.5X_i+0.5Z_i),
\]
and the treatment variable \(D_i\) is sampled from $D_i \sim \text{Bernoulli}\bigl(\pi(V_{i})\bigr).$

Because the CME is nonlinear in \(X_i\), the linear interaction model is biased. Recall that the kernel estimator is specified as follows:
\[
 Y_i   = f(X_i) + h(X_i) D_i + Z_i^{\top} \gamma(X_i) + \epsilon_i
\]
With a properly selected bandwidth, it can approximate the DGP effectively with \(f(X_i) = 1 + X_i\), \(h(X_i) = X_i^2\), and \(\gamma(X_i) = 0.5\). Bandwidth selection via cross-validation, which targets out-of-sample predictive accuracy, helps to find an optimal balance between bias and variance.

The AIPW-Lasso and DML estimators both model the outcome and the propensity score as functions of \(V\):
\[
\eta = \Bigl\{\,
\underbrace{\mathbb{E}(Y_i \mid D_i=1,V_i)}_{\mu(1,V_i)},\,
\underbrace{\mathbb{E}(Y_i \mid D_i=0,V_i)}_{\mu(0,V_i)},\,
\underbrace{\mathbb{E}(D_i \mid V_i)}_{\pi(V_{i})}
\Bigr\},
\]
and construct the orthogonal signal of the outcome as
\[
\Lambda_i(\hat{\eta}) = \hat{\mu}(1,V_i) \;-\; \hat{\mu}(0,V_i) 
\;+\; \frac{D_i\,[Y_i - \hat{\mu}(1,V_i)]}{\hat{\pi}(V_{i})}
\;-\; \frac{(1-D_i)\,[Y_i - \hat{\mu}(0,V_i)]}{1 - \hat{\pi}(V_{i})}.
\]

With this DGP, the true nuisance functions are $\mu(1,V_i) = X_i^2 + 1 + X_i + 0.5Z_i$, $\mu(0,V_i) = 1 + X_i + 0.5Z_i$, and $\pi(V_{i}) = \textrm{logit}^{-1}(0.5X_i+0.5Z_i)$. Although these functions are not complex, small samples can still challenge machine learning model fitting, leading to underfitting or convergence issues. Moreover, the cross-fitting procedure in DML---which reduces the training data in each fold---may further amplify these issues when data are limited.

\begin{figure}[!th]
\caption{Comparison of Estimators with DGP1}
\begin{subfigure}[b]{0.45\textwidth}
    \centering
    \includegraphics[width=\textwidth]{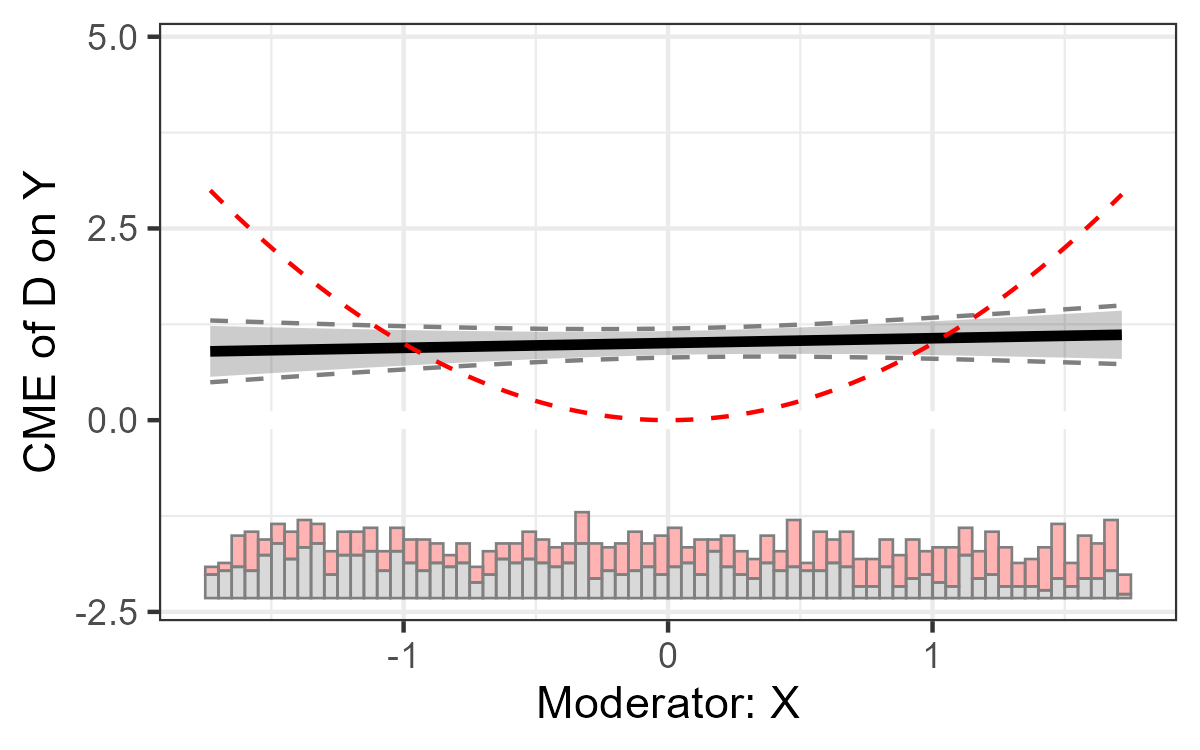}
    \caption{Linear ($n = 1{,}000$)}
    \label{fig:DGP1_lin1000}
\end{subfigure}
\hspace{0.02\textwidth}  
\begin{subfigure}[b]{0.45\textwidth}
    \centering
    \includegraphics[width=\textwidth]{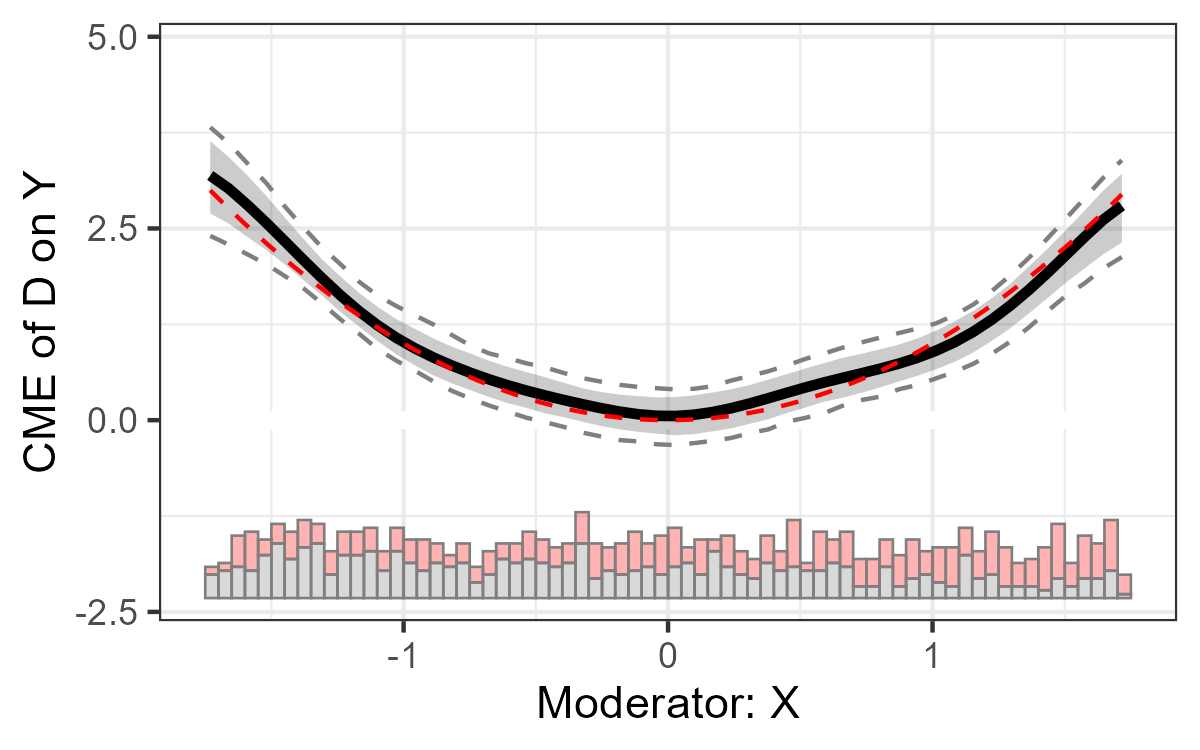}
    \caption{Kernel ($n = 1{,}000$)}
    \label{fig:DGP1_kcv1000}
\end{subfigure}\\
\begin{subfigure}[b]{0.45\textwidth}
    \centering
    \includegraphics[width=\textwidth]{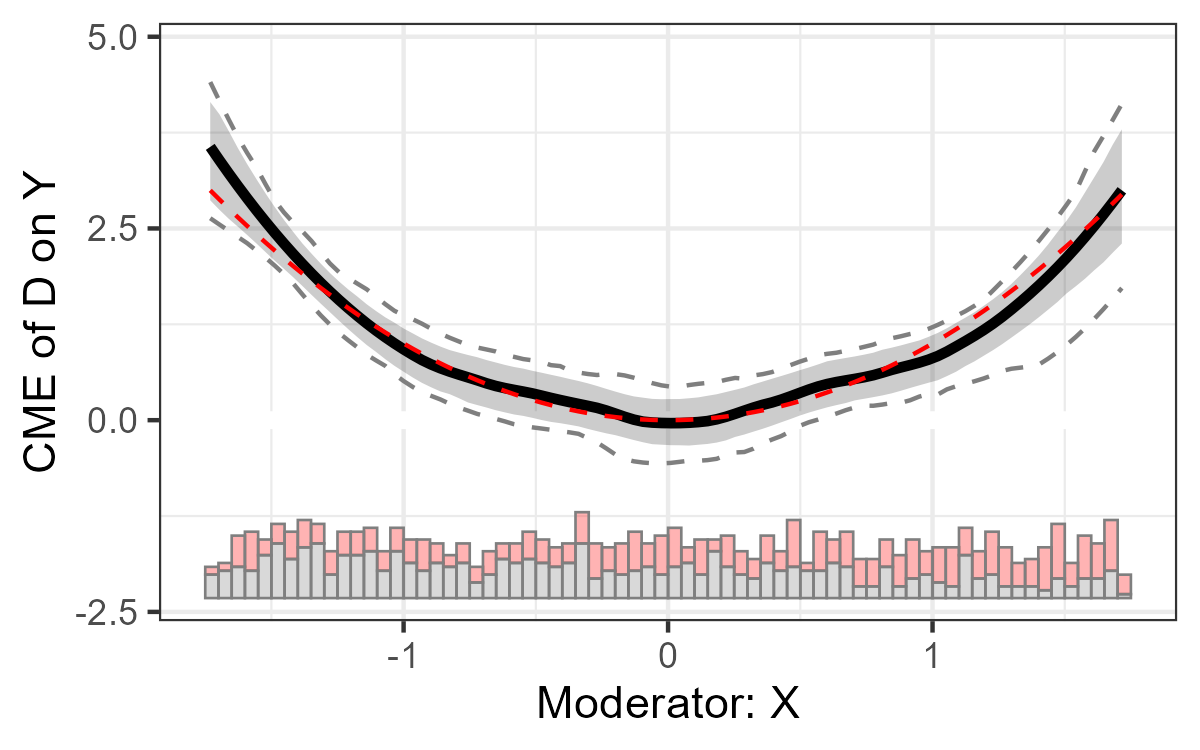}
    \caption{AIPW-Lasso ($n = 1{,}000$)}
    \label{fig:DGP1_aipw1000}
\end{subfigure}
\hspace{0.02\textwidth}  
\begin{subfigure}[b]{0.45\textwidth}
    \centering
    \includegraphics[width=\textwidth]{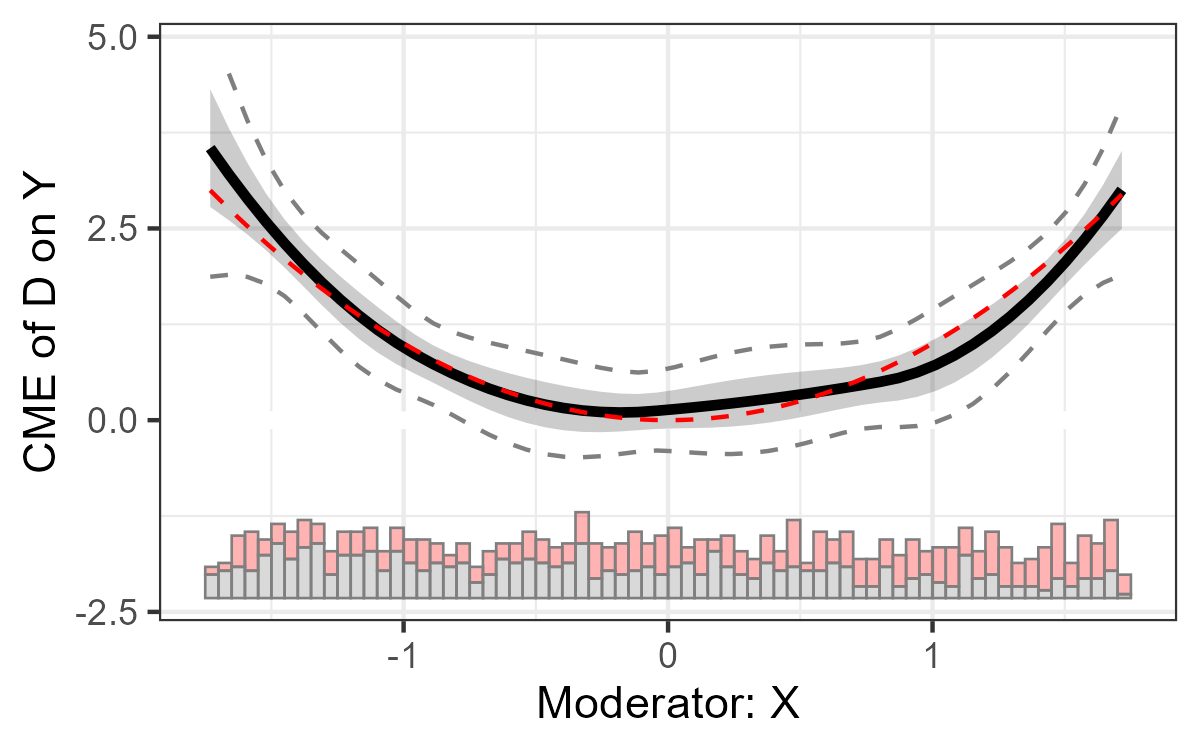}
    \caption{DML-NN ($n = 1{,}000$)}
    \label{fig:DGP1_nn1000}
\end{subfigure}
\\
\begin{subfigure}[b]{0.45\textwidth}
    \centering
    \includegraphics[width=\textwidth]{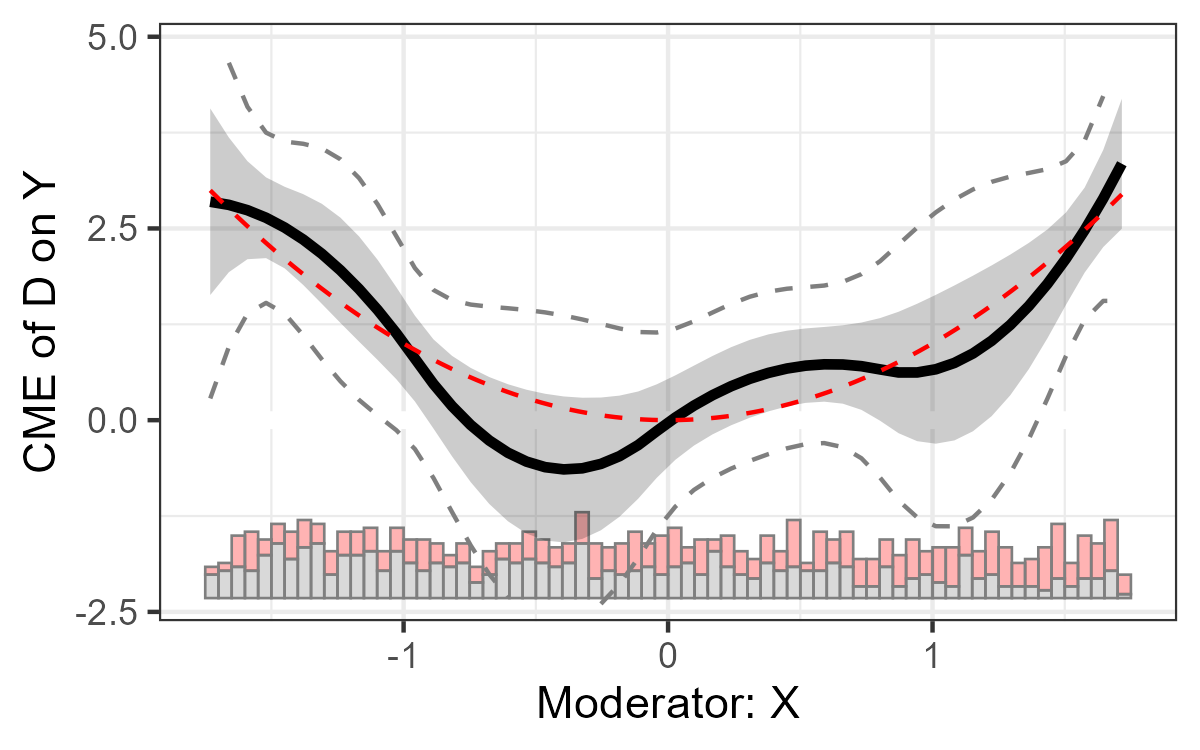}
    \caption{DML-HGB ($n = 1{,}000$)}
    \label{fig:DGP1_hgb1000}
\end{subfigure}
\hspace{0.02\textwidth}  
\begin{subfigure}[b]{0.45\textwidth}
    \centering
    \includegraphics[width=\textwidth]{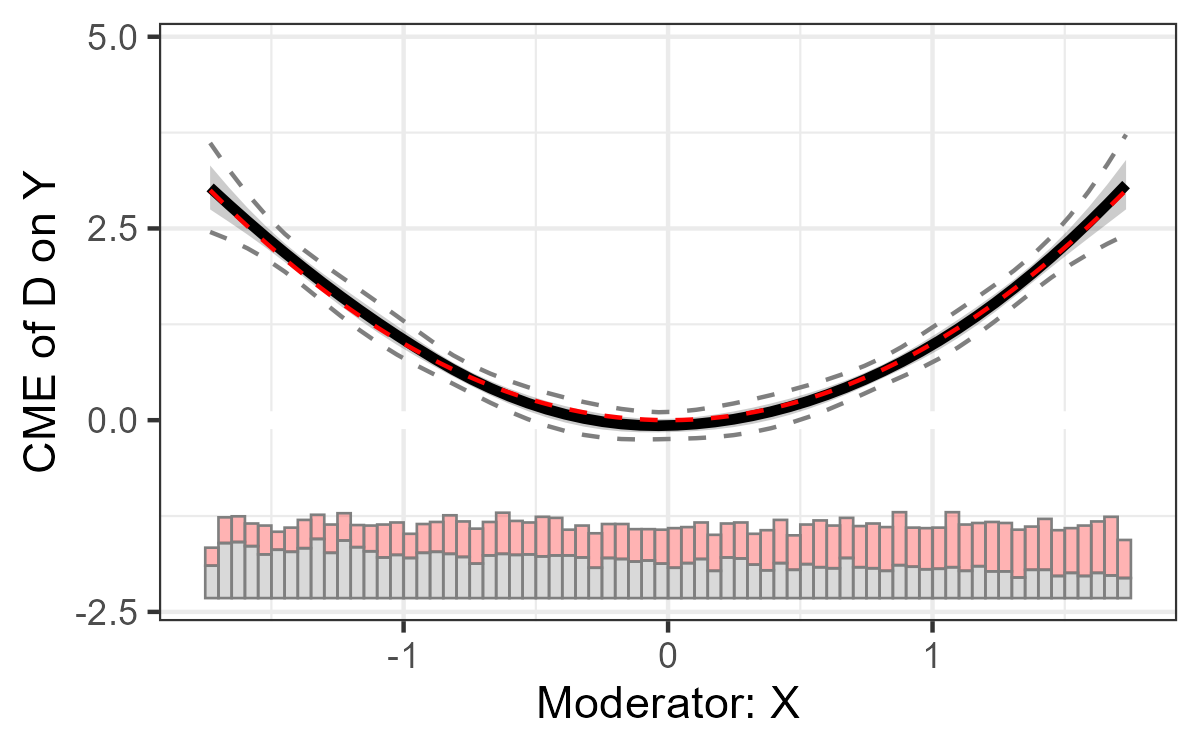}
    \caption{DML-HGB ($n = 10{,}000$)}
    \label{fig:DGP1_hgb10000}
\end{subfigure}
\label{fig:cme_estimation_dgp1}\\
{\footnotesize \textbf{Note}: In each figure: the red dashed line represents the true CME; the black solid line represents the estimated CME; the shaded area represents the pointwise confidence intervals, while the dashed gray lines represent the uniform confidence intervals.}
\end{figure}

In Figure~\ref{fig:cme_estimation_dgp1}, we compare the estimated CME to the true CME for a single simulation from the DGP. The first row presents model-based approaches, including the linear estimator (based on the linear interaction model) and the kernel estimator (with bandwidth selected via cross-validation). The second row shows results from the AIPW-Lasso estimator and the DML estimator using a neural network (DML-NN) for $n=1,000$. The third row presents the DML estimator using histogram gradient boosting (DML-HGB), showing results for both $n=1,000$ and $n=10,000$. At the bottom of each panel, we display the distribution of treated units (in red) and control units (in gray) along the moderator \(X\) using histograms.

Figure~\ref{fig:cme_estimation_dgp1} shows that the linear estimator exhibits substantial bias, while the kernel estimator closely approximates the true CME. AIPW-Lasso performs comparably to the kernel method, offering more stable estimation and narrower confidence intervals. DML-NN yields similar point estimates but with slightly wider confidence intervals. At \(n = 1{,}000\), DML-HGB performs poorly, with uniform confidence intervals that nearly always cover zero. However, at \(n = 10{,}000\), its performance improves markedly, and the estimated CME closely aligns with the truth. The DML estimator using a random forest (DML-RF), whose results are omitted here for brevity, performs similarly to DML-HGB.
These results suggest that both kernel and AIPW-Lasso estimators are well suited for estimating nonlinear CMEs when covariates enter the outcome model linearly and the sample size is moderate.

\begin{figure}[!t]
    \centering
    \caption{Performance of Different Methods: DGP1}
    \label{fig:method_comparison_dgp1}
    \begin{subfigure}[b]{0.48\textwidth}
        \centering
        \includegraphics[width=\textwidth]{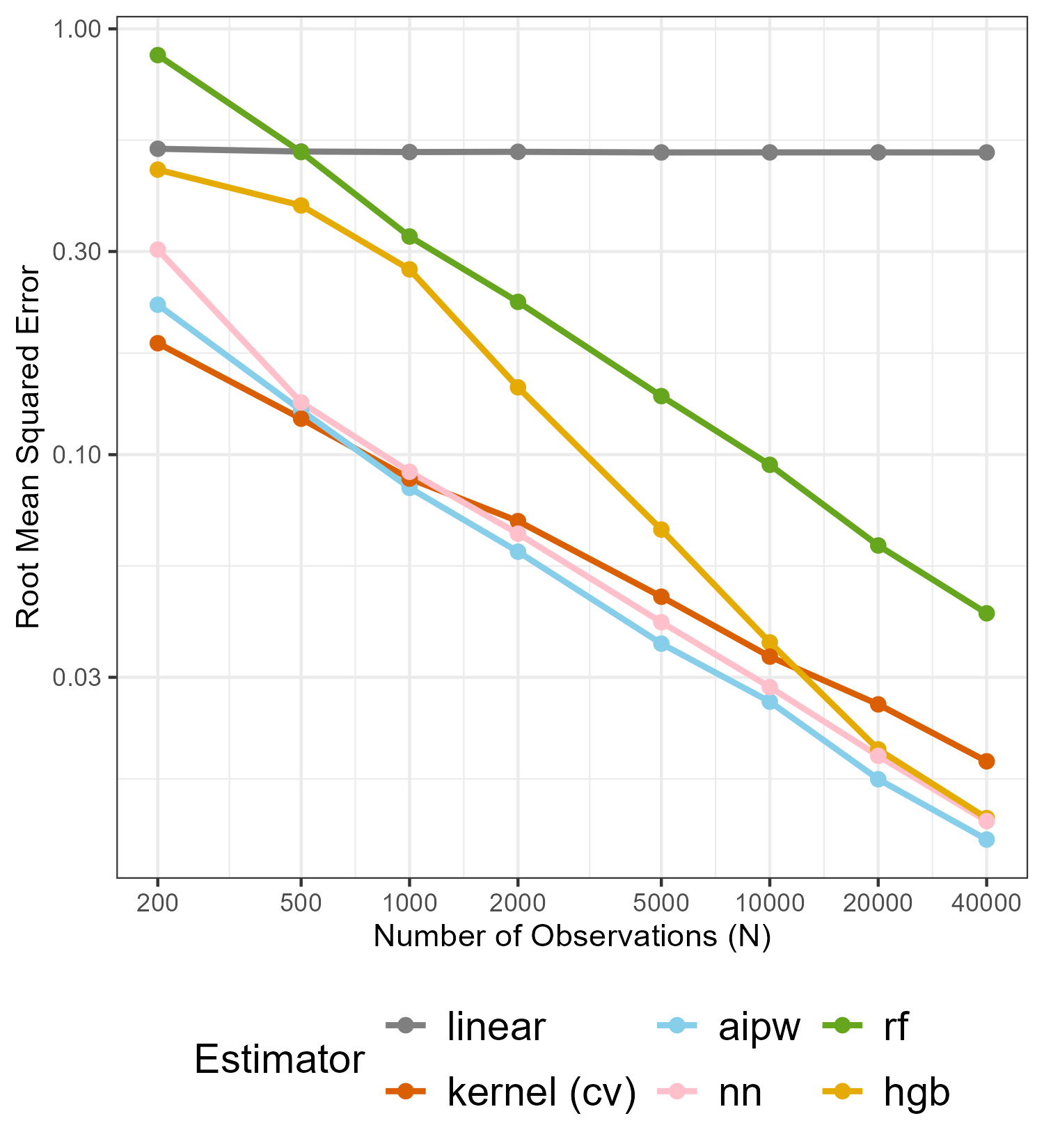}
        \caption{RMSE}
    \end{subfigure}
    \begin{subfigure}[b]{0.48\textwidth}
        \centering
        \includegraphics[width=\textwidth]{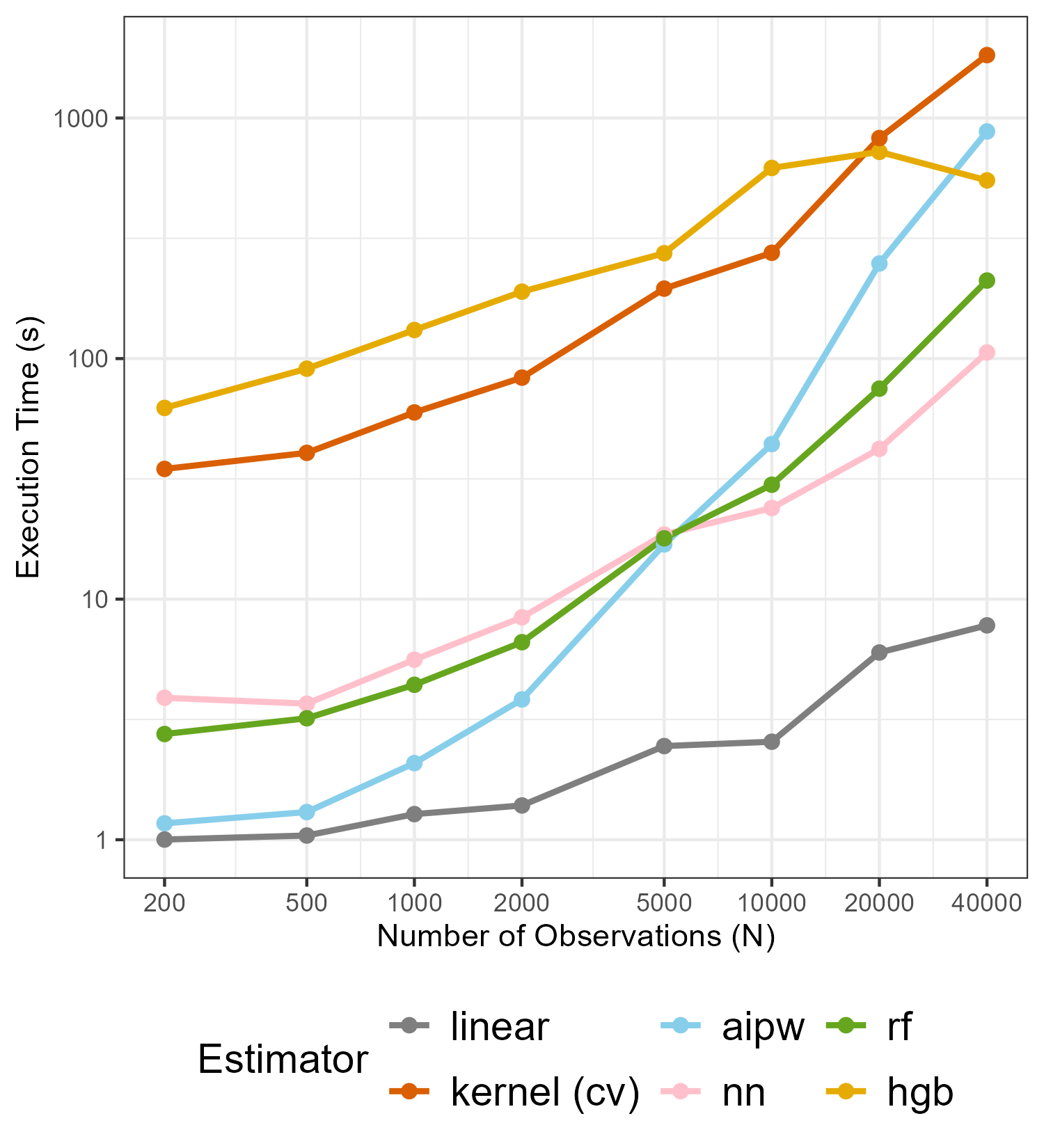}
        \caption{Execution time}
    \end{subfigure}
    \begin{minipage}{\linewidth}
{\footnotesize \textbf{Notes:} The above figures show how root mean squared error (left) and execution time (right) change as the sample size increases. Both axes in each figure are in log scale.}
    \end{minipage}    
\end{figure}

To evaluate the performance of each method in terms of accuracy and computational efficiency as the sample size increases, we compare two metrics in Figure~\ref{fig:method_comparison_dgp1}: the root mean squared error (RMSE) and execution time. RMSE measures the discrepancy between the true CME, \(\theta(X_i)\), and its estimate, \(\hat{\theta}(X_i)\), weighted by the probability density of \(X_i\), \(f_X(X_i)\). Since \(X_i\) is uniformly distributed in this simulation, the weights are constant over $X$. Execution time serves as a practical metric for computational efficiency.

Figure~\ref{fig:method_comparison_dgp1} presents results for the linear estimator, the kernel estimator, the AIPW-Lasso estimator, and DML methods using three different learners (NN, RF, and HGB). As expected, the linear estimator has the highest RMSE across all sample sizes due to substantial bias. The kernel estimator achieves relatively low RMSE but incurs higher computational costs because of bandwidth selection via cross-validation. The AIPW-Lasso estimator performs slightly better in RMSE and is generally less computationally demanding than the kernel method in small samples, although its runtime increases rapidly as the sample size grows. Among the DML estimators, DML-NN performs comparably to AIPW-Lasso and becomes more efficient in runtime when \(n > 5{,}000\). HGB achieves low RMSE once \(n\) exceeds 10{,}000, but its execution time grows substantially for larger \(n\). Random forest performs worse than both the NN and HGB in this setting. The kernel estimator is computationally more expensive than AIPW-Lasso and untuned DML methods, even in small samples, due to cross-validation.

With this DGP, the only source of nonlinearity is in the CME, while the covariate \(Z_i\) enters the outcome model linearly. Under these conditions, the kernel, AIPW-Lasso, and DML estimators can effectively capture the quadratic CME when the sample size is sufficiently large. However, with smaller samples, DML methods may suffer from instability and high variance, leading to worse performance. Therefore, when the sample size is small or moderate and computational efficiency is a key concern, AIPW-Lasso offers the best trade-off between accuracy and runtime. As the dataset grows, both AIPW-Lasso and DML methods become increasingly competitive in accuracy, albeit with greater computational cost.

\FloatBarrier

\subsection{Simulation Study 2: Nonlinear Covariate Effects}

We use the second simulation study to illustrate the advantages of the AIPW-Lasso and DML methods when covariates enter the outcome model nonlinearly. Based on the first simulation study, we modify only the outcome model as follows:
\[g(V_{i}) = 1 + X_i + e^{2Z_i + 2}.\] 
The CME, \(\theta(X_i)\), the propensity score, \(\pi(V_{i})\), and the distributions of \(X\) and \(Z_i\) remain unchanged. While the kernel estimator can effectively capture nonlinearity in the moderator itself, it struggles to accommodate nonlinearities in other covariates. Specifically, the kernel estimator restricts \(Z_i\) to enter linearly via \(\gamma(X_i)\), making it incapable of modeling terms such as \(e^{2Z_i + 2}\). By contrast, the AIPW-Lasso and DML methods can flexibly capture more complex effects of \(Z_i\) on \(Y_i\) through basis expansion or machine learning algorithms.

\begin{figure}[!t]
\caption{Comparison of Estimators with DGP2}
\label{fig:cme_estimation_dgp2}
\begin{subfigure}[b]{0.45\textwidth}
    \centering
    \includegraphics[width=\textwidth]{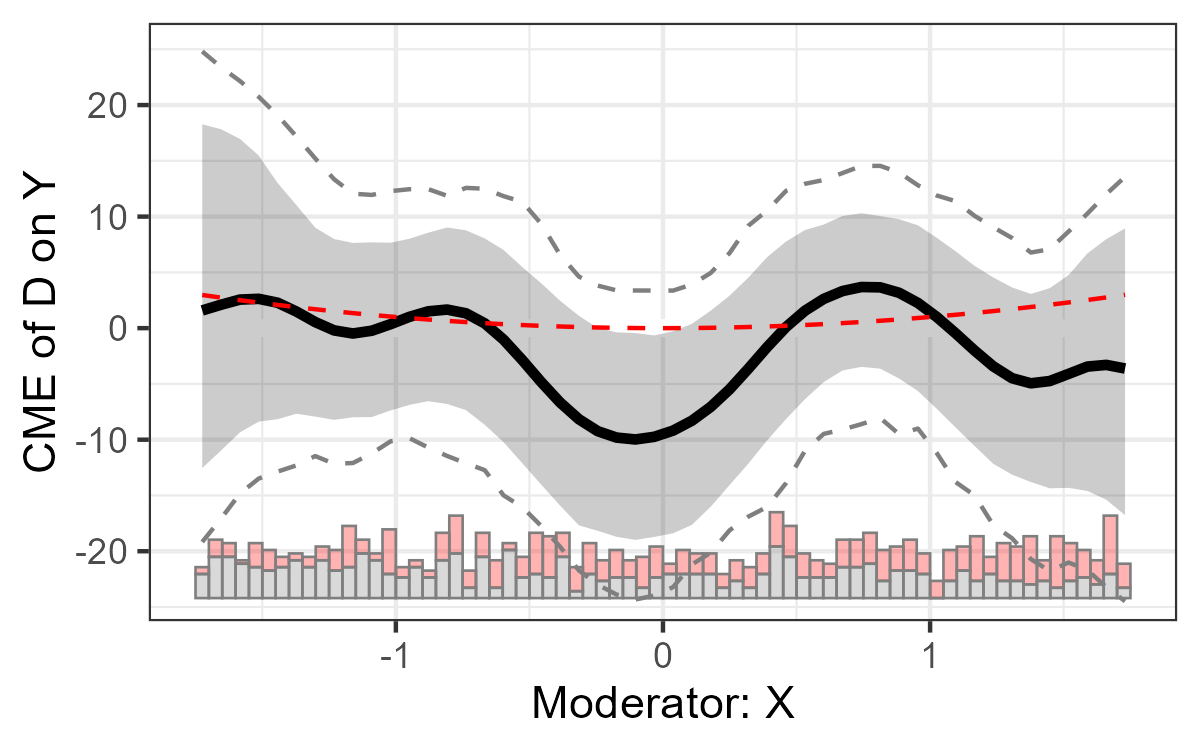}
    \caption{Kernel ($n = 1{,}000$)}
    \label{fig:DGP2_kcv1000}
\end{subfigure}
\hspace{0.02\textwidth}  
\begin{subfigure}[b]{0.45\textwidth}
    \centering
    \includegraphics[width=\textwidth]{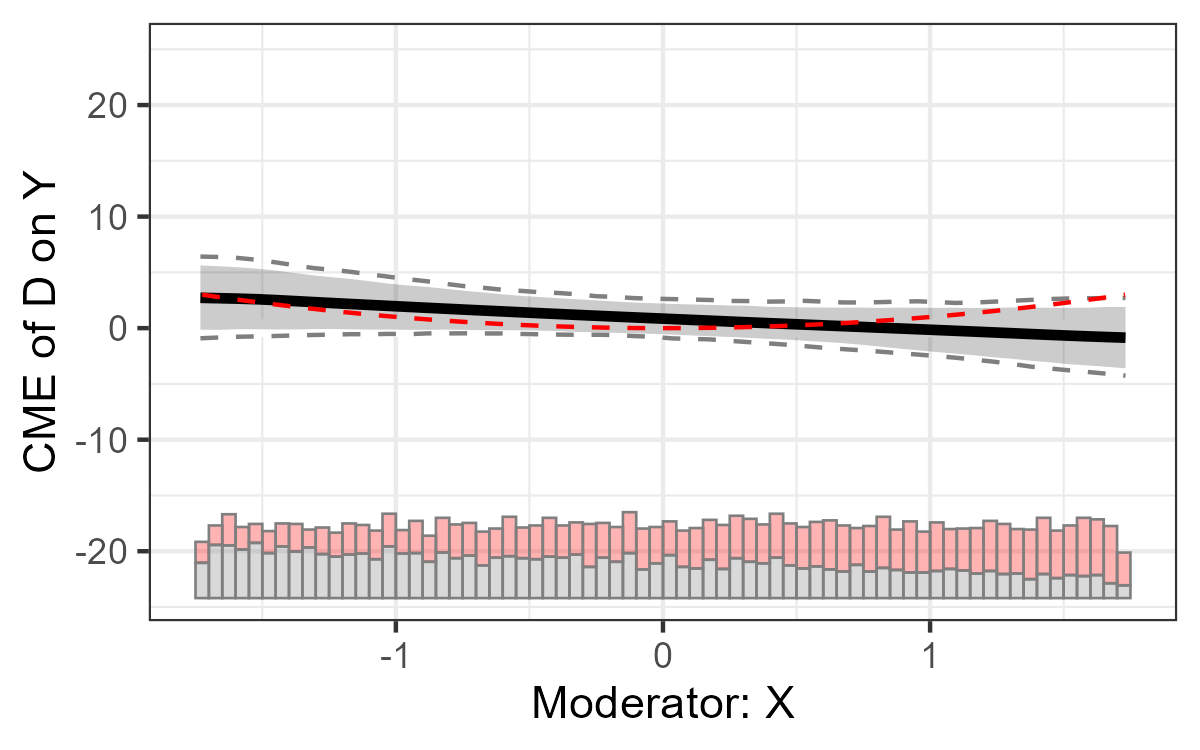}
    \caption{Kernel ($n = 10{,}000$)}
    \label{fig:DGP2_kcv10000}
\end{subfigure}\\
\begin{subfigure}[b]{0.45\textwidth}
    \centering
    \includegraphics[width=\textwidth]{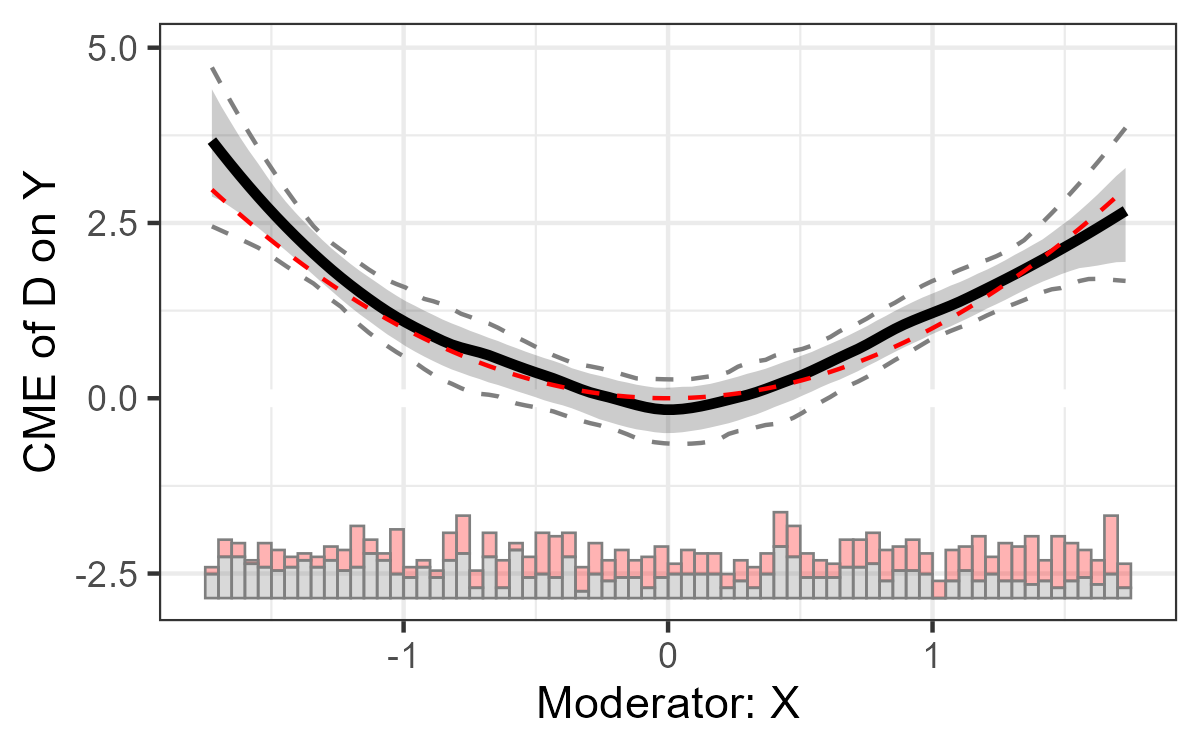}
    \caption{AIPW-Lasso ($n = 1{,}000$)}
    \label{fig:DGP2_aipw1000}
\end{subfigure}
\hspace{0.02\textwidth}  
\begin{subfigure}[b]{0.45\textwidth}
    \centering
    \includegraphics[width=\textwidth]{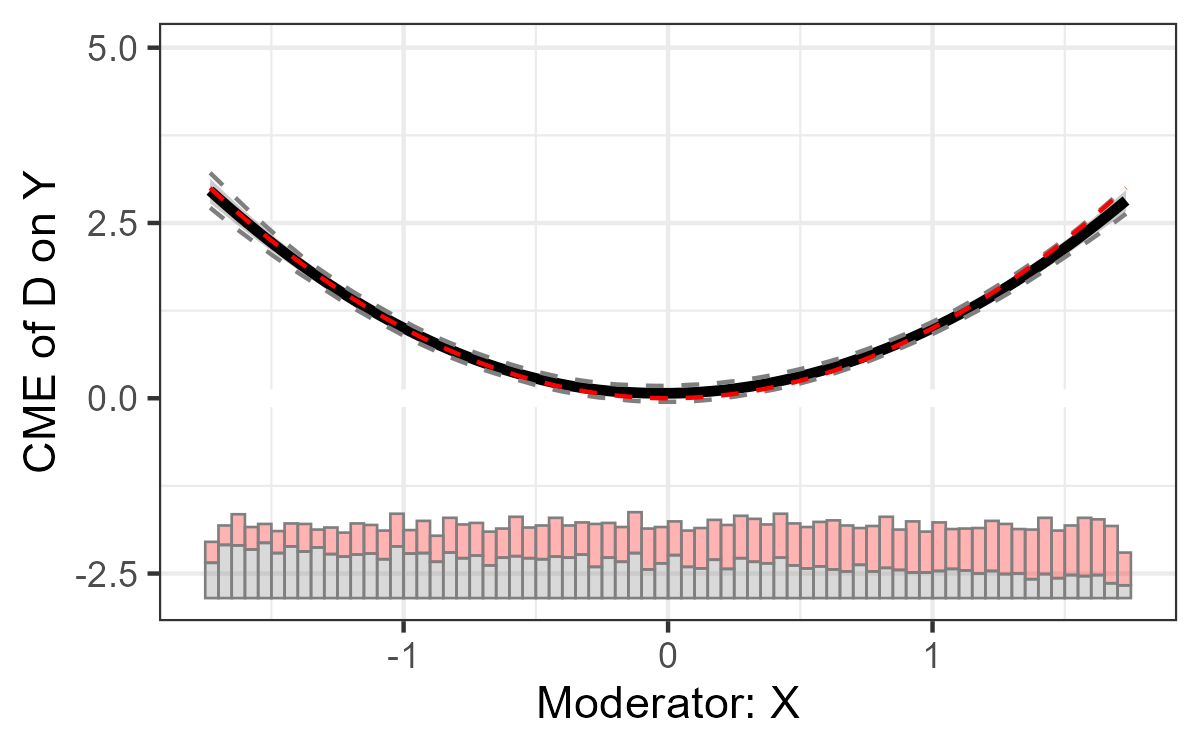}
    \caption{AIPW-Lasso ($n = 10{,}000$)}
    \label{fig:DGP2_aipw10000}
\end{subfigure}\\
\begin{subfigure}[b]{0.45\textwidth}
    \centering
    \includegraphics[width=\textwidth]{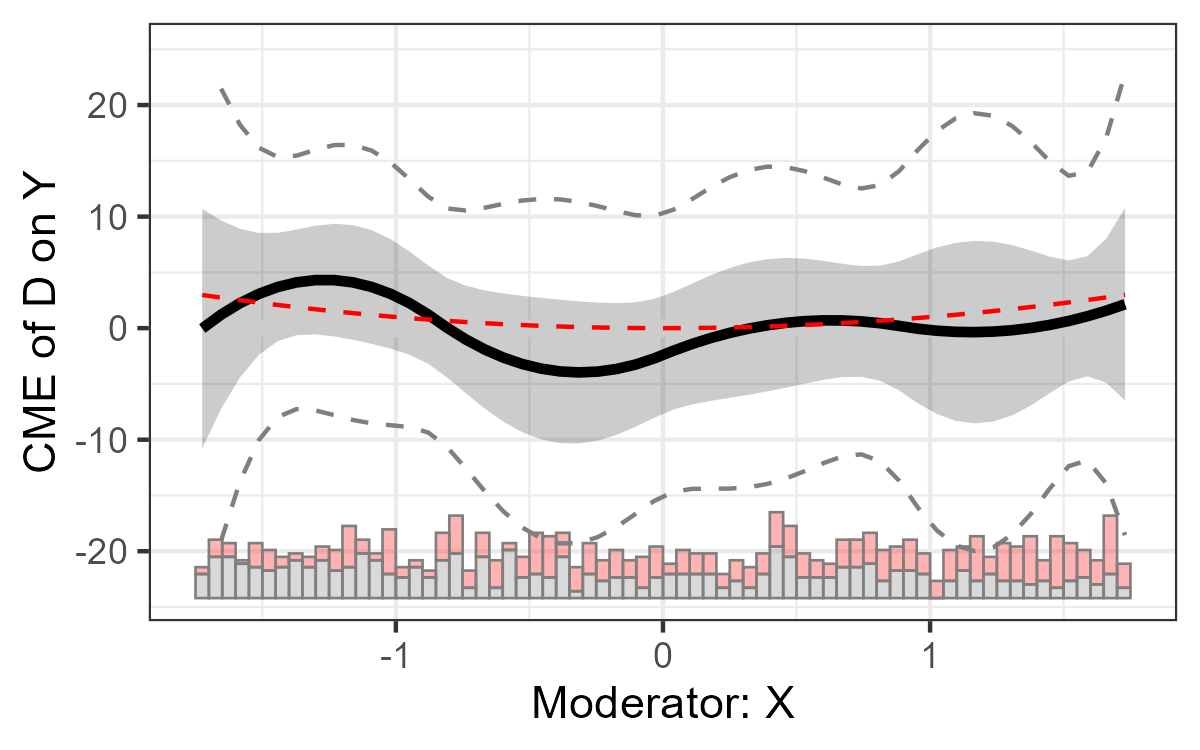}
    \caption{DML-NN ($n = 1{,}000$)}
    \label{fig:DGP2_nn1000}
\end{subfigure}
\hspace{0.02\textwidth}  
\begin{subfigure}[b]{0.45\textwidth}
    \centering
    \includegraphics[width=\textwidth]{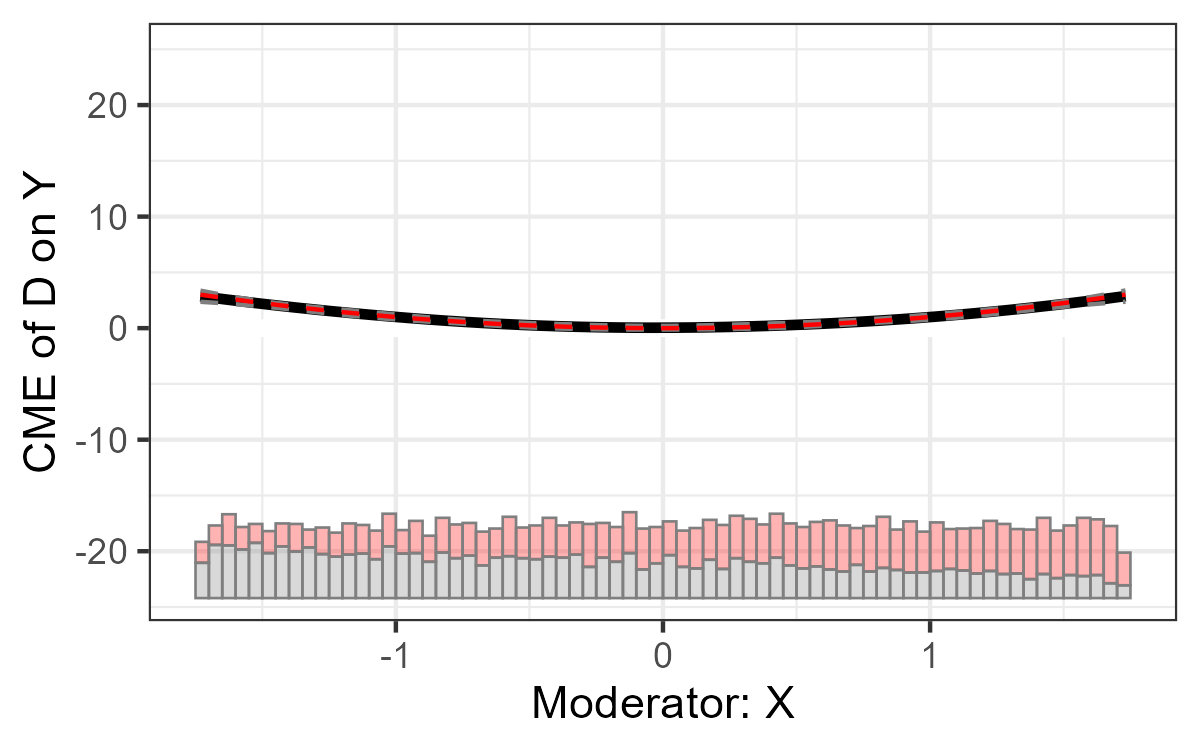}
    \caption{DML-NN ($n = 10{,}000$)}
    \label{fig:DGP2_nn10000}
\end{subfigure}\\
{\footnotesize \textbf{Note}: In each figure: the red dashed line represents the true CME; the black solid line represents the estimated CME; the shaded area represents the pointwise confidence intervals, while the dashed gray lines represent the uniform confidence intervals. At the bottom of each figure, we display histograms showing the distribution of treated units (in red) and control units (in gray) along the moderator \(X\).}
\end{figure}

In Figure~\ref{fig:cme_estimation_dgp2}, we compare the estimated CME with the true CME in a single simulated sample based on this DGP. The first row presents the kernel estimator (with bandwidth selected via cross-validation) for \(n = 1{,}000\) and \(n = 10{,}000\). The second row shows results from the AIPW-Lasso estimator for \(n = 1{,}000\) and \(n = 10{,}000\), using basis expansion and post-selection Lasso for variable selection. The third row presents results from DML-NN, with both the outcome and propensity score models fitted using NN (DML-RF and DML-HGB behave similarly). 

Unlike the first scenario, the kernel estimator fails to approximate the CME accurately at either sample size. For \(n = 1{,}000\), cross-validation selects a small bandwidth, resulting in highly noisy estimates. At \(n = 10{,}000\), the selected bandwidth is much larger, causing the kernel estimator to resemble the biased linear estimator and still fail to capture the nonlinearity introduced by \(e^{2Z_i + 2}\). In contrast, the AIPW-Lasso estimator performs remarkably well, even at \(n = 1{,}000\), likely because the exponential term is well approximated by the B-spline expansion of \(Z_i\). The DML method with the NN learner exhibits high variance at smaller \(n\), but its accuracy becomes comparable to that of AIPW-Lasso as \(n\) increases.

\begin{figure}[!h]
    \caption{Performance of Different Methods: DGP2}
    \label{fig:method_comparison_dgp2}
    \begin{subfigure}[b]{0.48\textwidth}
        \centering
        \includegraphics[width=\textwidth]{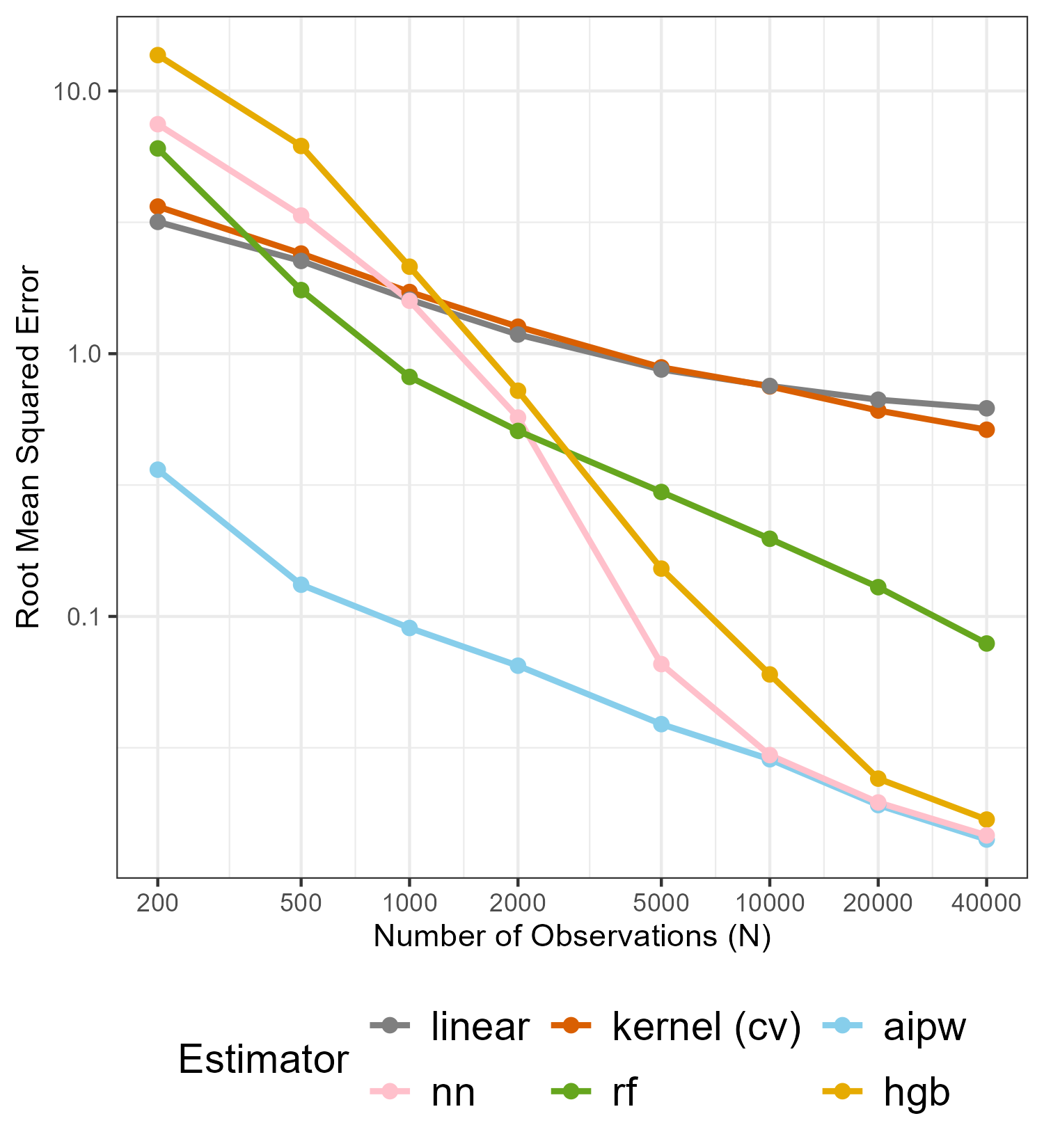}
        \caption{RMSE}
    \end{subfigure}
    \begin{subfigure}[b]{0.48\textwidth}
        \centering
        \includegraphics[width=\textwidth]{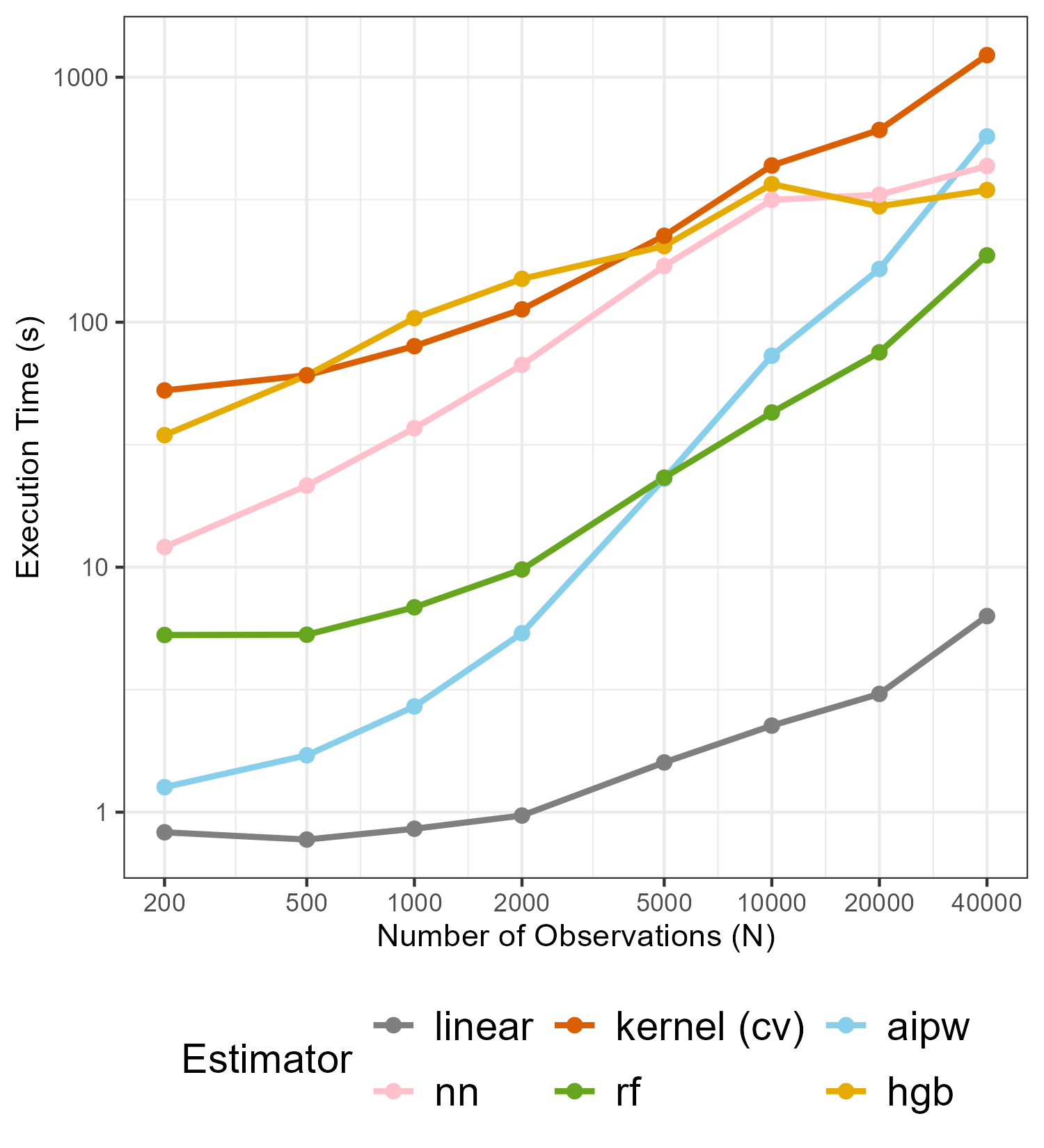}
        \caption{Execution time}
    \end{subfigure}
    \begin{minipage}{\linewidth}
    {\footnotesize\textbf{Notes:} The above figures show how root mean squared error (left) and execution time (right) change as the sample size increases. Both axes in each figure are in log scale.}
    \end{minipage}
\end{figure}

Figure~\ref{fig:method_comparison_dgp2} summarizes the RMSE and execution time of each estimator based on 200 simulation runs. We evaluate the linear estimator, the kernel estimator with cross-validation, the AIPW-Lasso estimator with basis expansion and post-selection Lasso, and DML methods using NN, RF, or HGB. Under this DGP, both the linear and kernel estimators exhibit substantial bias, even at large \(n\). By contrast, DML estimators improve significantly after approximately 3{,}000 observations, with NN and HGB achieving particularly low RMSE, albeit with higher computational cost than RF. Notably, the AIPW-Lasso estimator consistently attains low RMSE (as seen in Figure~\ref{fig:cme_estimation_dgp2}) with moderate computational overhead when the sample size is not large, although its runtime increases rapidly as \(n\) grows. 

In short, compared with the first study, the introduction of nonlinearity in \(g(V_i)\) through \(e^{2Z_i + 2}\) renders the kernel estimator ineffective at capturing this structure, resulting in biased CME estimates. In contrast, AIPW-Lasso and DML methods can flexibly accommodate nonlinearities in additional covariates. Therefore, when nonlinearities extend beyond the moderator, kernel methods become less suitable, and AIPW-Lasso or DML are more likely to provide greater accuracy and, in some cases, more efficient computation. AIPW-Lasso offers a strong advantage in accuracy at smaller sample sizes and remains computationally competitive at moderate sample sizes.

\FloatBarrier

\subsection{Simulation Study 3: High-Order Interactions and Discontinuities}

In our final simulation study, we evaluate the estimators under a substantially more complex data-generating process. This DGP is specifically designed to challenge the limits of each method by incorporating two particularly difficult features: (1) sharp discontinuities (piecewise-defined regions) in the outcome model and (2) high-order interactions in both the outcome and propensity score models.

These features are known to be difficult for estimators that rely on smooth approximation functions, such as the B-spline basis expansion used in our AIPW-Lasso implementation. In theory, tree-based DML learners like RF and HGB should be far more adept at capturing these structures, as their core algorithm is based on partitioning the covariate space. This simulation puts that theory to the test.

We simulate data with $p=5$ covariates ($X$ and four $Z$ variables), but we set $n=10{,}000$ and draw all covariates from a $\operatorname{Unif}(-\sqrt{3}, \sqrt{3})$ distribution. The true CME is kept simple, following the quadratic form $\theta(x)=x^2$.

The complexity is concentrated entirely in the nuisance functions. The baseline outcome model, $g(V_i)$, is constructed as a ``rule-based" piecewise function, which creates sharp discontinuities. The model starts with a highly nonlinear and discontinuous base function, $g_{base}(V_i) = 1 + X_i + f_2(Z_{i1})$, where $f_2$ is a piecewise function with three different cutoffs. The model then uses ``if-then" logic to replace this base function in different regions of the covariate space. For example, one functional form (involving $X_i^2$, $Z_{i1}Z_{i2}$, and $\sin(X_i+Z_{i1})$) is used if $Z_{i1} < 1 \text{ and }Z_{i2} < 0$, while a completely different function (involving $\text{abs}(X_iZ_{i1})$ and $Z_{i2}^3$) is used if $Z_{i1} < 1 \text{ and }Z_{i2} \geq0$. This process creates a complex patchwork of different nonlinear surfaces with hard breaks at the region boundaries. The propensity score model, $\pi(V_i)$, is similarly built by starting with a nonlinear baseline function (e.g., $X_i - X_i^2 + X_iZ_{i1} - 2\sin(X_i + Z_{i1})$) and then adding other high-order interaction terms (like $\cos(2X_i)$ or $X_iZ_{i1}Z_{i2}$) only if observations fall into specific regions defined by $Z_{i1}$ and $Z_{i2}$.

\begin{figure}[!h]
\caption{Estimated CME based on a single simulation following the Third DGP}
\begin{subfigure}[b]{0.44\textwidth}
    \centering
    \includegraphics[width=\textwidth]{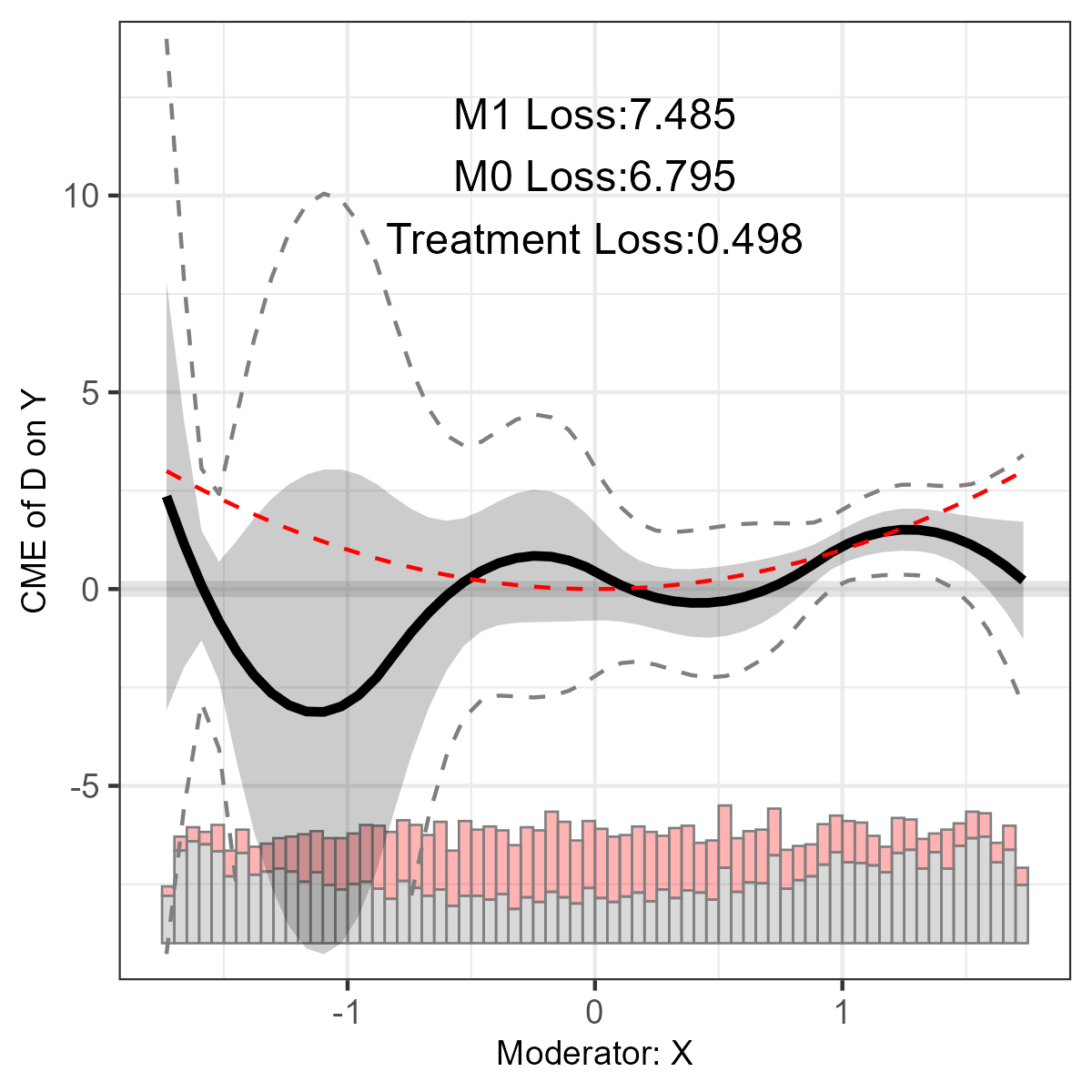}
    \caption{NN (tuned)}
\end{subfigure}
\hspace{0.01\textwidth}  
\begin{subfigure}[b]{0.44\textwidth}
    \centering
    \includegraphics[width=\textwidth]{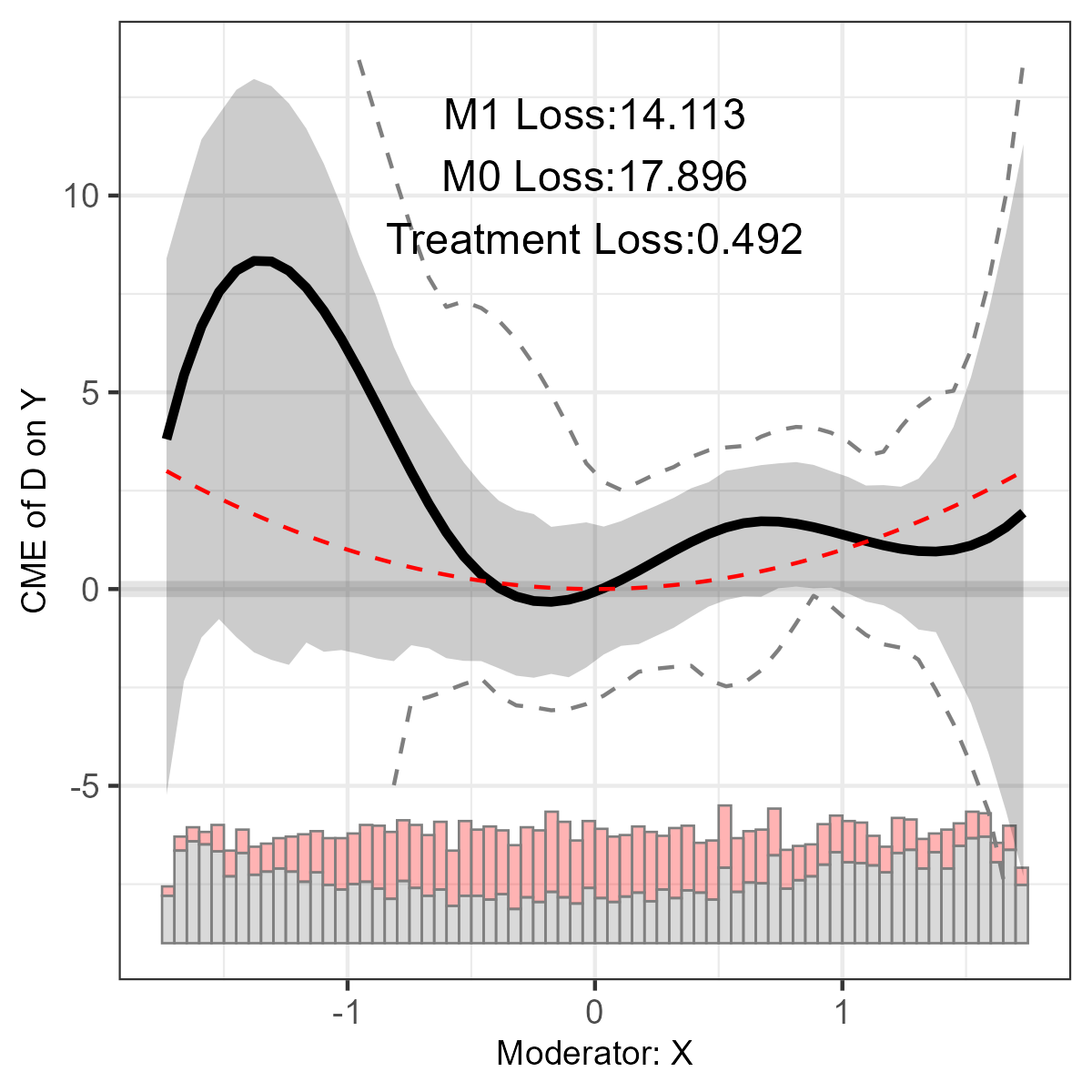}
    \caption{AIPW-Lasso}
\end{subfigure}
\\
\begin{subfigure}[b]{0.44\textwidth}
    \centering
    \includegraphics[width=\textwidth]{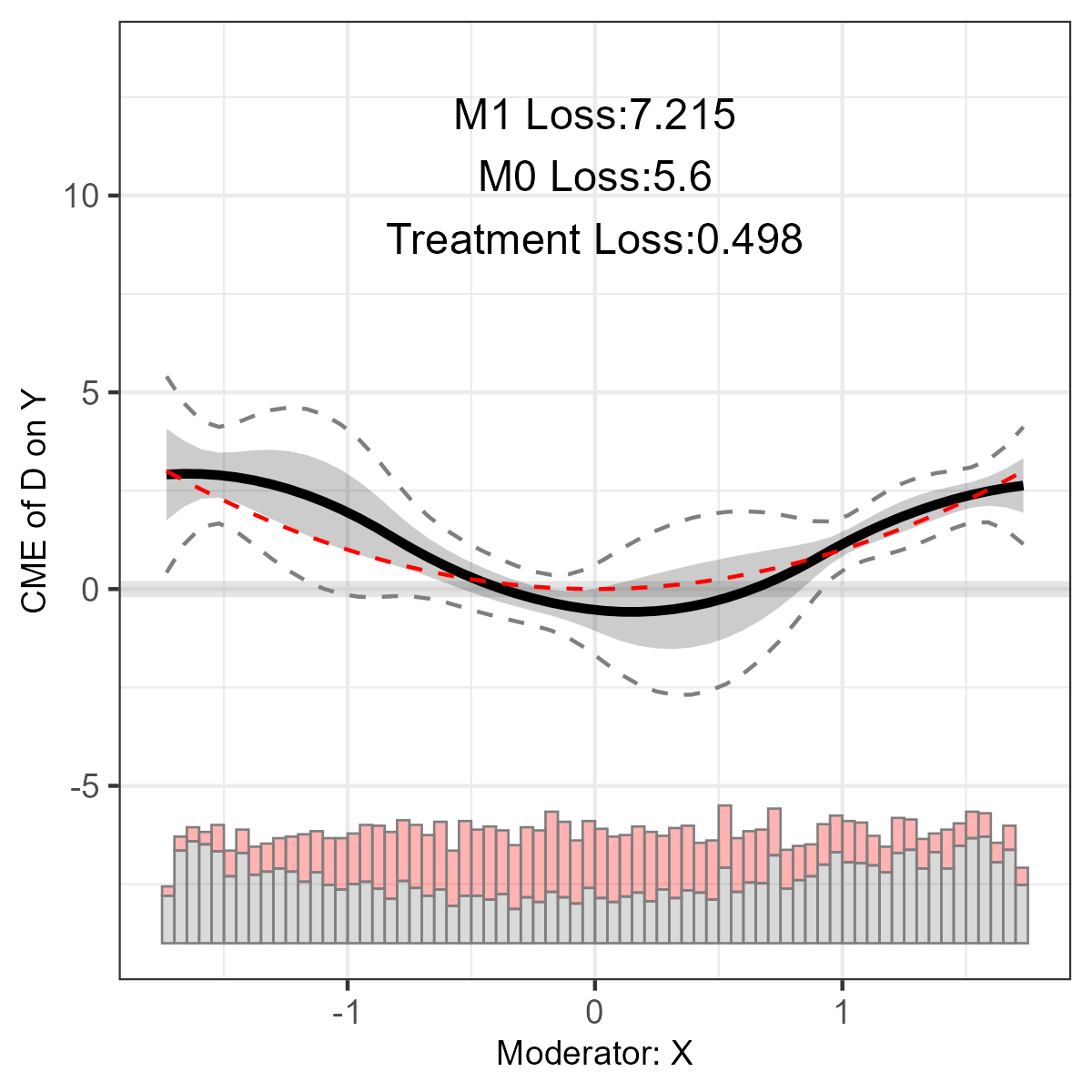}
    \caption{HGB (tuned)}
\end{subfigure}
\hspace{0.01\textwidth}  
\begin{subfigure}[b]{0.44\textwidth}
    \centering
    \includegraphics[width=\textwidth]{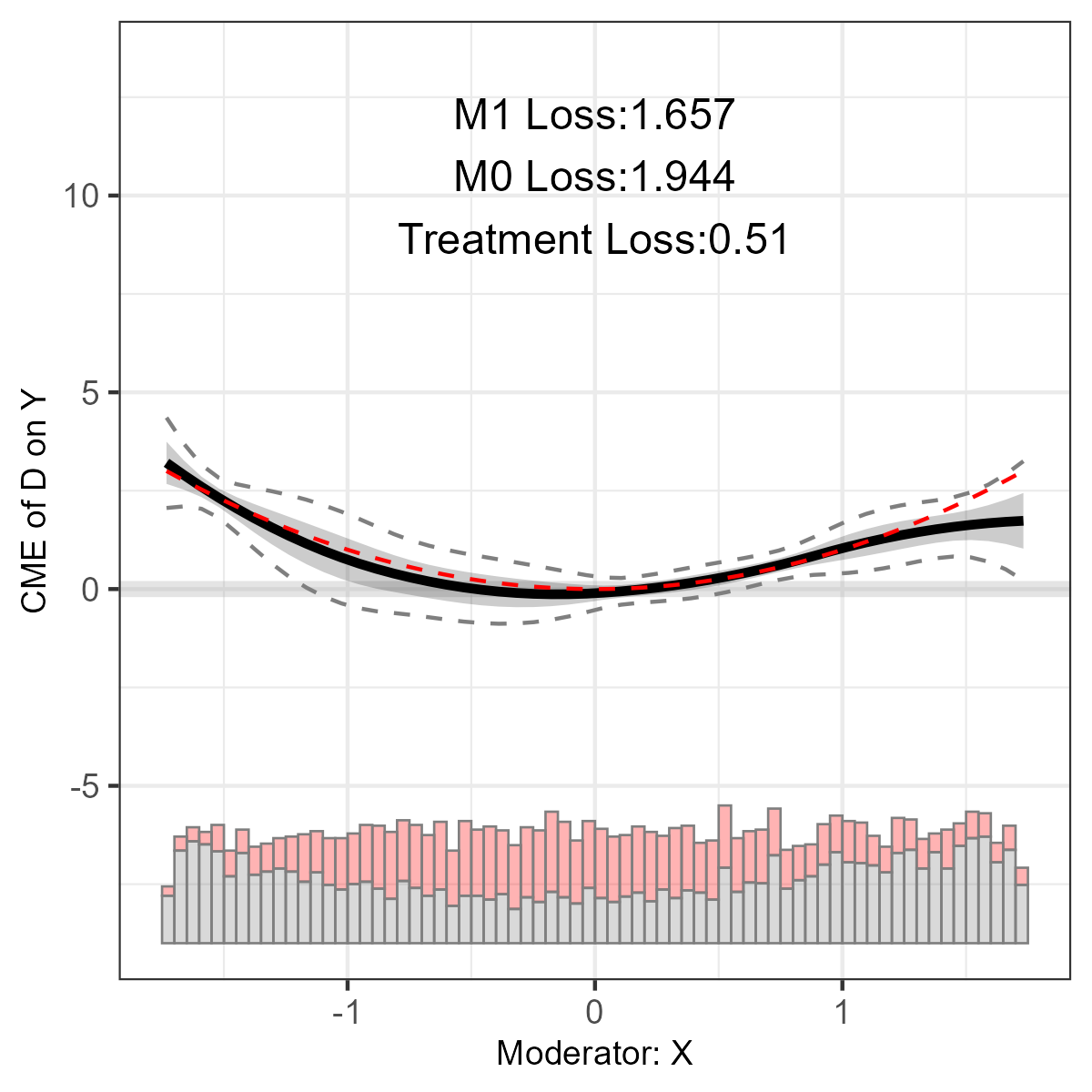}
    \caption{RF (tuned)}
\end{subfigure}
\label{fig:learner_cv_comparison_DGP4}\\
{\footnotesize \textbf{Note}: In each subfigure, the black solid line represents the estimated CME, while the red dashed line denotes the true CME. The shaded gray areas illustrate the pointwise confidence intervals, and the gray dashed lines depict the uniform confidence intervals. At the bottom of each subplot, a histogram displays the distribution of treated and control units across varying values of the moderator \(X\).}
\end{figure}

For DML learners, we use cross-validated hyperparameters; details are provided in Section D of the Online Appendix. 
In Figure~\ref{fig:learner_cv_comparison_DGP4}, we compare the performance of the AIPW-Lasso and the three DML methods with fine-tuned hyperparameters. We find that AIPW-Lasso, despite its strong performance in earlier, smoother settings, is unable to approximate the complex nuisance functions in this DGP and produces heavily biased CME estimates.

Interestingly, DML-NN also performs poorly. This suggests that standard neural network architectures, including the $(50,50)$ specification in the candidate pool, are too shallow to capture the sharp discontinuities of this DGP. This failure is reminiscent of the classic \textit{XOR} problem, where even simple nonlinear boundaries require sufficient network depth. Our DGP poses a substantially more demanding challenge, with multiple ``cliffs'' and high-order interactions. Neural networks are biased toward learning smooth functions and would likely require much deeper or more specialized architectures to approximate such highly piecewise structures.

In contrast, the tree-based DML learners---RF and HGB---perform markedly better. Their ability to partition the covariate space aligns well with the piecewise nature of the true functions, yielding substantially more accurate and reliable CME estimates.

\begin{figure}[!h]
    \caption{Performance of Different Methods: DGP4}
    \label{fig:method_comparison_dgp4}
    \begin{subfigure}[b]{0.44\textwidth}
        \centering
        \includegraphics[width=\textwidth]{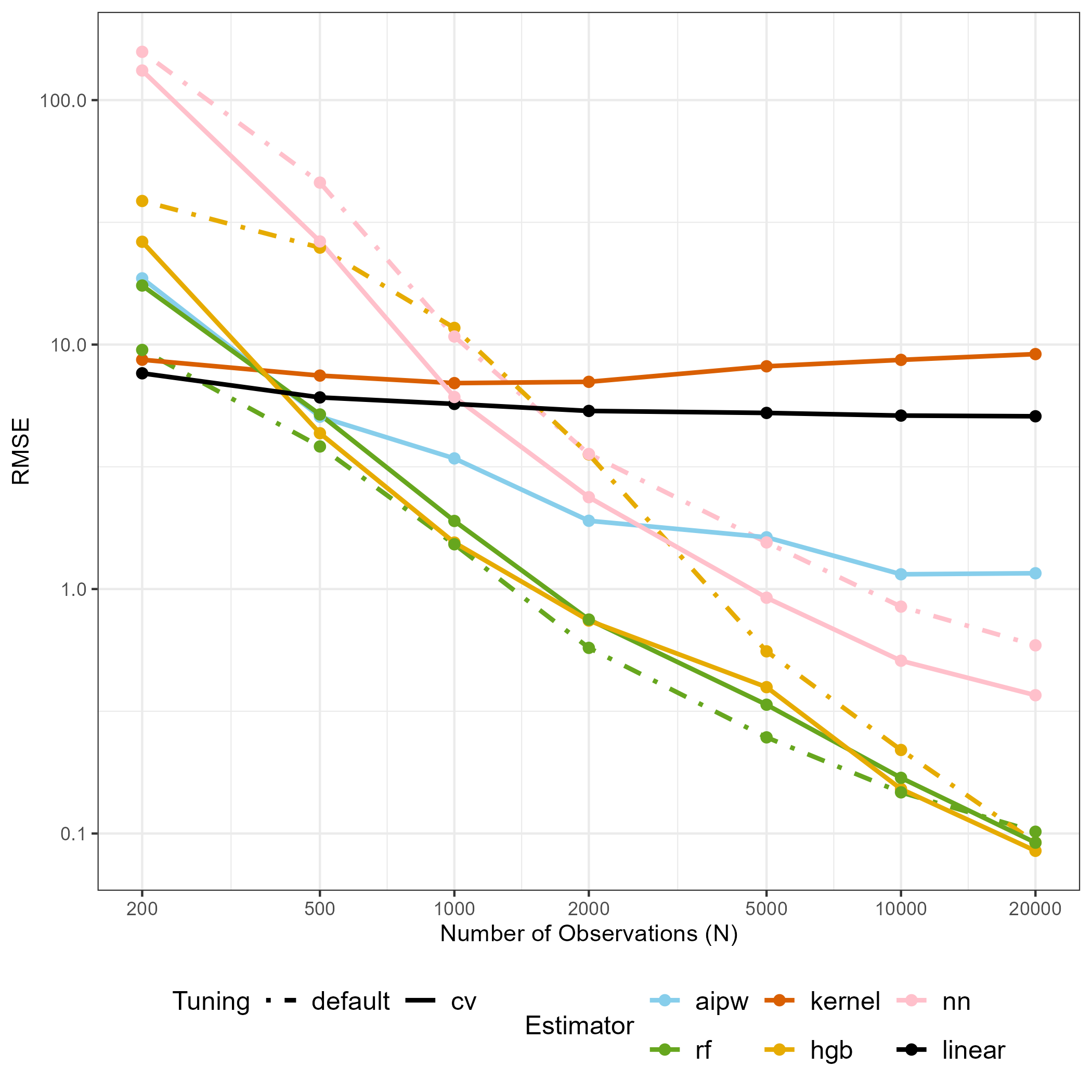}
        \caption{RMSE}
    \end{subfigure}\hspace{2em}
        \begin{subfigure}[b]{0.44\textwidth}
        \centering
        \includegraphics[width=\textwidth]{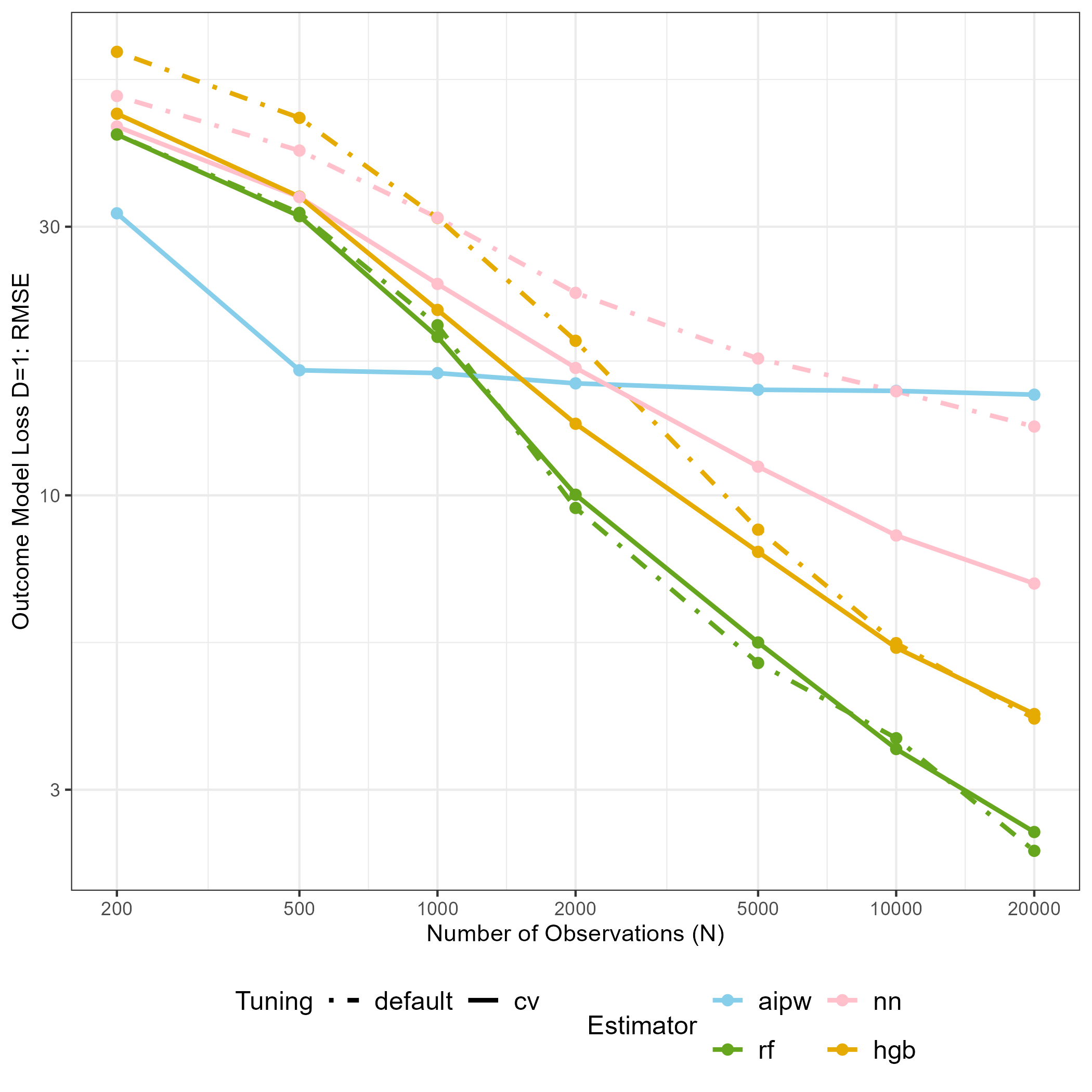}
        \caption{Loss Function (Outcome with $D=1$)}
    \end{subfigure}
    \begin{minipage}{\linewidth}
    {\footnotesize\textbf{Notes:} The above figures show how RMSE (left) and loss of the outcome model for treated units (right) change as the sample size increases. Both axes in each figure are in log scale.}
    \end{minipage}
\end{figure}

This pattern is confirmed in Figure~\ref{fig:method_comparison_dgp4}, which compares RMSE and nuisance loss functions across methods. In terms of RMSE, RF and HGB outperform AIPW-Lasso even at moderate sample sizes. The loss function plots clarify the source of this difference: the outcome model loss for AIPW-Lasso is substantially larger than that of RF and HGB and remains nearly flat as the sample size increases. This indicates that the B-spline basis expansion fails to capture the sharp, piecewise complexity of this DGP, regardless of sample size.

\subsection{Summary}

Across the simulation studies, three patterns emerge. First, the kernel estimator performs well when nonlinearity is confined to the moderator but is not robust to nonlinearities in other covariates. AIPW-Lasso provides a strong general benchmark in smooth settings, delivering low RMSE with moderate computational cost, particularly in small to medium samples.

Second, smooth approximation methods have clear limits. When nuisance functions contain discontinuities or rule-based interactions, AIPW-Lasso performs poorly, and DML becomes necessary. Among DML approaches, tree-based learners (RF and HGB) outperform NN in capturing piecewise structure.

Third, learner choice should match the form of complexity. Nuisance loss functions serve as useful diagnostics: out-of-sample losses for the outcome and propensity models track CME accuracy and help guide estimator selection and hyperparameter tuning.


\clearpage

\section{Conclusion}

Understanding conditional treatment effects—how the effect of a treatment $D$ on an outcome $Y$ varies with a moderating variable $X$—is essential in social science research. Despite significant advances, current methods face notable challenges, including unclear estimands, misinterpretation of statistical results, lack of sufficient overlap in empirical data, rigid assumptions on functional forms, and difficulty with complex settings like discrete outcomes. This Element introduces a robust methodological framework that overcomes these limitations and ensures valid statistical inference for the key estimand, the conditional marginal effect (CME).

The linear interaction model is still widely used in empirical research. Despite its popularity, this model often suffers from inadequate common support, misspecifications, and multiple testing issues. To mitigate these issues, in Section 2, we advocate for diagnostic tools such as inspecting raw data and the binning estimator. Moreover, we discuss methods such as uniform confidence intervals via bootstrapping to address multiple comparisons and improve inferential validity. The semiparametric kernel estimator proposed by \citet{hainmueller2019much} relaxes functional form assumptions. We further improve its robustness by incorporating adaptive kernels and fully moderated specifications.

In Section 3, we advance beyond classical approaches by introducing augmented inverse propensity score weighting (AIPW). The AIPW estimator for the CME combines outcome modeling and propensity score weighting, providing double robustness—maintaining consistency if either the propensity score model or outcome model is correctly specified. It takes three steps: (i) estimating the outcome and propensity score models, (ii) constructing the signals for treatment effects; (iii) smoothing over the moderator $X$. Compared with inverse propensity score weighting (IPW) alone, AIPW improves estimation stability and reduces variance. To accommodate high-dimensional covariates and relax functional form assumptions, we further introduce basis expansions and Lasso regularization techniques, which significantly stabilize estimation and improve precision. Finally, we extend AIPW to continuous treatments through partialing-out Lasso (PO-Lasso). 

In Section 4, we present double/debiased machine learning (DML) as a powerful extension suitable for high-dimensional and complex nuisance parameters. DML achieves robust estimation through Neyman orthogonality, which ensures that small estimation errors in nuisance functions do not bias CME estimates to first order. This framework extends the double robustness of AIPW-Lasso and PO-Lasso by leveraging flexible machine learning methods, such as neural networks, random forests, and gradient boosting, to model complex relationships between covariates and the outcome and treatment. The combination of orthogonalization, which insulates estimators from regularization bias, and cross-fitting, which prevents overfitting by splitting the data, ensures valid inference at a $\sqrt{n}$convergence rate.  We illustrate DML for binary and continuous treatments with empirical examples from political science, including orthogonal signal construction, residualization strategies, and application of the \texttt{interflex} package. We also show that DML can accommodate empirical applications with discrete outcomes.

Moreover, in Section 5, we provide Monte Carlo evidence comparing kernel-based methods, AIPW-Lasso, and DML estimators. Simulation studies indicate that while kernel estimators effectively handle simple nonlinearities, they falter with complex covariate dependencies. In contrast, AIPW-Lasso and DML successfully model intricate nonlinearities, particularly in higher-dimensional settings, though their performance crucially depends on sample size and effective hyperparameter tuning. Tuning, while computationally costly, often significantly improves accuracy of the DML estimators, especially with complex learners like neural networks. However, simpler learners such as random forests show only marginal gains or even slight declines after tuning. 

Finally, we offer the following practical recommendations for empirical researchers interested in estimating and interpreting conditional relationships:  

\begin{itemize}\itemsep0.5em
    \item Clearly define the quantity of interest, e.g., the CME, and state key assumptions—such as unconfoundedness, overlap, and functional form—before proceeding with the analysis.
    \item Assess data quality by checking for missing values and outliers, and evaluate the overlap assumption. If overlap is violated, consider trimming and/or clipping the data to improve balance.
    \item If the linear interaction model must be used for estimating the CME—for transparency, ease of computation and interpretation, or other reasons—supplement it with diagnostic tests for the rigid functional form, such as the binning estimator. 
    \item Use flexible modeling strategies to estimate the CME.
    \begin{itemize}\itemsep0em
        \item For experimental data, the kernel estimator is typically sufficient. Efficiency may be improved by incorporating basis expansions and post-Lasso selection.
        \item For observational data, we recommend doubly robust estimators in the DML family. AIPW-Lasso (or PO-Lasso for continuous treatments) is itself a DML estimator whose nuisance components are estimated via basis expansion with Lasso regularization. In relatively small samples and when nuisance functions are reasonably smooth, AIPW-Lasso often works sufficiently well, as our simulations illustrate. When sample size allows and nuisance structure calls for it---for example, when out-of-sample nuisance losses remain high under AIPW-Lasso---DML with flexible learners such as neural networks, random forests, or histogram gradient boosting can capture structure that basis expansions miss. When using these learners, cross-validate hyperparameters.
        \item For continuous treatments, the PO-Lasso and DML estimators implemented in \texttt{interflex} rely on the partially linear regression model (PLRM) and recover Definition~\ref{def:cme2} exactly when the PLRM holds; under misspecification they recover a best-linear projection of the derivative onto the residualized treatment, which we recommend as the operative target unless $f_{D\mid V}$ can be reliably estimated.
    \end{itemize}
    \item Visualize estimated CME to facilitate intuitive understanding of effect heterogeneity.
    \item Interpret results with care. Specifically, (i) avoid overusing causal language with respect to the moderator—CME captures the effect of $D$ along levels of the moderator $X$, but it does not represent the causal effect of $X$ itself. (ii) When making inferences about the sign or magnitude of the CME over an interval, use uniform confidence intervals to ensure proper coverage.
\end{itemize}

All the estimation strategies and diagnostic tools discussed in this Element can be implemented using the \texttt{interflex} package in \texttt{R}.


\clearpage




\clearpage
\setcounter{page}{1}
\setcounter{table}{0}
\setcounter{figure}{0}
\setcounter{equation}{0}
\setcounter{footnote}{0}
\renewcommand{\theexample}{A\arabic{example}}
\renewcommand{\theassumption}{A\arabic{assumption}}
\renewcommand\thetheorem{A\arabic{theorem}}
\renewcommand\thetable{A\arabic{table}}
\renewcommand\thefigure{A\arabic{figure}}
\renewcommand{\thepage}{A-\arabic{page}}
\renewcommand{\theequation}{A\arabic{equation}}
\renewcommand{\thefootnote}{A\arabic{footnote}}

\appendix  

\titleformat{\section}
  {\normalfont\Large\bfseries}
  {\thesection.}{0.5em}{}

\section{Proofs}

\subsection{Section 1 Proofs}

We show that the Stratified Difference-in-Means (SDIM) estimator and the IPW estimator of the CME are numerically equivalent when the covariate vector $V_i$ is discrete and the propensity score is estimated by within-cell frequencies:
$\hat{\theta}_{\mathrm{SDIM}}(x) = \hat{\theta}_{\mathrm{IPW}}(x)$.

\noindent\textbf{Proof.}
Fix the moderator value $x$. Let $V_i=(X_i, Z_i)$. Since $X_i$ is fixed to $x$, the variation in $V_i$ within this subpopulation is determined entirely by $Z_i$. Let $\mathcal{Z}_x$ denote the set of observed values of $Z$ for units with $X_i=x$. For each $z \in \mathcal{Z}_x$, define the cell $v=(x,z)$.

Define cell counts:
$n_{1v} := \#\{i: X_i=x, Z_i=z, D_i=1\}$,
$n_{0v} := \#\{i: X_i=x, Z_i=z, D_i=0\}$,
and $n_v := n_{1v}+n_{0v}$.
Let $N_x := \#\{i: X_i=x\} = \sum_{z \in \mathcal{Z}_x} n_v$.
We assume strict overlap in the sample: for every $v$ such that $n_v > 0$, we have $n_{1v} > 0$ and $n_{0v} > 0$.

Define within-cell sample means:
\[
\hat{\mu}^1(v) = \frac{1}{n_{1v}}\sum_{i \in I_{1v}} Y_i, \qquad
\hat{\mu}^0(v) = \frac{1}{n_{0v}}\sum_{i \in I_{0v}} Y_i,
\]
where $I_{dv} = \{i: X_i=x, V_i=v, D_i=d\}$. The SDIM estimator is:
\[
\hat{\theta}_{\mathrm{SDIM}}(x)
= \sum_{z \in \mathcal{Z}_x} \frac{n_v}{N_x} \bigl(\hat{\mu}^1(v)-\hat{\mu}^0(v)\bigr).
\]

The IPW estimator for the subpopulation $X=x$ is:
\[
\hat{\theta}_{\mathrm{IPW}}(x)
= \frac{1}{N_x}\sum_{i:X_i=x}\left[\frac{D_i Y_i}{\hat{\pi}(V_i)} - \frac{(1-D_i)Y_i}{1-\hat{\pi}(V_i)}\right].
\]
With discrete covariates, the nonparametric maximum likelihood estimator for the propensity score is the sample frequency: $\hat{\pi}(v) = n_{1v}/n_v$. Substituting this into the IPW formula and grouping terms by cell $v$:
\[
\hat{\theta}_{\mathrm{IPW}}(x)
= \frac{1}{N_x}\sum_{z \in \mathcal{Z}_x}\left[
\sum_{i \in I_{1v}}\frac{Y_i}{n_{1v}/n_v} - \sum_{i \in I_{0v}}\frac{Y_i}{1 - n_{1v}/n_v}
\right].
\]
Simplifying the fractions:
\[
\frac{1}{n_{1v}/n_v} = \frac{n_v}{n_{1v}} \quad \text{and} \quad \frac{1}{1 - n_{1v}/n_v} = \frac{1}{n_{0v}/n_v} = \frac{n_v}{n_{0v}}.
\]
Thus, the inner sums become:
\[
\sum_{i \in I_{1v}} Y_i \left(\frac{n_v}{n_{1v}}\right) = n_v \left( \frac{1}{n_{1v}}\sum_{i \in I_{1v}} Y_i \right) = n_v \hat{\mu}^1(v),
\]
\[
\sum_{i \in I_{0v}} Y_i \left(\frac{n_v}{n_{0v}}\right) = n_v \left( \frac{1}{n_{0v}}\sum_{i \in I_{0v}} Y_i \right) = n_v \hat{\mu}^0(v).
\]
Substituting these back into the expression for $\hat{\theta}_{\mathrm{IPW}}(x)$:
\[
\hat{\theta}_{\mathrm{IPW}}(x)
= \frac{1}{N_x} \sum_{z \in \mathcal{Z}_x} \bigl( n_v \hat{\mu}^1(v) - n_v \hat{\mu}^0(v) \bigr)
= \sum_{z \in \mathcal{Z}_x} \frac{n_v}{N_x} \bigl( \hat{\mu}^1(v) - \hat{\mu}^0(v) \bigr)
= \hat{\theta}_{\mathrm{SDIM}}(x).
\]

\subsection{Section 3 Proofs}

\paragraph{Identification based on IPW}

\begin{remark}
Consider a binary treatment $D_i\in\{0,1\}$ and covariates $V_i=(X_i,Z_i)$. Let
$\pi(v)=\Pr(D_i=1\mid V_i=v)$.
Assume (i) \emph{consistency/SUTVA}: $Y_i = D_i Y_i(1) + (1-D_i)Y_i(0)$,
(ii) \emph{unconfoundedness}: $(Y_i(0),Y_i(1)) \perp D_i \mid V_i$,
and (iii) \emph{overlap}: $0<\pi(V_i)<1$ a.s. Then for any $v$,
\[
\mathbb{E}\!\Bigl[\tfrac{D_i}{\pi(V_i)}\,Y_i \,\big|\, V_i=v\Bigr]
= \mathbb{E}[Y_i(1)\mid V_i=v],
\qquad
\mathbb{E}\!\Bigl[\tfrac{1-D_i}{1-\pi(V_i)}\,Y_i \,\big|\, V_i=v\Bigr]
= \mathbb{E}[Y_i(0)\mid V_i=v].
\]
\end{remark}

\noindent\textbf{Proof.}
Fix $v$. By overlap, $\pi(v)\in(0,1)$, and since $\pi(V_i)$ is a function of $V_i$,
\[
\mathbb{E}\!\left[\frac{D_i}{\pi(V_i)}Y_i \,\big|\, V_i=v\right]
= \frac{1}{\pi(v)}\,\mathbb{E}[D_iY_i\mid V_i=v].
\]
Moreover,
\[
\mathbb{E}[D_iY_i\mid V_i=v]
= \mathbb{E}[D_iY_i\mid V_i=v,D_i=1]\Pr(D_i=1\mid V_i=v),
\]
because $D_iY_i=0$ when $D_i=0$. When $D_i=1$, consistency implies $Y_i=Y_i(1)$,
so
\[
\mathbb{E}[D_iY_i\mid V_i=v,D_i=1]
= \mathbb{E}[Y_i(1)\mid V_i=v,D_i=1].
\]
By unconfoundedness, $Y_i(1)\perp D_i\mid V_i$, hence
$\mathbb{E}[Y_i(1)\mid V_i=v,D_i=1]=\mathbb{E}[Y_i(1)\mid V_i=v]$.
Therefore,
\[
\mathbb{E}[D_iY_i\mid V_i=v]
= \mathbb{E}[Y_i(1)\mid V_i=v]\;\pi(v),
\]
and substituting back yields
\[
\mathbb{E}\!\left[\frac{D_i}{\pi(V_i)}Y_i \,\big|\, V_i=v\right]
= \mathbb{E}[Y_i(1)\mid V_i=v].
\]

The control case is analogous:
\[
\mathbb{E}\!\left[\frac{1-D_i}{1-\pi(V_i)}Y_i \,\big|\, V_i=v\right]
= \frac{1}{1-\pi(v)}\,\mathbb{E}[(1-D_i)Y_i\mid V_i=v],
\]
and $(1-D_i)Y_i=0$ when $D_i=1$. When $D_i=0$, consistency gives $Y_i=Y_i(0)$, so
\[
\mathbb{E}[(1-D_i)Y_i\mid V_i=v]
= \mathbb{E}[Y_i(0)\mid V_i=v,D_i=0]\,(1-\pi(v))
= \mathbb{E}[Y_i(0)\mid V_i=v]\,(1-\pi(v)),
\]
where the last equality uses unconfoundedness. Dividing by $1-\pi(v)$ gives the result.

\paragraph{Identification based on AIPW}

\begin{remark}
Define the augmented inverse-propensity weighting (AIPW) signal
\[
\Lambda(V_i)
:= \mu_{1}(V_i)
+ \frac{D_i}{\pi(V_i)} \bigl(Y_i - \mu_{1}(V_i)\bigr)
- \mu_{0}(V_i)
- \frac{1 - D_i}{1 - \pi(V_i)} \bigl(Y_i - \mu_{0}(V_i)\bigr).
\]
Assume consistency/SUTVA, unconfoundedness $(Y_i(0),Y_i(1))\perp D_i\mid V_i$, and overlap $0<\pi(V_i)<1$ a.s.
If either (i) the propensity score is correct, $\pi(V_i)=\Pr(D_i=1\mid V_i)$, or
(ii) the outcome regressions are correct, $\mu_d(V_i)=\mathbb{E}[Y_i\mid D_i=d,V_i]$ for $d\in\{0,1\}$,
then
\[
\mathbb{E}\bigl[\Lambda(V_i)\mid V_i=v\bigr]
= \mathbb{E}\bigl[Y_i(1)-Y_i(0)\mid V_i=v\bigr].
\]
\end{remark}

\noindent\textbf{Proof.}

\textbf{Case 1: The outcome regressions are correct.}
Suppose $\mu_d(V_i)=\mathbb{E}[Y_i\mid D_i=d,V_i]$ for $d\in\{0,1\}$. Then
\[
\mathbb{E}\bigl[Y_i-\mu_d(V_i)\mid V_i=v, D_i=d\bigr]=0.
\]
Taking conditional expectations given $V_i=v$,
\begin{align*}
\mathbb{E}[\Lambda(V_i)\mid V_i=v]
&= \mu_1(v)-\mu_0(v)
+ \frac{1}{\pi(v)}\mathbb{E}\!\left[D_i\bigl(Y_i-\mu_1(v)\bigr)\mid V_i=v\right]
- \frac{1}{1-\pi(v)}\mathbb{E}\!\left[(1-D_i)\bigl(Y_i-\mu_0(v)\bigr)\mid V_i=v\right].
\end{align*}
Moreover,
\[
\mathbb{E}\!\left[D_i\bigl(Y_i-\mu_1(v)\bigr)\mid V_i=v\right]
= \Pr(D_i=1\mid V_i=v)\,\mathbb{E}[Y_i-\mu_1(v)\mid V_i=v, D_i=1]
= \pi(v)\cdot 0 = 0,
\]
and similarly the control residual term is zero. Hence
\[
\mathbb{E}[\Lambda(V_i)\mid V_i=v]=\mu_1(v)-\mu_0(v).
\]
By consistency, when $D_i=1$, $Y_i=Y_i(1)$, so $\mu_1(v)=\mathbb{E}[Y_i(1)\mid V_i=v, D_i=1]$,
and by unconfoundedness $\mathbb{E}[Y_i(1)\mid V_i=v, D_i=1]=\mathbb{E}[Y_i(1)\mid V_i=v]$.
Thus $\mu_1(v)=\mathbb{E}[Y_i(1)\mid V_i=v]$. Analogously, $\mu_0(v)=\mathbb{E}[Y_i(0)\mid V_i=v]$.
Therefore,
\[
\mathbb{E}[\Lambda(V_i)\mid V_i=v]
= \mathbb{E}[Y_i(1)-Y_i(0)\mid V_i=v].
\]

\textbf{Case 2: The propensity score is correct.}
Assume $\pi(V_i)=\Pr(D_i=1\mid V_i)$. Condition on $V_i=v$ and use
$\mathbb{E}[D_i\mid V_i=v]=\pi(v)$ and $\mathbb{E}[1-D_i\mid V_i=v]=1-\pi(v)$.
First, the $\mu_1$ terms satisfy
\[
\mathbb{E}\!\left[\mu_1(v)-\frac{D_i}{\pi(v)}\mu_1(v)\,\Big|\,V_i=v\right]
= \mu_1(v)\left(1-\frac{\mathbb{E}[D_i\mid V_i=v]}{\pi(v)}\right)=0,
\]
and similarly the $\mu_0$ terms satisfy
\[
\mathbb{E}\!\left[-\mu_0(v)+\frac{1-D_i}{1-\pi(v)}\mu_0(v)\,\Big|\,V_i=v\right]=0.
\]
Hence
\[
\mathbb{E}[\Lambda(V_i)\mid V_i=v]
= \mathbb{E}\!\left[\frac{D_i}{\pi(v)}Y_i-\frac{1-D_i}{1-\pi(v)}Y_i\,\Big|\,V_i=v\right].
\]
By the IPW identification result (which follows from consistency, unconfoundedness, and overlap),
\[
\mathbb{E}\!\left[\frac{D_i}{\pi(v)}Y_i\,\Big|\,V_i=v\right]=\mathbb{E}[Y_i(1)\mid V_i=v],
\qquad
\mathbb{E}\!\left[\frac{1-D_i}{1-\pi(v)}Y_i\,\Big|\,V_i=v\right]=\mathbb{E}[Y_i(0)\mid V_i=v].
\]
Therefore,
\[
\mathbb{E}[\Lambda(V_i)\mid V_i=v]
= \mathbb{E}[Y_i(1)-Y_i(0)\mid V_i=v].
\]
Combining the two cases proves double robustness.

\paragraph{Identification based on Outcome Modeling (On the Treated)}

\begin{remark}
Consider a binary treatment $D_i\in\{0,1\}$ and covariates $V_i=(X_i,Z_i)$.
Assume (i) consistency/SUTVA: $Y_i = D_iY_i(1) + (1-D_i)Y_i(0)$,
(ii) unconfoundedness: $(Y_i(0),Y_i(1))\perp D_i\mid V_i$,
and (iii) overlap in $X$: $0<\Pr(D_i=1\mid X_i)<1$ a.s.
Define $\mu_0(V_i):=\mathbb{E}[Y_i\mid D_i=0,V_i]$.
Then
\[
\mathbb{E}\!\Biggl[ \frac{(Y_i-\mu_{0}(V_i))\,D_i}{\Pr(D_i=1 \mid X_i)} \,\Bigg|\, X_i\Biggr]
=\mathbb{E}\!\Bigl[Y_i(1)-Y_i(0) \mid D_i=1, X_i\Bigr].
\]
\end{remark}

\noindent\textbf{Proof.}
Fix $x$. Since $\Pr(D_i=1\mid X_i=x)$ depends only on $x$,
\[
\mathbb{E}\!\Biggl[\frac{(Y_i-\mu_{0}(V_i))\,D_i}{\Pr(D_i=1 \mid X_i)} \,\Bigg|\, X_i=x\Biggr]
=\frac{1}{\Pr(D_i=1 \mid X_i=x)}\,
\mathbb{E}\!\Bigl[(Y_i-\mu_{0}(V_i))\,D_i \mid X_i=x \Bigr].
\]
Because $D_i$ is binary,
\[
\mathbb{E}\!\Bigl[(Y_i-\mu_{0}(V_i))\,D_i \mid X_i=x \Bigr]
=\mathbb{E}\!\bigl[Y_i-\mu_0(V_i)\mid D_i=1,X_i=x\bigr]\Pr(D_i=1\mid X_i=x).
\]
By consistency, conditional on $D_i=1$ we have $Y_i=Y_i(1)$, hence
\[
\mathbb{E}\!\bigl[Y_i-\mu_0(V_i)\mid D_i=1,X_i=x\bigr]
=\mathbb{E}\!\bigl[Y_i(1)-\mu_0(V_i)\mid D_i=1,X_i=x\bigr].
\]

Next, for any $v$,
\[
\mu_0(v)=\mathbb{E}[Y_i\mid D_i=0,V_i=v]
=\mathbb{E}[Y_i(0)\mid D_i=0,V_i=v]
=\mathbb{E}[Y_i(0)\mid V_i=v]
=\mathbb{E}[Y_i(0)\mid D_i=1,V_i=v],
\]
where the second equality uses consistency (when $D_i=0$) and the third uses unconfoundedness.
Therefore,
\[
\mathbb{E}[Y_i(1)-\mu_0(V_i)\mid D_i=1,V_i]
=\mathbb{E}[Y_i(1)-Y_i(0)\mid D_i=1,V_i].
\]
Taking conditional expectations given $D_i=1,X_i=x$ and using iterated expectations,
\[
\mathbb{E}[Y_i(1)-\mu_0(V_i)\mid D_i=1,X_i=x]
=\mathbb{E}[Y_i(1)-Y_i(0)\mid D_i=1,X_i=x].
\]

Substituting back,
\[
\mathbb{E}\!\Biggl[\frac{(Y_i-\mu_{0}(V_i))\,D_i}{\Pr(D_i=1 \mid X_i)} \,\Bigg|\, X_i=x\Biggr]
=\mathbb{E}[Y_i(1)-Y_i(0)\mid D_i=1,X_i=x],
\]
as claimed.

\paragraph{Identification based on IPW (On the Treated)}

\begin{remark}
Consider a binary treatment $D_i\in\{0,1\}$ and covariates $V_i=(X_i,Z_i)$.
Assume (i) consistency/SUTVA: $Y_i = D_iY_i(1)+(1-D_i)Y_i(0)$,
(ii) unconfoundedness: $(Y_i(0),Y_i(1))\perp D_i\mid V_i$,
and (iii) overlap: $0<\pi(V_i)<1$ a.s., and $0<\Pr(D_i=1\mid X_i)<1$ a.s.,
where $\pi(V_i):=\Pr(D_i=1\mid V_i)$.
Then
\[
\mathbb{E}\!\Biggl[ \frac{Y_i\bigl(D_i-\pi(V_i)\bigr)}{\Pr(D_i=1 \mid X_i)\bigl(1-\pi(V_i)\bigr)} \,\Bigg|\, X_i\Biggr]
=\mathbb{E}\!\Bigl[Y_i(1)-Y_i(0) \mid D_i=1, X_i\Bigr].
\]
\end{remark}

\noindent\textbf{Proof.}
Define $\tau(v):=\mathbb{E}[Y_i(1)-Y_i(0)\mid V_i=v]$.
By unconfoundedness, $\mathbb{E}[Y_i(1)-Y_i(0)\mid D_i=1,V_i=v]=\tau(v)$.
Hence for any $x$,
\[
\mathbb{E}[Y_i(1)-Y_i(0)\mid D_i=1,X_i=x]
=\mathbb{E}[\tau(V_i)\mid D_i=1,X_i=x].
\]
By iterated expectations and Bayes' rule,
\begin{align*}
\mathbb{E}[\tau(V_i)\mid D_i=1,X_i=x]
&=\int \tau(x,z)\, f(z\mid D_i=1,X_i=x)\,dz \\
&=\int \tau(x,z)\,\frac{f(z\mid X_i=x)\pi(x,z)}{\Pr(D_i=1\mid X_i=x)}\,dz \\
&=\frac{1}{\Pr(D_i=1\mid X_i=x)}\,
\mathbb{E}\!\left[\tau(V_i)\,\pi(V_i)\ \big|\ X_i=x\right].
\end{align*}

Next, by the IPW identity (from consistency, unconfoundedness, and overlap),
\[
\tau(V_i)
=\mathbb{E}\!\Biggl[\frac{Y_iD_i}{\pi(V_i)}-\frac{Y_i(1-D_i)}{1-\pi(V_i)}\ \Bigg|\ V_i\Biggr].
\]
Therefore,
\begin{align*}
\mathbb{E}\!\left[\tau(V_i)\pi(V_i)\mid X_i=x\right]
&=\mathbb{E}\!\left[
\mathbb{E}\!\left[\left(\frac{Y_iD_i}{\pi(V_i)}-\frac{Y_i(1-D_i)}{1-\pi(V_i)}\right)\pi(V_i)\ \Bigg|\ V_i\right]
\Bigg|\ X_i=x\right] \\
&=\mathbb{E}\!\left[ Y_iD_i - \frac{Y_i(1-D_i)\pi(V_i)}{1-\pi(V_i)}\ \Bigg|\ X_i=x\right].
\end{align*}
Combining the two steps,
\[
\mathbb{E}[Y_i(1)-Y_i(0)\mid D_i=1,X_i=x]
=\mathbb{E}\!\left[\frac{Y_iD_i}{\Pr(D_i=1\mid X_i=x)}
-\frac{Y_i(1-D_i)\pi(V_i)}{\Pr(D_i=1\mid X_i=x)(1-\pi(V_i))}\ \Bigg|\ X_i=x\right].
\]
Finally, note
\[
Y_iD_i - \frac{Y_i(1-D_i)\pi(V_i)}{1-\pi(V_i)}
=\frac{Y_i\bigl(D_i-\pi(V_i)\bigr)}{1-\pi(V_i)},
\]
so
\[
\mathbb{E}[Y_i(1)-Y_i(0)\mid D_i=1,X_i=x]
=\mathbb{E}\!\left[\frac{Y_i\bigl(D_i-\pi(V_i)\bigr)}{\Pr(D_i=1\mid X_i=x)(1-\pi(V_i))}\ \Bigg|\ X_i=x\right],
\]
which is the desired result.

\paragraph{Identification based on AIPW (On the Treated)}

\begin{remark}
Assume (i) \emph{consistency/SUTVA}: $Y_i=D_iY_i(1)+(1-D_i)Y_i(0)$,
(ii) \emph{unconfoundedness}: $(Y_i(0),Y_i(1))\perp D_i\mid V_i$,
and (iii) \emph{overlap}: $0<\pi(V_i)<1$ a.s. and $0<\Pr(D_i=1\mid X_i)<1$ a.s.,
where $\pi(V_i):=\Pr(D_i=1\mid V_i)$.
Define the signal
\[
\Xi(V_i)
:= \frac{1}{\Pr(D_i=1\mid X_i)}\Bigl(Y_i-\mu_{0}(V_i)\Bigr)
\left(D_i-\frac{\pi(V_i)(1-D_i)}{1-\pi(V_i)}\right).
\]
If either $\pi(V_i)$ is the true propensity score or $\mu_0(V_i)=\mathbb{E}[Y_i\mid D_i=0,V_i]$ is the true control regression, then for any $x$,
\[
\mathbb{E}\bigl[\Xi(V_i)\mid X_i=x\bigr]
= \mathbb{E}\bigl[Y_i(1)-Y_i(0)\mid D_i=1,X_i=x\bigr].
\]
\end{remark}

\noindent\textbf{Proof.}
Fix $x$ and let $p(x):=\Pr(D_i=1\mid X_i=x)$.
Let $\tau(v):=\mathbb{E}[Y_i(1)-Y_i(0)\mid V_i=v]$.
By unconfoundedness, $\mathbb{E}[Y_i(1)-Y_i(0)\mid D_i=1,V_i=v]=\tau(v)$. Thus:
\[
\mathbb{E}[Y_i(1)-Y_i(0)\mid D_i=1,X_i=x]
=\mathbb{E}[\tau(V_i)\mid D_i=1,X_i=x].
\]
Using Bayes' rule (or the definition of conditional expectation), we can rewrite the target as:
\[
\mathbb{E}[\tau(V_i)\mid D_i=1,X_i=x]
=\frac{1}{p(x)}\,\mathbb{E}\bigl[\tau(V_i)\Pr(D_i=1|V_i)\mid X_i=x\bigr].
\]
It therefore suffices to show that the numerator of $\Xi(V_i)$ satisfies:
\[
\mathbb{E}\left[\Bigl(Y_i-\mu_0(V_i)\Bigr)\left(D_i-\frac{\pi(V_i)(1-D_i)}{1-\pi(V_i)}\right)\ \Bigg|\ X_i=x\right]
=\mathbb{E}\bigl[\tau(V_i)\Pr(D_i=1|V_i)\mid X_i=x\bigr].
\]

\textbf{Case 1: The propensity score is correct.}
Assume $\pi(V_i)=\Pr(D_i=1\mid V_i)$. Conditioning on $V_i$, the expectation of the term inside the square brackets is:
\begin{align*}
&\mathbb{E}\left[\Bigl(Y_i-\mu_0(V_i)\Bigr)D_i \mid V_i\right]
- \frac{\pi(V_i)}{1-\pi(V_i)}\mathbb{E}\left[\Bigl(Y_i-\mu_0(V_i)\Bigr)(1-D_i) \mid V_i\right].
\end{align*}
Note that $\mu_0(V_i)$ is arbitrary here (potentially misspecified).
Consider the second term. By consistency, if $D_i=0$, then $Y_i = Y_i(0)$. Thus:
\[
\mathbb{E}[(Y_i-\mu_0(V_i))(1-D_i)\mid V_i]
= (1-\pi(V_i)) \bigl( \mathbb{E}[Y_i(0)\mid V_i] - \mu_0(V_i) \bigr).
\]
Consider the first term. If $D_i=1$, $Y_i=Y_i(1)$. Thus:
\[
\mathbb{E}[(Y_i-\mu_0(V_i))D_i\mid V_i]
= \pi(V_i) \bigl( \mathbb{E}[Y_i(1)\mid V_i] - \mu_0(V_i) \bigr).
\]
Subtracting the second term (scaled by $\frac{\pi}{1-\pi}$) from the first:
\begin{align*}
&\pi(V_i) \bigl( \mathbb{E}[Y_i(1)|V_i] - \mu_0(V_i) \bigr) - \frac{\pi(V_i)}{1-\pi(V_i)} (1-\pi(V_i)) \bigl( \mathbb{E}[Y_i(0)|V_i] - \mu_0(V_i) \bigr) \\
&= \pi(V_i) \Bigl( \mathbb{E}[Y_i(1)|V_i] - \mu_0(V_i) - \mathbb{E}[Y_i(0)|V_i] + \mu_0(V_i) \Bigr) \\
&= \pi(V_i) \tau(V_i).
\end{align*}
The misspecified $\mu_0$ cancels out perfectly. Taking the expectation over $V_i|X_i=x$ matches the target.

\textbf{Case 2: The outcome regression is correct.}
Assume $\mu_0(V_i)=\mathbb{E}[Y_i\mid D_i=0,V_i]$, but allow $\pi(V_i)$ to be misspecified.
The term of interest is:
\[
(Y_i-\mu_0(V_i))D_i \;-\; \frac{\pi(V_i)}{1-\pi(V_i)}(Y_i-\mu_0(V_i))(1-D_i).
\]
Conditional on $V_i$, the expected value of the second part is:
\[
-\frac{\pi(V_i)}{1-\pi(V_i)}\Pr(D_i=0\mid V_i)\,\mathbb{E}[Y_i-\mu_0(V_i)\mid V_i,D_i=0].
\]
Since $\mu_0$ is the true conditional mean for controls, $\mathbb{E}[Y_i - \mu_0(V_i) \mid V_i, D_i=0] = 0$. Thus, the entire second term vanishes, regardless of the value of $\pi(V_i)$.
The remaining term is:
\[
\mathbb{E}[(Y_i-\mu_0(V_i))D_i\mid V_i].
\]
Using consistency and unconfoundedness:
\[
\mathbb{E}[(Y_i(1)-\mu_0(V_i)) \cdot 1 \mid V_i, D_i=1] \Pr(D_i=1|V_i).
\]
Since $\mu_0(V_i) = \mathbb{E}[Y_i(0)|V_i]$, this becomes:
\[
\Pr(D_i=1\mid V_i) \bigl( \mathbb{E}[Y_i(1)|V_i] - \mathbb{E}[Y_i(0)|V_i] \bigr) = \Pr(D_i=1|V_i) \tau(V_i).
\]
This matches the target moment exactly. Thus, the estimator is consistent if either nuisance parameter is correctly specified.

\subsection{Section 4 Proofs}

\paragraph{Signals and CME (Continuous Treatment)}

To prove: $\E[\Lambda(\eta)\mid V=v] = \E[\partial_d Y(D)\mid V=v]$.

\noindent\textbf{Proof}

By the tower rule, $\E[\Lambda(\eta)\mid X=x] =  \E\Bigl[
\E[\Lambda(\eta)\mid V]\,\Big|\;X=x\Bigr],$ so it suffices to show that $\E[\Lambda(\eta)\mid V=v]= \E[\partial_d Y(D)\mid V=v].$ We want to show:
\begin{equation*}
\E[\Lambda(\eta)\mid V=v] =  \E[\partial_d Y(D)\mid V=v].
\end{equation*}

Expanding on the Neyman orthogonal score:
\begin{align*}
\E\bigl[\Lambda(\eta)\mid V=v\bigr] & = \E\Bigl[
\partial_{d}\mu(D,v) \mid V = v \Bigr] - \E\Bigl[\partial_d \log f(D|v)\,\bigl[Y -\mu(D,v)\bigr]\,\Big|\;V=v\Bigr]
\end{align*}

Note that we are integrating over the joint distribution of $Y$ and $D$ given $V=v$. We verify the second term is zero.
\begin{align*}
&\E\Bigl[-\,\partial_{d}\log f(D|v)\,\bigl(Y - \mu(D,v)\bigr)\;\Big|\;V=v\Bigr]  =  \\
&\int_{d}\!\!\int_{y}\Bigl(-\,\partial_{d}\log f(d|v)
\,\bigl[y-\mu(d,v)\bigr]\Bigr)\; f_{Y\mid D,V}(y\mid d,v)\,f_{D\mid V}(d\mid v)\;dy\,dd.
\end{align*}
By definition of the conditional mean, $\int y\,f_{Y\mid D,V}(y\mid d,v)\,dy=\mu(d,v).$ It follows that $\int [y-\mu(d,v)]\,f_{Y\mid D,V}(y\mid d,v)\,dy=0.$ Hence:
\begin{equation*}
\E\Bigl[-\,\partial_{d}\log f(D|v)\,\bigl(Y - \mu(D,v)\bigr)\;\Big|\;V=v\Bigr] = 0.
\end{equation*}
Therefore,
\begin{equation*}
\E[\Lambda(\eta)\mid V=v] = \E[\partial_d \mu(D,v) \mid V=v] = \E[\partial_d Y(D)\mid V=v].
\end{equation*}

\paragraph{Score Definition}
To verify the Neyman orthogonality for the AIPW score, we show that the G\^{a}teaux derivative of the expected score with respect to the nuisance functions is zero at the true nuisance parameters.

For each unit $i$, define the AIPW score:

\begin{equation*}
\Lambda_i(\eta) = \mu_1(V_i) - \mu_0(V_i) + \frac{D_i}{\pi(V_i)}\bigl[Y_i - \mu_1(V_i)\bigr] - \frac{1 - D_i}{1 - \pi(V_i)}\bigl[Y_i - \mu_0(V_i)\bigr].
\end{equation*}

where $\eta$ collectively denotes the nuisance functions $\eta = (\mu_1,\mu_0,\pi)$. At the true nuisance parameters $\eta_0 = (\mu_1^0,\mu_0^0,\pi^0)$, this score identifies the pointwise CATE.

To show Neyman orthogonality, we must verify:
\begin{equation*}
\left.
\frac{\partial}{\partial \eta}
\mathbb{E}\bigl[\Lambda_i(\eta)\bigr]
\right|_{\eta = \eta_0}
= 0.
\end{equation*}

We will look at the partial derivatives of $\mathbb{E}[\Lambda_i(\eta)]$ with respect to each nuisance function $\mu_1, \mu_0, \text{and } \pi$. In each case, we show the derivative is zero when evaluated at $\eta_0$.

\paragraph{Derivative with Respect to $\mu_1$}

Consider a small directional perturbation $h_{\mu_1}(v)$ of $\mu_1$. The G\^{a}teaux derivative in that direction is:

\begin{equation*}
\left.\frac{d}{dt}\right|_{t=0}
\mathbb{E}\Bigl[
\Lambda_i\bigl(\mu_1 + t\,h_{\mu_1},\,\mu_0,\pi\bigr)
\Bigr].
\end{equation*}

Within $\Lambda_i$, $\mu_1$ appears in two places: $+\mu_1(V_i)$ and inside the term $\frac{D_i}{\pi(V_i)}[Y_i - \mu_1(V_i)]$. Taking the derivative yields:

\begin{equation*}
\text{Partial w.r.t.\ }\mu_1:
\quad
h_{\mu_1}(V_i)\;-\;\frac{D_i}{\pi(V_i)}\bigl[h_{\mu_1}(V_i)\bigr]
=
h_{\mu_1}(V_i)\Bigl(1 -\frac{D_i}{\pi(V_i)}\Bigr).
\end{equation*}

Since $h_{\mu_1}(V_i)$ is a function of $V_i$:
\begin{equation*}
\mathbb{E}\Bigl[h_{\mu_1}(V_i)\Bigl(1 -\frac{D_i}{\pi(V_i)}\Bigr)\Bigr] = \mathbb{E}\Bigl[
h_{\mu_1}(V_i)\,
\mathbb{E}\bigl[1 - \frac{D_i}{\pi(V_i)} \bigm|\,V_i\bigr]
\Bigr].
\end{equation*}

At the true $\eta_0$, $\pi^0(v) = \Pr(D_i=1 \mid V_i=v)$. Therefore:

\begin{equation*}
\mathbb{E}\Bigl[\,1 - \frac{D_i}{\pi^0(V_i)} \Bigm|\,V_i=v\Bigr]
=
1 -\frac{\mathbb{E}[D_i \mid V_i=v]}{\,\pi^0(v)\,}
=
1 -\frac{\pi^0(v)}{\,\pi^0(v)\,}
= 0.
\end{equation*}

By an identical argument, the derivative w.r.t. $\mu_0$ also vanishes.

\paragraph{Derivative with Respect to $\pi$}

Now consider a small perturbation $h_\pi(v)$ of $\pi$. Within $\Lambda_i$, $\pi$ appears in the denominators:
\begin{equation*}
\frac{D_i}{\,\pi(V_i)\,}\quad \text{and} \quad \frac{1 - D_i}{\,1 - \pi(V_i)\,}.
\end{equation*}
Focusing on just the parts that depend on $\pi$, the derivative of $\frac{D_i}{\,\pi(V_i)\,}\bigl[Y_i - \mu_1(V_i)\bigr]$
and $-\frac{1 - D_i}{\,1 - \pi(V_i)\,}\bigl[Y_i - \mu_0(V_i)\bigr]$ sums to:

\begin{equation*}
h_\pi(V_i)\,\left(
-\,\frac{D_i}{\,\pi(V_i)^2\,}\bigl[Y_i - \mu_1(V_i)\bigr]
\;+\;
\frac{1 - D_i}{\,[1 - \pi(V_i)]^2}\bigl[Y_i - \mu_0(V_i)\bigr]
\right).
\end{equation*}

Factor out $h_\pi(V_i)$ and condition on $V_i=v$:

\begin{align*}
\mathbb{E}\Bigl[
h_\pi(V_i)\,
\Bigl(
-\,\frac{D_i}{\,\pi(V_i)^2\,}(Y_i - \mu_1(V_i))
\;+\;
\frac{1-D_i}{\,[1- \pi(V_i)]^2}\,(Y_i - \mu_0(V_i))
\Bigr)
\Bigr] = \\
\mathbb{E}\Bigl[
h_\pi(V_i)\,
\mathbb{E}\Bigl[
-\,\frac{D_i}{\,\pi^0(v)^2\,}(Y_i - \mu_1^0(v))
\;+\;
\frac{1-D_i}{\,[1-\pi^0(v)]^2\,}(Y_i - \mu_0^0(v))
\Bigm|\,V_i=v
\Bigr]
\Bigr].
\end{align*}

Define $\Gamma(v)$ as the sum of two expectations:

\begin{equation*}
\Gamma(v) = \underbrace{\mathbb{E}\Bigl[-\frac{D_i}{\pi^0(v)^2}(Y_i - \mu_1^0(v)) \Bigm| V_i=v \Bigr]}_{T_1(v)} \quad+\quad \underbrace{\mathbb{E}\Bigl[\frac{1 - D_i}{[1 - \pi^0(v)]^2}(Y_i - \mu_0^0(v)) \Bigm| V_i=v \Bigr]}_{T_2(v)}.
\end{equation*}

We will show each of $T_1(v)$ and $T_2(v)$ is zero under the true nuisance parameters.

\subparagraph{$T_1(v)$ Vanishes}

\begin{equation*}
T_1(v) = -\frac{1}{\pi^0(v)^2} \mathbb{E}\Bigl[ D_i(Y_i - \mu_1^0(v)) \Bigm| V_i=v \Bigr].
\end{equation*}

Expand the inner expectation:

\begin{equation*}
\mathbb{E}[D_i(Y_i - \mu_1^0(v)) \mid V_i=v] = \mathbb{E}[D_i Y_i \mid V_i=v] - \mu_1^0(v)\mathbb{E}[D_i \mid V_i=v].
\end{equation*}

Evaluate at the true parameters:
\begin{align*}
\mathbb{E}[D_i Y_i \mid V_i=v] &= \Pr(D_i=1\mid V_i=v)\cdot \mathbb{E}[Y_i\mid D_i=1, V_i=v] = \pi^0(v)\mu_1^0(v), \\
\mathbb{E}[D_i \mid V_i=v] &= \pi^0(v).
\end{align*}
Thus:

\begin{align*}
\mathbb{E}[D_i Y_i \mid V_i=v] &= \pi^0(v)\mu_1^0(v) \\
\mathbb{E}[D_i(Y_i - \mu_1^0(v)) \mid V_i=v] & = \pi^0(v)\mu_1^0(v) - \mu_1^0(v)\pi^0(v) = 0.
\end{align*}

Therefore $T_1(v) = -\frac{1}{\pi^0(v)^2} \cdot 0 = 0$.

\subparagraph{$T_2(v)$ Vanishes}

\begin{equation*}
    T_2(v) = \frac{1}{[1 - \pi^0(v)]^2} \mathbb{E}\Bigl[ (1 - D_i)(Y_i - \mu_0^0(v)) \Bigm| V_i=v \Bigr].
\end{equation*}

Similarly, expand the inner expectation:
\begin{equation*}
\mathbb{E}[(1 - D_i)(Y_i - \mu_0^0(v)) \mid V_i=v] = \mathbb{E}[(1 - D_i)Y_i \mid V_i=v] - \mu_0^0(v)\mathbb{E}[(1 - D_i) \mid V_i=v].
\end{equation*}

Evaluate at the true parameter:
\begin{align*}
\mathbb{E}[(1 - D_i)Y_i \mid V_i=v] & = (1 - \pi^0(v))\mu_0^0(v)\\
\mathbb{E}[(1 - D_i) \mid V_i=v] & = 1 - \pi^0(v).
\end{align*}

Thus,
\begin{equation*}
\mathbb{E}[(1 - D_i)(Y_i - \mu_0^0(v)) \mid V_i=v]  = (1 - \pi^0(v))\mu_0^0(v) - \mu_0^0(v)(1 - \pi^0(v)) = 0.
\end{equation*}
Thus, $T_2(v) = \frac{1}{[1-\pi^0(v)]^2}\cdot 0 = 0$.

Putting the two parts together: $\Gamma(v) = T_1(v) + T_2(v) = 0 + 0 = 0$ which means for every $v$:
\begin{equation*}
\mathbb{E}\Bigl[ -\frac{D_i}{\pi^0(v)^2}(Y_i - \mu_1^0(V_i)) + \frac{1-D_i}{[1-\pi^0(v)]^2}(Y_i - \mu_0^0(v)) \Bigm| V_i=v \Bigr] = 0.
\end{equation*}
Consequently,
\begin{equation*}
\mathbb{E}\Bigl[ h_\pi(V_i) \Bigl( -\frac{D_i}{\pi^0(V_i)^2}(Y_i - \mu_1^0(V_i)) + \frac{1-D_i}{[1-\pi^0(V_i)]^2}(Y_i - \mu_0^0(V_i)) \Bigr) \Bigr] = 0,
\end{equation*}
for any directional perturbation $h_\pi(\cdot)$. This completes the proof that the derivative of $\mathbb{E}[\Lambda_i(\eta)]$ w.r.t. $\pi$ is zero at $\eta_0$.


By combining this argument with the analogous derivatives w.r.t. $\mu_1$ and $\mu_0$, we conclude that
\begin{equation*}
\left. \frac{\partial}{\partial \eta}\,\mathbb{E}[\Lambda_i(\eta)] \right|_{\eta=\eta_0} = 0
\end{equation*}
for all directions of perturbation. Hence, the AIPW score is Neyman-orthogonal to first-order errors in $\eta$, which underlies its double robustness property and the extension to DML settings.


\clearpage

\section{Implementation Details for the Lasso Methods}

This section provides additional technical details on basis expansion and variable selection for Lasso methods that are omitted from the main text for brevity. We describe how flexible basis functions are constructed for \(V=(X,Z)\), how interactions are incorporated, and how Lasso and post-selection estimation are implemented in the outcome and propensity-score models. We also present supplementary simulation results that illustrate the performance of these flexible specifications.

\subsection{Basis Expansion}

One way to increase the flexibility of the outcome and propensity-score models is to expand the set of regressors \(V = (X, Z)\) beyond their raw forms. Specifically, each covariate—whether the moderator \(X\) or an additional covariate \(Z_j \in Z\)—can be replaced with a set of basis functions that capture nonlinearities, and interactions among them can also be incorporated. Common approaches include polynomial expansions (e.g., second degree or higher) and B\mbox{-}spline expansions, which partition the support of a variable into segments while enforcing smoothness across boundaries. By enriching the model space, the estimated relationships can better adapt to complex patterns that might otherwise be inadequately captured by purely linear terms.

Concretely, suppose \(V \in \mathbb{R}^{1+p}\), where \(p\) is the dimension of the covariates. For each variable \(L\) in \(V\), including \(X\), we can generate a set of transformed variables \(\{\phi_k(L)\}\), where each function represents a polynomial or B\mbox{-}spline basis expansion. Interaction terms can be constructed by multiplying these expansions for \(X\) with those for \(Z\), allowing the model to capture effect modification across covariates. We denote the full set of expansions by \(\psi(V)\).

Once expanded, these regressors can be incorporated into both the outcome model \(\mathbb{E}[Y \mid D, X, Z]\) and the propensity-score model \(\mathbb{E}[D \mid X, Z]\). In the outcome model, basis expansions enable more flexible relationships between \(X\), \(Z\), and \(Y\). If the true relationship is nonlinear, involves threshold effects, or exhibits heterogeneous slopes across different regions of \(X\) or \(Z\), the expanded basis helps approximate these complexities. In the propensity-score model, these expansions allow the probability of treatment to vary nonlinearly with the covariates. With a sufficiently rich expansion—and appropriate regularization, such as Lasso—these flexible specifications can reduce bias by better approximating the unknown data-generating process.

A practical rule of thumb is to prefer flexible yet parsimonious bases. For polynomial expansions, prefer low-degree polynomials (quadratic or cubic) and avoid high-degree polynomials. For B\mbox{-}splines, use cubic (degree 3) or natural splines for better boundary behavior, and employ a small number of interior knots (e.g., 3--6) placed at empirical quantiles \citep{de1978practical, hastie2009elements}. In principle, these hyperparameters can be selected by cross-validation or information criteria. Because we pair these choices with Lasso regularization tuned by cross-validation in the next section, we do not additionally cross-validate them here, to conserve computational resources. In the remaining sections and simulations, we use a B\mbox{-}spline expansion with \texttt{spline\_degree} = 3 and \texttt{spline\_df} = 6 (placing three internal knots at 25th, 50th, and 75th percentiles). We also include interaction terms between the expanded basis of $X$ and the expanded basis of each $Z$, as well as interactions between the expanded bases of different $Z$ variables.

We illustrate the benefit of basis expansions using another simulated example. We apply outcome modeling, IPW, and AIPW, showing results for each, with and without basis expansion.

\begin{example}[A Simulated Example with Complex Nonlinear Relationships]\label{ex:sim.nonlinear}
We simulate \(V = (X_i, Z_{i1}, Z_{i2}, Z_{i3})\) with \(X_i \sim \mathrm{Unif}[-2,2]\) and each \(Z_{ij} \sim \mathrm{Unif}[0,1]\) for \(j=1,2,3\). The treatment is generated by
\[
D_i \sim \mathrm{Bernoulli}\Bigl(
\mathrm{logit}^{-1}\bigl(-1 + 0.5 Z_{i1} + |X_i| - X_i \cdot Z_{i2} +
Z_{i3}^2 \bigr)\Bigr),
\]
and the outcome is generated by
\[
Y_i = 1 + X_i + D_i  - D_i X_i^2  + Z_{i1}^2 + \sin(Z_{i2}) + 0.5\, Z_{i1}\exp(Z_{i3})
+ \varepsilon_i,
\]
where \(\varepsilon_i \sim \mathcal{N}(0,1)\). Under this DGP, the true CME is \(\theta(x) = 1 - x^2\). Unconfoundedness holds: \(D_i \indep (Y_i(0), Y_i(1)) \mid V_i\). However, the researcher is unaware of the functional form of either the propensity-score model or the outcome model.
\end{example}

\begin{figure}[!pth]
\caption{With and Without Basis Expansion}\label{fig:AIPW_expansion}
\begin{minipage}[t]{1\textwidth}
    \centering
    \begin{subfigure}[b]{\textwidth}
        \centering
        \includegraphics[width=0.45\textwidth]{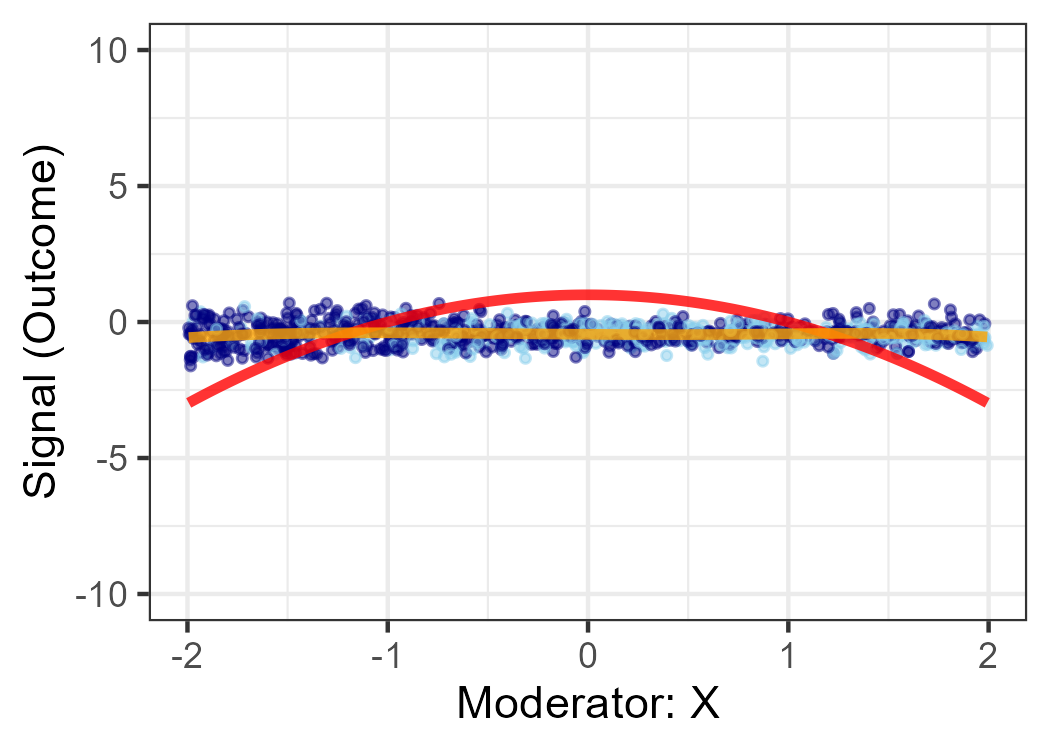}   
        \hspace{0.02\textwidth}
        \includegraphics[width=0.45\textwidth]{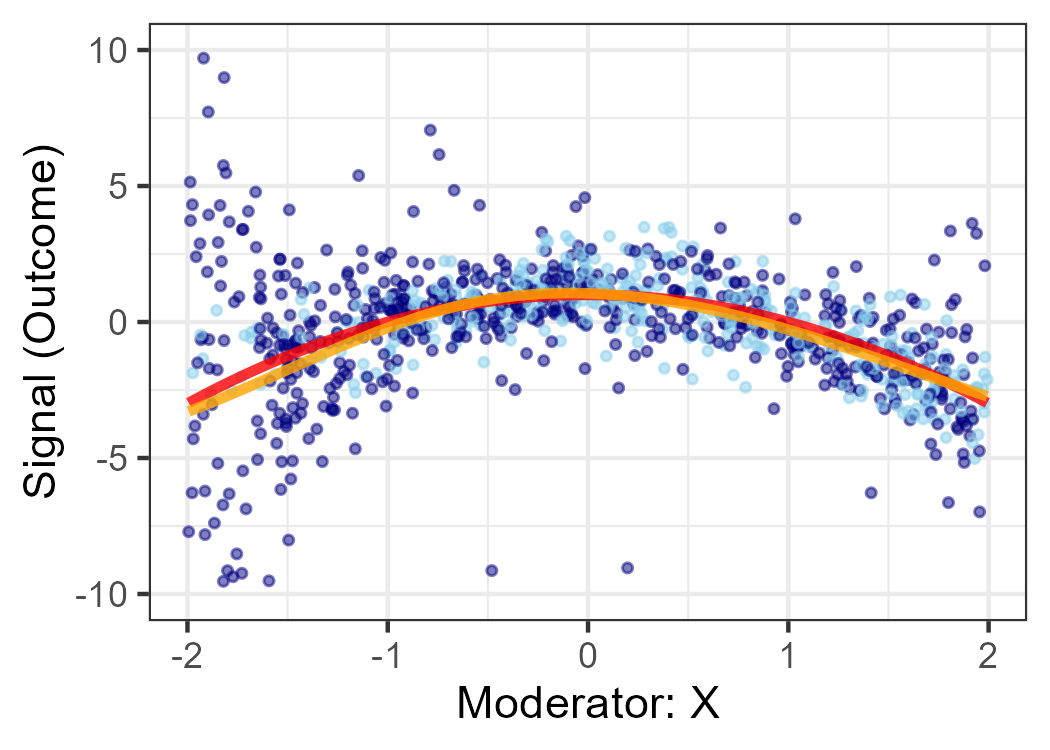}
        \caption{Outcome Modeling: without and with basis expansion}
    \end{subfigure}    
    \vspace{0.02\textwidth}    
    \begin{subfigure}[b]{\textwidth}
        \centering
        \includegraphics[width=0.45\textwidth]{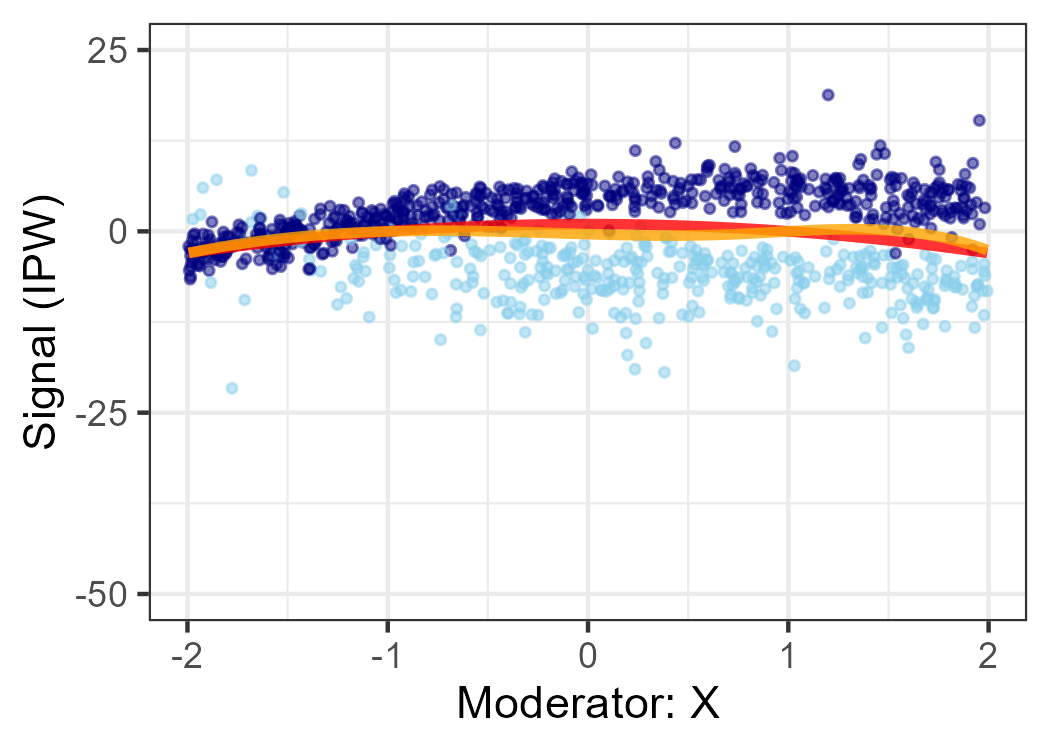}   
        \hspace{0.02\textwidth}
        \includegraphics[width=0.45\textwidth]{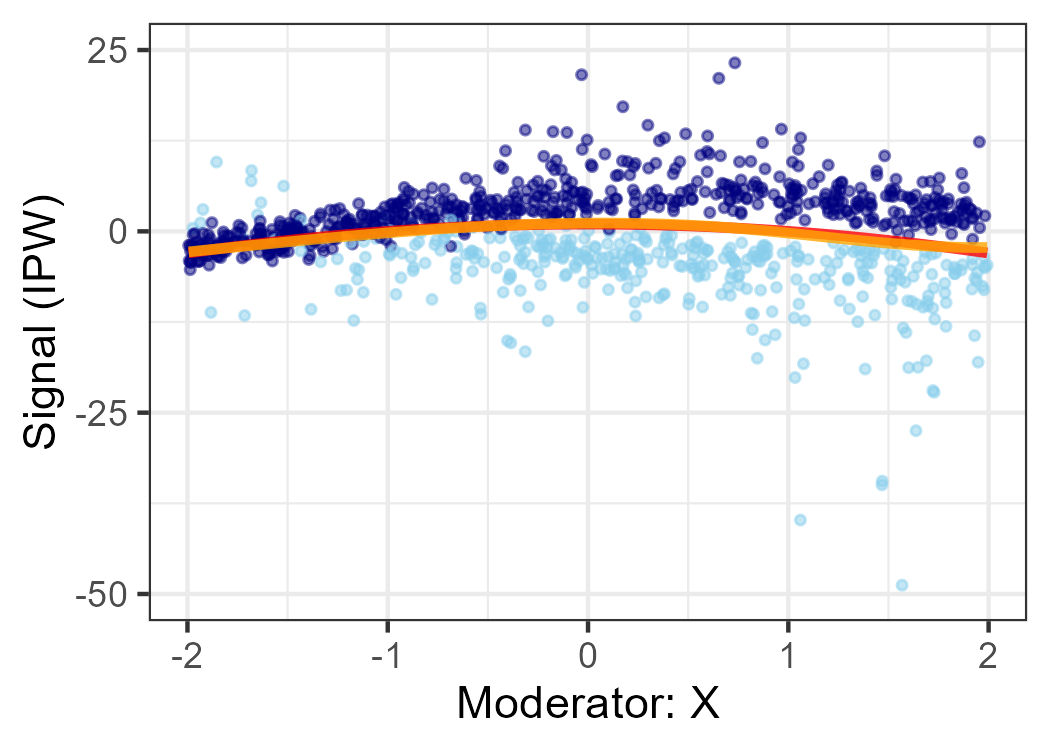}
        \caption{IPW: without and with basis expansion}
    \end{subfigure}  
    \vspace{0.02\textwidth}   
    \begin{subfigure}[b]{\textwidth}
        \centering
        \includegraphics[width=0.45\textwidth]{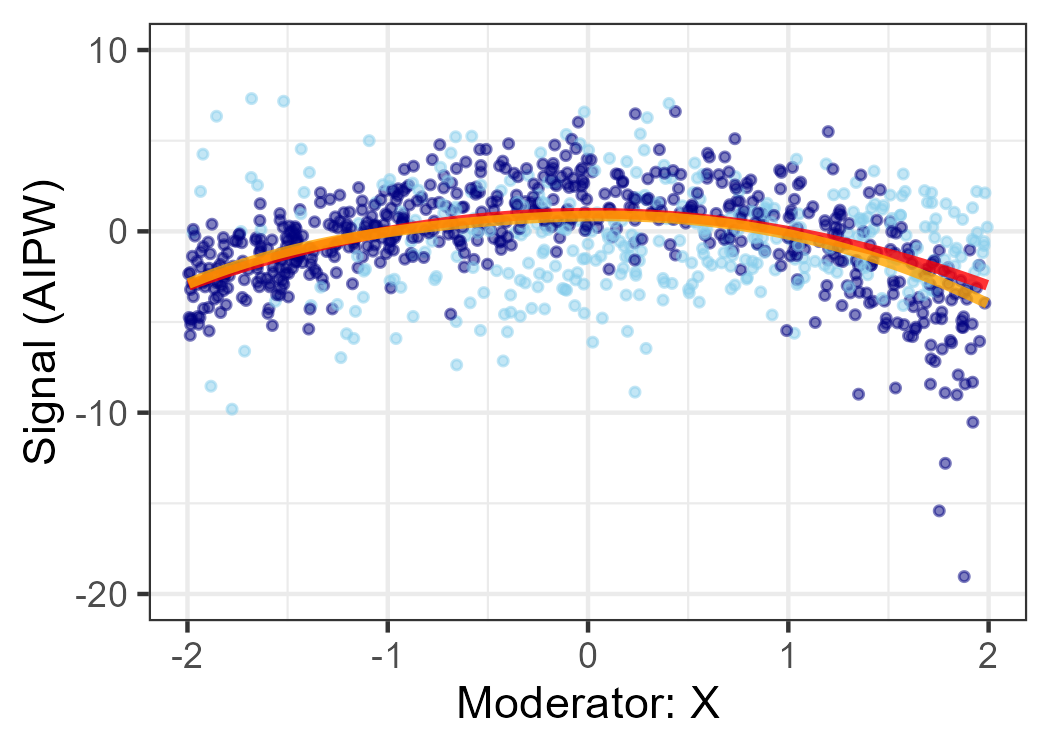}   
        \hspace{0.02\textwidth}
        \includegraphics[width=0.45\textwidth]{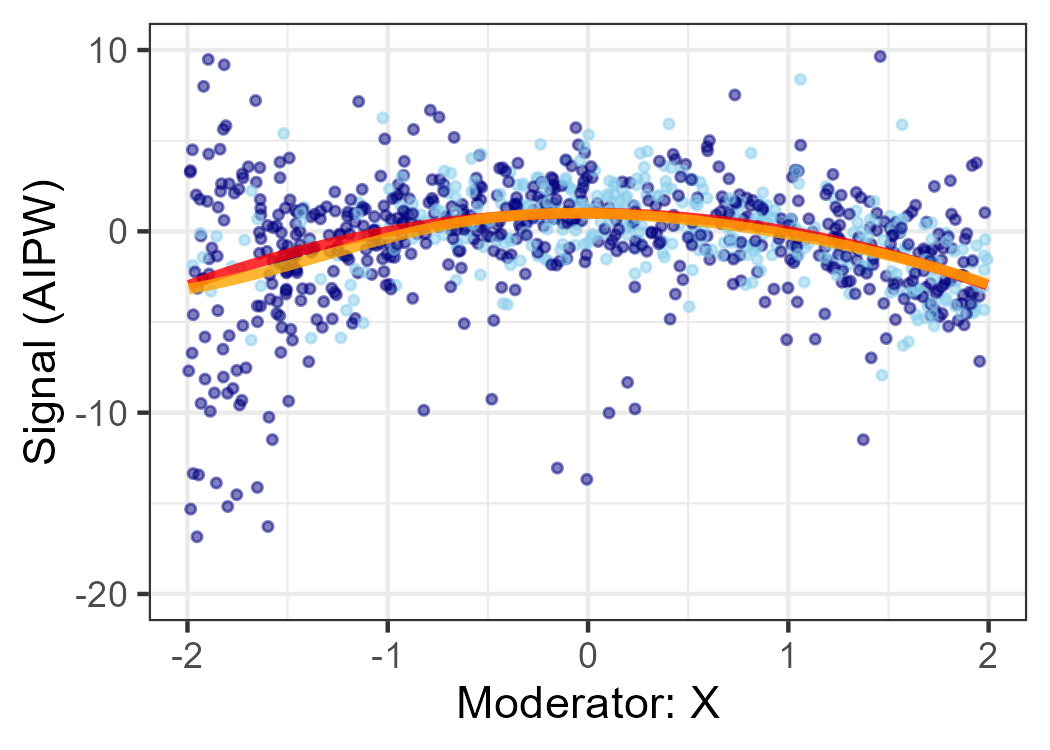}
        \caption{AIPW: without and with basis expansion}
    \end{subfigure}  
\end{minipage}
{\footnotesize \textit{Note}: Dark blue and light blue dots represent signals for the treatment and control groups, respectively. The red line represents the true CME, while the orange line represents the estimated CME. In each subfigure, the left and right panels show results without and with basis expansion, respectively.}
\end{figure}

Figure~\ref{fig:AIPW_expansion} shows the results. The left panel in each subfigure presents the signals and CME estimates without basis expansion, and the right panel includes basis expansions on both \(X\) and \(Z\). The left panels in Figure~\ref{fig:AIPW_expansion} demonstrate that (i) the linear outcome model fails to capture the nonlinearity; (ii) the IPW estimator also misses the true shape and occasionally produces large outliers in the estimated signals; and (iii) the AIPW estimator performs somewhat better even with a linear outcome model, likely because the two models compensate for each other’s deficiencies.

The right panels of Figure~\ref{fig:AIPW_expansion} show that both the outcome and AIPW estimators capture the inverse-\(U\) shape of the CME, \(\theta(x) = 1 - x^2\), more accurately and outperform their counterparts without basis expansions. However, the IPW estimator remains sensitive to propensity scores near 0 or 1, which results in large inverse weights and unstable estimates. In contrast, the AIPW signals are far more stable and concentrated within a narrow band, yielding more precise CME estimates.

Although basis expansions can enhance model flexibility, they also introduce more parameters to estimate, potentially leading to greater variance and a higher risk of overfitting, particularly when the number of covariates is large. Next, we show that regularized methods, such as Lasso, are essential for managing this high-dimensional setting and ensuring stable, reliable inference.

\subsection{Variable Selection}

When the number of covariates or expanded basis functions is large, conventional OLS or logistic regression may overfit and yield unstable estimates. To address this limitation, one can apply Lasso to achieve both regularization and variable selection. Lasso imposes an \(\ell_1\)-penalty on the regression coefficients, shrinking many of them toward zero and effectively selecting a sparse subset of relevant predictors. In its canonical form for a linear model with response \(\{y_i\}_{i=1}^n\) and predictors \(\{x_i\}_{i=1}^n \in \mathbb{R}^p\), the Lasso estimate \(\hat{\beta}_{\lambda}\) is defined as
\[
\hat{\beta}_{\lambda} = \arg\min_{\beta \in \mathbb{R}^p} \Biggl\{ \frac{1}{n} \sum_{i=1}^n \bigl(y_i - x_i^\top \beta\bigr)^2 + \lambda \sum_{j=1}^p |\beta_j| \Biggr\}.
\]
The \(\ell_1\)-penalty shrinks many coefficients to zero, and the trade-off between sparsity and goodness of fit is governed by the penalty parameter \(\lambda \geq 0\). A straightforward application of Lasso in this context is to estimate \emph{separate} Lasso regressions for each outcome model (one for the treated and one for the control subsample) and for the propensity-score model. Let 
\[
I_1=\{\,i: D_i=1\,\}, \quad \text{and} \quad I_0=\{\,i: D_i=0\,\},
\]
and, with some abuse of notation, let \(X_i\) denote the vector of predictors—including basis expansions and their interactions—for observation \(i\). Then one solves:
\begin{align*}
\hat{\beta}_{1,\lambda_1} &= \arg\min_{\beta_1} \left\{ \frac{1}{|I_1|} \sum_{i \in I_1} \bigl(Y_i - X_i^\top \beta_1\bigr)^2 + \lambda_1 \|\beta_1\|_1 \right\} 
\quad \Longrightarrow \quad 
\hat{\mu}^{1}(V_i) = X_i^\top \hat{\beta}_{1,\lambda_1}, \\
\hat{\beta}_{0,\lambda_0} &= \arg\min_{\beta_0} \left\{ \frac{1}{|I_0|} \sum_{i \in I_0} \bigl(Y_i - X_i^\top \beta_0\bigr)^2 + \lambda_0 \|\beta_0\|_1 \right\} 
\quad \Longrightarrow \quad 
\hat{\mu}^{0}(V_i) = X_i^\top \hat{\beta}_{0,\lambda_0},
\end{align*}
and, using all \(n\) observations for the propensity score,
\begin{align*}
\hat{\gamma}_{\lambda_p} &= \arg\min_{\gamma} \left\{ -\frac{1}{n} \sum_{i=1}^n \left[ D_i X_i^\top \gamma - \log\bigl(1 + \exp(X_i^\top \gamma)\bigr) \right] + \lambda_p \|\gamma\|_1 \right\} \\
&\Longrightarrow \quad \hat{\pi}(V_i) = \frac{1}{1 + e^{-X_i^\top \hat{\gamma}_{\lambda_p}}}.
\end{align*}
Here, \(\lambda_{1}\), \(\lambda_{0}\), and \(\lambda_{p}\) are penalty parameters for each model that can be selected via cross-validation. The estimated coefficients are then used to construct the outcome or propensity-score models and, subsequently, to build various signals as discussed previously.

\subsection{Post-Selection Lasso.}
A well-known issue with Lasso is that standard errors for its coefficients are not readily available, and inference is not straightforward. Because Lasso produces sparse solutions, one can define the model selected by Lasso as the set of predictors with nonzero coefficients:
\[
\widehat{\mathcal{A}}_1=\left\{j: \hat{\beta}_{1,\lambda_1,j} \neq 0\right\},\quad \widehat{\mathcal{A}}_0=\left\{j: \hat{\beta}_{0,\lambda_0,j} \neq 0\right\},\quad \widehat{\mathcal{A}}_p=\left\{j: \hat{\gamma}_{\lambda_p,j} \neq 0\right\}.
\]
Post-selection coefficients are obtained by refitting the model using the appropriate \emph{unpenalized} estimator—OLS for linear models and logistic regression for logit models—on the variables selected by Lasso. Specifically, the post-selection coefficients for \(\beta_{1,\lambda_{1}}\), denoted \(\tilde{\beta}_{1,\lambda_{1}}\), are estimated by running a standard OLS regression on the treated units using only the predictors in \(\widehat{\mathcal{A}}_1\); similarly, one obtains \(\tilde{\beta}_{0,\lambda_{0}}\) and \(\tilde{\gamma}_{\lambda_{p}}\). The various signals are then constructed based on these refined estimates.

\citet{belloni2012sparse} and \citet{belloni2013least} provide results showing that predictions based on post-selection coefficients perform at least as well as those based on penalized coefficients. Under the conditions that guarantee valid predictions by Lasso, the post-selection coefficients often yield slightly improved performance, and in other cases, they perform comparably.

\begin{example}[Many Covariates and Complex Nonlinear Relationships]\label{ex:sim.complex}
We simulate a sample of 1,000 units with \(V_i = (X_i, Z_{i1}, Z_{i2}, \dots, Z_{i8})\), where \(X_i \sim \mathrm{Unif}[-2,2]\) and each \(Z_j \sim \mathrm{Unif}[0,1]\) for \(j = 1, 2, \dots, 8\). The treatment and outcome are generated in the same way as in Example~\ref{ex:sim.nonlinear}, meaning that covariates \(Z_{i4},\dots,Z_{i8}\) are redundant. The true CME is \(\theta(x) = 1 - x^2\). Unconfoundedness holds: \(D_i \indep (Y_i(0), Y_i(1)) \mid V_i\).
\end{example}

\begin{figure}[!th]
\caption{AIPW with Basis Expansion: w/o and w/ Variable Selection}
\begin{subfigure}[b]{0.47\textwidth}
    \centering
    \includegraphics[width=\textwidth]{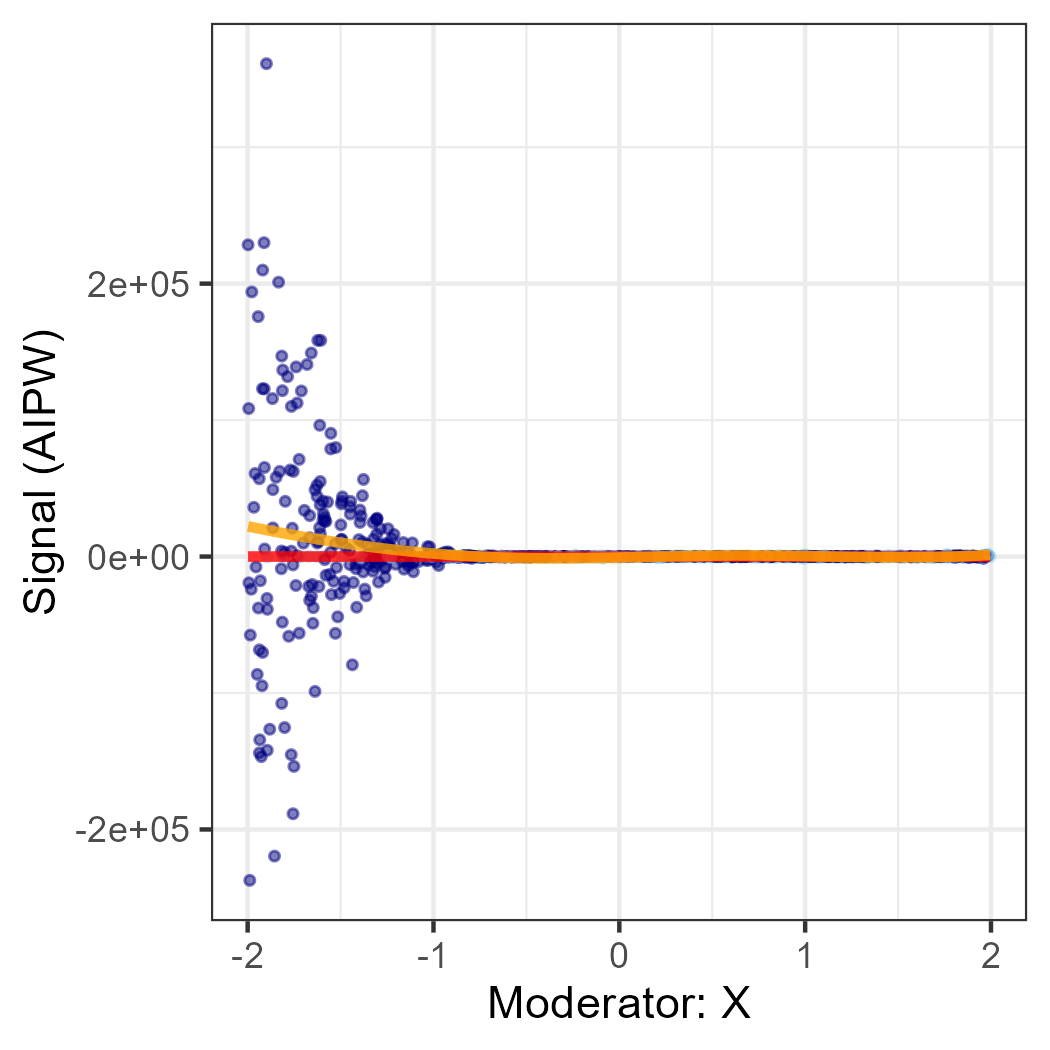}
    \caption{\hspace{1em} Without variable selection}
\end{subfigure}
\hspace{0.02\textwidth}
\begin{subfigure}[b]{0.47\textwidth}
    \centering
    \includegraphics[width=\textwidth]{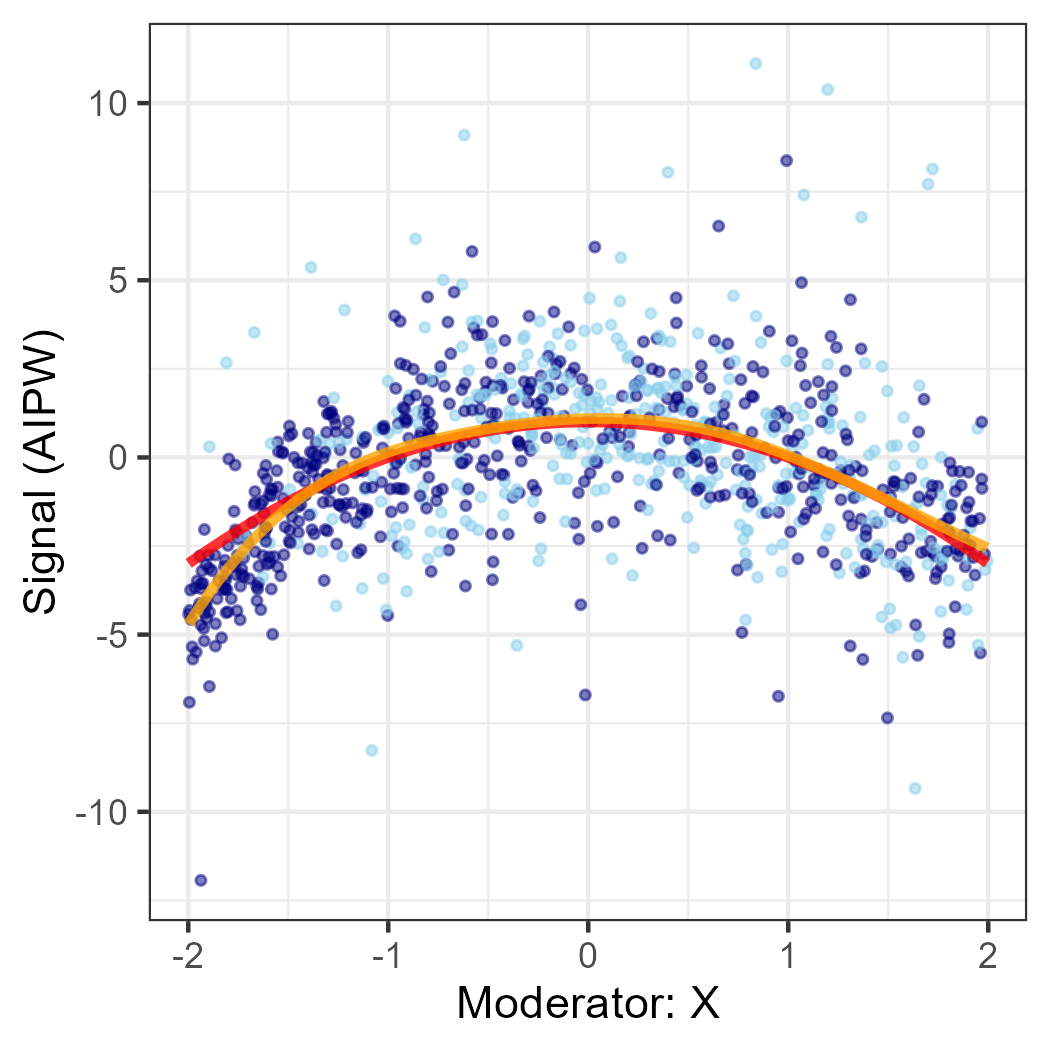}
    \caption{\hspace{1em} With variable selection}
\end{subfigure}
\label{fig:AIPW_selection}\\
{\footnotesize \textit{Note}: Dark blue and light blue dots represent AIPW signals for the treatment and control groups, respectively. The red line represents the true CME, while the orange line represents the estimated CME.}
\end{figure}

Figure~\ref{fig:AIPW_selection} presents the AIPW signals and CME estimates without and with Lasso selection. Because the additional variables \(Z_{i4},\dots,Z_{i,10}\) are included despite having little influence on the treatment or outcome, identifying the relevant covariates becomes nontrivial. Figure~\ref{fig:AIPW_selection} illustrates how Lasso improves estimation by selecting a sparse, informative subset of predictors, leading to drastically more stable signals and more accurate CME estimates. Without the variable selection, the basis expansions and interaction terms over $(X,Z)$ make the estimated signals noisy and the CME estimation unstable.

\clearpage
\section{Regularity Conditions for Function-Valued DML}

This section collects the formal assumptions governing the asymptotic theory of the sieve-based DML estimator for a function-valued parameter $\theta_0(x)$ in Section 4. The framework closely follows \citet{semenova2021debiased}. When $\theta_0$ is function-valued, DML methods can still achieve desirable asymptotic properties, but the set of required assumptions expands to govern the choice of sieve dimension $d$. To ensure that $\hat{\theta}(x)$ retains the $\sqrt{N}$-rate and satisfies both pointwise and uniform asymptotic normality, we must manage the trade-off between approximation error and estimation variance. In particular, the sieve dimension must grow large enough to render approximation bias negligible, yet slowly enough relative to $N$ to prevent variance from dominating and to allow uniform inference.

\begin{assumption}[Identification and Approximation]
Suppose that:

\renewcommand\labelenumi{(\theenumi)}
\begin{enumerate}
\item Eigenvalues bounded. Let $Q = \mathbb{E}[p(X)p(X)^\prime]$ denote the population covariance matrix of the basis functions. There exist constants $0 < C_{min} < C_{max} < \infty$ such that uniformly in $d$,
\[
C_{min} \le \lambda_{min}(Q) \le \lambda_{max}(Q) \le C_{max}.
\]

\item Approximation error. Write
\[
g(x) = p(x)^\prime \beta_0 + r_g(x),
\]
where $r_g(x)$ denotes the approximation error. Both the mean-squared magnitude $\mathbb{E}[r_g(X)^2]$ and the uniform magnitude $\sup_{x\in\mathcal{X}} |r_g(x)|$ are controlled as in Assumption 3.3 of \citet{semenova2021debiased}.
\end{enumerate}
\end{assumption}

This condition ensures that the sieve basis is well-conditioned and that $\theta_0(x)$ can be approximated sufficiently well as $d$ increases.\\

When we approximate $\theta_0(x)$ by projecting the orthogonal signal onto $d$ basis functions, each observation contributes to estimating $d$ coefficients. If $d$ grows too quickly relative to $N$, the moment-based estimator exhibits high variance. The following restriction controls the admissible growth rate of the basis.

\begin{assumption}[Growth Condition]
Suppose that:

\renewcommand\labelenumi{(\theenumi)}
\begin{enumerate}
\item Basis growth. Define
\[
\xi_d = \sup_{x\in\mathcal{X}} \|p(x)\|.
\]
Then
\[
\frac{\xi_d^2 \log N}{N} = o(1).
\]

\item Lipschitz basis regularity. Define the normalized basis
\[
\alpha(x) = \frac{p(x)}{\|p(x)\|}.
\]
There exist $m>2$ and constants $\xi_d^L$ such that
\[
\xi_d^L = \sup_{x \neq x'} 
\frac{\|\alpha(x)-\alpha(x')\|}{\|x-x'\|} < \infty,
\]
and
\[
(\xi_d^L)^{\frac{2m}{m-2}} \cdot \frac{\log N}{N} \le 1,
\qquad
\log \xi_d^L \le \log d.
\]
\end{enumerate}
\end{assumption}

The Lipschitz requirement ensures that the normalized basis does not vary too abruptly over $\mathcal{X}$, which is crucial for empirical-process arguments supporting uniform convergence. Stronger growth control is therefore required for uniform inference than for pointwise results.\\

The next assumption governs nuisance estimation and stochastic fluctuations of the orthogonal signal.
\begin{assumption}[Small Bias and Error]
Suppose that:

\renewcommand\labelenumi{(\theenumi)}
\begin{enumerate}
\item Bounded conditional variance. Let $U = Y - g(X)$. Then
\[
\sup_{x\in\mathcal{X}} \mathbb{E}[U^2 \mid X=x] \le 1.
\]

\item Small bias in nuisance estimation. Let $\hat{\eta}_k$ denote the cross-fitted nuisance estimators. With high probability, $\hat{\eta}_k$ lies in a shrinking neighborhood $T_N$ around $\eta_0$, such that the projection bias terms $B_N$ and $\Lambda_N$, as defined in \citet{semenova2021debiased}, satisfy $B_N = o(1)$ and $\Lambda_N = o(1)$.

\item Tail bounds. There exists $m>2$ such that
\[
\sup_{x\in\mathcal{X}} \mathbb{E}[|U|^m \mid X=x] \le 1.
\]
\end{enumerate}
\end{assumption}

The bounded variance and tail conditions ensure sufficiently strong concentration inequalities to support uniform convergence. The small-bias restriction guarantees that orthogonalization and cross-fitting render nuisance estimation errors asymptotically negligible.\\

We now state the pointwise limit theory for the sieve-based DML estimator.

\begin{theorem}[Pointwise $\sqrt{N}$-Consistency and Asymptotic Normality]
Under Assumptions C.1–C.3, for any fixed $x \in \mathcal{X}$,
\[
\sqrt{N}\big(\hat{\theta}(x) - \theta_0(x)\big)
=
\frac{1}{\sqrt{N}} \sum_{i=1}^N \psi_x(W_i)
+ o_p(1),
\]
where $\psi_x(W_i)$ is an influence function with $\mathbb{E}[\psi_x(W_i)] = 0$ and $\mathbb{E}[\psi_x(W_i)^2] = \sigma^2(x) < \infty$.

Consequently,
\[
\sqrt{N}\big(\hat{\theta}(x) - \theta_0(x)\big)
\;\leadsto\;
N\!\big(0, \sigma^2(x)\big).
\]
\end{theorem}

This result establishes $\sqrt{N}$-consistency and asymptotic normality at any fixed $x$. Under the stronger growth and moment conditions in Assumptions C.2 and C.3, the convergence extends uniformly over $\mathcal{X}$, enabling valid construction of simultaneous confidence bands via multiplier bootstrap. Formal proofs are provided in \citet{semenova2021debiased}.

\clearpage

\section{Fine-tuning DML}

For many machine learning methods, model tuning is crucial to ensure strong performance. In this simulation study, we compare how different machine learners—NN, RF, and HGB—perform under default hyperparameters versus hyperparameters fine-tuned via cross-validation, using a more complex DGP. Although default settings are convenient and sometimes adequate, they often fail to capture the underlying complexity of the data, resulting in underfitting or overfitting depending on the context. In DML, where both the outcome and propensity score models must be estimated, it is especially unlikely that a single set of hyperparameters will perform well for both. Cross-validation allows for more targeted tuning by optimizing out-of-sample predictive performance separately for each model component. By comparing default and cross-validated configurations side by side, we highlight the trade-off between computational cost—since cross-validation can be time-intensive—and the potential for improved accuracy through more careful tuning.

In this setting, we consider a nonlinear outcome model and a nonlinear propensity score model, each depending on a single moderator \(X\) and four additional covariates \(z_{i1}\), \(z_{i2}\), \(z_{i3}\), and \(z_{i4}\). Specifically, the outcome variable \(Y_i\) is then given by
\[
Y_i = \theta(X_i)\,D_i + g(V_i) + \epsilon_i,
\]
in which \(D_i \sim \operatorname{Bernoulli}(\pi(V_i))\) and \(\epsilon_i \sim \mathcal{N}(0,1)\) and
\[
\begin{aligned}
g(V_i) =\;& 1 + X_i + \exp(2X_i + 2) + 3\,\sin(X_i + Z_{i1}) + 2 X_i Z_{i2} \\[5pt]
&+ Z_{i3}\,\mathbf{1}\{Z_{i3} > 0\} - 4\,\frac{\sin(Z_{i4})}{Z_{i4}} + 2\,Z_{i3}^2 Z_{i4}.\\
\end{aligned}
\]
The propensity score model is given by:
\[
\pi(V_i) = \text{logit}^{-1}\{0.25 + X_i - X_i^2 - z_{i1}^2 + 2X_i z_{i1} z_{i2} - \text{logit}^{-1}(z_{i1} + z_{i3}) + 2z_{i2}^2 \cos(z_{i4})\}
\]
This highly nonlinear specification poses a challenge for all estimation methods and underscores the importance of hyperparameter tuning in flexible learners. As in previous studies, the true CME is quadratic: $\theta(X_i)=X_i^2$.

\paragraph*{Tuning NN.} Our NN model begins with the default parameter settings in \texttt{scikit-learn}’s \texttt{MLPRegressor} and \texttt{MLPClassifier}, which include a single hidden layer of 100 neurons, ReLU activation, the Adam optimizer, an L2 regularization term (\texttt{alpha}) of 0.0001, and a maximum of 1,000 training iterations. To allow for greater flexibility, we focus our hyperparameter search on four key components: layer architecture, activation function, solver, and regularization strength. Specifically, we vary the hidden layer configurations among \texttt{(50,)}, \texttt{(100,)}, and \texttt{(50, 50)}, allowing for both simpler single-layer networks and deeper architectures capable of capturing more complex relationships. We compare the \texttt{tanh} and \texttt{relu} activation functions to account for different gradient behaviors and consider both stochastic gradient descent (\texttt{sgd}) and Adam (\texttt{adam}) as solvers, given their distinct convergence properties. Finally, we tune the L2 penalty term \texttt{alpha} over \(\{10^{-4}, 10^{-3}, 10^{-2}\}\), balancing the trade-off between bias and variance. Although a broader grid could yield further refinements, we limit the search space to ensure adequate coverage of key parameters while controlling the computational burden of cross-validation in repeated simulation runs.

The code snippet below demonstrates how to implement this hyperparameter tuning using the \texttt{interflex} package:

\begin{lstlisting}[language=R]
param_grid_nn <- list(
  hidden_layer_sizes = list(
    list(50L),
    list(100L),
    list(50L, 50L)
  ),
  activation = c("tanh", "relu"),
  solver = c("sgd", "adam"),
  alpha = c(1e-4, 1e-3, 1e-2)
)

sol.nn.cv <- interflex(data = sim_data, Y="Y", D="D", X="X", Z=Z.list, estimator="DML", CV = TRUE, model.t="nn", model.y="nn",param.grid.y = param_grid_nn, param.grid.t = param_grid_nn)
\end{lstlisting}

\paragraph*{Tuning RF.} Our random forests model uses the default parameter settings in \texttt{scikit-learn}’s \texttt{RandomForestRegressor} and \texttt{RandomForestClassifier}, which include 100 estimators (\texttt{n\_estimators} = 100) and unrestricted tree depth. In our grid search, we broaden the search space by allowing \texttt{n\_estimators} to be 100 or 300, striking a balance between computational overhead and variance reduction through ensemble averaging. We also vary \texttt{max\_depth} among \{\texttt{NULL}, 5, 10\} so that the model can adapt its complexity based on data size and noise levels. To control the formation of splits, we choose \texttt{min\_samples\_split} in \{2, 10\} and \texttt{min\_samples\_leaf} in \{1, 5\}, thereby mitigating the risk of overfitting by requiring more observations in node splits and leaves. Furthermore, we compare \texttt{max\_features} set to 1.0 (all features considered) versus 0.8 (a subsample of features at each split) to foster diversity among the trees. Finally, we explore both \texttt{bootstrap} = \texttt{TRUE} (standard bootstrap sampling) and \texttt{FALSE} (sampling without replacement) to gauge their effect on variance reduction.

\paragraph*{Tuning HGB.} Our HGB model starts from the default parameter settings in \texttt{scikit-learn}’s \texttt{HistGradientBoostingRegressor} and \texttt{HistGradientBoostingClassifier}, which use a learning rate of 0.1 and a maximum of 100 boosting iterations. We tune several other key hyperparameters to enhance flexibility while avoiding overfitting. Specifically, we let \texttt{learning\_rate} range over \{0.01, 0.1\}, spanning conservative to more aggressive updates, and allow up to either 100 or 200 boosting iterations (\texttt{max\_iter}). We examine two distinct values for \texttt{max\_leaf\_nodes} (31 and 127) to regulate tree complexity, and vary \texttt{max\_depth} among \{\texttt{NULL}, 3, 5\} to impose a bound on tree depth only when potentially beneficial. In addition, \texttt{min\_samples\_leaf} takes values \{5, 20\} to guard against overfitting by avoiding extremely small leaves, while \texttt{l2\_regularization} in \{0.0, 1.0\} manages the trade-off between bias and variance. Finally, we tune \texttt{max\_features} in \{1.0, 0.8\}, controlling the fraction of features available at each split.\medskip

\begin{figure}[!t]
\caption{Estimated CME based on a single simulation following the Fourth DGP}
\begin{subfigure}[b]{0.23\textwidth}
    \centering
    \includegraphics[width=\textwidth]{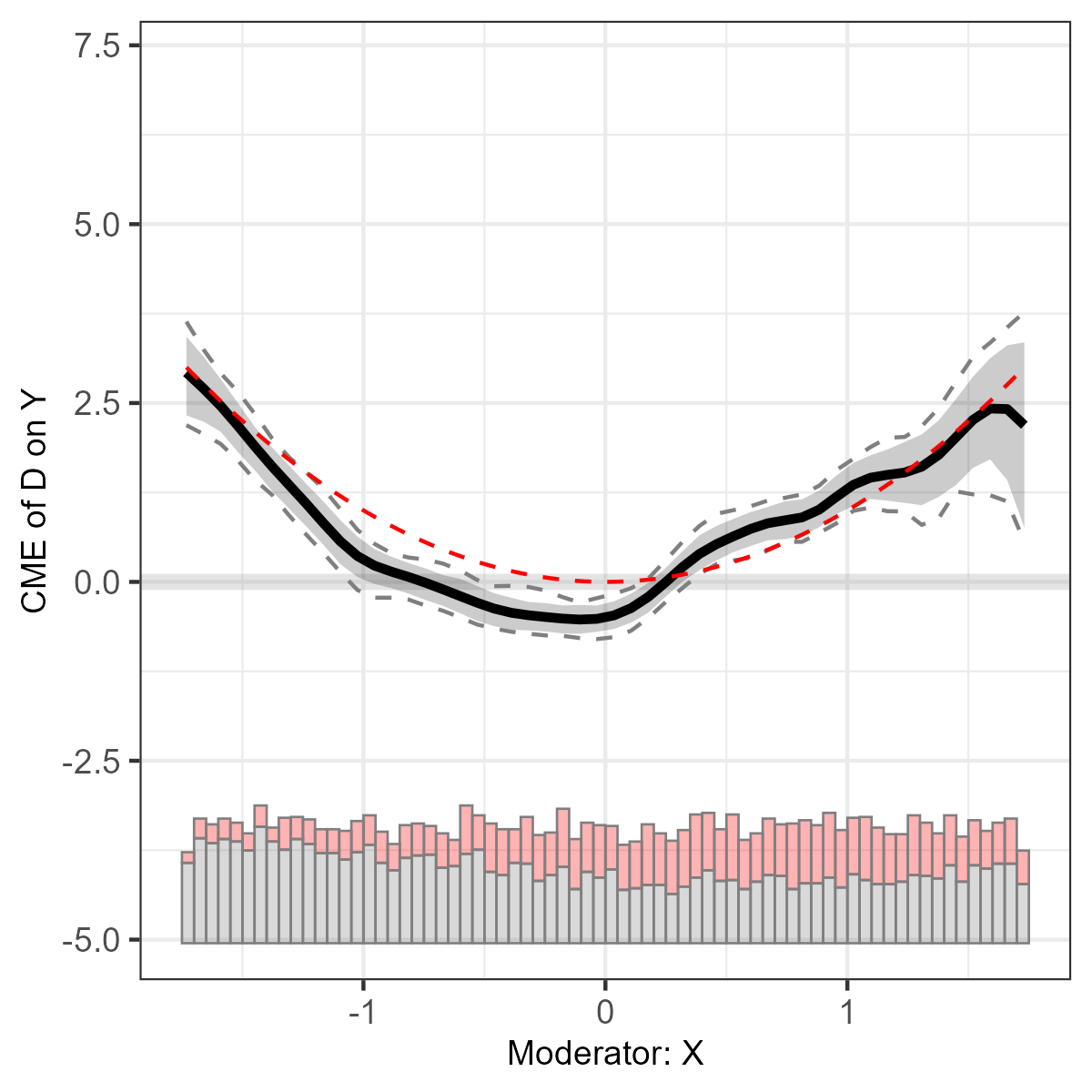}
    \caption{Kernel}
\end{subfigure}
\hspace{0.01\textwidth}  
\begin{subfigure}[b]{0.23\textwidth}
    \centering
    \includegraphics[width=\textwidth]{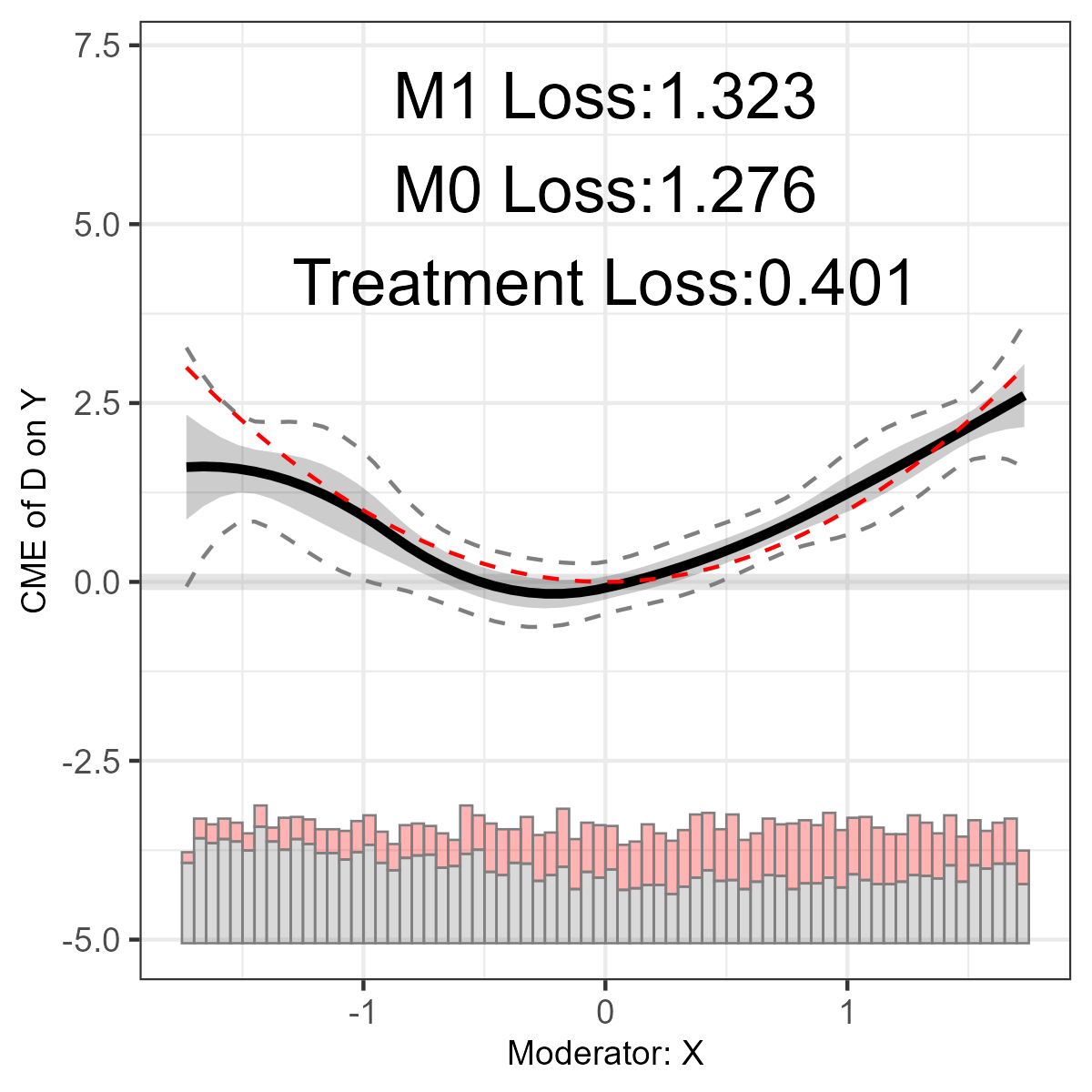}
    \caption{NN (default)}
\end{subfigure}
\hspace{0.01\textwidth}  
\begin{subfigure}[b]{0.23\textwidth}
    \centering
    \includegraphics[width=\textwidth]{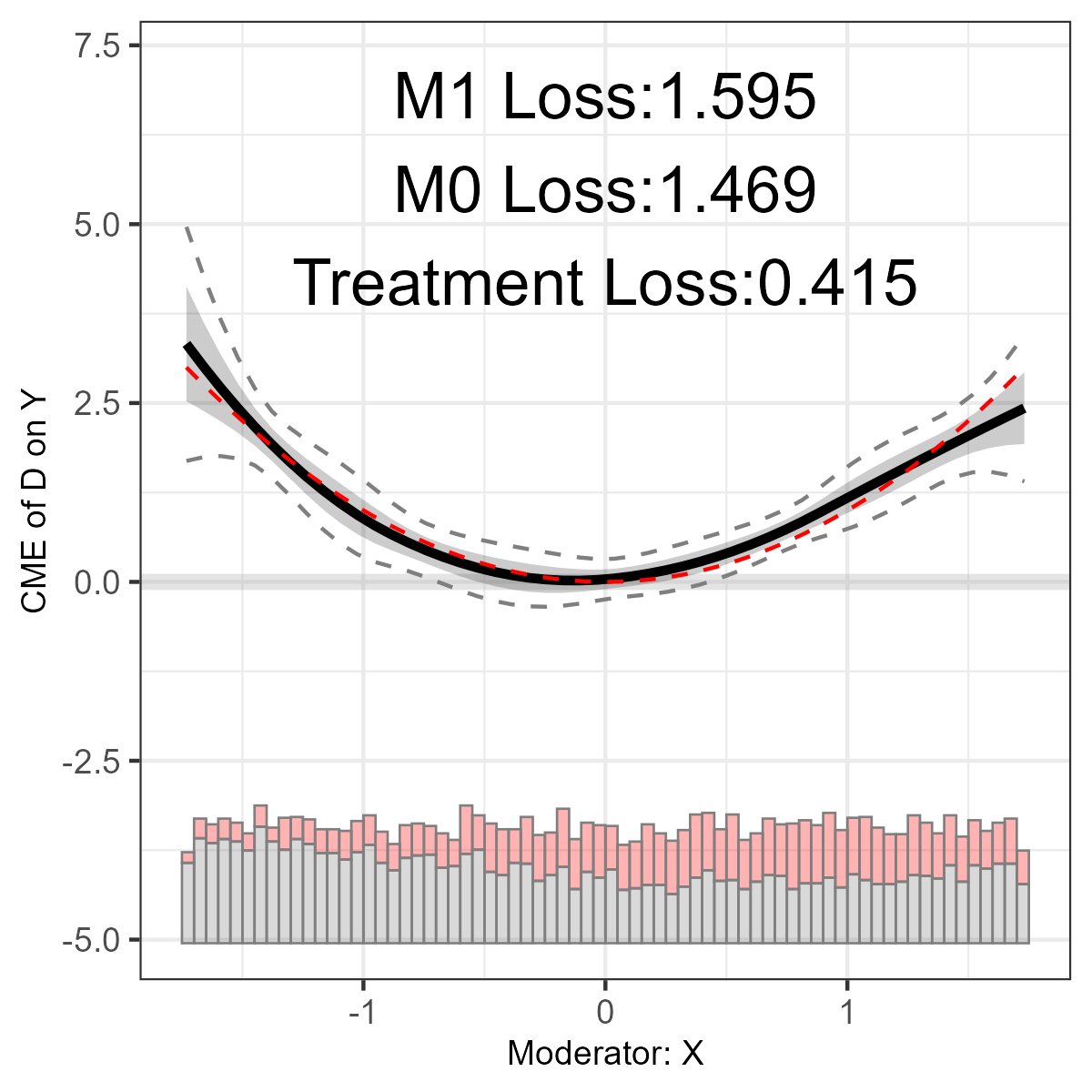}
    \caption{HGB (default)}
\end{subfigure}
\hspace{0.01\textwidth}  
\begin{subfigure}[b]{0.23\textwidth}
    \centering
    \includegraphics[width=\textwidth]{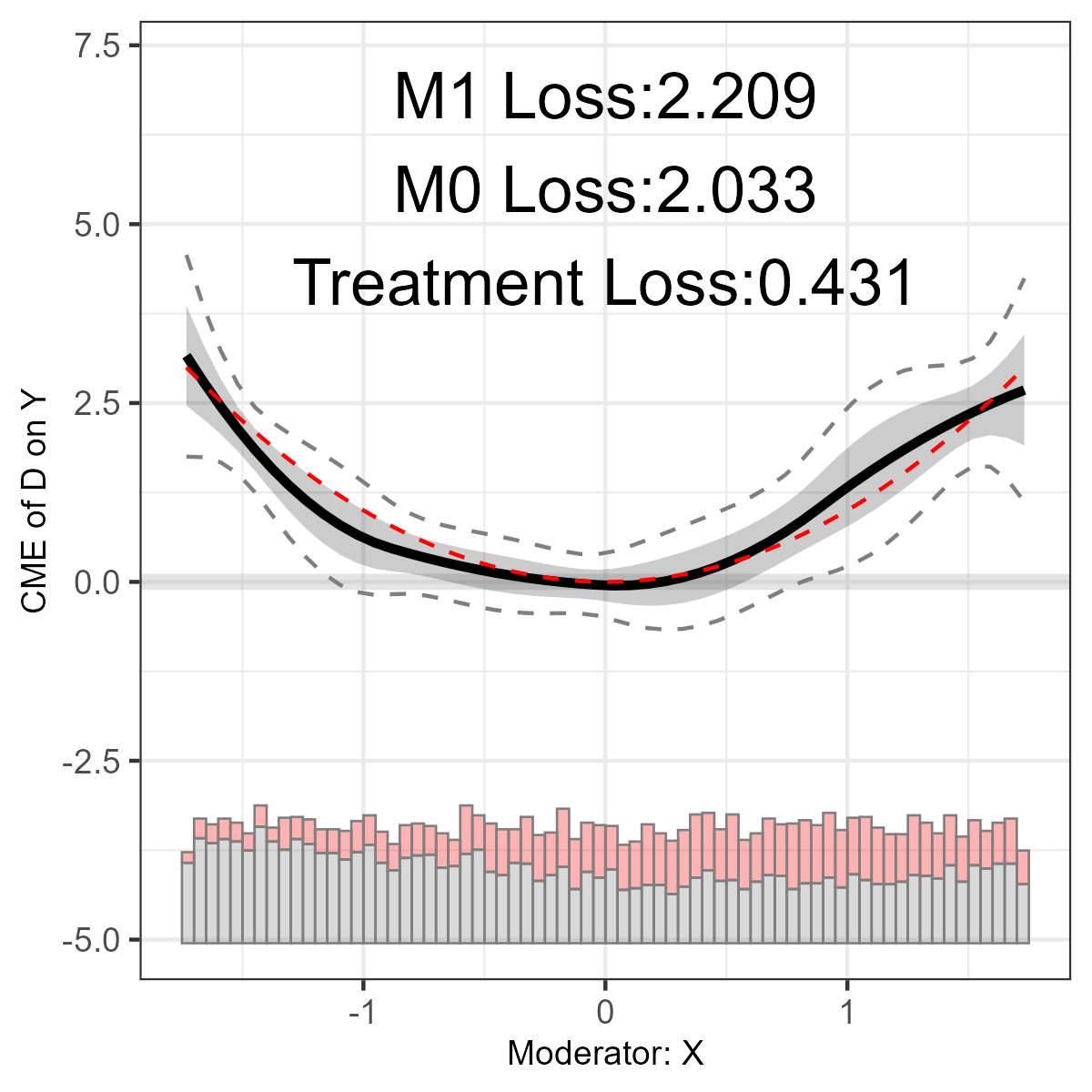}
    \caption{RF (default)}
\end{subfigure}
\\
\begin{subfigure}[b]{0.23\textwidth}
    \centering
    \includegraphics[width=\textwidth]{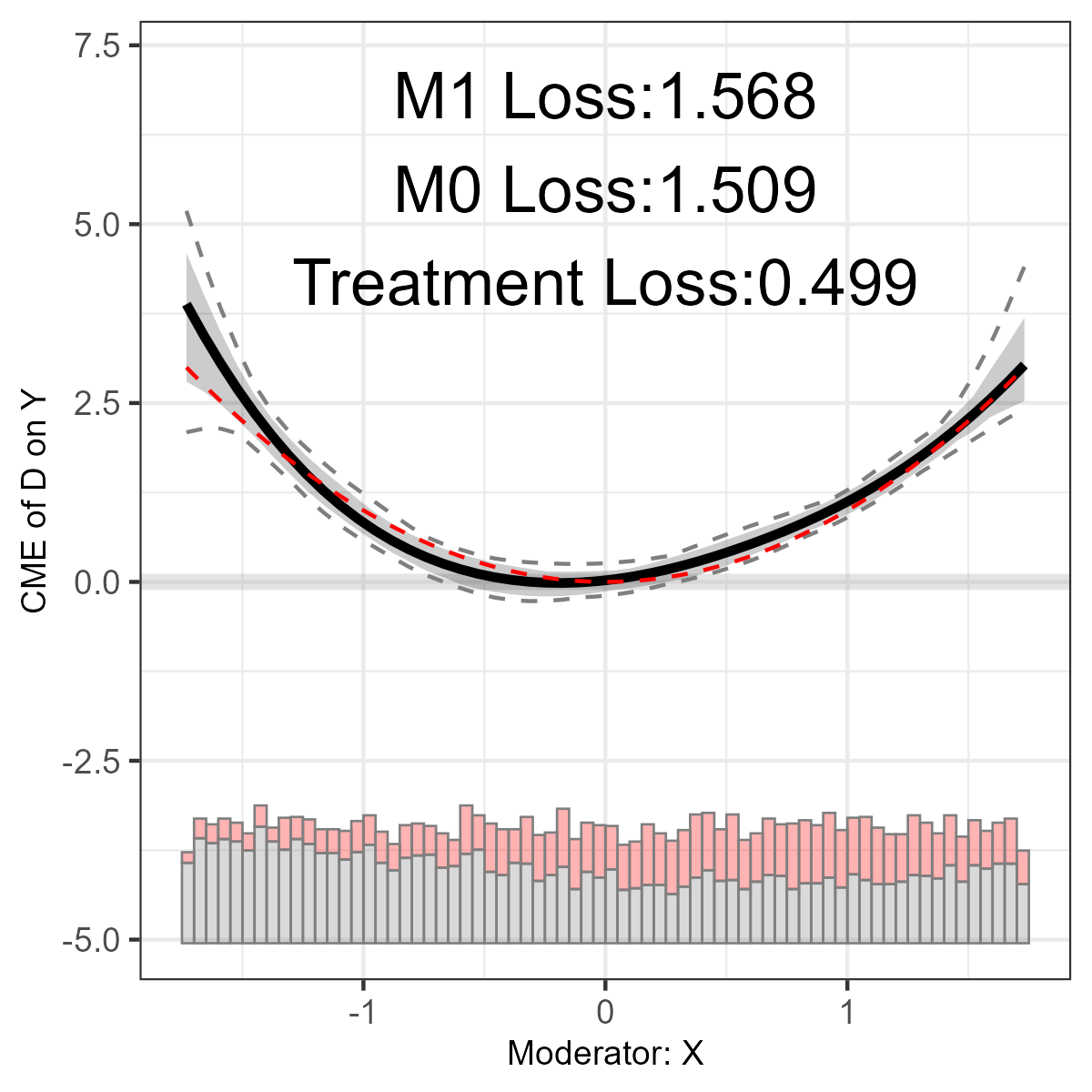}
    \caption{AIPW-Lasso}
\end{subfigure}
\hspace{0.01\textwidth}  
\begin{subfigure}[b]{0.23\textwidth}
    \centering
    \includegraphics[width=\textwidth]{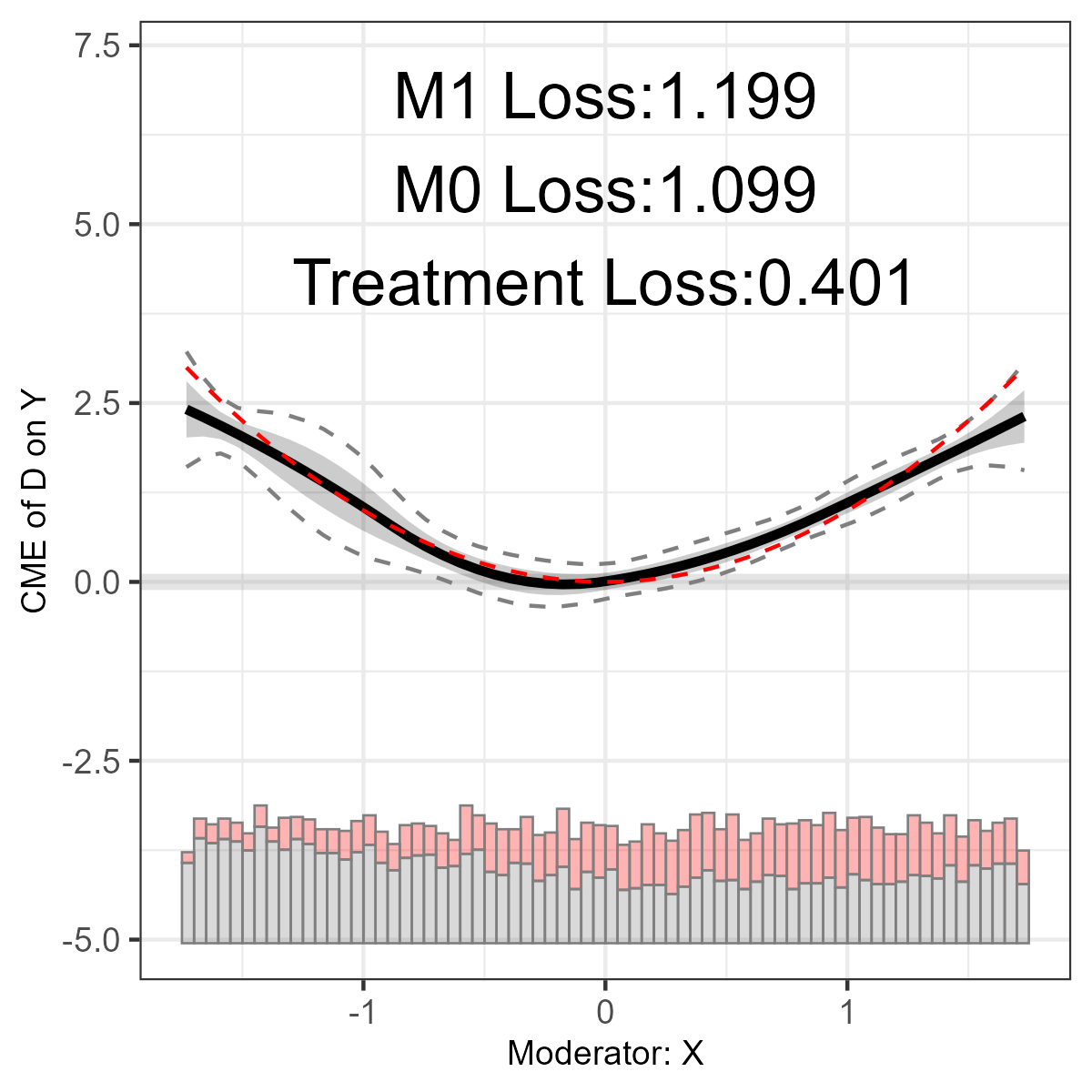}
    \caption{NN (tuned)}
\end{subfigure}
\hspace{0.01\textwidth}  
\begin{subfigure}[b]{0.23\textwidth}
    \centering
    \includegraphics[width=\textwidth]{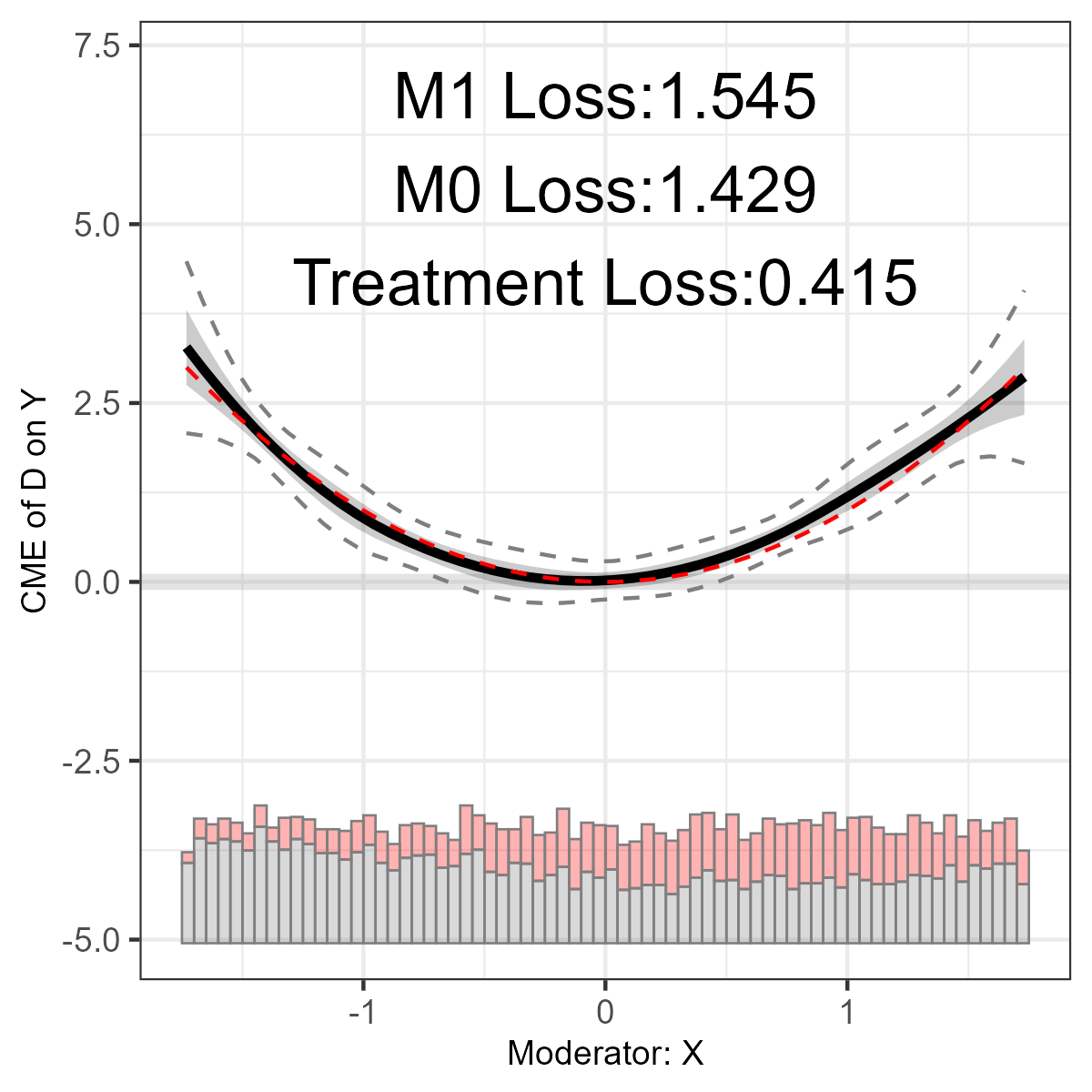}
    \caption{HGB (tuned)}
\end{subfigure}
\hspace{0.01\textwidth}  
\begin{subfigure}[b]{0.23\textwidth}
    \centering
    \includegraphics[width=\textwidth]{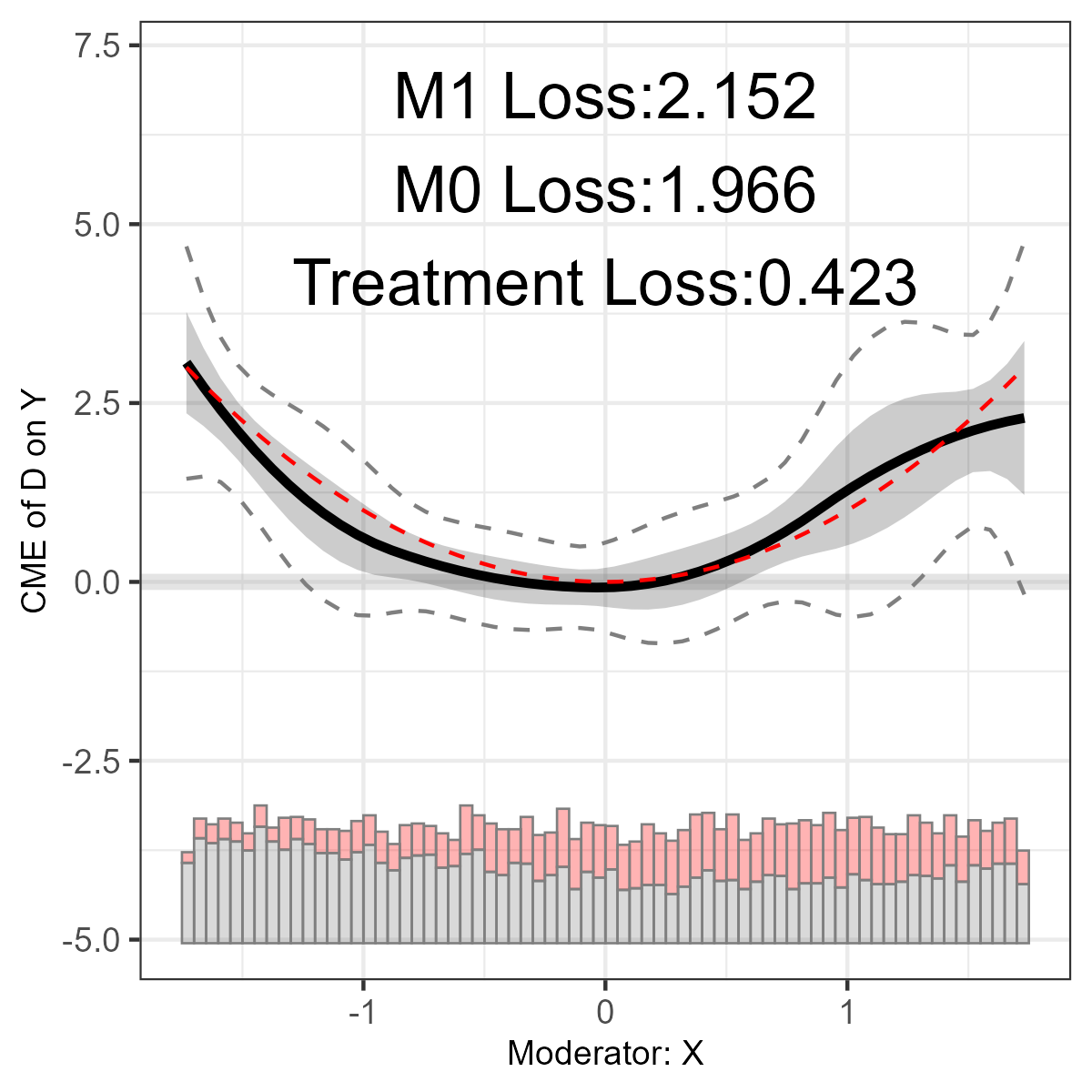}
    \caption{RF (tuned)}
\end{subfigure}
\label{fig:learner_cv_comparison}\\
{\footnotesize \textbf{Note}: In each subfigure, the black solid line represents the estimated CME, while the red dashed line denotes the true CME. The shaded gray areas illustrate the pointwise confidence intervals, and the gray dashed lines depict the uniform confidence intervals. At the bottom of each subplot, a histogram displays the distribution of treated and control units across varying values of the moderator \(X\).}
\end{figure}

In Figure~\ref{fig:learner_cv_comparison}, we compare the estimated CME to the true CME under two different hyperparameter settings (default vs. fine-tuned). Panels (b), (c), and (d) display estimates obtained using each learner’s default hyperparameters, while panels (f), (g), and (h) show estimates based on cross-validated hyperparameters. We also show the estimates given by the kernel estimator in (a), and the AIPW-Lasso estimator in (e), for comparison.

A key practical question is how to select the best learner and hyperparameter set without knowing the true CME. As shown at the top of each panel in Figure~\ref{fig:learner_cv_comparison}, the cross-fitted loss functions for the nuisance models, $\mu_1$, $\mu_0$, and $\pi$, provide a useful diagnostic. Lower loss---root mean squared error for $\mu$ or cross-entropy for $\pi$---indicates better out-of-sample fit for the corresponding nuisance component.\footnote{For the AIPW-Lasso estimator, we use out-of-sample fitting errors from the linear outcome regression and the logistic treatment assignment model.} Although not a guarantee, learners with lower average nuisance losses are generally expected to yield more accurate and reliable CME estimates. This pattern is evident in the figure: tuned models, which perform better visually, also report lower nuisance losses than their default counterparts (e.g., panel (b) versus (f)).

Notably, the NN learner highlights the importance of tuning hyperparameters to the data-generating process. For the outcome model, the selected hyperparameters are $\texttt{activation = tanh}$, $\texttt{alpha = 1e-4}$, $\texttt{hidden\_layer\_sizes = (50,50)}$, and $\texttt{solver = adam}$; in contrast, the propensity score model performs best with $\texttt{activation = tanh}$, $\texttt{alpha = 0.01}$, $\texttt{solver = sgd}$, and $\texttt{hidden\_layer\_sizes = (50,50)}$. This divergence in optimal specifications underscores the value of cross-validation, as the outcome and propensity score functions may differ in complexity and thus require different degrees of flexibility. Moreover, comparing subfigure (b) (default hyperparameters) with subfigure (f) (cross-validated hyperparameters) shows that DML with NN learners yields more accurate CME estimates when hyperparameters are properly tuned rather than set to defaults.

In contrast to the NN results, the HGB learner exhibits only modest improvement when moving from default hyperparameters to those selected via cross-validation. For the outcome model, the chosen hyperparameters are \(\texttt{l2\_regularization} = 1\), \(\texttt{learning\_rate} = 0.1\), \(\texttt{max\_depth} = 5\), \(\texttt{max\_features} = 1\), \(\texttt{max\_iter} = 200\), \(\texttt{max\_leaf\_nodes} = 31\), and \(\texttt{min\_samples\_leaf} = 20\). The propensity score model retains the same values for the penalty term, learning rate, and minimum samples per leaf but uses \(\texttt{max\_leaf\_nodes} = 127\) and \(\texttt{max\_iter} = 100\). Although these configurations differ meaningfully---particularly in tree size---between the outcome and propensity models, the resulting CME estimates show only a slight numerical improvement over the default settings, as illustrated in Figure~\ref{fig:learner_cv_comparison},(c) and (g). This pattern suggests that the default HGB parameters already capture the underlying signal reasonably well, with cross-validation providing incremental rather than substantial gains.

Using a similar procedure for the RF learner, cross-validation selects \texttt{bootstrap = TRUE}, \texttt{max\_depth = 10}, \texttt{max\_features = 0.8}, \texttt{min\_samples\_leaf = 1}, \texttt{min\_samples\_split = 2}, and \texttt{n\_estimators = 300} for the outcome model, while the propensity score model uses the same values except for \texttt{min\_samples\_leaf = 5}. These settings imply slightly deeper trees, a broader feature subset, and more trees than the defaults, which may improve variance reduction. However, comparing Figure~\ref{fig:learner_cv_comparison},(d) and (h), these adjustments do not yield noticeable improvements in the estimated CME, and overall performance remains inferior to both the NN and HGB learners. This result suggests that the random forests may be less well suited to capturing the nonlinear structure in this DGP, or that more specialized tuning, or larger sample sizes, are required to achieve performance comparable to the other methods.

\begin{figure}[!t]
    \caption{Performance of Different Methods: DGP3}
    \label{fig:method_comparison_dgp3}
    \begin{subfigure}[b]{0.45\textwidth}
        \centering
        \includegraphics[width=\textwidth]{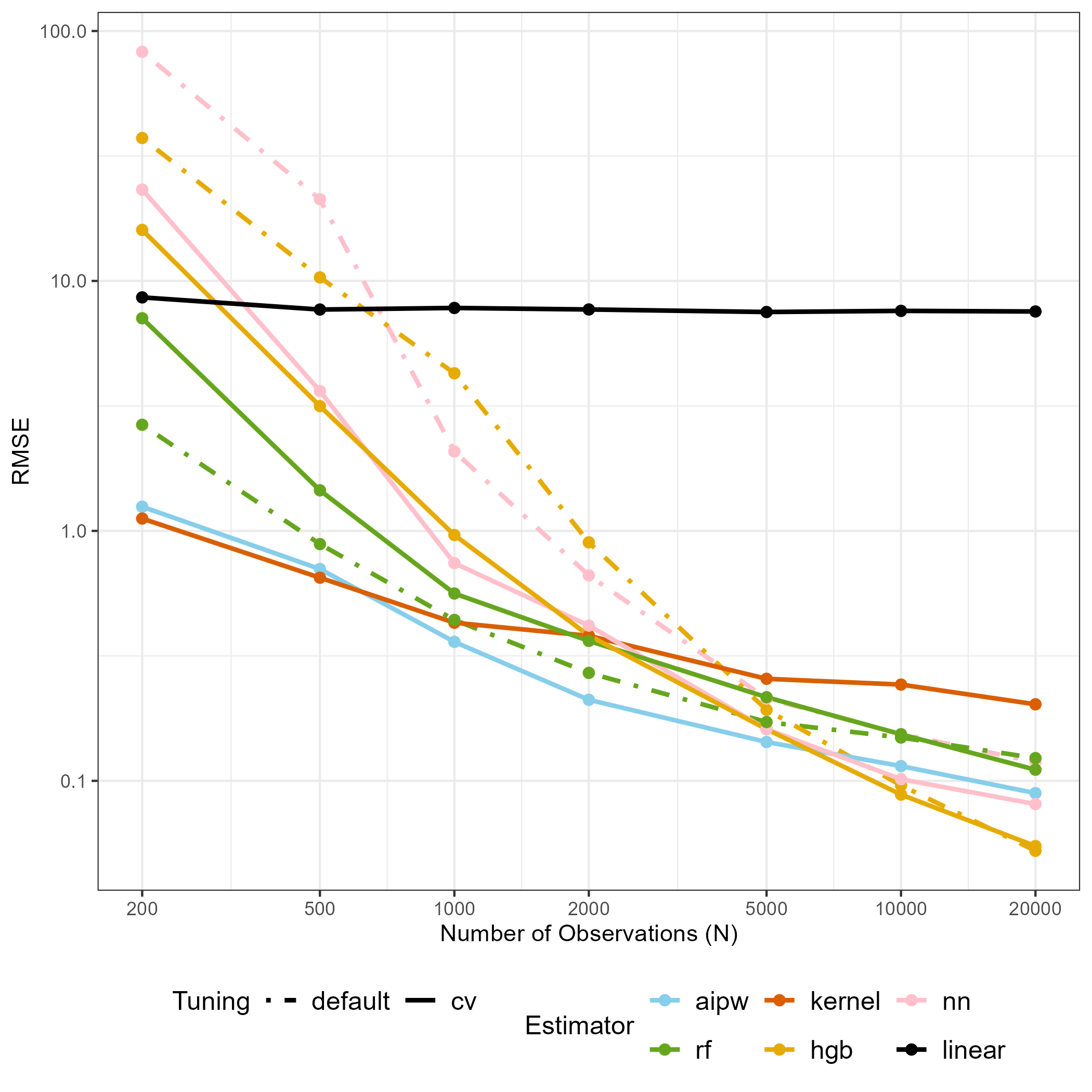}
        \caption{RMSE}
    \end{subfigure}
    \begin{subfigure}[b]{0.45\textwidth}
        \centering
        \includegraphics[width=\textwidth]{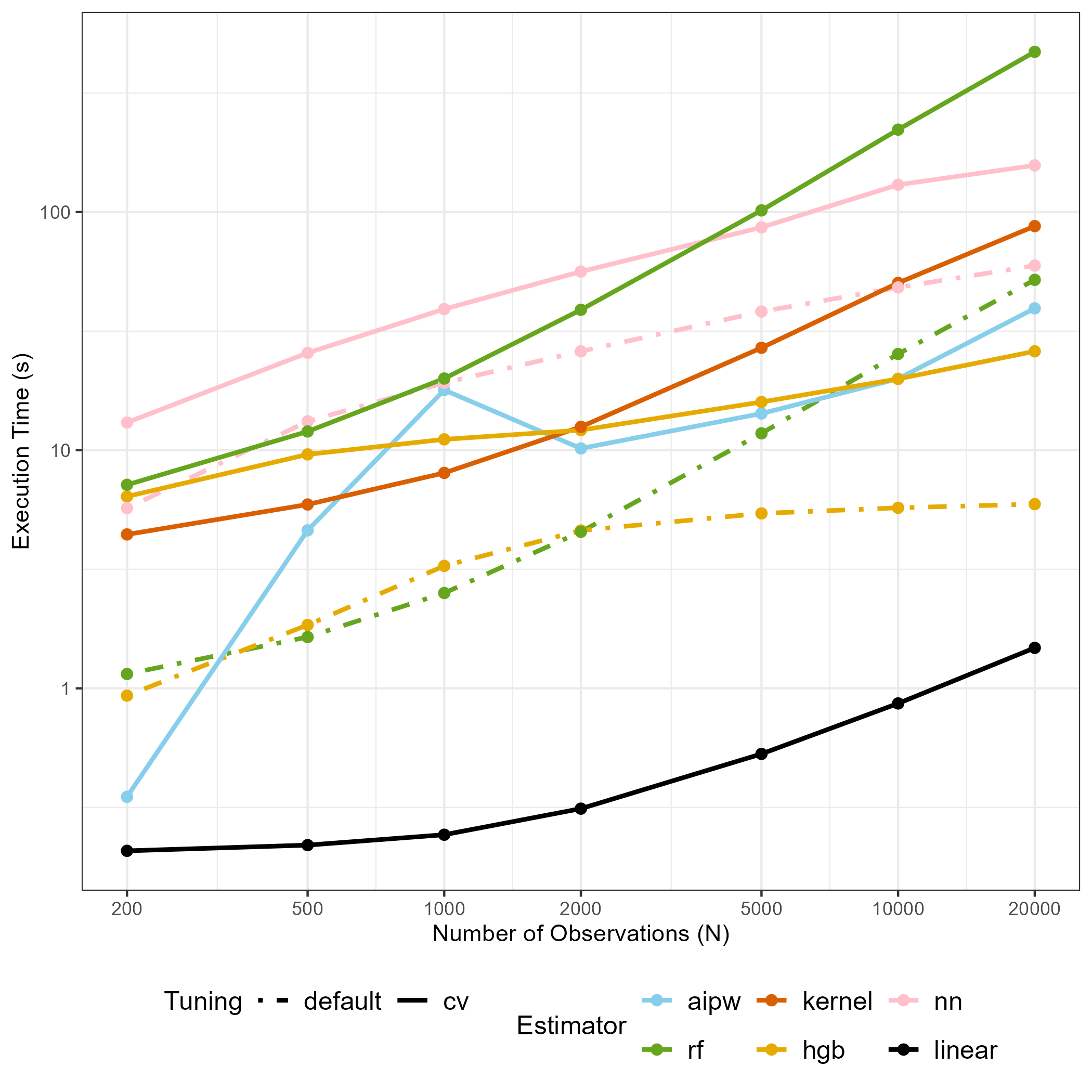}
        \caption{Execution time}
    \end{subfigure}
        \begin{subfigure}[b]{0.45\textwidth}
        \centering
        \includegraphics[width=\textwidth]{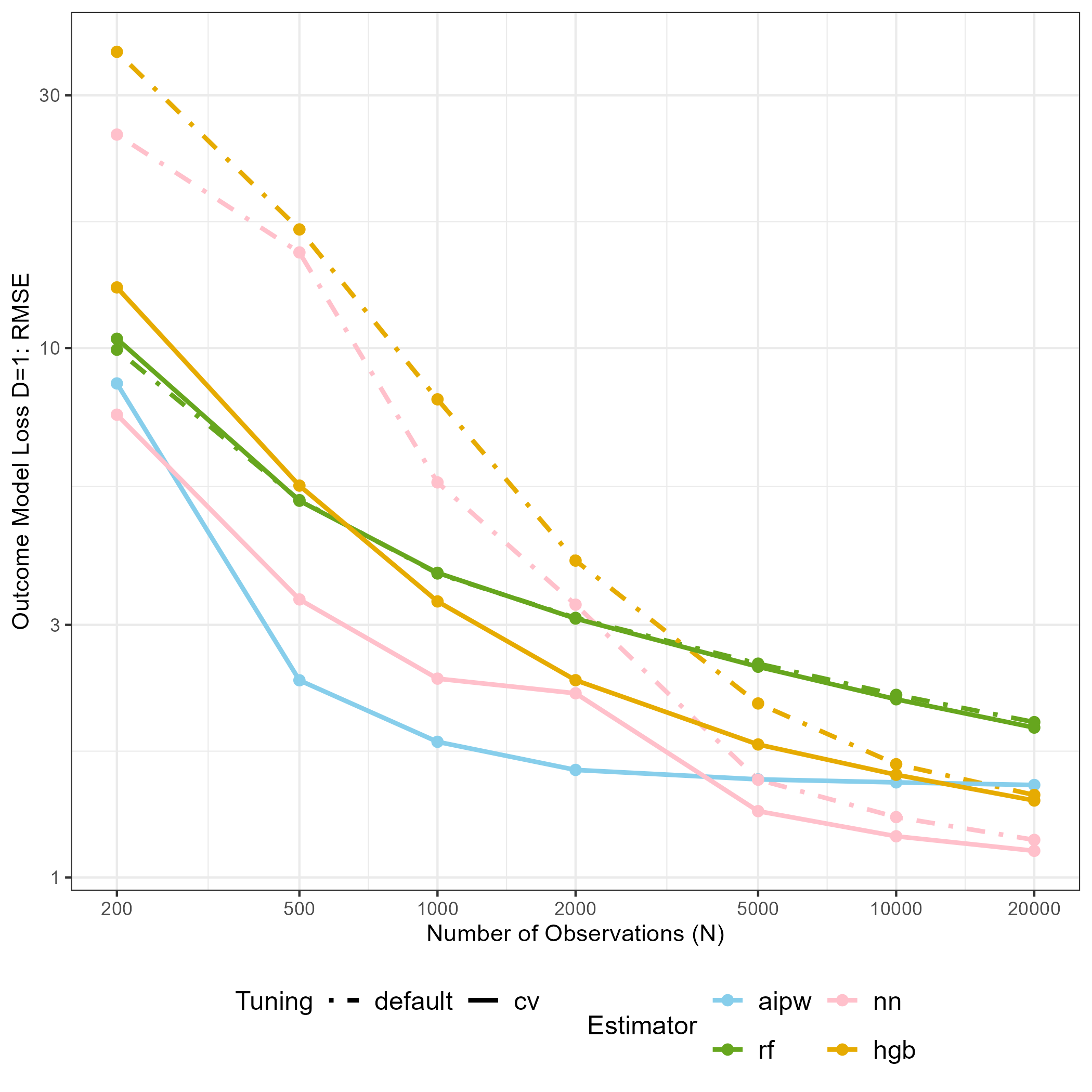}
        \caption{Loss Function (Outcome with $D=1$)}
    \end{subfigure}
    \begin{subfigure}[b]{0.45\textwidth}
        \centering
        \includegraphics[width=\textwidth]{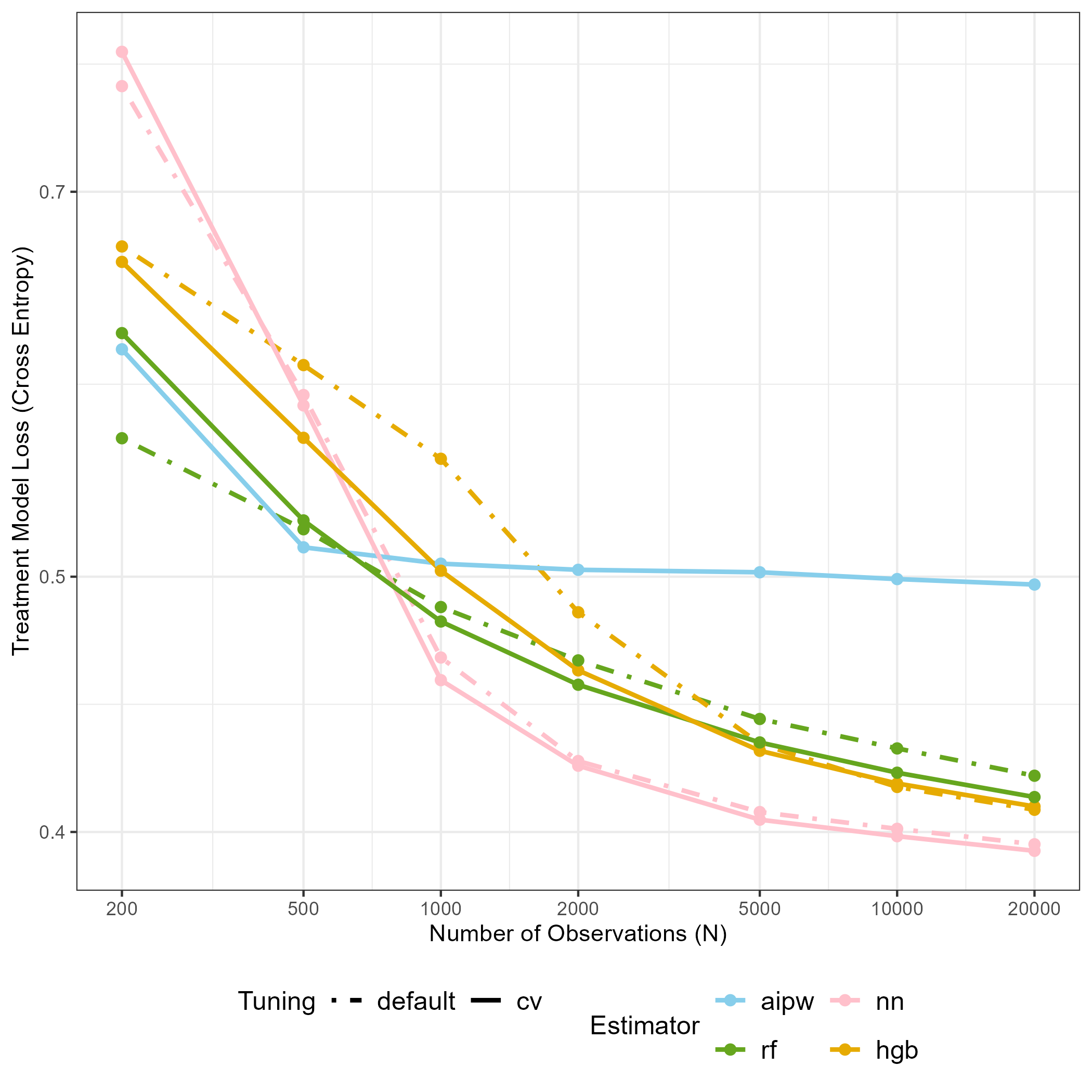}
        \caption{Loss Function (Treatment)}
    \end{subfigure}
    \begin{minipage}{\linewidth}
    {\footnotesize\textbf{Notes:} The above figures show how RMSE (top-left), execution time (top-right), loss of the outcome model for treated units (bottom-left), and loss of the treatment model (bottom-right) change as the sample size increases. Both axes in each figure are in log scale.}
    \end{minipage}
\end{figure}

In Figure~\ref{fig:method_comparison_dgp3}, we compare RMSE, loss functions, and execution time across the kernel estimator, AIPW-Lasso, and DML with different ML learners. The comparison includes both default hyperparameters and those selected via cross-validation, based on 100 simulations of the previously described DGP.

We observe the following patterns. First, all DML methods outperform the kernel estimator once the sample size $n$ is sufficiently large. Second, for NN and HGB, cross-validating hyperparameters consistently yields lower RMSE, especially when the sample size is modest. This likely reflects that fine-tuning mitigates overfitting in moderate samples. By contrast, RF shows little improvement from cross-validation, suggesting that gains in out-of-sample predictive performance do not necessarily translate into better CME estimates. Third, the AIPW-Lasso estimator performs competitively: it outperforms the kernel estimator even at moderate sample sizes and only slightly underperforms the tuned NN and HGB DML methods when $n$ is very large. Finally, the loss function plots (bottom row) tell a similar story. For all methods, outcome and treatment losses decline as the sample size increases, and fine-tuning hyperparameters (e.g., comparing ``nn'' to ``nn-cv'') reduces loss in line with the RMSE improvements. 

These plots also show that AIPW-Lasso often achieves lower loss than DML methods at small to moderate sample sizes. Moreover, cross-validation substantially increases execution time, highlighting a trade-off between computational cost and improved model fit.

\clearpage


\bibliographystyle{apsr}
\bibliography{refs}
\end{document}